\documentclass[aps,prx,twocolumn,superscriptaddress,longbibliography]{revtex4-1}
\usepackage{float}
\usepackage{graphicx}% Include figure files
\usepackage{dsfont}
\usepackage[latin1]{inputenc}
\usepackage{version} % for comment
\usepackage{color}
\usepackage{amssymb}
\usepackage{amsmath}
\usepackage{braket}
\usepackage{bm}
\usepackage{upgreek}
\usepackage{colortbl}
\usepackage{graphicx}% Include figure files
\usepackage{dcolumn}% Align table columns on decimal point
\usepackage{bm}% bold math
\usepackage{bigints}
\usepackage{xcolor}  % for writting parts in color, e.g. in red
\usepackage[pagebackref=false]{hyperref} % choice of color for hyperref links
\usepackage{enumitem}

\setcitestyle{super}

\definecolor{jrp}{rgb}{1,0,0}

\definecolor{rim}{rgb}{0,1,0}

\definecolor{pr}{rgb}{0.7,0,0}

\definecolor{mjg}{rgb}{.08,.05,.8}

\definecolor{yyl}{rgb}{.8,.05,.08}

\newcommand{\delete}[1]{{}} 
\colorlet{shadecolor}{gray!40}

\hypersetup{
    bookmarks=true,         % show bookmarks bar?
    unicode=false,          % non-Latin characters in Acrobat?s bookmarks
    pdftoolbar=true,        % show Acrobat?s toolbar?
    pdfmenubar=true,        % show Acrobat?s menu?
    pdffitwindow=false,     % window fit to page when opened
    pdfstartview={FitH},    % fits the width of the page to the window
    pdftitle={OTOC2_main},    % title
    pdfauthor={X. Mi},     % author
    pdfkeywords={keyword1} {key2} {key3}, % list of keywords
    pdfnewwindow=true,      % links in new window
    colorlinks=true,       % false: boxed links; true: colored links
    linkcolor=red,          % color of internal links (change box color with linkbordercolor)
    citecolor=blue,        % color of links to bibliography
    filecolor=magenta,      % color of file links
    urlcolor=cyan,           % color of external links         
}

\makeatletter

\makeindex

\begin{document}

\title{Constructive interference at the edge of quantum ergodic dynamics}
% xxx= edge, onset, threshold
%\title{Emergence of constructive large-loop interference in quantum many-body observables} 
\author{Google Quantum AI and Collaborators$^{\hyperlink{authorlist}{\dagger}}$}

\begin{abstract}

Quantum observables in the form of few-point correlators are the key to characterizing the dynamics of quantum many-body systems \cite{Feynman_1982, Lloyd96, Altman_PRXQ_2021}. In dynamics with fast entanglement generation, quantum observables generally become insensitive to the details of the underlying dynamics at long times due to the effects of scrambling. In experimental systems, repeated time-reversal protocols have been successfully implemented to restore sensitivities of quantum observables \cite{baum85multiple}. Using a 103-qubit superconducting quantum processor, we characterize ergodic dynamics using the second-order out-of-time-order correlators \cite{LarkinOvchinnikov, Shenker_JHEP_2014, Hosur_JHEP_2016, Maldacena2016,  Swingle_PRA_2016, AleinerIoffe_AP_2016, Roberts_JHEP_2017, Garttner2017, Nahum_PRX_2018, Keyserlingk_PRX_2018, ParticleConservingOTOCPollman, Khemani_PRX_2018, Yao_Nature_2019, Mi_OTOC_2021}, OTOC$^{(2)}$. In contrast to dynamics without time reversal, OTOC$^{(2)}$ are observed to remain sensitive to the underlying dynamics at long time scales. Furthermore, by inserting Pauli operators during quantum evolution and randomizing the phases of Pauli strings in the Heisenberg picture, we observe substantial changes in OTOC$^{(2)}$ values. This indicates that OTOC$^{(2)}$ is dominated by constructive interference between Pauli strings that form large loops in configuration space. The observed interference mechanism endows OTOC$^{(2)}$ with a high degree of classical simulation complexity, which culminates in a set of large-scale OTOC$^{(2)}$ measurements exceeding the simulation capacity of known classical algorithms. Further supported by an example of Hamiltonian learning through OTOC$^{(2)}$, our results indicate a viable path to practical quantum advantage.
\end{abstract}

\maketitle

Identifying complex correlations between the many-body degrees of freedom in a quantum system is a central goal for the simulation of quantum dynamics. Even spectroscopic questions can be formulated in terms of few-point dynamical correlations. As entanglement grows with system size or evolution time, the resulting dynamics are often ergodic. Consequently, the sensitivities toward details of quantum dynamics decay exponentially for most quantum observables, limiting their utility in revealing many-body correlations. Numerical or analytical studies of correlations are also hindered by the difficulty of identifying subtle contributing processes, which undermine common simplifying assumptions. Moreover, the linearity of the Schr\"odinger equation precludes the use of classical techniques based on sensitivity to initial conditions -- methods that have proven effective in detecting the butterfly effect and characterizing chaos. 

\begin{figure}
\centering
\includegraphics[width=0.99\columnwidth]{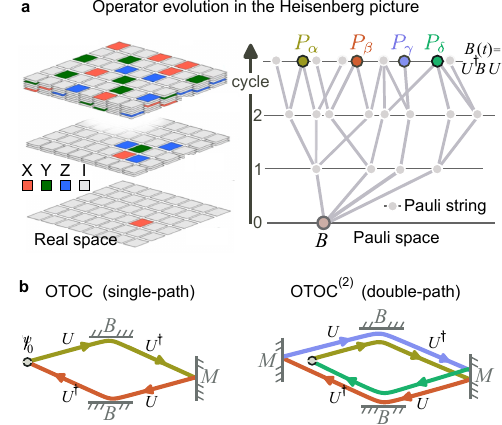}
\caption{{\bf OTOCs as interferometers.} {\bf a}, When dynamical protocols involve echoing, the Heisenberg picture of the operator evolution is the natural framework for studying dynamics. {\bf b}, OTOC and OTOC$^{(2)}$ can be viewed as time interferometers, which highlights their capability of refocusing on desired details and echoing out unwanted dynamics. See text for the definition of parameters. }
\label{fig:1}
\end{figure}

As a solution toward the above challenge, experimental protocols that employ \textit{refocusing} to echo out nearly all evolution have become essential for probing highly entangled dynamics. These protocols have proven indispensable in quantum metrology and sensing,\cite{Monika_PRL_2016,li2023improving}, as well as in studies of chaos, black holes, and thermalization,\cite{Hayden2007,Sekino_JHEP_2008,Shenker_JHEP_2014,Maldacena2016,Xu_PRX_2019}. Dynamical sequences which include time reversal are most naturally described in the Heisenberg picture of operator evolution (Fig.~\ref{fig:1}). The sequence can be conceptualized as an interference problem, where correlations reflect coherent interference across many-body trajectories. Computing an observable can thus be expressed as a sum over distinct trajectories. In this conceptual framework, each time reversal corresponds to the addition of two interference arms and also additional cross-terms contributing to experimental observables, which are formally known as out-of-time-order correlators (OTOCs) \cite{LarkinOvchinnikov, Shenker_JHEP_2014, Hosur_JHEP_2016, Maldacena2016,  Swingle_PRA_2016, AleinerIoffe_AP_2016, Roberts_JHEP_2017, Garttner2017, Nahum_PRX_2018, Keyserlingk_PRX_2018, ParticleConservingOTOCPollman, Khemani_PRX_2018, Yao_Nature_2019, Mi_OTOC_2021}.

In our work, we perform a family of OTOC experiments and leverage the interference framework to understand how different paths and their combinations reveal quantum correlations inaccessible without time reversal or via numerical methods. More specifically, we use the unique programmability of a digital quantum processor to change the number of interference arms (Fig.~\ref{fig:2}) and insert either noisy (Fig.~\ref{fig:3}) or coherent (Fig.~\ref{fig:5}) phase-shifters in each arm. In response, we find that OTOCs are significantly more sensitive toward these perturbations compared to observables in the absence of time reversal. Furthermore, we discover that this sensitivity is enhanced as the order $k$ of OTOC$^{(k)}$ (i.e. the number of interference arms) increases. In particular, OTOC$^{(2)}$ reveals constructive interference between Pauli strings that is invisible in lower-order observables.

To understand how repeated time-reversal restores sensitivity toward quantum dynamics, we first consider measuring the Pauli operator $M \in \{ X, Y, Z \}$ of a qubit $q_\text{m}$ located in a square lattice of qubits and initialized in an eigenstate of $M$. The measurement at a time $t$ is equivalent to the time-ordered correlator\,(TOC), $\braket{M (t) M}$, where $M (t) = U^\dagger(t) M U(t)$ denotes the time-evolved $M$ in the Heisenberg picture, $U$ is a many-body unitary, and $\braket{...}$ denotes expectation value over a particular initial state. As observed in previous experiments \cite{Kaufman_Science_2016, Pan_Science_2022_Gauge, Zhang_Science_2023_Metamaterials, Andersen_Nature_2025}, $\braket{M (t) M}$ decays exponentially over time when $U$ is ergodic. This stems from the scrambling of quantum information from the initial state of $q_\text{m}$ into the system's exponentially large Hilbert space. 

The above decay can, however, be {\it partially} refocused through an evolution outlined in Fig.~\ref{fig:2}a: Here, the dynamics $U$ is replaced with the nested echo sequence $U_{k} (t) = B(t) [M B(t)]^{k - 1}$, where $B(t) = U^\dagger(t) B U(t)$ is the time-evolved operator of another Pauli $B$ acting on qubit(s) $q_\text{b}$ some distance away from $q_\text{m}$, and $k \geq 1$ is an integer. The action of $U_k$ may be understood as dispersing the information injected by $M$, modifying it by $B$, reversing it back to $M$, and repeating this process $k - 1$ times. Since $U_k^\dagger (t) = U_k (t)$, the expectation value (denoted as $\mathcal{C}^{(2k)}$ in this case) may be written as:
\begin{equation}
    \mathcal{C}^{(2k)} = \braket{U_k^\dagger (t) M U_k (t) M} = \braket{(B(t)M)^{2k}}.
\label{eqn:c_def}\end{equation}

The quantity $\mathcal{C}^{(2)}$ coincides with the well-known out-of-time-order correlator \cite{LarkinOvchinnikov, Shenker_JHEP_2014, Hosur_JHEP_2016, Maldacena2016,  Swingle_PRA_2016, AleinerIoffe_AP_2016, Roberts_JHEP_2017, Garttner2017, Nahum_PRX_2018, Keyserlingk_PRX_2018, ParticleConservingOTOCPollman, Khemani_PRX_2018, Yao_Nature_2019, Mi_OTOC_2021, Braumuller2022}, OTOC. We therefore refer to $\mathcal{C}^{(2k)}$ as OTOC$^{(k)}$ or $k^\text{th}$-order OTOC. Eqn.~\ref{eqn:c_def} yields two key insights: First, if the information originating from $q_\text{m}$ has not yet reached $q_\text{b}$, $B(t)$ commutes with $M$ and the information coming back to $q_\text{m}$ is identical to its initial value. We therefore expect the existence of a wavefront across which $\mathcal{C}^{(2k)}$ decays. By increasing the separation between $q_\text{m}$ and $q_\text{b}$, this front may be pushed later in time, thus allowing large signals to be measured where TOCs are $\sim$0. Second, provided $U$ is not a Clifford sequence, information starting from $M$ and returning to $M$ can take multiple different paths in configuration space. Correlations between Paulis strings within $B(t)$ may therefore manifest through constructive interference between different paths for $\mathcal{C}^{(2k)}$ with $k \geq 2$.

\section{Sensitivity of OTOCs toward quantum dynamics}

\begin{figure*}
	\centering
 	\includegraphics[width=2\columnwidth]{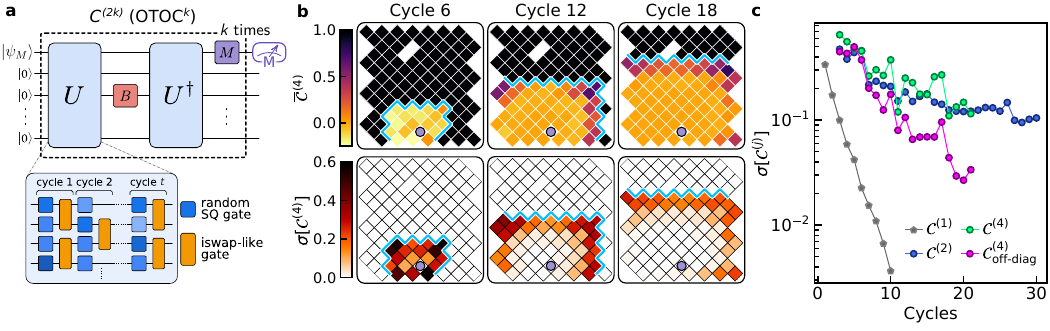}
 	\caption{\textbf{Sensitivity of OTOCs toward microscopic details of quantum dynamics.} {\bf a}, Top: Quantum circuit schematic for measuring OTOCs of different orders, OTOC$^{(k)}$. Here, $\ket{\psi_M}$ is an eigenstate of the measurement operator $M$ (realized as $Z$ in this work). The operator $B$ is realized as $X$. Bottom: Implementation of the unitary $U$ as $t$ cycles of single- and two-qubit gates. Each single-qubit (SQ) gate is $e^{-i \frac{\theta}{2} (\cos(\phi) X + \sin(\phi) Y)}$, where $\theta/\pi \in \{0.25, 0.5, 0.75\}$ and $\phi/\pi$ is chosen randomly from the interval $[-1,1]$. Each iSWAP-like gate is equivalent to an iSWAP followed by a CPHASE gate with a conditional phase of $\approx 0.35$~rad. {\bf b}, The mean ($\overline{\mathcal{C}}^{(4)}$) and standard deviation ($\sigma [ \mathcal{C}^{(4)} ]$) of OTOC$^{(2)}$  ($\mathcal{C}^{(4)}$), measured over 100 circuit instances for $t$ = 6, 12 and 18 cycles. The color at each qubit site indicates data collected with $B$ applied to the given qubit. Purple dot indicates the fixed location of $q_\text{m}$. Cyan lines represent the lightcone of $q_\text{m}$. {\bf c}, Standard deviation of four quantities, TOC ($\mathcal{C}^{(1)}$), OTOC ($\mathcal{C}^{(2)}$), OTOC$^{(2)}$ ($\mathcal{C}^{(4)}$) and the off-diagonal component of OTOC$^{(2)}$ ($\mathcal{C}^{(4)}_\text{off-diag}$). In the cases of $\mathcal{C}^{(2)}$, $\mathcal{C}^{(4)}$ and $\mathcal{C}^{(4)}_\text{off-diag}$, $q_\text{m}$ has the same fixed location as panel \textbf{b} whereas $q_\text{b}$ is gradually moved further from $q_\text{m}$ as circuit cycles increase, such that an OTOC mean of $\overline{\mathcal{C}}^{(2)} \approx 0.5$ is maintained. $\mathcal{C}^{(1)}$ corresponds to $\braket{Z (t) Z}$ measured at a qubit close to the center of the lattice.}
 	\label{fig:2}
 \end{figure*} 

We begin by characterizing the sensitivity of OTOC$^{(2)}$ toward the microscopic details of quantum dynamics. Fig.~\ref{fig:2}a schematically shows our quantum circuits, which are composed of random single-qubit and fixed two-qubit gates.  The experiment is done by first fixing the choices of $q_\text{m}$ and $q_\text{b}$. A circuit instance $i$ is then generated by varying the random parameters of the single-qubit gates that interleave the deterministic set of two-qubit gates. For a fixed number of circuit cycles $t$ within $U$, the quantity $\mathcal{C}^{(2k)} (t, q_\text{m}, q_\text{b}, i)$ is repeatedly measured until the statistical noise of the measurement is less than $<$10$\%$ of its average value. This protocol is then repeated by varying $t$, $q_\text{m}$ and $q_\text{b}$, as well as the circuit instance $i$, which is sampled 50 to 250 times. Lastly, all experimental $\mathcal{C}^{(4)}$ or $\mathcal{C}^{(2)}$ values are normalized by a global rescaling factor obtained through error-mitigation strategies (SI, Sections~II.E.1 and II.F.1).

The upper panel of Fig.~\ref{fig:2}b displays the values of $\overline{\mathcal{C}}^{(4)} (t, q_\text{b})$ for different circuit cycles and choices of $q_\text{b}$, where the overline denotes averaging over circuit instances. The location of $q_\text{m}$ is fixed throughout these measurements. The information front introduced above is clearly visible in the experimental data: For each $t$, there is a boundary for $q_\text{b}$ beyond which $\mathcal{C}^{(4)}$ is $\sim$1. This boundary defines the light-cone of $q_\text{m}$, corresponding to the set of qubits that have been entangled with $q_\text{m}$. Moreover, we find that the circuit-to-circuit fluctuation of $\mathcal{C}^{(4)} (t, q_\text{b})$, defined as the standard deviation $\sigma$ of its value over circuit instances, is of the same order of magnitude as the average value near the information front (bottom panel of Fig.~\ref{fig:2}b). This observation indicates that $\mathcal{C}^{(4)}$ is indeed highly sensitive toward the details of the underlying evolution $U$, an effect that we will later utilize to demonstrate its application in Hamiltonian learning.

To study more systematically the decay of OTOC sensitivity over time, we measure the standard deviations of various OTOCs as functions of $t$ (Fig.~\ref{fig:2}c). We observe that the standard deviations of $\mathcal{C}^{(2)}$, $\mathcal{C}^{(4)}$ and the off-diagonal part of OTOC$^{(2)}$ ($\mathcal{C}^{(4)}_\text{off-diag}$, see Section~\ref{sec: interference} for definition) decay algebraically and remain $>$0.01 beyond $t = 20$. The standard deviation of a TOC, which does not possess the echo-like structure of OTOCs, is found to decay exponentially over time and becomes $<$0.01 at $t = 9$. The stark contrast between TOC and OTOCs indicates that the interferometric nature of the latter is crucial for enhancing sensitivities to quantum dynamics. In Section~IV of the SI, we provide further studies of OTOC$^{(k)}$ fluctuations using one-dimensional Haar random circuits, finding that these observables decay as a power-law consistent with the two-dimensional experimental data.

\section{Large-loop Interference in OTOC$^{(2)}$} \label{sec: interference}

 \begin{figure*}
 	\centering
 	\includegraphics[width=2\columnwidth]{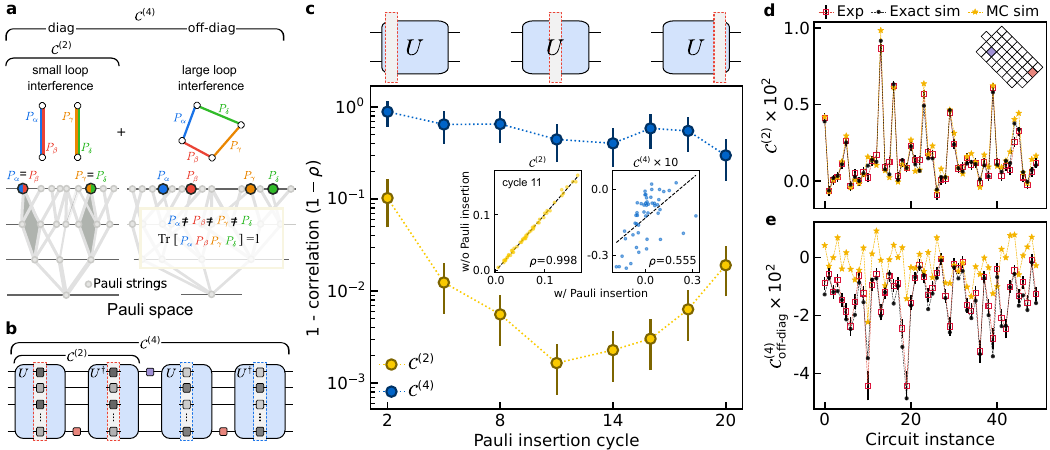}
 	\caption{\textbf{Quantum interference and classical simulation complexity of OTOC$^{(2)}$.} {\bf a}, In the Heisenberg picture, the time-evolved $B(t)$ branches into a superposition of multi-qubit Pauli strings. For $\mathcal{C}^{(2)}$ in which only two copies of $B (t)$ are present, the final strings $P_\alpha$ and $P_\beta$ need to be identical to contribute. For $\mathcal{C}^{(4)}$, the strings ($P_\alpha$, $P_\beta$, $P_\gamma$, $P_\delta$) contribute a ``diagonal'' component $\mathcal{C}^{(4)}_\text{diag}$ when $P_\alpha = P_\beta$ and $P_\gamma = P_\delta$, or an ``off-diagonal'' component $\mathcal{C}^{(4)}_\text{off-diag}$ when $P_\alpha \neq P_\beta \neq P_\gamma \neq P_\delta$. {\bf b}, Protocol for probing quantum interference: Random Pauli operators are inserted at one circuit cycle, which changes the signs of Pauli string coefficients. {\bf c}, Relative signal change,  characterized by $1 - \rho$, as a function of the cycle at which Paulis are inserted. $\rho$ refers to the Pearson correlation between experimental data from 50 different 40-qubit circuits (t = 22 cycles), obtained with and without Pauli insertion (see insets for data at cycle 11). Error bars denote standard errors estimated from resampling the experimental data. {\bf d}, Comparison of experimental $\mathcal{C}^{(2)}$ values against exactly simulated $\mathcal{C}^{(2)}$ for a set of 40-qubit circuit instances. Values computed using cached Monte-Carlo (CMC) heuristic algorithms are shown for comparison, achieving an SNR of 5.3 similar to the quantum processor (SNR = 5.4). Inset shows the circuit geometry (red: $q_\text{m}$; blue: $q_\text{b}$) used for the experiments in panels \textbf{c}, \textbf{d} and \textbf{e}. {\bf e}, Experimental $\mathcal{C}^{(4)}_\text{off-diag}$ values on the same set of 40 qubits, alongside exact and CMC simulations. $\mathcal{C}^{(4)}_\text{off-diag}$ is measured by subtracting the Pauli-averaged $\mathcal{C}^{(4)}$ from the non-averaged $\mathcal{C}^{(4)}$. Here the experimental SNR is 3.9 whereas the SNR from CMC is 1.1. Error bars on experimental data are based on an empirical error model discussed in SI Section~II.F.3 and II.F.4.}
 	\label{fig:3}
 \end{figure*} 

The experiments performed in the preceding section, which change all details of $U$ between OTOC measurements, is analogous to shifting the arms of an interferometer entirely (Fig.~\ref{fig:1}b). In this section, we demonstrate that higher order OTOCs also become increasingly more sensitive toward the phases of interferometer arms, a hallmark of interference phenomena. Conceptually, many-body interference may be understood by first noting that Eqn.~\ref{eqn:c_def} may, in the ergodic limit, be expressed as:
 \begin{equation}
     \mathcal{C}^{(2k)} = \text{Tr} [(B (t) M) ^ {(2k)}]\,/ 2^N.
 \label{eqn:trace} \end{equation}
For a {\it given circuit instance}, $B (t)$ can be decomposed in the basis of the $4^N$ Pauli strings $\{ P_n \}$ of our $N$-qubit system:
 \begin{equation}
     B (t) = \sum_{n = 1}^{4^N} b_{n} (t) P_n,
 \label{eqn:pauli_sum} \end{equation}
 where $\{ b_{n} \}$ is a set of real-valued time-dependent coefficients. The time evolution of $B$ may be visualized by the bottom left panel of Fig.~\ref{fig:3}a, where $B$ is seen to branch out in Pauli space over time due to the action of non-Clifford gates in our circuits, creating the so-called operator entanglement \cite{ZanardiPRA_2001,Mi_OTOC_2021,garcia2023resource}. In the schematic, the trajectories of some Pauli strings are also seen to recombine during time evolution. This mechanism, which affects both $\mathcal{C}^{(4)}$ and $\mathcal{C}^{(2)}$, will be referred to as small-loop interference.

The mechanism of large-loop interference is related to how Pauli strings at the end of the time evolution contribute to the experimental observable. It is only present in $\mathcal{C}^{(4)}$, as schematically shown in the upper panels of Fig.~\ref{fig:3}a. Since $M P_n M = \pm P_n$, Eqn.~\ref{eqn:trace} may be written for $\mathcal{C}^{(4)}$ as:
\begin{equation}
     \mathcal{C}^{(4)} = \sum_{\alpha, \beta, \gamma, \delta} c_{\alpha\beta\gamma\delta} \text{Tr} [P_\alpha P_\beta P_\gamma P_\delta].
\label{eqn:trace_pauli_sum} \end{equation}
Here each $c_{\alpha\beta\gamma\delta}$ is also a real-valued coefficient. Each Pauli string in this expression is represented as a colored segment in the diagrams within the top panels of Fig.~\ref{fig:3}a. The length of this segment qualitatively represents the Hamming distance between the Pauli string and the identity. Multiplying two Pauli strings joins them at one end and forms a new Pauli string connecting the two new terminal points. 

For the trace in Eqn.~\ref{eqn:trace_pauli_sum} to be non-zero, the product of the four Pauli strings must be the identity, i.e. the diagram must form a loop. This condition is satisfied through two distinct means: If $\alpha = \beta$ and $\gamma = \delta$, we obtain the so-called diagonal contribution $\mathcal{C}^{(4)}_\text{diag}$ to $\mathcal{C}^{(4)}$ in which the loops enclose zero area. If $\alpha \neq \beta \neq \gamma \neq \delta$, we obtain the off-diagonal contribution $\mathcal{C}^{(4)}_\text{off-diag}$, named after the fact that each term may be thought of as an off-diagonal element in the $4^{4N} \times 4^{4N}$ density matrix formed by the Paulis. Diagrammatically, $\mathcal{C}^{(4)}_\text{off-diag}$ consists of a superposition of operator loops, each of which contains three unconstrained Pauli strings and therefore encloses an arbitrarily large area. For comparison, large-loop interference is absent in $\mathcal{C}^{(2)}$ since only Pauli strings with $\alpha = \beta$ contribute, similar to $\mathcal{C}^{(4)}_\text{diag}$.

To characterize the effects of quantum interference, we insert random Pauli operators at different cycles within each application of $U$ and $U^\dagger$ (Fig.~\ref{fig:3}b). The inserted Paulis randomize the signs of the coefficients $c_{\alpha\beta\gamma\delta}$ in Eqn.~\ref{eqn:trace_pauli_sum} without changing their amplitudes. This is analogous to shifting the phase around an ordinary interference loop without changing its intensity. By ensemble-averaging over random Paulis and probing the changes in $\mathcal{C}^{(4)}$ or $\mathcal{C}^{(2)}$, the contribution of quantum interference to each observable can then be quantified. Figure~\ref{fig:3}c shows experimentally measured $1 - \rho$, where $\rho$ is the Pearson correlation between circuits with and without inserted Paulis. The inserted Paulis produce substantial changes in $\mathcal{C}^{(4)}$, indicating that the large-loop interference effects (i.e. $\mathcal{C}^{(4)}_\text{off-diag}$) are a dominant contribution to $\mathcal{C}^{(4)}$. In contrast, the signal change for $\mathcal{C}^{(2)}$ is much weaker overall, owing to the presence of only small-loop interference. The larger residual effect of Pauli insertion near the edges of $U$ is attributed to the fact that quantum gates closer to the $B$ and $M$ operators have more weight in the resulting signals. We also note that $1 - \rho$ is slightly reduced at later insertion cycles for the case of $\mathcal{C}^{(4)}$, which may arise from external decoherence processes that tend to reduce the visibility of quantum interference. Lastly, by subtracting the $\mathcal{C}^{(4)}$ of each circuit after Pauli insertion from its original value, the off-diagonal contribution $\mathcal{C}^{(4)}_\text{off-diag}$ can be experimentally extracted.

We find that the observed interference effects in $\mathcal{C}^{(2)}$ and $\mathcal{C}^{(4)}$ are closely connected to their complexities with respect to approximate classical simulation algorithms, since the degree of quantum interference sets the level of allowed classical approximation. For OTOC, we find that it is sometimes well-approximated by combining exact wave-function evolution and Monte-Carlo simulation that ignores the effects of small interference loops. Two such algorithms, cached Monte-Carlo (CMC) and tensor networks Monte-Carlo (TNMC), are described in Section~III.B of the SI. Figure~\ref{fig:3}d shows experimental $\mathcal{C}^{(2)}$ values from a set of circuits comprising 40 qubits, along with values of $\mathcal{C}^{(2)}$ computed using CMC. To quantify the accuracy of each data set, we define a signal-noise-ratio (SNR, see Methods) against exactly simulated $\mathcal{C}^{(2)}$ values shown in the same figure. The SNR achieved by CMC (5.3) is seen to be similar to the experimental SNR of 5.4. For the off-diagonal OTOC$^{(2)}$ ($\mathcal{C}^{(4)}_\text{off-diag}$), classical algorithms are far less accurate and achieve a much lower SNR (1.1) compared to experiment (3.9), as shown in Fig.~\ref{fig:3}e. In Section~III.C of the SI, we review all classical simulation algorithms attempted as part of this work and show that none succeeds in approximating $\mathcal{C}^{(4)}_\text{off-diag}$.

\begin{figure}
 	\centering
 	\includegraphics[width=1\columnwidth]{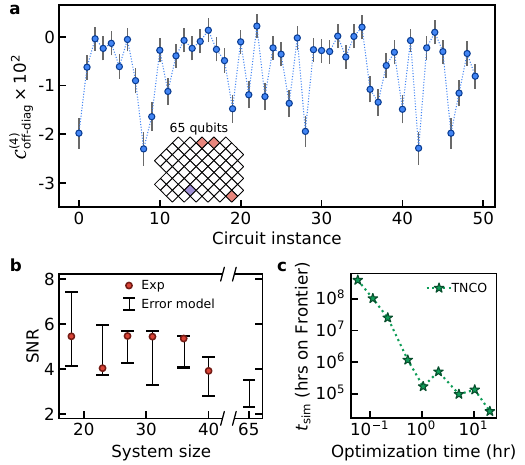}
\caption{\textbf{Measuring OTOC$^{(2)}$ in the classically challenging regime.} {\bf a}, $\mathcal{C}^{(4)}_\text{off-diag}$ measured on a set of 65-qubit circuits each having $t = 23$ cycles. The qubit geometry is indicated in the inset, where $B$ acts simultaneously on three different qubits. {\bf b}, Experimental SNRs for circuits measured with system sizes ranging from 18 to 40 qubits. Error bars correspond to the 95\% confidence interval of an empirical error model (see SI Section~II.F.3 and II.F.4). Error bars in panel {\bf a} are based on the same empirical error model. {\bf c}, Estimated time to compute $\mathcal{C}^{(4)}_\text{off-diag}$ of a single circuit in {\bf a} on the Frontier supercomputer using TN contraction. The estimate is obtained by running a specially designed optimization algorithm \cite{boixo2017simulation,gray2021hyper,Kalachev_arxiv_2022, Pan_TN_PRL_2022} on 20 Google Cloud virtual machines (totaling 1200 CPUs) up to a period of 24 hours ($x$-axis). Estimates using a publicly available library \texttt{cotengra} \cite{gray2021hyper} lead to costs that are ten times higher after the same optimization time.}
\label{fig:4}
\end{figure}

\begin{figure}
 	\centering
 	\includegraphics[width=1\columnwidth]{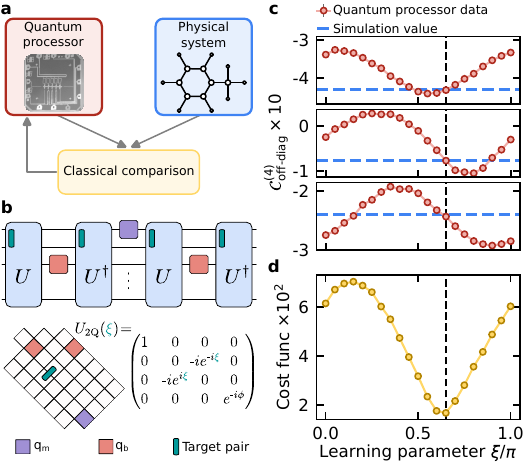}
\caption{\textbf{Application to Hamiltonian learning.} 
{\bf a}, Scheme for applying OTOC$^{(2)}$ toward Hamiltonian learning: OTOC$^{(2)}$ measured in a physical system of interest are compared with quantum simulation of OTOC$^{(2)}$ using a parameterized Hamiltonian of the same system. Hamiltonian parameters are then optimized to minimize the difference between the two data sets. {\bf b}, Demonstrating a one-parameter learning experiment: Here a collection of classically simulated  $\mathcal{C}^{(4)}_\text{off-diag}$ values from 20 circuit instances having a 34-qubit geometry (bottom left panel) are treated as data from a physical system of interest. The goal is to learn a particular phase $\xi / \pi = 0.6$ of the two-qubit gate unitary $U_\text{2Q}$ belonging to one pair of qubits (green bar in the top and bottom left panels). {\bf c}, Experimentally measured $\mathcal{C}^{(4)}_\text{off-diag}$ (quantum processor data) as a function of $\xi$ for three different circuit instances. Blue lines indicate the ideal values of $\mathcal{C}^{(4)}_\text{off-diag}$ from classical simulation, which intersect all three data sets close to the target value of $\xi$ (vertical dashed line). {\bf d}, An optimization cost function, corresponding to the RMS difference between quantum processor and classical simulation data of all 20 circuit instances, as a function of $\xi$. The cost function is minimized at the target value of $\xi$.}
\label{fig:5}
\end{figure}

\section{Toward practical quantum advantage}

The combination of large sensitivity and high classical simulation complexity makes higher-order OTOCs such as OTOC$^{(2)}$ a prime candidate for achieving the long-standing goal of practical quantum advantage. To illustrate this potential, we perform two additional experiments which show: (i) OTOC$^{(2)}$ can be accurately resolved in regimes that are currently intractable with classical supercomputers. (ii) A specific example wherein OTOC$^{(2)}$ is used to accomplish a practical task.

We begin by demonstrating (i): Figure~\ref{fig:4}a shows a set of $\mathcal{C}^{(4)}_\text{off-diag}$ measurements performed with a 65-qubit geometry with $B$ applied simultaneously to three different qubits, chosen in order to maximize the effective quantum volume (e.g. number of two-qubit gates falling within the lightcones of $B$ and $M$ operators \cite{KECHEDZHI2024431}). To estimate the accuracy of these measurements, we next characterize experimental error (i.e. SNR) across six different system sizes up to 40 qubits (Fig.~\ref{fig:4}b). Here we observe that the SNR degrades only weakly as the system size increases and is also captured by the confidence interval of an empirical error model detailed in the SI (Section~II.F.3 and II.F.4). Based on the same error model, the SNR of the 65-qubit dataset is projected to range from 2 to 3. To achieve this accuracy with classical simulation, heuristic classical algorithms are insufficient (see Section~III.C of the SI) and tensor networks (TN) contraction is the most effective approach. In Fig.~\ref{fig:4}c, we show the estimated cost of simulating $\mathcal{C}^{(4)}_\text{diag}$ via TN contraction on the Frontier supercomputer, which converges to $\sim$3.2 years. This is a factor of $\sim$13000 longer than the experimental data collection time of 2.1 hours per circuit, indicating that this experiment is currently in the beyond-classical regime of quantum computation.

To apply OTOCs toward real-world applications, we consider a physical system of interest characterized by a Hamiltonian with a set of unknown parameters. The physical system supplies a collection of OTOC$^{(2)}$ data, which is compared against quantum simulation of the same Hamiltonian. The unknown parameters are then optimized until the quantum-simulated data match real-world experimental data (Fig.~\ref{fig:5}a). The slowly decaying signal size and sensitivity of OTOC$^{(2)}$, as demonstrated in Fig.~\ref{fig:2} and Fig.~\ref{fig:3}, make it a particularly suitable candidate toward accomplishing this task, known as Hamiltonian learning~\cite{Bairey_PRL_2019,obrien21quantum,Cotler_PRA_2023,schuster23learning}.

To demonstrate the proposed scheme in practice, we construct a one-parameter learning example shown in Fig.~\ref{fig:5}b: A set of $\mathcal{C}^{(4)}_\text{off-diag}$ values from 20 random circuit instances, produced by classical simulation to mimic the role of the ``physical system'' in Fig.~\ref{fig:5}a, are provided. All details of $U$, except the phase $\xi$ of one two-qubit gate located along the passage between $q_\text{m}$ and $q_\text{b}$, are also given. To learn the unknown parameter $\xi$, we measure $\mathcal{C}^{(4)}_\text{off-diag}$ on the quantum processor while varying $\xi$. Results for three circuit instances are shown in Fig.~\ref{fig:5}c, where we see smooth oscillations of experimental signals that are distinct between different instances. Importantly, all oscillations intersect the classically simulated values of $\mathcal{C}^{(4)}_\text{off-diag}$ at the target value of $\xi$. This is further reflected in Fig.~\ref{fig:5}d, where we have constructed a cost function between the classically simulated and experimentally measured values of $\mathcal{C}^{(4)}_\text{off-diag}$. The cost function is seen to display a global minimum at the target $\xi$ value.

\section{Conclusion}
In this work, we have shown that OTOCs possess quantum interference effects which endow them with large sensitivity to details of quantum dynamics and, in the case of OTOC$^{(2)}$, also high levels of classical simulation complexity. As such, OTOCs are viable candidates for realizing practical quantum advantage, a major milestone sought by recent experiments\cite{Kim2023, dwave_advantage_arxiv_2024,haghshenas2025digital}. Generally, practical quantum advantage can be formulated as the task of measuring the expectation values of low-rank observables, e.g., energy or correlations\,\cite{Altman_PRXQ_2021, Daley2022_review}, such that the observable:

\vspace{-2mm}
\begin{enumerate}[label=(\roman*)]
    \item can be experimentally measured with a proper accuracy, in our case with signal-to-noise ratio (SNR) above unity. More formally, the observable is in BQP\,\cite{Vazirani_Complexity_1997}.

    \item lies beyond the reach of both exact classical simulation and heuristic methods that trade accuracy for efficiency\,\cite{Tindall_PRXQ_2024,Begusic_SciAdv_2024,KECHEDZHI2024431,tindall2025,mauron2025}.
\end{enumerate}

\vspace{-2mm}
\noindent  Satisfying both defines a 'Goldilocks zone' for quantum advantage. To demonstrate practical quantum advantage, one more criterion is required:

\begin{enumerate}[label=(\roman*),start=3]
    \item the observable should yield practically relevant information about the quantum system.
\end{enumerate}
\vspace{-2mm}

Here, by measuring a many-body observable with SNR $>$ 2 and showing that it is beyond the reach of currently known classical simulation algorithms, we have made progress toward (i) and (ii). At the same time, the proof-of-principle for (iii) is demonstrated with a dynamic learning problem. While the random circuits used in the dynamic learning demonstration remains a toy model for Hamiltonians that are of practical relevance, the scheme is readily applicable to real physical systems. One such example is solid-state nuclear magnetic resonance systems, where dipolar couplings between spin pairs can be inverted without complete knowledge of their strength,\cite{Rhim_PRB_1971}. Comparing experimental data from such systems with quantum simulation outcomes may allow more accurate estimates of these couplings. We leave this exciting real-world application for future work.

\section{Methods}

{\bf Quantum processor and RCS benchmark.} The quantum processor originally comprises 105 frequency-tunable superconducting transmon qubits connected by tunable couplers. Upon cooldown, we find that two qubits were inoperable due to broken coupler bias lines. They are therefore excluded from the experiment. Details of the processor, including gate and readout errors, qubit coherence times, anharmonicities and frequencies, can be found in Section~II.A of the SI. Due to improved $T_1$ (median: 106 $\mu$s) compared to our earlier quantum processors \cite{Arute2019,Acharya_2023_M2,Morvan2024_RCS}, a median two-qubit gate (iSWAP-like) error of 0.15\% is achieved for this device. As a system-wide benchmark, we performed a random circuit sampling (RCS) experiment and estimated the overall fidelity to be $0.001$ at 40 circuit cycles, which doubles the circuit volume compared to previous records and corresponds to $\sim$10$^{25}$ years of simulation run-time on the Frontier supercomputer using tensor network (TN) contraction algorithms (SI, Section~II.B).

{\bf Signal-to-noise ratio.} The SNR is defined as follows: The set of circuit-specific $\mathcal{C}^{(2)}$ (or $\mathcal{C}^{(4)}_\text{off-diag}$) values first have their mean subtracted and are then rescaled by their variance, i.e. $\mathcal{C}^{(s)} = \left( \mathcal{C} - \overline{\mathcal{C}} \right) / \sigma (\mathcal{C})$ where $\mathcal{C}$ represents either $\mathcal{C}^{(2)}$ or $\mathcal{C}^{(4)}_\text{off-diag}$. This is done to both the set of experimentally measured ($\{ \mathcal{C}_\text{exp} \}$) and the set of numerically simulated quantities ($\{ \mathcal{C}_\text{sim} \}$). The SNR is then $1/\sqrt{\overline{(\mathcal{C}^{(s)}_\text{exp} - \mathcal{C}^{(s)}_\text{sim})^2}}$, where the overline denotes averaging over all circuit instances. This definition of SNR is directly related to the Pearson correlation coefficient $\rho$ as SNR $= 1/ \sqrt{2(1 - \rho)}$.

{\bf Initial states.} For measurements of $\mathcal{C}^{(2k)}$ with respect to a single initial state, an auxiliary qubit may be entangled with $q_\text{m}$ after preparing the system in the desired initial state, then read out at the end \cite{Mi_OTOC_2021,DTC_Nature_2022}. In our present work, we simplify this scheme by initializing $q_\text{m}$ in the $\ket{0}$ state which is an eigenstate of the measurement operator $M = Z$. Measuring $q_\text{m}$ in the $Z$ basis directly yields $\mathcal{C}^{(2k)}$.

It is also possible to measure $\mathcal{C}^{(2k)}$ of an infinite-temperature initial state, i.e. a maximally mixed state with the density matrix $I / 2^N$, where $I$ is the identity matrix and $N$ is the number of qubits. This is done by averaging over different initial states. In Section~II.E.4 of the SI, we show measurements of infinite-temperature $\mathcal{C}^{(2)}$ after averaging over initial states in a set of 66-qubit circuits. A sizable circuit-to-circuit fluctuation is still found in this case, indicating that the signal size of $\mathcal{C}^{(2k)}$ remains large even for infinite-temperature states.

{\bf Choice of circuit geometries.} The various circuit geometries studied throughout this work are chosen with the following considerations: 1) For each system size, we place $q_\text{b}$ and $q_\text{m}$ at locations close to the opposing edges of a lattice. 2) The total number of circuit cycles for each system size is adjusted to be the maximum beyond which the average signal size for $\mathcal{C}^{(4)}_\text{off-diag}$ is reduced below 0.01. These two considerations are implemented such that each small-scale geometry (e.g. the six different geometries used in Fig.~\ref{fig:4}c) mimic the geometry and overall signal size of the 65-qubit circuits in Fig.~\ref{fig:4}a.

{\bf Classical simulation costs of 40-qubit circuits.} Each 40-qubit circuit instance in Fig.~\ref{fig:3}d and Fig.~\ref{fig:3}e required 3 hours to simulate exactly on a Google Cloud virtual machine with 11.5 Tb of RAM and 416 CPUs. Matching the accuracy (i.e. SNR$\sim$5) of the quantum experiment in Fig.~\ref{fig:3}d using cached MC simulation required 1 billion cache size (see Section~III.B.2 of the SI for definition) and 6 days of simulation time (per circuit) on a single NVIDIA H100 GPU to gather sufficient statistics. 

{\bf Circuit-to-circuit fluctuations of $\mathcal{C}^{(2k)}$.} In Section~IV of the SI, we analyze the moments of $\mathcal{C}^{(2)}$ and $\mathcal{C}^{(4)}$ theoretically and numerically for the 1D bricklayer circuit of two-qubit Haar random gates. In the case of $\mathcal{C}^{(2)}$ we use a combination of perturbation theory and large scale matrix product state simulations to relate the power-law circuit-to-circuit fluctuations to the small interference loops introduced in the main text. In the case of $\mathcal{C}^{(4)}$ and $\mathcal{C}^{(4)}_\text{off-diag}$, similar analysis is challenging and instead we verify the polynomial scaling of fluctuations with exact numerics. We also demonstrate the existence of a dynamical quantum phase transition in the spectrum of the operator $MB(t)$ (Eqn.~\ref{eqn:trace}). The properties of this transition are captured in an analytically tractable random matrix model which allows us to compute moments $k\geq1$. This quantum criticality is consistent with the classically intractable power-law correlations in $\mathcal{C}^{(2k)}$. 

{\bf Sign problem in computing $\mathcal{C}^{(4)}$.} In Section~\ref{sec: interference}, we demonstrated a significant independent contribution in Eqn.~\ref{eqn:trace_pauli_sum} from non-paired trajectories in the space of Pauli strings ($\mathcal{C}^{(4)}_\text{off-diag}$ in Fig.~\ref{fig:3}a) which carry coefficients with random signs. In Section~III.C.3 of the SI, We further argue theoretically that this interference effect is not an artifact of the Pauli representation, but is intrinsic to $\mathcal{C}^{(4)}$. We accomplish this for the circuit-averaged $\mathcal{C}^{(4)}$ which is strictly easier to compute classically. We use a mapping between the average $\mathcal{C}^{(4)}$ and a model of a magnet which possesses the structure of the symmetric group of order 4, that reflects universal symmetries of the random circuit ensemble \cite{Nahum_PRX_2018, Keyserlingk_PRX_2018, ParticleConservingOTOCPollman, Khemani_PRX_2018, Mi_OTOC_2021}. We show numerically that the sign problem in this model is severe. We therefore argue that the severe sign problem is an inevitable feature that presents a barrier for classical sampling algorithms to compute $\mathcal{C}^{(4)}$.

\onecolumngrid
\newpage
\vspace{1em}
\begin{flushleft}
{\hypertarget{authorlist}{${}^\dagger$} \small Google Quantum AI and Collaborators}
% {* \small Google Quantum AI and Collaborators}

\bigskip
{\small
\renewcommand{\author}[2]{#1$^\textrm{\scriptsize #2}$}
\renewcommand{\affiliation}[2]{$^\textrm{\scriptsize #1}$ #2 \\}

\newcommand{\xGoogle}{\affiliation{1}{Google Research}}

\newcommand{\xPrinceton}{\affiliation{2}{Department of Physics, Princeton University, Princeton, NJ}}

\newcommand{\xBerkeleyChem}{\affiliation{3}{Department of Chemistry, University of California, Berkeley, Berkeley, CA}}

\newcommand{\xLBNLChem}{\affiliation{4}{Chemical Sciences Division, Lawrence Berkeley National Laboratory, Berkeley, CA}}

\newcommand{\xCIFAR}{\affiliation{5}{CIFAR Azrieli Global Scholars Program, 661 University Ave, Toronto, ON, Canada}}

\newcommand{\xUMass}{\affiliation{6}{Department of Electrical and Computer Engineering, University of Massachusetts, Amherst, MA}}

\newcommand{\xNVIDIA}{\affiliation{7}{NVIDIA Corporation, 2788 San Tomas Expressway, Santa Clara, 95051, CA, USA}}

\newcommand{\xUCSB}{\affiliation{8}{Department of Physics, University of California, Santa Barbara, CA}}

\newcommand{\xStorrs}{\affiliation{9}{Department of Physics, University of Connecticut, Storrs, CT}}

\newcommand{\xAuburnECE}{\affiliation{10}{Department of Electrical and Computer Engineering, Auburn University, Auburn, AL}}

\newcommand{\xCaltechIQI}{\affiliation{11}{Institute for Quantum Information and Matter, Caltech, Pasadena, CA}}

\newcommand{\xCaltechCMS}{\affiliation{12}{Computational and Mathematical Sciences, Caltech, Pasadena, CA}}

\newcommand{\xHarvard}{\affiliation{13}{Department of Chemistry, Harvard University, Cambridge, MA}}

\newcommand{\xNASA}{\affiliation{14}{Quantum Artificial Intelligence Laboratory, NASA Ames Research Center, Moffett Field, California 94035, USA}}

\newcommand{\xKBR}{\affiliation{15}{KBR, 601 Jefferson St., Houston, TX 77002, USA}}

\newcommand{\xUCSBCS}{\affiliation{16}{Department of Computer Science, University of California, Santa Barbara, CA}}

\newcommand{\xMITLab}{\affiliation{17}{Research Laboratory of Electronics, Massachusetts Institute of Technology, Cambridge, MA}}

\newcommand{\xMITEE}{\affiliation{18}{Department of Electrical Engineering and Computer Science, Massachusetts Institute of Technology, Cambridge, MA}}

\newcommand{\xMITPhysics}{\affiliation{19}{Department of Physics, Massachusetts Institute of Technology, Cambridge, MA}}

\newcommand{\xUCRPA}{\affiliation{20}{Department of Physics and Astronomy, University of California, Riverside, CA}}

\newcommand{\xDartmouthPhys}{\affiliation{21}{Department of Physics and Astronomy, Dartmouth College, Hanover, NH}}

\newcommand{\xCaltechWBI}{\affiliation{22}{Walter Burke Institute for Theoretical Physics, Caltech, Pasadena, CA}}

\newcommand{\xMPP}{\affiliation{23}{Max Planck Institute for the Physics of Complex Systems, Dresden, Germany}}

\newcommand{\Google}{1}
\newcommand{\Princeton}{2}
\newcommand{\BerkeleyChem}{3}
\newcommand{\LBNLChem}{4}
\newcommand{\CIFAR}{5}
\newcommand{\UMass}{6}
\newcommand{\NVIDIA}{7}
\newcommand{\UCSB}{8}
\newcommand{\Storrs}{9}
\newcommand{\AuburnECE}{10}
\newcommand{\CaltechIQI}{11}
\newcommand{\CaltechCMS}{12}
\newcommand{\Harvard}{13}
\newcommand{\NASA}{14}
\newcommand{\KBR}{15}
\newcommand{\UCSBCS}{16}
\newcommand{\MITLab}{17}
\newcommand{\MITEE}{18}
\newcommand{\MITPhysics}{19}
\newcommand{\UCRPA}{20}
\newcommand{\DartmouthPhys}{21}
\newcommand{\CaltechWBI}{22}
\newcommand{\MPP}{23}

\author{Dmitry A.~Abanin}{\Google,\! \Princeton},
\author{Rajeev Acharya}{\Google},
\author{Laleh Aghababaie-Beni}{\Google},
\author{Georg Aigeldinger}{\Google},
\author{Ashok Ajoy}{\BerkeleyChem,\! \LBNLChem,\! \CIFAR},
\author{Ross Alcaraz}{\Google},
\author{Igor Aleiner}{\Google},
\author{Trond I.~Andersen}{\Google},
\author{Markus Ansmann}{\Google},
\author{Frank Arute}{\Google},
% ...

\bigskip

\xGoogle
\xPrinceton
\xBerkeleyChem
\xLBNLChem
\xCIFAR
\xUMass
\xNVIDIA
\xUCSB
\xStorrs
\xAuburnECE
\xCaltechIQI
\xCaltechCMS
\xHarvard
\xNASA
\xKBR
\xUCSBCS
\xMITLab
\xMITEE
\xMITPhysics
\xUCRPA
\xDartmouthPhys
\xCaltechWBI
\xMPP
}

\vspace{3mm}

\vspace{3mm}
\begin{footnotesize}
{\bf Author contributions---}The Google Quantum AI team conceived and designed the experiment. The theory and experimental teams at Google Quantum AI developed the data analysis, modeling and metrological tools that enabled the experiment, built the system, performed the calibrations, and collected the data. The Berkeley, Dartmouth, and Google Quantum AI teams designed protocols for the Hamiltonian learning experiment. All authors wrote and revised the manuscript and the Supplementary Information.\par 
\end{footnotesize}

\vspace{3mm}
\begin{footnotesize}
{\bf Acknowledgements---}We are grateful to A.~Ashkenazi, S.~Brin, S.~Pichai and R.~Porat for their executive sponsorship of the Google Quantum AI team, and for their continued engagement and support.\par 
\end{footnotesize}

\vspace{3mm}
\begin{footnotesize}
{\bf Corresponding author---}Correspondence to Hartmut Neven (neven@google.com).
\end{footnotesize}

% \vspace{3mm}
% \begin{footnotesize}
% {\bf Code availability---}The Python simulation code used in theoretical analysis are available for download at https://doi.org/10.5281/zenodo.5570676.
% \end{footnotesize}

\vspace{3mm}
\begin{footnotesize}
{\bf Competing interests---}The authors declare no competing interests.
\end{footnotesize}

\end{flushleft}

\twocolumngrid

\bibliography{otoc2.bib}
\bibliographystyle{naturemag}

\end{document}

\title{Supplementary Information for ``Constructive interference at the edge of quantum ergodic dynamics''}
\author{Google Quantum AI and collaborators}

\maketitle

\tableofcontents

\section{Outline}
In this supplementary text, we provide further details on the experimental procedures used in the main text, solidify the theory behind quantum advantage in computing OTOC$^{(2)}$, and show additional experimental results and classical computation methods designed as part of this study.

The quantum advantage of OTOC$^{(2)}$ is rooted in three separate claims; for the reader wishing to directly validate these, we now highlight the important parts of the supplementary material to visit:
\begin{enumerate}
    \item The claim that our experiment faithfully reproduces OTOC$^{(k)}$: see Section~\ref{sec:otoc2_error_mitigation}.
    \item The claim that OTOC$^{(2)}$ is beyond the capabilities of classical computers to simulate to the same precision: see Section~\ref{sec:classical_otoc2}.
    \item The claim that the signal size of OTOC$^{(k)}$ only decays as a power law in ergodic dynamics: see Section~\ref{sec:otoc_theory}.
\end{enumerate}

\begin{table*}[!]
    \centering
    \begin{tabular}{c|c|c|c|c|c}\hline
         Method & Section & References & Scaling & SNR$_{\mathrm{OTOC}}>1$ & SNR$_{\mathrm{OTOC}^2}>1$ \\ \hline \hline
         Exact tensor network contraction & \ref{sec:tn}, \ref{sec:tn_estimates_otoc2} & \cite{Mi_OTOC_2021, Morvan2024_rcs} & Exponential & \cmark & \cmark \\ \hline
         Tensor network with gate removal & \ref{sec:tngr} & - & Exponential & \cmark & \cmark \\ \hline
         Neural quantum state methods & \ref{sec:nn_wf} & \cite{Carleo_2017,Westerhout_2020,Lin_2022} & Exponential & \xmark & \xmark \\ \hline
         Matrix product state methods & \ref{sec:mps_mpo} & \cite{H_mery_2019,_stlund_1995,Vidal_2003} & Exponential & \xmark & \xmark \\ \hline
         Clifford-based expansions & \ref{sec:clifford_expansion} & \cite{Mi_OTOC_2021, kechedzhi2024effective} & Exponential & \cmark & - \\ 
         & & & Polynomial & \xmark & - \\ \hline
         Pauli pathing truncation & \ref{sec:pauli_truncation} & \cite{begusic2024, angrisani2025} & Exponential & \xmark & \xmark \\ \hline
         %Reduced density operator methods & \ref{sec:rdo} & \cite{cree2022approximate} & Polynomial & Y & - \\ \hline
         Vanilla Monte Carlo & \ref{sec:mc_otoc2} & - & Polynomial & \xmark & \xmark \\ \hline
         Cached Monte Carlo & \ref{sec:cachedMC}, \ref{sec:cached_mc_otoc2} & - & Polynomial & \cmark & \xmark \\ \hline
         Tensor-network Monte Carlo & \ref{sec:tnmc}, \ref{sec:mc_otoc2} & - & Polynomial & \cmark & \xmark \\
         \hline
    \end{tabular}
    \caption{\textbf{Table of classical computation methods tested in this work.} We link to the different sections where each method can be found, as well as any previous literature upon which our implementations are based, and whether the method (as used in our study) scales exponentially or polynomially in the system size. We finally note whether we were able to achieve an SNR $>1$ for OTOC$^{(1)}$ and OTOC$^{(2)}$ systems (when tested), as a rough measure of whether we observe any correlation between a given approximation and the target OTOC.
    Note that the performance metric mentioned here does not correspond to competing with our beyond-classical claims, but on the other hand the performance of any of the above methods can potentially be further improved.
    For details on actual performance see the referenced sections.}
    \label{tab:method_summary}
\end{table*}

Our supplemental begins in Section~\ref{sec:expt}, where we give the technical details of our experiment at a level sufficient to reproduce our results.
We detail the quantum processors used (Section~\ref{sec:quantum_processor}), and perform a random circuit sampling (RCS) experiment on our next generation Willow hardware, setting a new benchmark for beyond-classical performance at this task.
We further outline the circuits implemented for the OTOC$^{(2)}$ experiment (Section~\ref{sec:circuit_implementation}) and the error mitigation methods used (Section~\ref{sec:otoc2_error_mitigation} and~\ref{sec:snr_otoc2}).

As part of the experimental section, we also give a dataset estimating the first-order OTOC, i.e. OTOC$^{(1)}$ on large systems, using the same gate ensemble as in the main text. We demonstrate the formation of a diffusive OTOC front in the looped iSWAP pattern (as used in the main text), as compared to the gate order used in standard RCS experiments which demonstrate exponentially-decaying OTOC$^{(1)}$ and OTOC$^{(2)}$ signals.
This allows us to demonstrate sensitivity to circuit parameters in experiments using up to $105$ qubits; the full Willow chip.
However, as we will explain in later sections of the supplemental, a large part of this sensitivity is not due to quantum interference, and can be accurately estimated using a Monte Carlo approximation in the Pauli-Liouville representation of the OTOC. This allows classical simulation to compete with quantum experiments in terms of accuracy. On the other hand, we find that despite this, no current implementation of classical algorithms can reproduce a set of OTOC$^{(1)}$ data at a system size of 95 qubits, indicating that simulating OTOC$^{(1)}$ remains a practically difficult task. Lastly, we demonstrate that even in the future event of a better classical algorithm that succeeds in approximating OTOC$^{(1)}$ at large scale, we may still be able to combine classical simulation and quantum simulation to attain better accuracy than either alone, therefore maintaining quantum advantage through such a hybrid approach. However, due to the strong competition of classical methods for simulating OTOC$^{(1)}$, and their potential to be further improved, we have focused on OTOC$^{(2)}$ throughout discussions in the main text.

In Section~\ref{sec:classical}, we outline multiple classical algorithms with which we have attempted to simulate the experimental OTOC$^{(1)}$ and OTOC$^{(2)}$ results.
We summarize the set of classical algorithms used in Table~\ref{tab:method_summary}.
The algorithms used can be roughly binned into exact algorithms with an exponential classical cost (typically in the system size), and heuristic algorithms with a polynomial classical cost but potentially unbounded error bars. We find a range of performance of these methods across OTOC$^{(1)}$ and OTOC$^{(2)}$ simulations, which is difficult to summarize. Nonetheless, as an indication that a classical method has had some success with OTOC simulation, we state whether our implementation was able to achieve an SNR greater than $1$ for OTOC$^{(1)}$ and/or OTOC$^{(2)}$.

For ease of presentation, we first study the performance of our various classical techniques on simulating OTOC$^{(1)}$ circuits in Section~\ref{sec:classical_otoc}.
Algorithms known in the literature based on matrix product state (MPS) (Section~\ref{sec:mps_mpo}) or neural quantum state (NQS) techniques (Section~\ref{sec:nn_wf}), as well as algorithms based on Pauli pathing truncation (Section~\ref{sec:pauli_truncation}) or Clifford-based expansions (Section~\ref{sec:clifford_expansion}) that work well for expectation value estimation, struggle to estimate even OTOC$^{(1)}$.
However, as part of this work, we have developed a set of algorithms based on Monte-Carlo sampling of Pauli paths.
In Section~\ref{sec:classical_otoc} we outline the ``vanilla'' Monte Carlo algorithm which is surprisingly accurate for the ensemble of random circuits considered in this work, despite its simplicity. Then, in Section~\ref{sec:cachedMC} and Section~\ref{sec:tnmc}, we outline two extensions that can be tuned to arbitrarily high accuracy (at potentially exponential cost). Neither of these algorithms has been able to improve over the error-mitigated device performance of the quantum chip presented in Section~\ref{sec:otoc_experiments}. However, we believe that the gap is sufficiently small that further improvements might allow for classical simulation of this data. That a Markovian approximation should work so well for operator spreading is a direct consequence of our use of random circuits, which have a scrambling effect, preventing the formation of large loops in a path-integral formalism.

In Section~\ref{sec:classical_otoc2}, we extend the above approaches to attempt a classical calculation of OTOC$^{(2)}$ for our 65-qubit circuit.
In Section~\ref{sec:mc_otoc2} and~\ref{sec:cached_mc_otoc2}, we show that our extension to OTOC$^{(2)}$ breaks the Monte Carlo approaches that worked so successfully for OTOC$^{(1)}$.
In lieu of good poly-time heuristics, we instead consider the cost of tensor network contraction in Section~\ref{sec:tn_estimates_otoc2}, which yield the classical estimates stated in the main text.
As a possible heuristic improvement on this, in Section~\ref{sec:tngr} we consider reducing the cost of tensor network contraction by removing gates from the circuit. This reduces the tensor network contraction to a heuristic as the error from gate removal is possible to estimate without access to the true results. In order to make any reasonable attempt at estimating the accuracy, we develop an {\it ad-hoc} method using Monte-Carlo simulation of the lower-order OTOC$^{(2)}$ to estimate the SNR loss from removing gates. Within the uncertainty of such an estimate, this highly heuristic protocol for OTOC$^{(2)}$ simulation breaks roughly even with the quantum processor in terms of time to solution. However, as the accuracy of this method has very little guarantee, we retain our claim that our OTOC$^{(2)}$ simulation is beyond-classical.

We finish the supplemental by theoretically justifying our findings of large fluctuations in OTOC circuits, and giving further evidence that the hardness encountered for estimating OTOC$^{(2)}$ circuits is likely broader than the methods considered here.
In Section~\ref{sec:otoc_theory} we develop two physical models for OTOCs in random circuits that demonstrate inverse-polynomially sized fluctuations for the values of $k$ used in this text.
The first model (Section~\ref{sec:permutation_basis}) analyses OTOC$^{(k)}$ in one-dimensional spin chains with Haar-random gates, demonstrating polynomial fluctuations via perturbation theory in the inverse of the local Hilbert space dimension $d$, and via tensor network calculations for $d=2$.
The second model (Section~\ref{sec:phasetransition}) is an exactly-solvable random-matrix theory in which one can find the spectrum of the correlation operator $B(t)M$ analytically.
We show that the spectrum undergoes a gap closing transition as the evolution becomes more and more scrambling.
We also verify numerically that a similar transition occurs in local random circuits.
On general grounds, we expect that the existence of the phase transition leads to strong circuit-to-circuit fluctuations of OTOC$^{(k)}$.
This validates \mbox{part 3)} of our beyond-classical claim.
Finally, in 
%Section~\ref{sec:average_otoc2_permutation} 
Section~\ref{sec:cached_mc_otoc2} 
we demonstrate evidence for a significant sign problem in the estimation of even the mean OTOC$^{(2)}$ in Haar-random circuits.
This adds to the validation of part 2) of our beyond-classical claim that was already demonstrated in Section~\ref{sec:classical_otoc2}.
The size of the sign problem found contrasts with a similar calculation of the variance of OTOC$^{(1)}$; a sign problem was noted previously in Ref.~\cite{Mi_OTOC_2021}, but we demonstrate in Section~\ref{sec:permutation_basis} that this is not as large, as evidenced by our success in calculating circuit-specific OTOC$^{(1)}$ in Section~\ref{sec:classical_otoc}.

\section{Experimental details and additional data}\label{sec:expt}

% I think it's better to start a section immediately with a subsection
\subsection{Quantum processor details}\label{sec:quantum_processor}

The quantum processor used in this work consists of 105 frequency-tunable superconducting transmon qubits, connected by tunable couplers. Two of the couplers were inoperable and consequently, our experiment is conducted with a two-dimensional grid of 103 qubits.

\begin{figure}[t!]
	\centering
	\includegraphics[width=1\columnwidth]{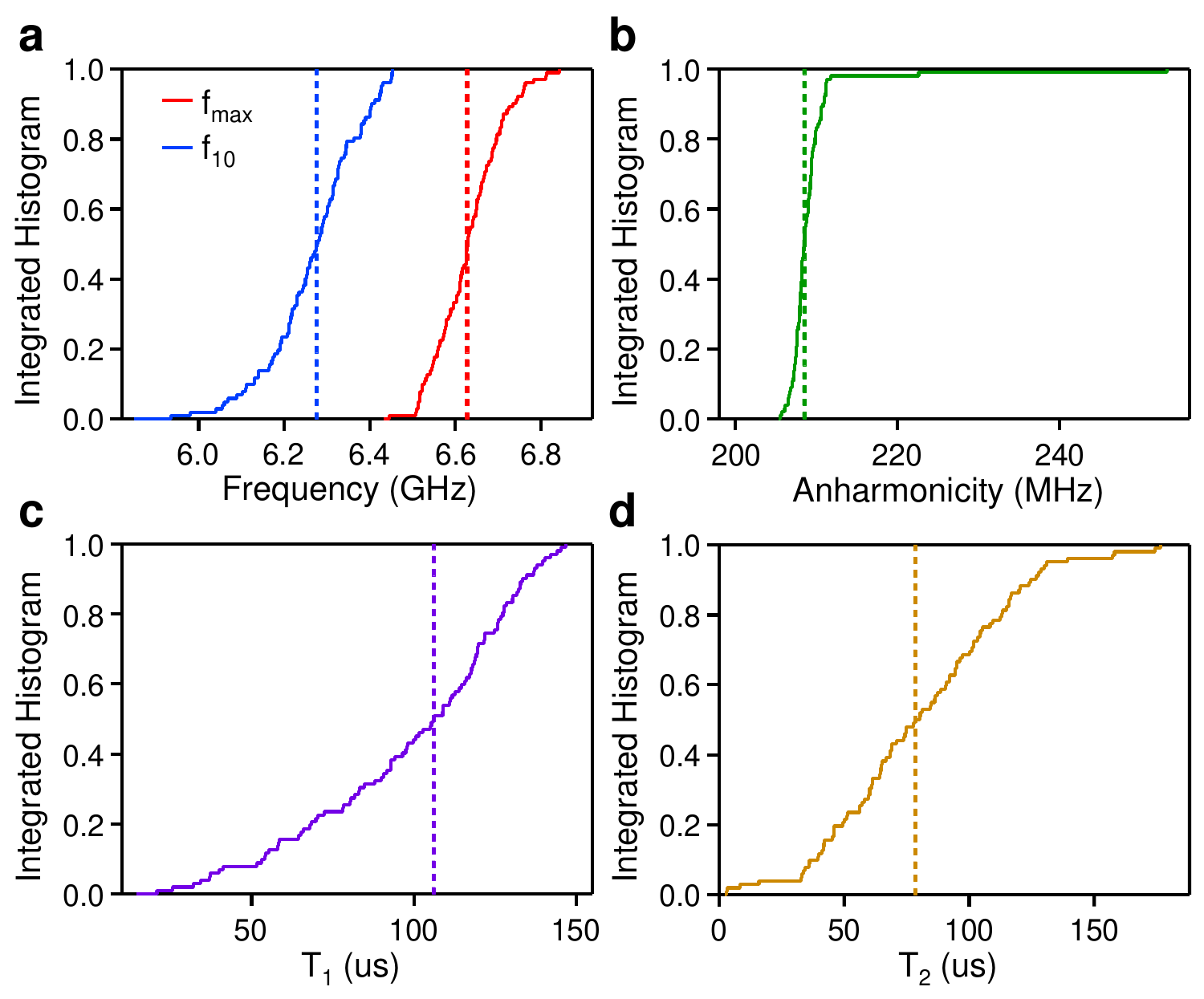}
	\caption{\textbf{Quantum processor details.} Integrated histograms of maximum qubit frequencies $f_\text{max}$ (panel {\bf a}), qubit frequencies at idle $f_{10}$ (panel {\bf a}), qubit anharmonicities (panel {\bf b}), qubit $T_1$ (panel {\bf c}) and qubit $T_2$ (panel {\bf d}).}
	\label{fig:s1}
\end{figure}

The transmon qubits used in this work are similar in design to our recent work on quantum error correction \cite{y1_paper_2024}. The main difference between the two processors is the shaping of ground planes near the qubits, which resulted in improved $T_1$ (median = 106 $\mu$s). Other qubit-related metrics, such as maximum allowed frequencies ($f_\text{max}$), frequencies at idle positions of the qubits ($f_{10}$), qubit anharmonicities and qubit $T_2$ measured using Carr-Purcell-Meiboom-Gill (CPMG) pulse sequences, are shown as integrated histograms in Fig.~\ref{fig:s1}.

\begin{figure}[t!]
	\centering
	\includegraphics[width=1\columnwidth]{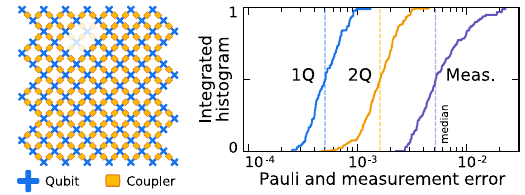}
	\caption{\textbf{Gate and measurement errors.} Left: Layout of the 103-qubit system used for this work. Semi-transparent symbols denote broken qubits and couplers in the original 105-qubit quantum processor. Right: Integrated histogram of simultaneous operation errors.}
	\label{fig:error_hist}
\end{figure}

The median single-qubit gate fidelity is measured to be 99.95\% using Clifford randomized benchmarking. The median two-qubit (iSWAP-like) gate fidelity is 99.85\%, obtained using cross-entropy benchmarking (XEB). The median readout fidelity is 99.5\%. The single-qubit gate, two-qubit gate, and readout times are $\SI{30}{ns}$, $\SI{29}{ns}$, and $\SI{600}{ns}$ respectively. All metrics are measured with all qubits in simultaneous operation, and the exact distribution of errors across the quantum processor are plotted in Fig.~\ref{fig:error_hist}.

The iSWAP-like gates are calibrated by bringing the frequencies of two qubits into resonance for a time \mbox{$t_\text{p} = \SI{21}{ns}$}, during which the coupler is ramped down in frequency to turn on a coupling with maximum amplitude $g_\text{max} / 2 \pi \approx \SI{18}{MHz}$. These parameters are chosen to synchronize the leakage and swap error channels \cite{quintana_prl_2019}. Additionally, effects of distortions in flux pulses are minimized by placing nearest-neighbor qubits close in frequency and allowing additional paddings between flux and microwave pulses \cite{Morvan2024_rcs}. Lastly, we use Floquet calibration techniques to characterize the conditional phase of the iSWAP-like gates \cite{Neill2021_ring}, which arises from the dispersive interaction between the $\ket{11}$ and $\ket{02}$ states and has a median of $\SI{0.35}{rad}$ across the processor.

\subsection{RCS quantum processor benchmark}

\begin{figure}[t!]
	\centering
	\includegraphics[width=1\columnwidth]{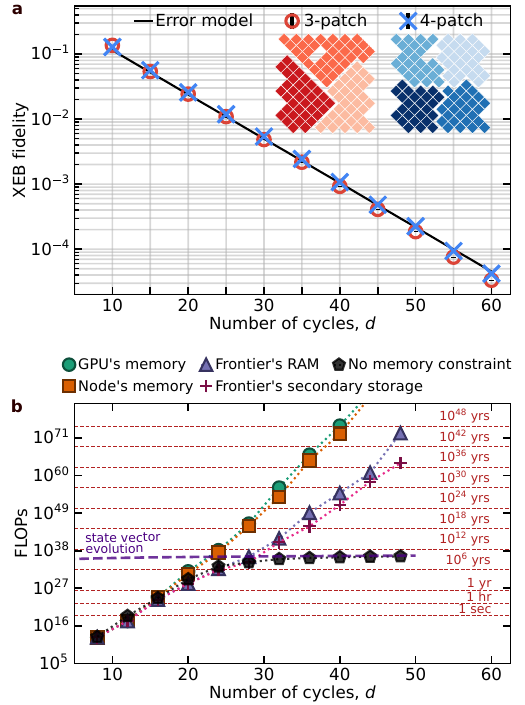}
	\caption{\textbf{RCS quantum processor benchmark.} {\bf a}, Cross-entropy bench-marking (XEB) of the 103-qubit quantum processor used in this work. The qubit grid is divided into three (red points) and four (blue points) disjoint simulatable patches which are measured in parallel (upper inset).Data correspond to the product of fidelities of the constituent patches, and solid line corresponds to digital error model prediction. {\bf b}, Simulation time for the computation of a single amplitude of the output of the 103-qubit RCS task at different depths.
	We first optimize the contraction path and slicing procedure for different memory constraints imposed on the contraction.
	We then estimate the time it would take to contract such a tensor network assuming $20\%$ FLOP efficiency on the Frontier supercomputer.}
	\label{fig:rcs_benchmark}
\end{figure}

As a system-wide benchmark, we perform random circuit sampling (RCS) experiments on our quantum processor. Calculating the XEB fidelity of the full 103-qubit processor is computationally intractable for classical computers. To estimate this fidelity, we employ a ``patched'' approach: we modify the random circuits by removing a small subset of two-qubit gates. 
This dissection effectively isolates sections of qubits, creating disjoint patches for which XEB fidelity can be efficiently calculated. 
In our experiments (Fig.~\ref{fig:rcs_benchmark}a), we divide our 103-qubit circuit into 3 and 4 patches (see inset of Fig.~\ref{fig:rcs_benchmark}a), and estimate the overall circuit fidelity by multiplying the XEB fidelities of its constituent patches.
To obtain each XEB fidelity data point, we averaged the results from $200$ different random circuits, each run with $50,000$ measurements. 
The estimated fidelity across different circuit depths shows close agreement with predictions from our digital error model, which includes single- and two-qubit gate and measurement errors.
We estimate the overall fidelity of 103-qubit RCS circuits to be $0.1\%$ at $40$ circuit cycles, which doubles the circuit volume compared to previous records~\cite{Morvan2024_rcs}, and have collected $100$ million sample bitstrings for a single circuit at this depth.

%We estimate the time it would take to classically perform a similar sampling task on a large supercomputer.
%These comparisons have become standard in the literature as a benchmark of the computational power of near term quantum computers through random circuit sampling~\cite{arute2019quantum,wu2021strong,zhu2022quantum,Morvan2024_rcs,decross2024computational}.
%We consider Frontier, the world's most performant supercomputer, with a peak floating point operations per second (FLOPS) of 1.71 ExaFLOPS.

To quantify the computational power of our processor, we estimate the classical resources required to perform a comparable RCS task.
Such comparisons have become a standard benchmark in the literature~\cite{arute2019quantum,wu2021strong,zhu2022quantum,Morvan2024_rcs,decross2024computational}.
% Approximating the classical resources required to perform a comparable RCS task become a standard benchmark in the literature~\cite{arute2019quantum,wu2021strong,zhu2022quantum,Morvan2024_rcs,decross2024computational}, providing a measure of the computational power of near-term quantum computers.
We consider Frontier, the world's most powerful supercomputer, with a peak performance of 1.71 ExaFLOPs.

Highly optimized tensor network contraction (TNC) algorithms are widely considered the most competitive classical approach for simulating near-term RCS experiments~\cite{gray2021hyper,huang2020classical,pan2022solving,kalachev2021classical,kalachev2021multi,Morvan2024_rcs} (see also Section~\ref{sec:classical}).
These methods involve mapping the quantum circuit to a tensor network, which is then contracted to compute probability amplitudes of specific output bitstrings.
This enables the sampling of bitstrings through rejection sampling~\cite{markov2018quantum}.

The computational cost of TNC is highly sensitive to the chosen contraction ordering.
To minimize this cost, we have implemented a highly efficient optimizer of contraction paths that considers techniques such as slicing to enforce memory constraints in the TNC~\cite{chen2018classical,huang2020classical,kalachev2021multi,kalachev2021classical,pan2022solving,Morvan2024_rcs} and caching and reuse of intermediate tensors~\cite{villalonga2019flexible,kalachev2021classical,kalachev2021multi,pan2022solving,Morvan2024_rcs}.
Further details on these techniques and implementation of the optimizer are provided in Section~\ref{sec:tn}.

In Fig.~\ref{fig:rcs_benchmark}b, we present estimated times for computing a single probability amplitude on Frontier, a key step in simulating RCS.
We analyze the scaling with circuit depth $d$ under various memory constraints, parameterized by the maximum allowed size $2^W$ (where the exponent $W$ is usually called the \emph{width}) of intermediate tensors generated during the contraction.
These constraints range from fitting within a single GPU ($W=31$), a single node ($W=33$), utilizing all of Frontier's RAM ($W=48$), utilizing all of Frontier's secondary storage ($W=54$), to even an idealized scenario with no memory constraints ($W=\infty$).

Consistent with previous studies~\cite{Morvan2024_rcs,decross2024computational}, we observe a crossover from exponential to linear scaling with depth in the absence of memory constraints.
However, imposing memory limits enforces an exponential scaling, with tighter constraints leading to dramatically increased computational costs.
Our estimates assume a $20\%$ FLOP efficiency, which is certainly optimistic for scenarios with loose memory constraints, where bandwidth limitations become significant.

Remarkably, even in the idealized scenario without memory constraints and assuming $20\%$ FLOP efficiency, the estimated time to compute a single output probability amplitude is about a billion years at depths $d\geq 32$.
This establishes RCS on our quantum processor as a task that vastly exceeds the capabilities of current classical supercomputers.

\subsection{Measurement circuit schematics}

\subsubsection{Time-ordered correlator}

\begin{figure}[t!]
	\centering
 	\includegraphics[width=1\columnwidth]{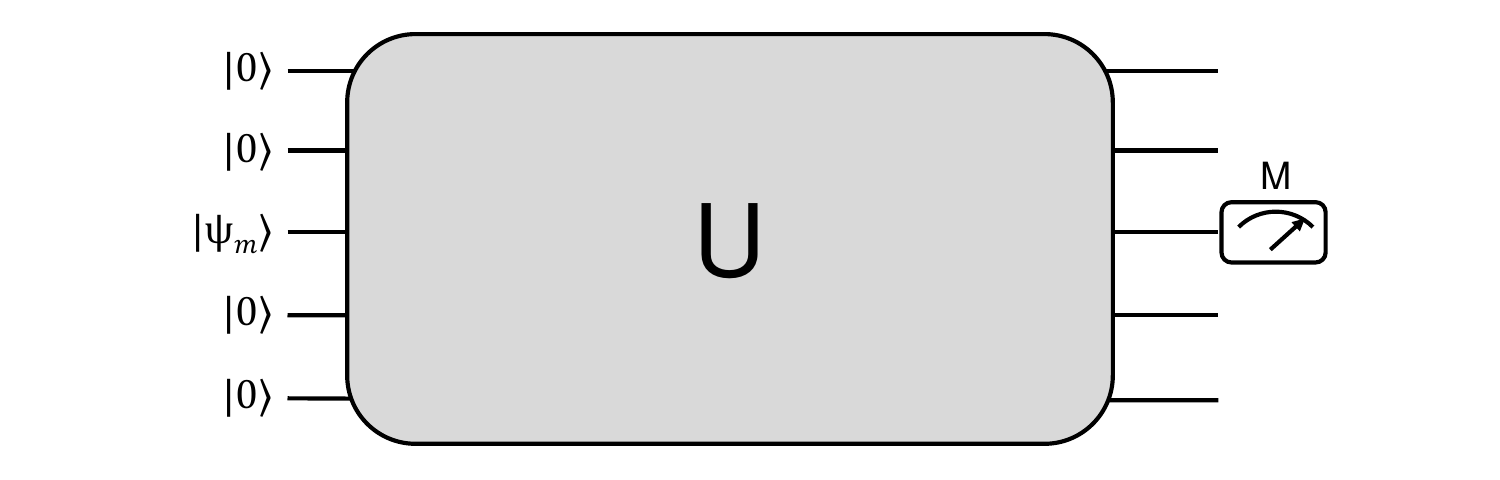}
	\caption{\textbf{Circuit diagram for measuring a time-ordered correlator.} Circuit diagram for measuring a time-ordered correlator, $\langle M(t)M \rangle$.}
	\label{fig:toc_diagram}
\end{figure} 

\begin{figure*}[t!]
	\centering
 	\includegraphics[width=2\columnwidth]{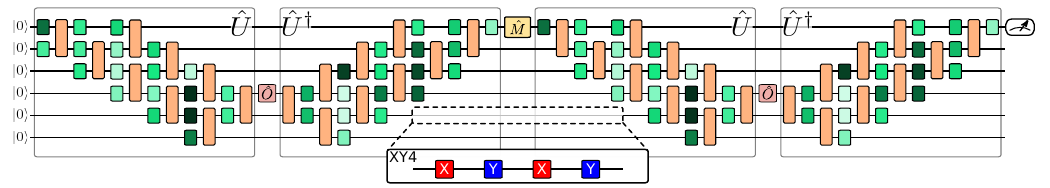}
	\caption{\textbf{Quantum circuit for OTOC$\boldsymbol{^{(2)}}$.} Detailed circuit structure of an OTOC$^{(2)}$ experiment. Quantum gates outside the lighcones of $q_\text{b}$ and $q_\text{m}$ are removed and replaced with dynamical decoupling (XY4) sequences.}
	\label{fig:s2}
\end{figure*} 

To measure the time-ordered correlator (TOC) shown in Fig.~2c of the main text, we employ the circuit schematic in Fig.~\ref{fig:toc_diagram}. We initialize the system in an eigenstate of the measurement operator $M$, $\ket{\psi_\text{m}}$, then time-evolve the system with the scrambling unitary $U$. In the end, we measure $\langle M \rangle$ at the measurement qubit $q_\text{m}$. Averaging over the two eigenstates of $M$ yields a measurement outcome equivalent to $\langle M(t)M \rangle$. Similar to OTOC$^{(k)}$, we have chosen $M$ to be $Z$ so $\ket{\psi_\text{m}}$ is either $\ket{0}$ or $\ket{1}$.

\subsubsection{OTOC$^{(k)}$}\label{sec:circuit_implementation}

\begin{figure}[t!]
	\centering
 	\includegraphics[width=1\columnwidth]{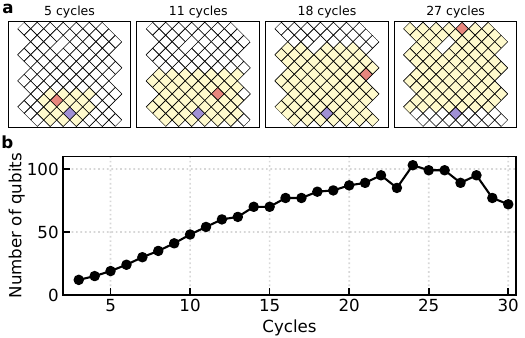}
	\caption{\textbf{Circuit geometries for measuring OTOC$\boldsymbol{^{(k)}}$ fluctuations.} {\bf a,} Circuit geometries used for measuring the fluctuations of $\mathcal{C}^{(2)}$, $\mathcal{C}^{(4)}$ and $\mathcal{C}^{(4)}_\text{off-diag}$ in Fig.~2c of the main text. Blue square indicates location of $q_\text{m}$. Red square indicates location of $q_\text{b}$. Yellow shaded squares indicate qubits that are within the lightcones of $q_\text{m}$ and $q_\text{b}$. {\bf b,} Number of qubits within the lightcones of $q_\text{m}$ and $q_\text{b}$ for each cycle.}
	\label{fig:moving_flies}
\end{figure} 

\begin{figure*}[t!]
	\centering
 	\includegraphics[width=2\columnwidth]{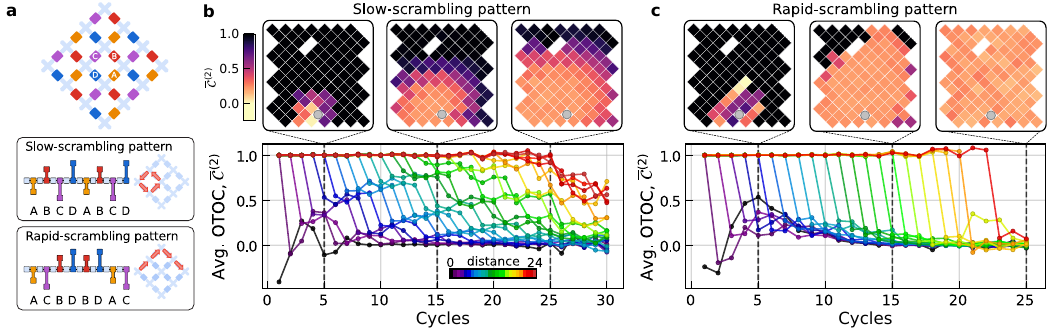}
	\caption{\textbf{Circuit structure and decay of OTOC$\boldsymbol{^{(1)}}$.} {\bf a}, Top panel: Schematic showing two-qubit gates applied in different cycles with different colors. Bottom panels: Two circuit patterns with different orders of applying the two-qubit gates. Red arrows indicate how an imaginary ``particle'' is propagated in each case.  {\bf b}, Top panels: OTOC measured with different qubits as $q_\text{b}$ and averaged over 8 circuit instances, $\overline{\mathcal{C}}^{(2)}$, for $d$ = 5, 10 and 25 cycles in the slow-scrambling pattern. Grey diamond indicates the fixed location of $q_\text{m}$. Bottom panel: Time-dependent  $\overline{\mathcal{C}}^{(2)}$ for qubits with different lightcone distances from $q_\text{m}$ (color legend). Qubits with the same lightcone distance from $q_\text{m}$ are averaged together. {\bf c}, Same data as {\bf b} but for the rapid-scrambling pattern.}
	\label{fig:loop_pattern}
\end{figure*} 

The key requirement for implementing OTOC and OTOC$^{(2)}$ measurements is inverting a random circuit ${U}$. While SQ gates are readily invertible by changing the phases of the microwave drive, the iSWAP-like gates are not as straightforward due to the conditional phase, the sign of which cannot be changed with virtual $Z$ gates alone. In our past works, this was achieved by increasing the length of the iSWAP-like gates, which reduces the conditional phase proportionally \cite{Mi_OTOC_2021}. However, this approach leads to higher incoherent errors in the iSWAP-like gates and lower fidelities. The inversion also remains imperfect due to the residual conditional phase.

In our current work, we find that the iSWAP-like gates can be inverted by the following transformation: ${G}_{12}^\dagger = {X}_2 {G}_{12} {X}_1 {Z}_2^{\varphi / \pi} $. Here ${G}_{12}$ is the iSWAP-like gate and $\varphi$ is its conditional phase, ${X}_1$ and ${X}_2$ are Pauli-$ X$ gates applied to qubits 1 and 2, and ${Z}_2$ is a Pauli-$ Z$ gate applied to qubit 2. The additional $x$-rotations introduced through this transformation are then combined with the existing random SQ gates in each circuit cycle into a single layer of single-qubit gates. As such, the iSWAP-like gates are exactly inverted without any overhead in SQ gate count.

When compiling the quantum circuits for OTOC and OTOC$^{(2)}$, we remove all gates that are outside either the lightcone of $q_\text{m}$ or the lightcone of $q_\text{b}$. This is because they either do not contribute to the observables or have limited impact. The resulting circuit structure for OTOC$^{(2)}$ is shown in Fig.~\ref{fig:s2} (the circuit structure for OTOC$^{(1)}$ is effectively half the OTOC$^{(2)}$ circuit). To reduce idle errors, we apply the XY4 dynamical decoupling sequence~\cite{gullion_1990} to qubits outside the lightcones.

Lastly, when measuring the fluctuation of OTOC$^{(k)}$ in Fig.~2c of the main text, we have adjusted the location of $q_\text{b}$ as circuit cycle increases in order to keep the measurement close to the wavefront of $q_\text{m}$ (i.e. the point where circuit-averaged OTOC is $\sim$0.5). For completeness, we show the choices of $q_\text{b}$ for different cycles in Fig.~\ref{fig:moving_flies}. The effective system size, corresponding to the number of qubits within the lightcones of $q_\text{m}$ and $q_\text{b}$, is also shown in the same figure for reference. We note that the decrease in the effective system size at cycles 29 and 30 is due to the $q_\text{b}$ chosen for those two cycles having more slowly growing lightcones.

\subsection{Scrambling circuit choice}

The employment of iSWAP-like gates leads to a class of maximally scrambling circuits which exhibit fast-decaying OTOCs in one-dimensional (1D) geometries \cite{Claeys_PRReseach_2020, Mi_OTOC_2021}. This presents a practical challenge for achieving experimental advantage, since the number of two-qubit gates $N_\text{2D}$ may not be sufficiently large to challenge tensor networks contraction algorithms before the signal size is too small to measure.  We discover that, in 2D setups, scrambling may be tuned by engineering circuit structures without changing the types of gates. Fig.~\ref{fig:loop_pattern}a shows two spatial patterns over which the iSWAP-like gates may be applied in 2D circuits: In the ``rapid-scrambling pattern'', a sequence of two-qubit gates forms a non-self-intersecting path connecting $q_\text{m}$ and $q_\text{b}$ (middle panel in Fig.~\ref{fig:loop_pattern}a). Here maximally scrambling behavior similar to 1D geometry is expected. In the ``slow-scrambling pattern'' (bottom panel in Fig.~\ref{fig:loop_pattern}a), gate sequences connecting $q_\text{m}$ and $q_\text{b}$ form loops instead. This pattern demonstrates slower scrambling and, as we will show below, qualitatively different decay of OTOCs.

Fig.~\ref{fig:loop_pattern}b and Fig.~\ref{fig:loop_pattern}c show, for the two different gate patterns, the spatial-temporal dependence of OTOCs after averaging over circuits with different single-qubit gates, $\overline{\mathcal{C}}^{(2)}$. We observe qualitatively different behaviors between the two gate patterns: In the slow-scrambling pattern, the decay of $\overline{\mathcal{C}}^{(2)}$ becomes slower as the distance between $q_\text{m}$ and $q_\text{b}$ increases. This effect, known as wavefront broadening, arises from operator spreading via a biased diffusion process having an average velocity slower than the lightcone \cite{Nahum2018-pw, von-Keyserlingk2018-ez}. In the case of the rapid-scrambling pattern, the wavefront resembles a sharp step function instead. This shape is characteristic of maximally scrambling circuits, in which ${B}(t)$ spreads ballistically at the velocity of the lightcone \cite{Claeys_PRReseach_2020}.

Throughout the main text and the rest of this supplement, we will focus on quantum circuits that adopts the slow-scrambling pattern. We note that the diffusive behavior of the slow-scrambling circuits is in fact more representative of generic nonintegrable dynamics in 2D. For example, Haar random circuits studied in Section~\ref{sec:otoc_theory} exhibit wavefront broadening similar to the slow-scrambling pattern shown here. 

\subsection{Experimental results on OTOC$^{(1)}$}\label{sec:otoc_experiments}

In the main text, we have largely focused on OTOC$^{(2)}$ owing to the presence of large-loop interference which makes classical simulation particularly difficult. In this section, we present a complete set of experimental measurements that demonstrate two key findings: i) While the small-loop interference of OTOC allows heuristic classical algorithms to compete with experimental results at small system scales ($\leq40$ qubits), the practical implementation cost of heuristic algorithms increases dramatically as system size grows. We present OTOC measurements with $\sim$100 qubits and show that without a classical supercomputer like Frontier, no practical implementation of currently known heuristic algorithms is capable of simulating such circuits. ii) Even in the event where a heuristic algorithm is capable of simulating OTOC with better accuracy than the quantum processor, the outcome of the heuristic algorithm may be combined with quantum processor results to achieve an accuracy higher than the classical simulation alone. These two findings indicate that while higher-order OTOCs are convenient choices of observables for demonstrating quantum advantage, practical tasks such as Hamiltonian could very well employ lower-order OTOCs since their simulation remains a daunting task for classical computers.

\subsubsection{Error-mitigation for OTOC$^{(1)}$}\label{sec:otoc_error_mitigation}

Before presenting the experimental results related to the two findings above, we first describe the error-mitigation scheme used for OTOC$^{(1)}$ measurements. Due to the presence of noise, quantum observables such as OTOC$^{(1)}$ and OTOC$^{(2)}$ are scaled down by a noise damping factor, $F_\text{noise}$, that is in principle circuit-dependent. In circuits with the ``rapid-scrambling'' gate patterns where each Pauli string has the same butterfly velocity, the noise-damping factor on OTOC is close to the value of Loschmidt echo fidelity corresponding to $\langle {M} \rangle$ of $q_\text{m}$ when the butterfly operator, ${B}$, is set to identity. This behavior was experimentally confirmed in our previous work \cite{Mi_OTOC_2021}.

\begin{figure}[t!]
	\centering
	\includegraphics[width=1\columnwidth]{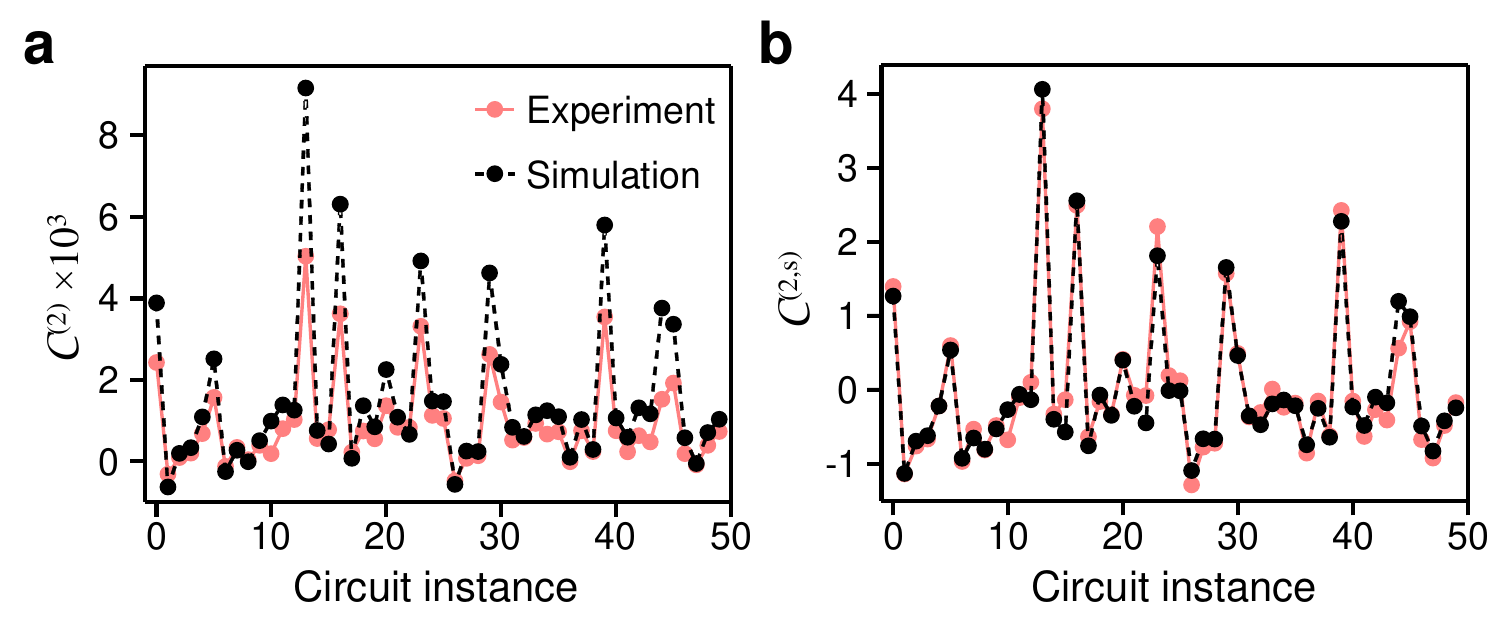}
	\caption{\textbf{Error mitigation through global re-scaling.} {\bf a}, Raw experimental values of $\mathcal{C}^{(2)}$ for the data shown in Fig.~3d of the main text along with values from exact simulation. {\bf b}, Re-scaled values of $\mathcal{C}^{(2)}$, $\mathcal{C}^{(2,s)}$, for the same circuit instances and comparison with exact simulation.}
	\label{fig:s3}
\end{figure}

In our current circuits which involve loops, $F_\text{noise}$ is no longer described by the echo fidelity due to the finite spread of butterfly velocities. On the other hand, we also find that $F_\text{noise}$ is largely insensitive to specifics of single-qubit gates and is therefore approximately a global scaling factor. To demonstrate this, we show in Fig.~\ref{fig:s3}a the raw measurements of
the OTOC$^{(1)}$, $\mathcal{C}^{(2)}$, for the data presented in Fig.~3d of the main text. Here it is observed that the circuit-dependent values of $\mathcal{C}^{(2)}$ from exact simulation are reproduced well in the noisy experiment but with an overall damping of amplitudes.

As a simple error-mitigation strategy, we define the accuracy metric, the signal-to-noise ratio (SNR), to be insensitive to the global scaling factor. As described in the main text, we define a set of ``rescaled'' OTOCs as $\mathcal{C}^{(2, s)} = (\mathcal{C}^{(2)} - \overline{\mathcal{C}}\,\!^{(2)}) / \sigma\! \left(\mathcal{C}^{(2)} \right)$, where $\overline{\mathcal{C}}\,\!^{(2)}$ and $\sigma\! \left(\mathcal{C}^{(2)} \right)$ are respectively the mean and standard deviation of $\mathcal{C}^{(2)}$ over 50 instances. 
The SNR is then defined as
\begin{equation}\label{eq:def_snr_rescaled}
    {\rm SNR} = \frac{1}{\sqrt{\overline{\big(\mathcal{C}_{\rm exp\vphantom{sim}}^{(2, s)} - \mathcal{C}_{\rm sim}^{(2, s)}\big)^2}}},
\end{equation}
with $\mathcal{C}_{\rm exp}^{(2, s)}$ and $\mathcal{C}_{\rm sim}^{(2, s)}$ being the OTOCs from experimental results and numerical simulations respectively.
Here, subtracting the mean is motivated by the fact that circuit-averaged OTOCs are efficiently simulated through classical population dynamics. Indeed, rescaling the entire data set by $\sigma\! \left(\mathcal{C}^{(2)} \right)$ normalizes the standard deviation to 1, thereby removing the global noise-damping factor from the signal. The rescaled OTOCs are shown in Fig.~\ref{fig:s3}b, where the experiment agrees well with the exact simulation.

While our definition of SNR is insensitive to the global scale factor, we note that in practice the noise-damping factor, $F_\text{noise}$, can be empirically determined through comparison with efficient classical simulations using the Monte-Carlo (MC) or hybrid TNMC algorithms (described in more details in Section~\ref{sec:classical}). Specifically, the former is used when presenting the $\mathcal{C}^{(2)}$ data in Fig.~2c of the main text, where the scaling factor is empirically derived from the ratio of $\sigma\! \left(\mathcal{C}^{(2)} \right)$ between the MC simulation and experimental Pauli-averaged OTOCs. The factor is then used to rescale the experimental OTOCs without the Pauli ensemble-averaging.

\subsubsection{OTOC$^{(1)}$ experiment with 95 qubits}\label{sec:otoc_95q_experiment}

\begin{figure*}[t!]
	\centering
	\includegraphics[width=2\columnwidth]{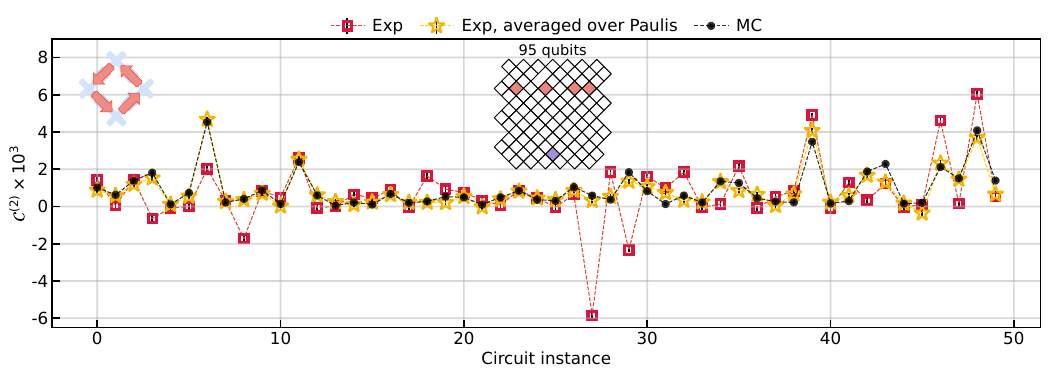}
	\caption{\textbf{OTOC measurements with 95 qubits.} Circuit-dependent $\mathcal{C}^{(2)}$ measured on a 95-qubit geometry shown on the left (blue: $q_\text{m}$, red: $q_\text{b}$). The total number of two-qubit gates is $N_\text{2D} = 1000$. Here the error bars indicate statistical shot noise $F/\sqrt{K}$ which is much smaller than data. Experimental results after averaging over random Paulis applied everywhere in the circuits (see Fig.~3b of the main text), and comparison with Monte Carlo simulation, are also shown.}
	\label{fig:otoc_95q}
\end{figure*} 

We now present a set of large-scale OTOC data in Fig.~\ref{fig:otoc_95q}, taken with a 95-qubit geometry comprising $N_\text{2D} = 1000$ iSWAP-like gates. In order to compare against heuristic classical algorithms, we also need to bound the accuracy of these results similar to what was done for $\mathcal{C}^{(4)}$ in the main text. In this case, it is in fact possible to modify the quantum circuits in a way that render them classically simulatable using Monte-Carlo methods. This is done by injecting random Pauli operators everywhere in the unitary $U$ and measure the value of $\mathcal{C}^{(2)}$ for each circuit after averaging over an ensemble of random Paulis. The averaging process effectively eliminates all quantum interference processes during the time-evolution of $B(t)$ in Pauli space, making the final amplitude of each Pauli string exactly tractable through Monte-Carlo simulation (Section~\ref{sec:mc_intro}). 

Fig.~\ref{fig:otoc_95q} shows the experimental values of $\mathcal{C}^{(2)}$ after Pauli-averaging, which deviates significantly from the data without Pauli averaging. In particular, we find that some circuit instances (e.\,g. $i = 6$) are enhanced by the averaging process while other instances are suppressed by the averaging process (e.\,g. 46 and 48). This relative fluctuation may be understood by the fact that the interference contribution to $\mathcal{C}^{(2)}$ can have either the same or the opposite sign as $\mathcal{C}^{(2)}$, and therefore subtracting its value may lead to either suppression or enhancement of the experimental signal. 

In the same figure, we have also shown results of Monte-Carlo simulation which agrees well with the Pauli-averaged experimental data up to a SNR of 3.5. Taking this as the accuracy of the non-averaged dataset, we find that our best attempt at simulating $\mathcal{C}^{(2)}$ values of these 95-qubit circuits using cached Monte-Carlo only results in SNR of $\sim\!0.7$ (Fig.~\ref{fig:OTOC_CMC_large_systems}b) when compared against experiment, suggesting that the simulation results likely have lower accuracy than the experimental data. Simulating each circuit in Fig.~\ref{fig:otoc_95q} using TN contraction algorithms is estimated to require $\sim\!3$ days on the Frontier supercomputer, in comparison to $\sim\!30$ minutes of data acquisition time on the quantum processor.

\subsubsection{Turning heuristic classical simulation into error-mitigation}

\begin{figure}[t!]
	\centering
	\includegraphics[width=1\columnwidth]{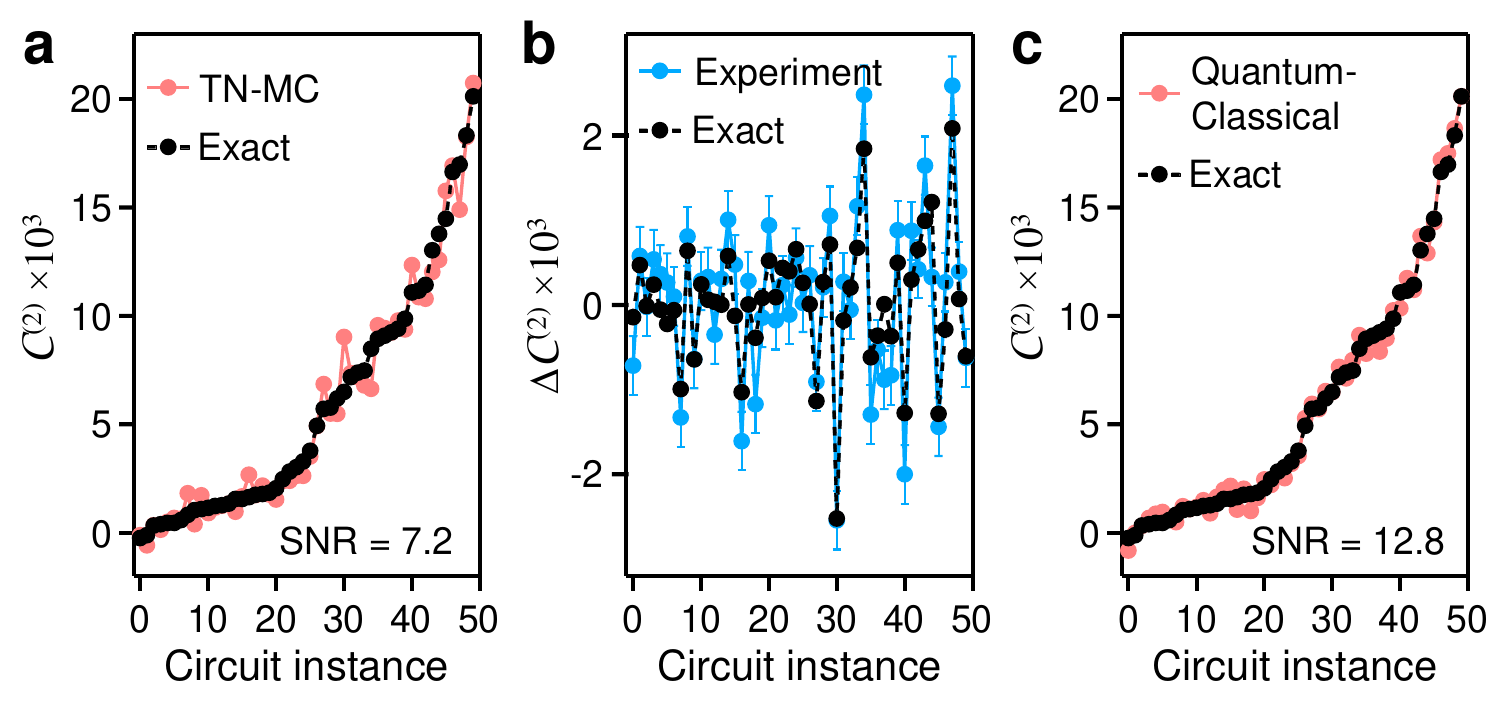}
	\caption{\textbf{Error mitigation through efficient classical simulation.} {\bf a}, Comparison of exactly simulated OTOCs and OTOCs simulated using the approximate TNMC algorithm. The circuits here include 31 qubits and a total of 24 cycles, with the ``cuts'' in the TNMC algorithm placed at circuit cycles 10 and 14. The circuit instances are sorted in the order of increasing OTOC values for clarity. {\bf b}, Difference between the TNMC and exact simulation results, $\mathcal{C}^{(2)}_{\rm off-diag}$. Experimentally measured $\mathcal{C}^{(2)}_{\rm off-diag}$ values are also shown for comparison, after re-scaling by a noise-damping factor $F_\text{noise}$ (see text). {\bf c}, Error-mitigated $\mathcal{C}^{(2)}$ obtained by adding the experimentally measured $\mathcal{C}^{(2)}_{\rm off-diag}$ in panel {\bf b} to the TNMC results in panel {\bf a}. Exact simulation results are also shown for comparison.}
	\label{fig:s4}
\end{figure} 

In this section, we demonstrate how quantum advantage may be maintained in some cases even when a heuristic classical algorithm is effective at capturing the experimental observable. Specifically, we focus on a hybrid tensor-network Monte-Carlo (TNMC) algorithm introduced in Section~\ref{sec:tnmc}. This classical algorithm combines TN contraction with MC approximation along specific ``cuts'' (i.e. cycles) of the unitary $U$. The simulation outcome of this algorithm is equivalent to $\mathcal{C}^{(2)}$ after averaging over Paulis injected along the same cuts. Fig.~\ref{fig:s4}a shows the results of implementing the TNMC algorithm on a set of 31 qubits, where a high SNR of 7.2 is achieved and exceeds the SNR of 6.0 achieved by the quantum processor for the same set of circuits.

Despite the better performance of the TNMC algorithm, we find that the accuracy of the quantum processor can be significantly boosted by using the TNMC results for error-mitigation. The procedure is outlined in Fig.~\ref{fig:s4}: We start by computing the TNMC results along select cuts (Fig.~\ref{fig:s4}a), denoted as $\mathcal{C}_\text{TNMC}^{(2)}$. Next, an experiment is performed which measures $\mathcal{C}^{(2)}$ with and without random Paulis inserted to the same cycles as the cut locations used in the TNMC algorithm. We denote the outcomes as $\mathcal{C}_\text{exp}^{(2)}$ and $\mathcal{C}_\text{TNMC, exp}^{(2)}$ respectively. Finally, the formula for the hybrid OTOC estimate is then:
\begin{equation*}
    \mathcal{C}^{(2)}_{\text{hybrid}} = \mathcal{C}^{(2)}_{\text{TNMC}} + \frac{\sigma(\mathcal{C}^{(2)}_{\text{TNMC}})}{\sigma(\mathcal{C}^{(2)}_{\text{TNMC, exp}})}
    \mathcal{C}^{(2)}_{\rm off-diag},
    %(\mathcal{C}^{(2)}_{\text{exp}} - \mathcal{C}^{(2)}_{\text{TNMC, exp}}).
\end{equation*}
with $\mathcal{C}^{(2)}_{\rm off-diag} = \mathcal{C}^{(2)}_{\text{exp}} - \mathcal{C}^{(2)}_{\text{TNMC, exp}}$.
The results, shown in Fig.~\ref{fig:s4}b, are in good agreements with the difference between exact $\mathcal{C}^{(2)}$ and $\mathcal{C}^{(2)}_\text{TNMC}$. The resulting SNR, which equals 12.8 in this particular case, is now higher than the $\text{SNR} = 7.2$ obtained with classical TNMC simulation alone.

\begin{figure}[t!]
	\centering
	\includegraphics[width=1\columnwidth]{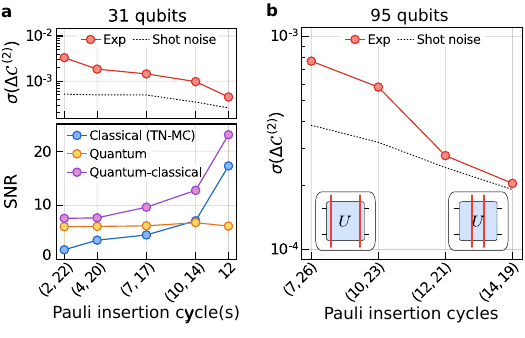}
	\caption{\textbf{Hybrid quantum-classical OTOC measurements.} {\bf a}, Top: Standard deviation of $\mathcal{C}^{(2)}_{\rm off-diag}$, $\sigma(\mathcal{C}^{(2)}_{\rm off-diag})$, over different choices of cycles for inserting random Paulis. Data are obtained with the same set of 31-qubit circuits as Fig.~\ref{fig:s4}. Dashed line indicates statistical shot noise based on the number of measurements in each case. Bottom: SNR (compared against exact simulation) for the classical TN-MC algorithm with different Pauli insertion cycles and hybrid quantum-classical results at the same cycles. The SNR obtained on the quantum device alone is also shown for comparison.  {\bf b}, Similar to the top panel of {\bf a} but measured with the 95-qubit circuits used in Fig.~\ref{fig:otoc_95q}. 180 million measurement shots are taken per circuit for the (14, 19) injection cycles.}
	\label{fig:hybrid_classical_quantum}
\end{figure} 

To demonstrate that the same observation holds irrespective of the specific implementation of the classical algorithm, we show in Fig.~\ref{fig:hybrid_classical_quantum}a SNRs obtained with the quantum processor alone, TNMC algorithm alone and the combination of quantum processor and TNMC algorithm. We find, for the set of 31-qubit circuits studied here, that the SNRs obtained by using both quantum and classical results outperforms the SNRs of quantum or classical results alone, regardless of the locations of the cuts. In particular, an SNR as high as 22 is obtained in the quantum-classical case for a cut placed in the middle of $U$ (cycle 12). These results indicate that as long as the difference of a classical approximation and exact quantum simulation is sufficiently large to be measured on the quantum processor above the statistical shot noise floor, quantum advantage may be maintained through the hybrid quantum-classical protocol outlined in this section.

It is instructive to apply the same protocol to the 95-qubit circuits of Fig.~\ref{fig:otoc_95q} and probe whether the difference $\mathcal{C}^{(2)}_{\rm off-diag}$ can be measured. Fig.~\ref{fig:hybrid_classical_quantum}b shows the experimentally measured standard deviations $\sigma(\mathcal{C}^{(2)}_{\rm off-diag})$ for four different implementations of a two-cut TNMC algorithm. We observe that $\sigma(\mathcal{C}^{(2)}_{\rm off-diag})$ decreases as the cuts are moved toward the middle of $U$, which has 31 cycles in total. This is consistent with Fig.~3c of the main text. Importantly, at cut locations (12, 21) which approximately divides $U$ into three equal parts and therefore is optimal for practical TNMC implementation, we can still resolve $\sigma(\mathcal{C}^{(2)}_{\rm off-diag})$ above the statistical shot noise floor. Based on the estimates in Section~\ref{sec:tnmc}, TNMC implemented with these cuts requires the Frontier supercomputer and is therefore practically infeasible is to attempt at the moment of writing. Based on these observations and the results from the previous section, we conclude that exceeding the accuracy of the 95-qubit OTOC measurements in Fig.~\ref{fig:otoc_95q} is not possible without extensive runtimes on classical supercomputers.

\subsubsection{Infinite-temperature OTOC$^{(1)}$}\label{sec:inf_temp}

Throughout the experiments reported in the main text, we have used a product state as the initial state of the system. In real systems of interest, oftentimes the OTOC being measured corresponds to that of infinite-temperature states. In other words, $\mathcal{C}^{(2)} = \text{Tr} [B (t) M]$. Here, we show that infinite-temperature OTOC can also be measured on our device.

\begin{figure}[t!]
	\centering
	\includegraphics[width=1\columnwidth]{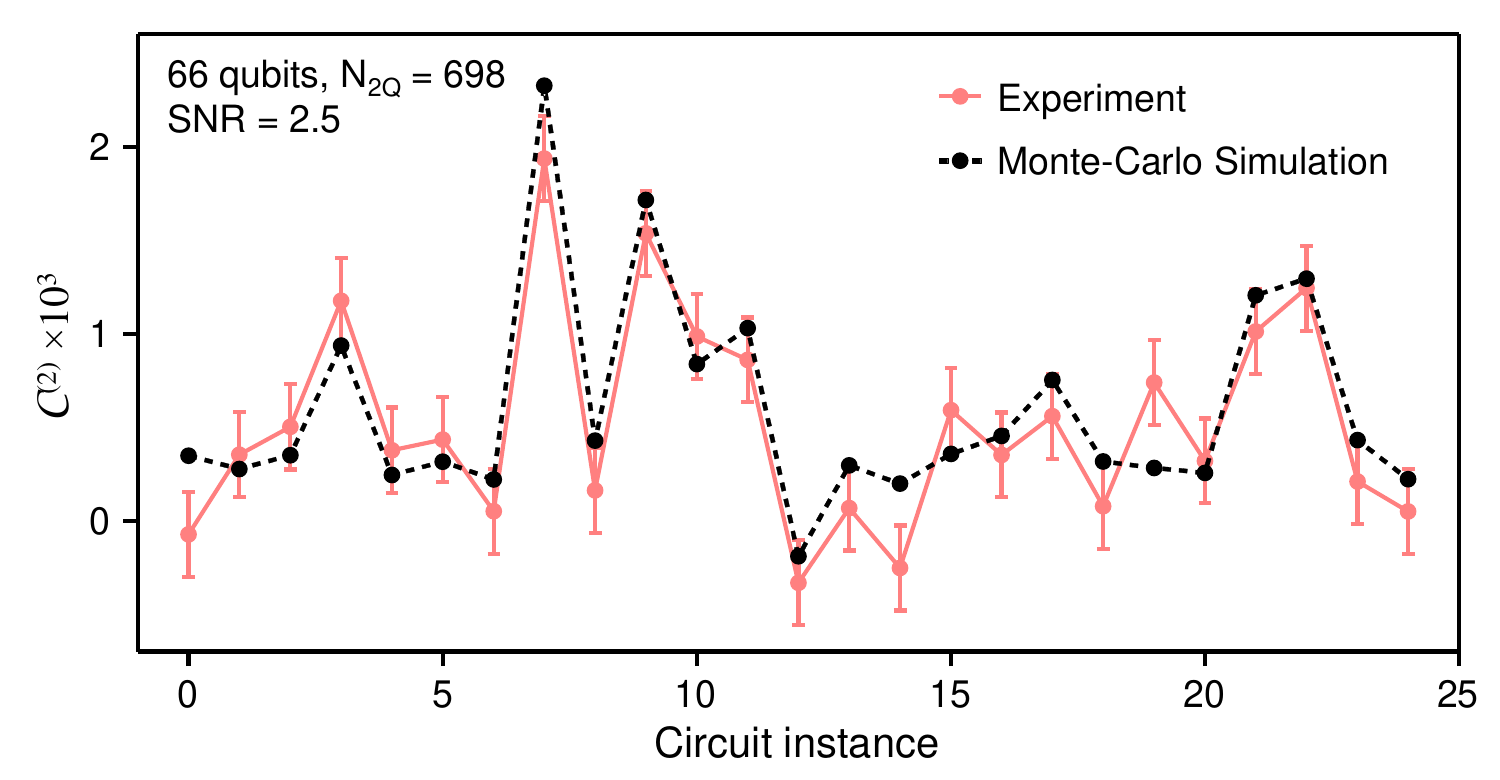}
	\caption{{\bf Infinite-temperature OTOC$\boldsymbol{^{(1)}}$.} Experimental measurements of $C^{(2)}$ on a set of 66-qubit circuits with $N_\text{2D} = 698$ iSWAP-like gates. Data shown are averaged over 900 different initial states (see text for details). Monte-Carlo simulation of the same circuits is shown for comparison, which agrees with experimental results with a SNR of 2.5.}
	\label{fig:s9}
\end{figure} 

The experimental protocol for measuring infinite-temperature OTOC consists of averaging over bitstring initial states of the form $\ket{00101101...}$ where each qubit is randomly initialized at the $\ket{0}$ or $\ket{1}$ state. When a sufficient number of initial states are used, the results are expected to converge to the infinite-temperature OTOC. To enable comparison with simulation, we fix the Rabi angles in the random single-qubit gate to be \mbox{$\theta / \pi = 0.5$} rather than randomly chosen from $\{ 0.25, 0.5, 0.75 \}$. This gate ensemble minimizes the chances of recombined trajectories in Pauli space (see Fig.~3a of the main text) based on small-scale numerical studies (not shown). As such, quantum interference effects are minimized and Monte-Carlo simulation is expected to capture the infinite-temperature OTOC relatively well.

To test this protocol, we perform initial-state-averaged measurements of $\mathcal{C}^{(2)}$ on a 66-qubit geometry with 698 iSWAP-like gates. This experiment is conducted on one of our earlier generations of quantum processors (see later discussion), which has higher gate errors compared to the current processor. Despite this, we find that the initial-state-averaged OTOCs agree well with Monte-Carlo simulation with a SNR of 2.5 (Fig.~\ref{fig:s9}), which is primarily limited by statistical shot noise. The scheme outlined can be straightforwardly extended to higher-order OTOCs, such as OTOC$^{(2)}$.

\subsection{Additional details for the OTOC$^{(2)}$ experiment}

In this Section we further discuss additional details for OTOC$^{(2)}$ regarding the error-mitigation protocol used for OTOC$^{(2)}$ circuits, as well as the procedure to estimate the SNR of the 65-qubit experiment in Fig.~4a of the main text.

\subsubsection{Error-mitigation for OTOC$^{(2)}$}\label{sec:otoc2_error_mitigation}

\begin{figure}[t!]
	\centering
	\includegraphics[width=1\columnwidth]{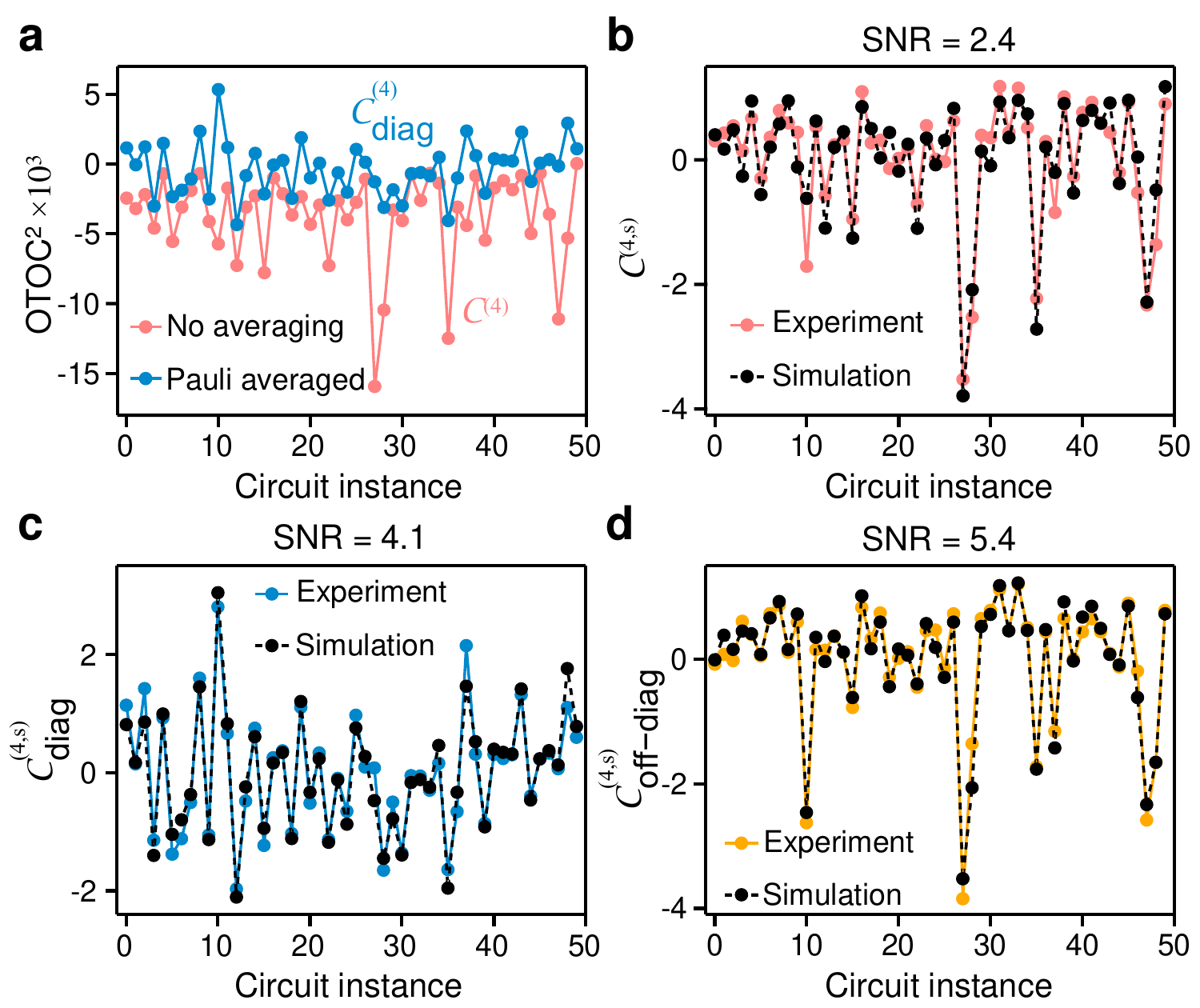}
	\caption{\textbf{Error mitigation of OTOC$\boldsymbol{^{(2)}}$.} {\bf a}, Experimentally measured values of OTOC$^{(2)}$ with ($\mathcal{C}^{(4)}_\text{diag}$) and without ($\mathcal{C}^{(4)}$) averaging over random Paulis inserted into the middle of ${U}$. Data shown are without any re-scaling. {\bf b}, $\mathcal{C}^{(4)}$ after re-scaling by its own standard deviation $\sigma (\mathcal{C}^{(4)})$, $\mathcal{C}^{(4,s)}$, and comparison with re-scaled exact simulation. {\bf c}, $\mathcal{C}^{(4)}_\text{diag}$ after re-scaling by its own standard deviation $\sigma (\mathcal{C}^{(4)}_\text{diag})$, $\mathcal{C}^{(4, s)}_\text{diag}$, and comparison with simulated values of $\mathcal{C}^{(4)}_\text{diag}$. Here $\sigma (\mathcal{C}^{(4)}_\text{diag})$ is a factor of $F_\text{noise} = 1.6$ larger in simulation than in experiment.  {\bf d}, Re-scaled off-diagonal contribution $\mathcal{C}^{(4)}_\text{off-diag} = \mathcal{C}^{(4)} - \mathcal{C}^{(4)}_\text{diag}$, $\mathcal{C}^{(4, s)}_\text{off-diag} = \left(\mathcal{C}^{(4)}_\text{off-diag} - \overline{\mathcal{C}^{(4)}}_{\rm off-diag}\right) / \sigma (\mathcal{C}^{(4)}_\text{off-diag})$, and comparison with simulation. Here $\sigma (\mathcal{C}^{(4)}_\text{off-diag})$ is a factor of $F_\text{noise} = 3.8$ larger in simulation than in experiment. In panels {\bf b} through {\bf d}, the mean of each data set has been subtracted, consistent with the treatment for OTOC, $\mathcal{C}^{(2)}$.}
	\label{fig:s5}
\end{figure} 

\begin{figure*}[t!]
	\centering
	\includegraphics[width=2\columnwidth]{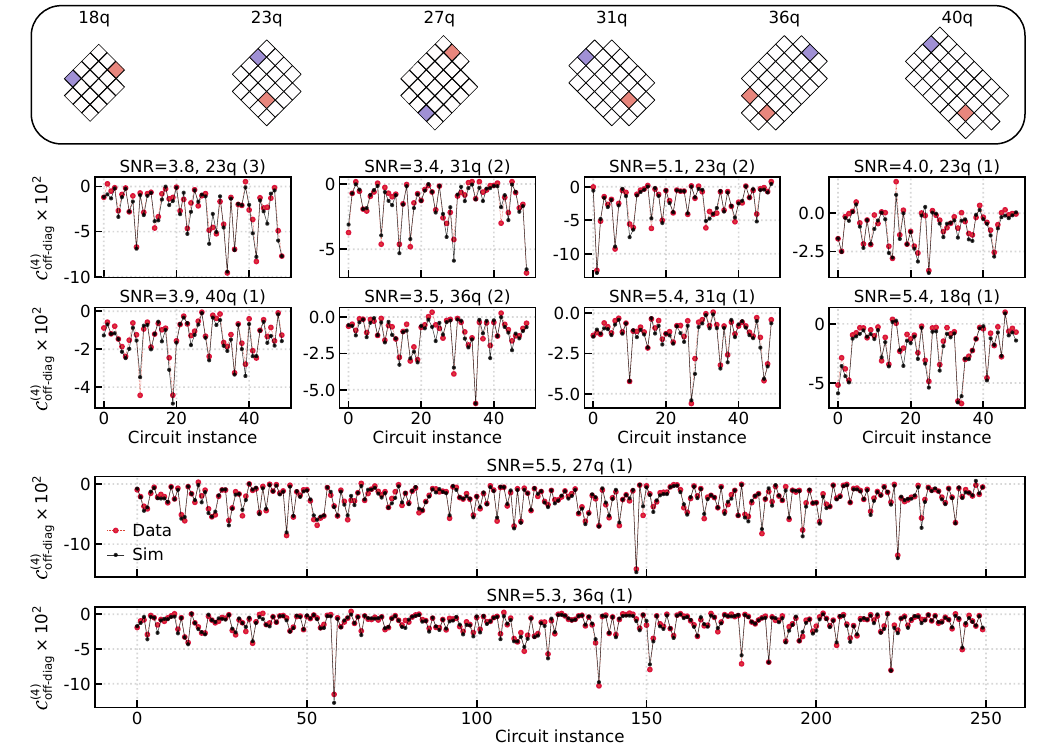}
	\caption{\textbf{Measurements of OTOC$\boldsymbol{^{(2)}}$ for different system sizes.} Experimentally measured vs exactly simulated values of $\mathcal{C}^{(4)}_\text{off-diag}$ for 6 different system sizes and geometries listed on the top panel. Some data are measured on older generation quantum processors with higher gate errors. The SNR, system and quantum processor used are listed as the title of each data panel. The quantum processor numbers represent the following: 1: Willow processor used in the main text. 2: Sycamore processor used in a previous work \cite{Morvan2024_rcs}. 3: Sycamore processor used in another previous work \cite{Acharya2023}. For the 27-qubit and 36-qubit geometries, we measured more circuits on Willow in order to observe the error scaling shown in Fig.~4b of the main text.}
	\label{fig:s7}
\end{figure*} 

\begin{figure}[t!]
	\centering
	\includegraphics[width=1\columnwidth]{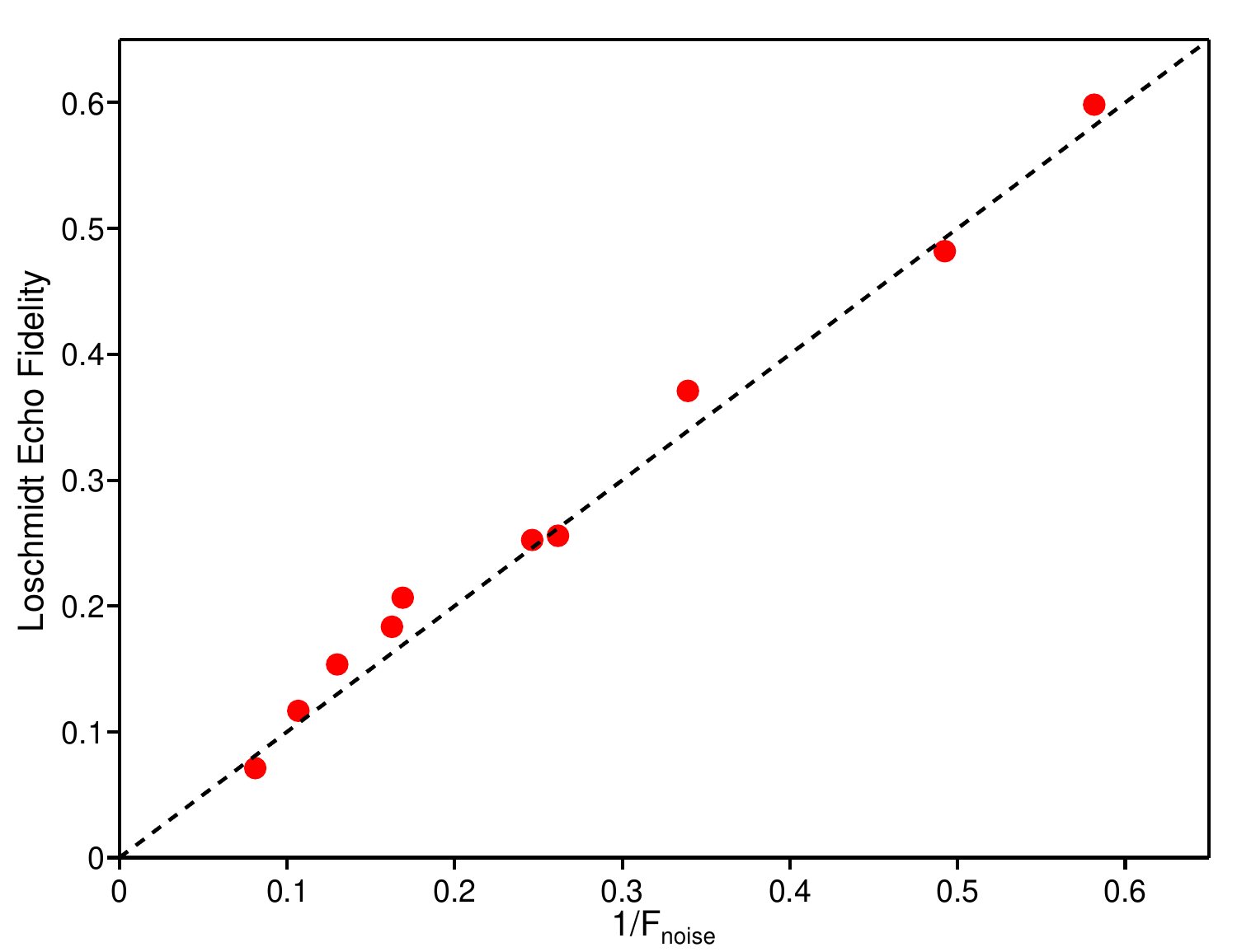}
	\caption{{\bf Estimating noise-damping factor from Loschmidt echo fidelities.} Loschmidt echo fidelity over the inverse noise-damping factor, $1/F_\text{noise}$, measured on the 10 sets of circuits shown in Fig.~\ref{fig:s7}. Dashed line shows the trend corresponding to echo fidelity =  $1/F_\text{noise}$.}
	\label{fig:s6}
\end{figure} 

Error-mitigation of OTOC$^{(2)}$ requires a different approach from OTOC$^{(1)}$. The zero-distance condition (which is satisfied by the dominant terms in 
\mbox{$\mathcal{C}^{(4)} = {\rm Tr}[\rho\, (M B(t))^4]$}) involves the trace of four Pauli strings, 
$\Tr[P_{\alpha_1}P_{\alpha_2}P_{\alpha_3}P_{\alpha_4}] = 1$. Unlike the case for
$\mathcal{C}^{(2)}$ where the zero-distance condition only involves two Pauli strings (and, therefore, being more restrictive), two terms satisfy the zero-distance condition for
$\mathcal{C}^{(4)}$:
either the Pauli strings are pairwise the same (three out of the four combinations), or $\alpha_1 \not= \alpha_2 \not= \alpha_3 \not= \alpha_4$. As a result, $\mathcal{C}^{(4)}$ cannot be easily reduced to classical probabilities, rendering MC techniques ineffective.

In Fig.~\ref{fig:s5}a, we show the raw OTOC$^{(2)}$ measurements for a set of 31-qubit circuits with ($\mathcal{C}^{(4)}_\text{diag}$) and without ($\mathcal{C}^{(4)}$) averaging over random Paulis inserted in the middle of ${U}$. In Fig.~\ref{fig:s5}b, we compare the re-scaled $\mathcal{C}^{(4, s)} = (\mathcal{C}^{(4)} - \overline{\mathcal{C}}^{(4)}) / \sigma(\mathcal{C}^{(4)})$ with numerical simulations, finding an SNR of 2.4. This modest SNR is understood to arise from the fact that noise affects the diagonal contribution $\mathcal{C}^{(4)}_\text{diag}$ and the off-diagonal contribution $\mathcal{C}^{(4)}_\text{off-diag}$ differently and rescaling by a single noise factor $F_\text{noise}$ leads to inaccuracies. To test this in experiment, we measure $\mathcal{C}^{(4)}_\text{diag}$ explicitly through averaging Paulis inserted in the middle of circuit evolution, finding an improved SNR of 4.1 for this component after the same global rescaling procedure (Fig.~\ref{fig:s5}c). Moreover, the off-diagonal contribution $\mathcal{C}^{(4)}_\text{off-diag}$, obtained after subtracting $\mathcal{C}^{(4)}_\text{diag}$ from $\mathcal{C}^{(4)}$, shows an SNR of 5.4 after global rescaling (Fig.~\ref{fig:s5}d). Comparing the values of the global re-scaling factors $F_\text{noise}$ obtained in both cases, we find that indeed they differ by more than a factor of 2, confirming the hypothesis above. Separating out the two contributions to $\mathcal{C}^{(4)}$ by measuring the signals after Pauli-averaging therefore serves as an effective error-mitigation strategy, beyond revealing large-loop interference as presented in the main text.

To test the robustness of the error-mitigation protocol, we conducted extensive measurements of $\mathcal{C}^{(4)}_\text{off-diag}$ across different system sizes. The results are shown in Fig.~\ref{fig:s7}. The six sets of data conducted on Processor 1 (Willow) are used for the SNR analysis in Fig.~4b and Fig.~4c of the main text (also see next section). We find that even on earlier generation (Sycamore) quantum processors where the gate errors are nearly twice the errors on Willow, we were able to routinely obtain SNRs above 3. In particular, the 36-qubit data set on Processor 2 have a noise damping factor ($F_\text{noise} = 0.08$) close to the 65-qubit Willow data set in Fig.~4a of the main text ($F_\text{noise} = 0.05$) and still yield a SNR of 3.5.

While the noise damping factor, $F_\text{noise}$, does not affect our SNR due to its definition, it remains desirable to estimate its magnitude for OTOC$^{(2)}$ in practical applications such as Hamiltonian learning. Unlike OTOC for which efficient classical simulation such as Monte-Carlo or TNMC exist and may be used to determine the noise-damping factor $F_\text{noise}$ through comparison with experimental data, no efficient classical simulation exists (to our knowledge) for $\mathcal{C}^{(4)}_\text{off-diag}$. Instead, we show that empirically $1 / F_\text{noise}$ is well-correlated with the Loschmidt echo fidelity. Here the echo fidelity is defined as the average of two different values of $\langle {M} \rangle$ measured at $q_\text{m}$ after setting either ${B}$ or ${M}$ to identity in Fig.~\ref{fig:s2}. The results are then further averaged over all instances. As shown in Fig.~\ref{fig:s6}, the echo fidelities are close to $1 / F_\text{noise}$ over the 10 small-scale experiments shown in Fig.~\ref{fig:s7}. The 65-qubit data set in Fig.~4a of the main text are therefore re-scaled by the empirically measured Loschmidt echo fidelity of 0.05. The observed correlation between $1 / F_\text{noise}$ and Loschmidt echo fidelity is further confirmed through noisy simulations of small-scale circuits shown in the next section.

\subsubsection{Validating error-mitigation with noisy numerics}\label{sec:digital_error_model}

In this section, we numerically check the error mitigation of the previous section using both simple depolarizing noise, and realistic noisy simulation including qubit decay and dephasing.

\begin{figure*}
    \centering
    \includegraphics[width=\textwidth]{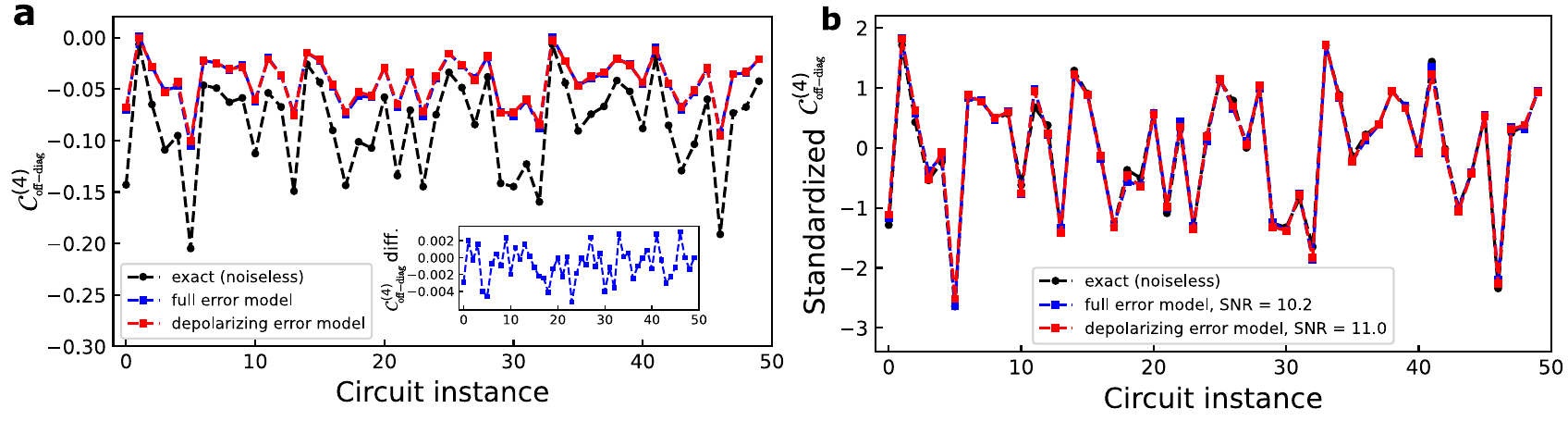}
    \caption{{\bf a,} Noisy simulations for $\mathcal{C}^{(4)}_{\rm off-diag}$ in 16-qubit circuits, based on the realistic error model and a local depolarizing error model. Noiseless simulation is also shown for comparison. (inset) Difference between the two noisy simulations. {\bf b,} Re-scaled signals used to compute the SNR for each error model.}
    \label{fig:realistic_error}
\end{figure*}

\begin{table}[b]
\centering
\begin{tabular}{c|c|c}
\hline
\hline
Error source & Mean & Std.~dev.~\\
\hline
$T_1$  & 80 $\mu$s  & 20 $\mu$s  \\
$T_{\phi}$  & 100 $\mu$s  & 25 $\mu$s   \\
$\epsilon_{1Q}$  & $2.6 \times 10^{-4}$  & $0.7 \times 10^{-4}$  \\
$\epsilon_{2Q}$  & $1.2 \times 10^{-3}$  & $0.4 \times 10^{-3}$  \\
$\epsilon_{\textrm{RO}}$  & $7.5 \times 10^{-3}$  & $-$  \\
$1/\Gamma_{\textrm{12}}$  & 1.5 $ms$  & $-$  \\
\hline
\hline
\end{tabular}
\caption{\textbf{Estimated error rates in our device.} We report the mean and standard deviation among qubits.}
\label{table:error-rates}
\end{table}

Realistic noise modeling in transmon devices typically requires accounting for several physical effects, including $T_1$ decay, dephasing, leakage, and qubit inhomogeneity~\cite{Acharya2023}.
Nevertheless, we expect the signals measured in this work to be largely insensitive to the exact details of the noise model.
This simplification arises from two main factors.
First, in our circuit implementation, we apply randomized compiling to the two-qubit gates and dynamical decoupling pulses during periods of qubit idling; both schemes effectively transform non-Pauli error channels (e.g.~coherent errors and amplitude damping) into Pauli channels.
Second, by design, our circuits involve highly scrambling dynamics. 
This implies that local errors propagate under subsequent gates into complicated, generic operators, regardless of the initial form of the error or its exact location. 

To validate the above expectations, we perform noisy simulations with a realistic error model developed previously for our device~\cite{Acharya2023}.
The model contains the following sources of error:
\begin{itemize}
    \item \emph{$T_1$ decay}---An amplitude damping channel (from state $\ket 1$ to $\ket 0$) is applied to each qubit after each circuit layer. The decay probability is $\tau/T_{1}(q)$, where $T_{1}(q)$ is the lifetime of qubit $q$ and $\tau = 30$ ns is the time duration of each layer.  
    \item \emph{Dephasing}---A dephasing channel is applied to each qubit after each circuit layer. The error probability is $\tau/T_{\phi}(q)$, where $T_{\phi}(q)$ is the (white noise) dephasing time. 
    \item \emph{Additional MW error}---A single-qubit depolarizing channel is applied after each single-qubit MW gate. The Pauli error probability is $\epsilon_{1Q}(q)$, for a gate acting on qubit $q$. This accounts for additional sources of gate error, including MW crosstalk or imperfect gate calibration, not captured by $T_{1}(q)$ and $T_{\phi}(q)$.
    \item \emph{Additional iSWAP error}---A two-qubit depolarizing channel is applied after each iSWAP gate, with Pauli error probability $\epsilon_{2Q}(q1,q2)$, for a gate acting on qubits $q_1$ and $q_2$.
    \item \emph{Readout error}---A bit-flip channel is applied to the measured qubit immediately prior to readout, with error probability $\epsilon_{\textrm{RO}}$. 
    \item \emph{Leakage}---A leakage transition from state $\ket{1}$ to $\ket{2}$ is applied to the each qubit after each circuit layer, with probability $\Gamma_{12}\tau$. If a leakage transition occurs, the subsequent gates acting on the qubit are removed (replaced by an identity operator), and the qubit is measured as the  $\ket{0}$ or $\ket{1}$ state with equal probability. \footnote{In practice, amplitude damping, dephasing, and leakage are represented using a combined decoherence channel. For more details, see \cite{Acharya2023}.} 
\end{itemize} 
From experimental measurements, we estimate the  average error rates for each noise source in our device, as well as the variation among qubits (Table \ref{table:error-rates}).
We model qubit inhomogeneity by sampling qubit-dependent error rates from a Gaussian distribution with the reported mean and standard deviation. 

Based on this error model, we compute the $\mathcal{C}^{(4)}_{\rm off-diag}$ for a set of 16-qubit circuits.
We also compute the signal for a simplified depolarizing model that omits qubit inhomogeneity and leakage. 
For the latter model, we replace each of the above error channels (except leakage) with a local depolarizing channel, whose Pauli error rate is equal to the average error rate of the original channel after Pauli twirling. 
As shown in Fig.~\ref{fig:realistic_error}(a), the subtracted signals for the two error models are in close agreement with each other, with a relative difference of $\sim 1\%$.
Moreover, after standardizing the signal across circuit instances, we find SNRs for the two noise models that differ by less than 10\% [Fig.~\ref{fig:realistic_error}(b)]. In either case, an SNR larger than 10 is observed, indicating the effectiveness of the error-mitigation protocol.

From these results, we conclude that an error model consisting of uniform, local depolarizing channels is sufficient to describe the noise in our device, and thus for our subsequent analysis we adopt this simplified error model. \\

\subsubsection{Residual errors in error-mitigated OTOC$\!^{(2)}$}\label{sec:otoc2_error_model}

In this section, we present a phenomenological model for the experimental errors that remain even after the error-mitigation protocols outlined in the previous sections. This model will be used to fit small-scale experimental data and estimate the bound on the SNR of the large 65-qubit experiment in the next section.\\

\noindent\textbf{Notation and Concepts ---} For fixed $t,q_\mathrm{m}$ and $q_\mathrm{b}$, the set of all possible circuit instances forms a probability space. From this set we draw $N_\mathrm{I}$ instances. We denote an instance average of an instance dependent variable $X(i)$ over a sample of $N_\mathrm{I}$ instances as
\begin{align}
    \overline{X} = \frac{1}{N_\mathrm{I}}\sum_{i=1}^{N_\mathrm{I}} X(i),
\end{align}
the variance as
\begin{align}
     \sigma[X]^2 = \frac{1}{N_\mathrm{I}}\sum_{i=1}^{N_\mathrm{I}} (X(i)-\overline{X})^2
\end{align}
and the correlation with another random variable $Y$ as
\begin{align}
    \rho[X,Y] = \frac{\overline{XY}-\overline{X}\,\overline{Y}}{\sigma[X]\sigma[Y]}.
\end{align}

In this particular section, we refer with $\mathcal{C}_\mathrm{nois}\equiv\mathcal{C}^{(4)}_\mathrm{off-diag,noise}$ to the difference between $\mathcal{C}^{(4)}$ and the Pauli-averaged $\mathcal{C}^{(4)}$, both measured on a quantum device affected by noise (Fig. 3 b). Similarly, $\mathcal{C}_\mathrm{ex}\equiv\mathcal{C}^{(4)}_\mathrm{off-diag,exact}$ denotes the corresponding result obtained from a noiseless simulation or an ideal quantum device.\\

\begin{figure}
    \centering
    \includegraphics{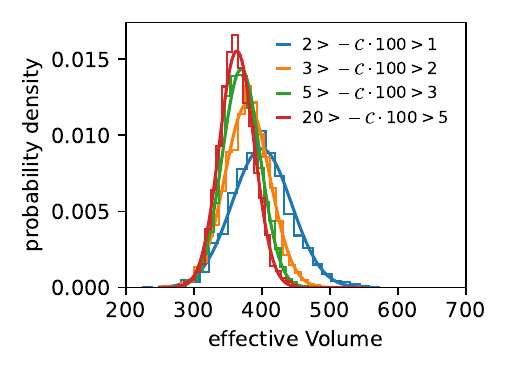}
    \caption{\textbf{Distribution of the effective volume for \mbox{$\boldsymbol{\mathcal{C}\equiv \mathcal{C}_\mathrm{ex}}$}.} Lines indicate fitted Gaussian distributions. Obtained via noisy simulation of $10000$ instances of a $21$ qubit-circuit with $4\times 68$ two-qubit gates. The simulations are performed by adding in total $491$ single qubit depolarizing channels after every two qubit gate with varying gate error rate $p$. The effective volume is obtained by fitting $\mathcal{C}_\mathrm{nois}=a(1-p)^{V_\mathrm{eff}}+c$. The average fitting error on the effective volume is about $20$ for the instances with $-0.02<\mathcal{C}_\mathrm{ex}<-0.01$ and about $10$ otherwise.}
    %\includegraphics{figures/SNR_study/21_2_3_mc_error_mitigation_estimate_SNR_via_gate_removal_different_decay2.pdf}
    %\caption{Noisy simulation results for the relative decay of the measurement for $5$ exemplary circuit instances for different error rates. The simulations are performed by adding single qubit depolarizing channels after every two qubit gate. The overall signal decay is measured via the ratio of the standard deviation. This shows that every instance decays with a different rate when subject to noise. Here, we used a circuit with $26$ qubits with $17$ cycles.}
    \label{fig:circuit_instance_depending_damping}
\end{figure}

\noindent\textbf{Physical Mechanism Responsible for the Multiplicative Error: Fidelity Fluctuations ---} In the presence of noise, a measurement obtained on the noisy quantum device $\mathcal{C}_\mathrm{nois}$ can be related to the exact result $\mathcal{C}_\mathrm{ex}$ by introducing an instance dependent damping factor $s$
\begin{align}
    \mathcal{C}_\mathrm{nois} = s \, \mathcal{C}_\mathrm{ex}.
\end{align}
This factor $s$ can be related to an instance dependent effective circuit volume $V_\mathrm{eff}$~\cite{kechedzhi2024effective} via the effective gate error $p$
\begin{align}
    s \sim (1-p)^{V_\mathrm{eff}}.
\end{align}
The distribution of $V_\mathrm{eff}$ is shown in Fig.~\ref{fig:circuit_instance_depending_damping}.

Next, we use $p\ll1$ and assume the effective volume to be Gaussian distributed $V_\mathrm{eff}\sim\mathcal{N}(\overline{V_\mathrm{eff}},\sigma_{V_\mathrm{eff}})$ over the circuit instances with $\sigma_{V_\mathrm{eff}}\ll 1/p$. Then $s$ follows a log-normal distribution with mean
\begin{align}
     \overline{s}=(1-p)^{\overline{V_\mathrm{eff}}}e^{\log(1-p)^2 \sigma_{V_\mathrm{eff}}^2/2}\approx e^{-p\overline{V_\mathrm{eff}}}
\end{align}
and variance
\begin{align}
     \sigma[s]^2&=\overline{s}^2\left( e^{\log(1-p)^2\sigma_{V_\mathrm{eff}}^2}-1\right)\nonumber \\
     &\approx  (\overline{s}\,p\,\sigma_{V_\mathrm{eff}})^2,\\
     &\approx  (\overline{s}\,\log(\overline{s})\,\sigma_{V_\mathrm{eff}}/\overline{V_\mathrm{eff}})^2. \label{eq:theory_fluctuations}
\end{align}
Hence, $s$ may be considered a random variable with the above mean and variance. This shows that replacing $s$ by its mean to estimate $\mathcal{C}_\mathrm{ex}$ by $\overline{s}\,\mathcal{C}_\mathrm{nois}$, we will introduce a multiplicative error of size
\begin{align}
    \sigma[\mathcal{C}_\mathrm{ex}-\overline{s}\,\mathcal{C}_\mathrm{nois}]^2 = \sigma[s]^2\overline{C_\mathrm{ex}^2}. \label{eq:derivation_mult_noise}
\end{align}
This perspective serves as the foundation for the error model introduced in the next section.\\

\noindent\textbf{Noise Model for Measurements ---}
The set of all possible circuit instances can be used to form a probability space. Therefore, the measured result $\mathcal{C}_{\mathrm{nois}}$ and the exact result $\mathcal{C}_{\mathrm{ex}}$ are random variables.

As discussed in the preceding section and empirically validated in Fig.~4 (b), the measurement error is observed to increase with the magnitude of the exact result, $|\mathcal{C}_{\mathrm{ex}}|$. This motivates to model the relation between $\mathcal{C}_{\mathrm{nois}}$ and $\mathcal{C}_{\mathrm{ex}}$ as
\begin{align}
    \mathcal{C}_{\mathrm{nois}} = s \mathcal{C}_{\mathrm{ex}}+\epsilon, \label{eq:model}
\end{align}
where we introduced two more random variables: the damping factor $s$ and the additive error $\epsilon$. Within the model, $s,\epsilon$ and $\mathcal{C}_{\mathrm{ex}}$ are independent and follow
\begin{align}
    s\sim\mathcal{N}(g,g\sigma_\mathrm{m}),\quad \epsilon \sim\mathcal{N}(c,\sigma_\mathrm{a}),\label{eq:model_part2}
\end{align}
where $g$, $\sigma_\mathrm{m}$, $c$, and $\sigma_\mathrm{a}$ are model parameters to be determined. 

By comparing to Eq.~\ref{eq:theory_fluctuations}, we find them to be related to the effective volume as
\begin{align}
 g=e^{-p \overline{V_\mathrm{eff}}}\quad \mathrm{and}\quad \sigma_\mathrm{m}=p \sigma_{V_\mathrm{eff}}.
\end{align}\\

\noindent\textbf{Predictions of the Noise Model --- }
For a dataset, which is fully described by Eq.~\eqref{eq:model} and Eq.~\eqref{eq:model_part2}, we can derive relationships between the model parameters and the statistical properties of $\mathcal{C}_\mathrm{nois},\mathcal{C}_\mathrm{ex}$. These relationships serve to validate the model's applicability and enable predictions based on it.

First, taking the sample average of Eq.~\eqref{eq:model} yields an expression for the mean additive error
\begin{align}
    c = \overline{\mathcal{C}_\mathrm{nois}}-g\overline{\mathcal{C}_\mathrm{ex}}. \label{eq:model_p_c}
\end{align}

Next, one can relate the mean damping factor $g$ to the correlation:
\begin{align}
\rho[\mathcal{C}_\mathrm{nois},\mathcal{C}_\mathrm{ex}] =  g\frac{\sigma[\mathcal{C}_\mathrm{ex}]}{\sigma[\mathcal{C}_\mathrm{nois}]}.\label{eq:model_g_pred}
\end{align}
From Eq.~\eqref{eq:model} we find
\begin{align}
    g^2 \sigma[\mathcal{C}_\mathrm{ex}]^2 =  \sigma[\mathcal{C}_\mathrm{nois}]^2-\sigma_a^2 -g^2\sigma_\mathrm{m}^2 \overline{(\mathcal{C}_\mathrm{ex})^2},
\end{align}
which we can substitute into Eq.~\eqref{eq:model_g_pred} to obtain an expression for the squared correlation
\begin{align}
    \rho[\mathcal{C}_\mathrm{nois},\mathcal{C}_\mathrm{ex}]^2 = 1 - \frac{\sigma_\mathrm{a}^2 + g^2\sigma_\mathrm{m}^2 \overline{(\mathcal{C}_\mathrm{ex})^2}}{\sigma[\mathcal{C}_\mathrm{nois}]^2}. \label{eq:model_pred_rho_without_g}
\end{align}
This expression can be analyzed to understand the different contributions of errors to the correlation and, therefore, the SNR $=1/\sqrt{2(1-\rho)}$. \\

\noindent\textbf{Maximum Likelihood Estimation and Model Validation ---}
In this section, we will probe the applicability of the model by assessing its self-consistency. For that, we use maximum likelihood estimation (MLE) to determine the model parameters $g,c,\sigma_\mathrm{m}$ and $\sigma_\mathrm{a}$. We then validate the model by comparing the MLE-derived parameters with the predictions of Eq.~\eqref{eq:model_p_c}, Eq.~\eqref{eq:model_g_pred} and Eq.~\eqref{eq:model_pred_rho_without_g}.

To obtain an MLE estimate of the noise model parameters we need the conditional probability $p(\mathcal{C}^{(4)}_\mathrm{nois}|\mathcal{C}_\mathrm{ex},\epsilon,s)$. It can be inferred from Eq.~\eqref{eq:model}: 
\begin{align}
        &p(\mathcal{C}^{(4)}_\mathrm{nois}|\mathcal{C}_\mathrm{ex},\epsilon,s) = \delta(s\mathcal{C}_\mathrm{ex}+\epsilon-\mathcal{C}_\mathrm{nois})
\end{align}
Integrating out $\epsilon$ and $s$ using Eq.~\eqref{eq:model_part2}, we find the likelihood function for a dataset of $N$ measurements $\{\mathcal{C}_{\mathrm{nois}}(i)\}_{i=1}^N$ and $\{\mathcal{C}_{\mathrm{ex}}(i)\}_{i=1}^N$
\begin{align}
        &p(\{\mathcal{C}_\mathrm{nois}(i)\}_i|\{\mathcal{C}_\mathrm{ex}(i)\}_i) \nonumber \\
        \propto&\exp\sum_i \left[  \frac{- \log (\sigma_\mathrm{a}^2 + [\mathcal{C}_\mathrm{ex}(i)]^2 g^2\sigma_\mathrm{m}^2)}{2}\right.\nonumber \\
        &\left.-\frac{(\mathcal{C}_\mathrm{nois}(i) - g \mathcal{C}_\mathrm{ex}(i) - c)^2}{2(\sigma_\mathrm{a}^2 + [\mathcal{C}_\mathrm{ex}(i)]^2 g^2 \sigma_\mathrm{m}^2)}\right]. \label{eq:likelihood}
\end{align}
Then, the model parameters $g,c,\sigma_\mathrm{m},\sigma_\mathrm{a}$ are set the values maximizing the likelihood function Eq.~\ref{eq:likelihood}. The covariance matrix of the parameters is estimated using the inverse Hessian of the log-likelihood.\\

\begin{figure}
    \centering
    \includegraphics{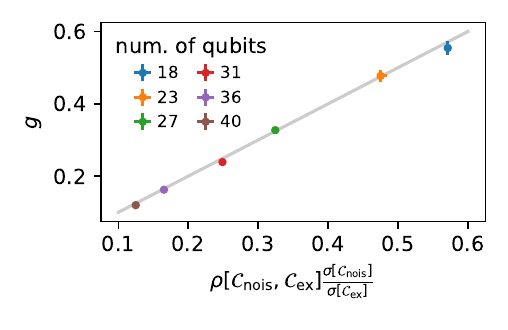}
    \caption{\textbf{Comparison of $\boldsymbol{g}$ obtained via MLE to the model prediction $\boldsymbol{\rho[\mathcal{C}^{(4)}_\mathrm{off-diag},\mathcal{C}_\mathrm{ex}]}$, Eq.~\eqref{eq:model_g_pred}, using the experimental data for 50 instances (250 for 27 and 36 qubits)}. Errors on $g$ are estimated from the likelihood fit, errors on the sample estimate are based on the shot noise of the experimental data.}
    \label{fig:check_model_consistency_g}
\end{figure}

\begin{figure}
    \centering
    \includegraphics{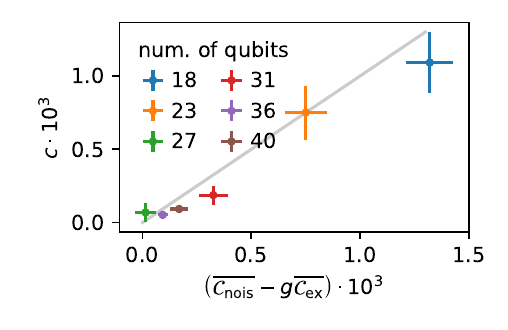}
    \caption{\textbf{Results for $\boldsymbol{c}$ from the MLE compared to the sample estimate $\boldsymbol{\overline{\mathcal{C}_\mathrm{nois}}-g\,\overline{\mathcal{C}_\mathrm{ex}}}$, Eq.~\eqref{eq:model_p_c} with $\boldsymbol{g}$ from Eq.~\eqref{eq:model_g_pred}, using the measured data for 50 instances (250 for 27 and 36 qubits).}  Errors on $c$ are estimated from the likelihood fit, errors on the sample estimate are based on the shot noise of the experimental data.}
    \label{fig:check_model_consistency_c}
\end{figure}

\begin{figure}
    \centering
    \includegraphics{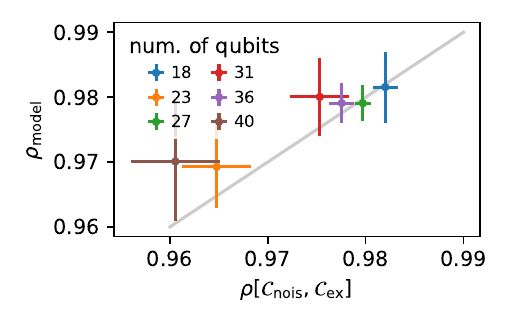}
    \caption{\textbf{Results for the sample result $\boldsymbol{\rho[\mathcal{C}_\mathrm{nois},\mathcal{C}_\mathrm{ex}]}$ and $\boldsymbol{\rho_\mathrm{model}=\sqrt{1 - \frac{\sigma_\mathrm{a}^2 + g^2\sigma_\mathrm{m}^2 \overline{(\mathcal{C}_\mathrm{ex})^2}}{\sigma[\mathcal{C}_\mathrm{nois}]^2}}}$ with the MLE-derived $\boldsymbol{g,\,\sigma_\mathrm{m},\,\sigma_\mathrm{a}}$, see Eq.~\eqref{eq:model_pred_rho_without_g}.} We use the measured data for $50$ instances ($250$ for $27$ and $36$ qubits).  Errors on the model result for $\rho$ are estimated from the likelihood fit and the shot noise of the experimental data, errors on the sample estimate $\rho[\mathcal{C}_\mathrm{nois},\mathcal{C}_\mathrm{ex}]$ are based on the shot noise of the experimental data.}
    \label{fig:check_model_consistency_rho}
\end{figure}

\noindent\textbf{Model Validation ---}
We assess the model's consistency by comparing the MLE-derived parameters with the predictions of Eq.~\eqref{eq:model_g_pred} (for $g$), Eq.~\eqref{eq:model_p_c} (for $c$) and Eq.~\eqref{eq:model_pred_rho_without_g} (for $\rho$), as shown in Fig.~\ref{fig:check_model_consistency_g}, Fig.~\ref{fig:check_model_consistency_c} and Fig.~\ref{fig:check_model_consistency_rho} respectively.
We observe a good agreement between the MLE results and the model predictions: The deviation between the MLE result for $g$ and the model prediction Eq.~\eqref{eq:model_g_pred} is within the error bars, and the model prediction Eq.~\eqref{eq:model_p_c} agrees well with the MLE result for $c$.

 \subsubsection{SNR estimate for large-scale OTOC$^{(2)}$}\label{sec:snr_otoc2}
 
  \begin{figure}
    \centering
    \includegraphics[width=1\columnwidth]{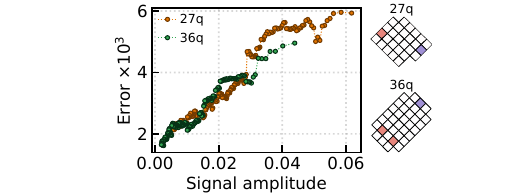}
    \caption{{\bf Scaling of experimental error vs signal size.} Experimental error on $\mathcal{C}^{(4)}_\text{off-diag}$ as a function of the signal amplitude of $\mathcal{C}^{(4)}_\text{off-diag}$ for two different system sizes (27 and 36 qubits, with geometries shown on the right; see Fig.~\ref{fig:s7} for raw data). Here a total of 250 circuit instances are measured for each system size and the results are sorted according to $|\mathcal{C}^{(4)}_\text{off-diag}|$. A rolling window analysis with a size of 50 circuits is then done to obtain the experimental error, i.e. the root-mean-square (RMS) difference between exactly simulated and experimental $\mathcal{C}^{(4)}_\text{off-diag}$, against signal amplitude, i.e. the RMS value of exactly simulated $\mathcal{C}^{(4)}_\text{off-diag}$. The number of two-qubit gates is $N_\text{2Q} = 284$ for 27q and $N_\text{2Q} = 496$ for 36q.}
    \label{fig:error_scaling_exp}
\end{figure}

 Having built an error model for OTOC$^2$, we now attempt to give an estimate on the SNR expected from the 65-qubit experiment in Fig.~4a of the main text. We first note that since exact simulation results are not available for such circuits, the distributions of error terms $s(i)$ and $\epsilon(i)$ are not possible to characterize directly. To proceed, we make the observation that both the SNR (Fig.~\ref{fig:s7}) and the scaling of errors vs signal amplitude (Fig.~\ref{fig:error_scaling_exp}) are quite similar between the 27-qubit and 36-qubit geometry, despite a nearly factor of 2 difference in the number of two-qubit gates and noise damping factor $F_\text{noise}$. This may be related to the fact that due to similar signal sizes (i.e. $\sigma(\mathcal{C}^{(4)}_\text{off-diag})$) used in all of our circuit geometries, quantum volume fluctuation and therefore the distribution of errors are approximately constant over different geometries. We therefore use the multiplicative errors $\sigma_\text{m}$ and additive errors $\sigma_\text{a}$ obtained from fitting the 27-qubit (Willow) experimental data to infer the SNRs of all other system sizes.

Once $\sigma_\text{m}$ and $\sigma_\text{a}$ are fixed, the SNR of a given experimental data set may therefore be inferred by a simple simulation consisting of the following steps:
\begin{enumerate}
    \item Scale each experimentally measured circuit-dependent value of $\mathcal{C}^{(4)}_\text{off-diag} (i)$ by a random multiplicative factor $s(i)$ drawn from a normal distribution with mean of 1 and standard deviation of $\sigma_\text{m}$.
    \item Add a random value drawn from a normal distribution with mean of 0 and standard deviation of $\sigma_\text{a}$ to the circuit-dependent values from step 1. 
    \item Add a random value draw from a normal distribution with mean of 0 and standard deviation of $\sigma_\text{s}$ to the circuit-dependent values from step 2. Here $\sigma_\text{s}$ is the statistical shot noise estimated based on the number of measurements conducted for each circuit.
    \item Obtain a single SNR value after comparing the circuit-dependent values from step 3 to the original set of $\mathcal{C}^{(4)}_\text{off-diag} (i)$.
    \item Repeat steps 1 through 4 a total of 1 million times so as to produce a distribution of SNRs. Bounds on the SNR are then obtained from the percentiles of this distribution.
\end{enumerate}

Using this procedure and error parameters ($\sigma_\text{m}$ and $\sigma_\text{a}$) obtained from the 27-qubit data set, we find that the experimental SNRs of the other 5 system sizes all fall within the 95\% confidence interval of the SNR, as shown in Fig.~4c of the main text~\footnote{The parameters $\sigma_\text{m}$ and $\sigma_\text{a}$ are obtained in the error model of the previous section, where errors are added to the exact values to estimate noisy values. The same parameters can be used to estimate exact values from noisy values, as long as the correlation remains high.}. This agreement therefore motivates the usage of the same error parameters $\sigma_\text{m}$ and $\sigma_\text{a}$ to project the SNR of the 65-qubit data, which is found to have a SNR with confidence interval $[2.3, 3.5]$. Owing to potential unknown systematic errors not captured by this empirical error model, we give an estimate of 2 to 3 for the SNR.

 \subsubsection{Interference effects vs system size for OTOC$^{(2)}$}
 
 \begin{figure}
    \centering
    \includegraphics[width=\columnwidth]{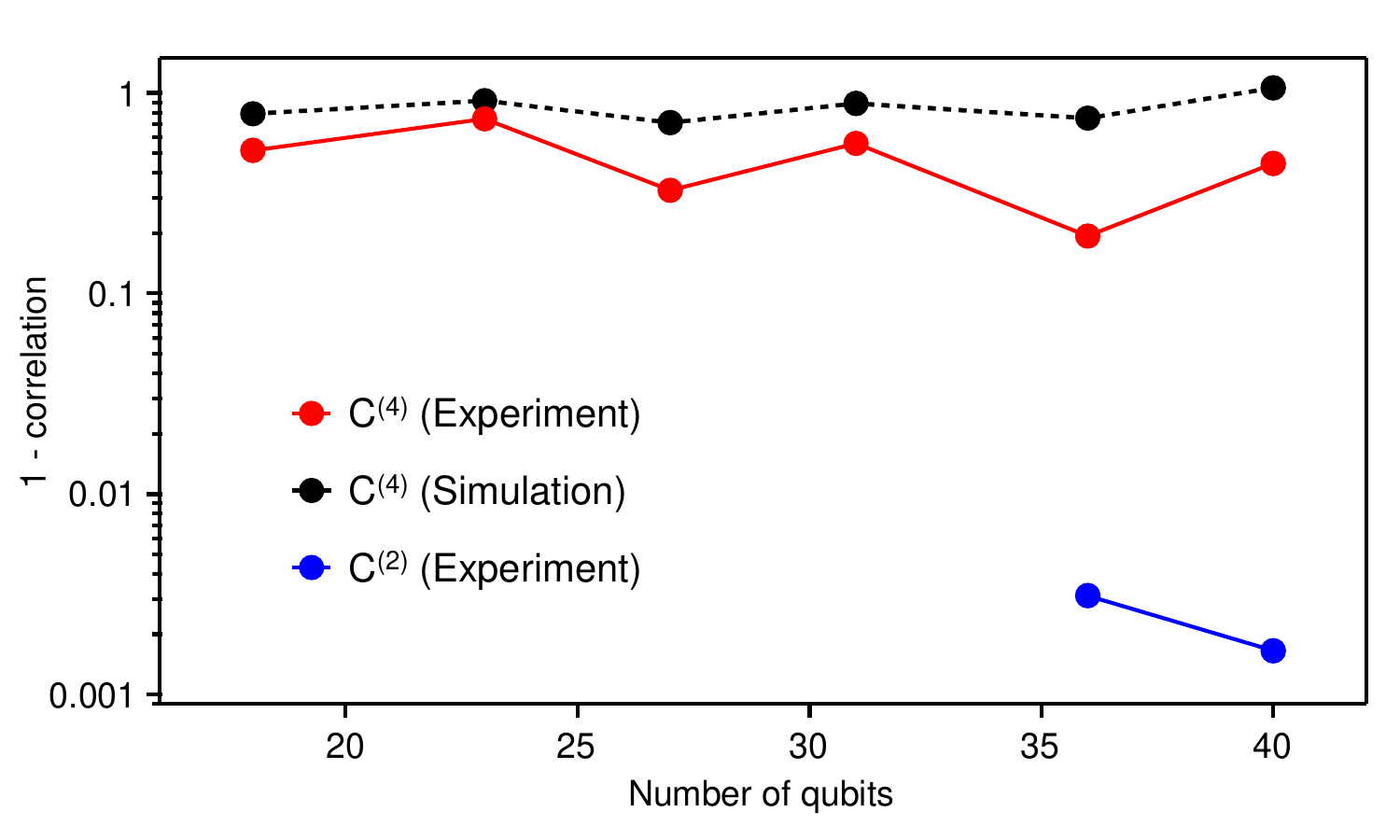}
    \caption{\textbf{Signal change as a result of inserting Paulis into the middle of $\boldsymbol{U}$ for different system sizes.} Here correlation refers to Pearson correlation between Pauli-averaged and non-averaged signals. Results are shown for both $\mathcal{C}^{(4)}$ and $\mathcal{C}^{(2)}$. For $\mathcal{C}^{(4)}$, exact simulation results are also shown. The $\mathcal{C}^{(4)}$ circuits used here are the same as the circuits used in Fig.~4c of the main text}
    \label{fig:s10}
\end{figure}

In Fig.~3c of the main text, we have contrasted the interference effects in $\mathcal{C}^{(4)}$ against $\mathcal{C}^{(2)}$ by fixing the system size at 40 and varying the location of Pauli insertion. In Fig.~\ref{fig:s10}, we show similar experimental data as Fig.~4c of the main text but vary system size instead. For $\mathcal{C}^{(2)}$, it is seen that even for 31 qubits, Paulis injected in the middle of $U$ have little impact on the observed signals, with correlation nearing 0.997. On the other hand, $\mathcal{C}^{(4)}$ manifests a much stronger signal change which barely decays as system size increases. In fact, the small decay in the case of $\mathcal{C}^{(4)}$ is not present in the noiseless simulation of the same circuits, indicating it originates from the effect of noise which reduces interference effects more at larger system sizes. This result complements the finding in Fig.~3c of the main text and indicates that large-loop interference remains a significant contribution to $\mathcal{C}^{(4)}$ even at large system sizes.

\section{Classical algorithms}\label{sec:classical}

%{\color{green}Tom: This needs to be updated to better represent the structure of the classical algorithms section. Also some of the emphasis on the quantity of work done should IMO be removed, I think this is already clear from the paper, and we should show by doing rather than saying.}

%We dedicated significant effort, approximately ten person-years, to estimating the time required for both known and newly designed classical algorithms to calculate the out-of-time-order correlators measured on our quantum processors. Over 1 trillion repetitions of quantum circuits were run to improve the accuracy of the experiments and benchmark them against these algorithms. 
The classical algorithms that were conceived during the course of this work are described in this section. We categorize these algorithms into two types: exact (e.\,g., tensor network simulation) and heuristic algorithms that attempt to approximate the true result with an uncontrolled error  $\epsilon$. For each algorithm, we describe its implementation and estimate its best-in-class performance on various trial systems. We also provide details of the analysis of the circuit cutting and gate-truncated tensor network contraction algorithms used for the time estimates in the main text.

We first comment on the choice of the algorithms we analyze. There exist numerous classical algorithms that leverage the physical properties of out-of-equilibrium quantum dynamics to reduce computational costs. Several of these methods have successfully replicated measurements of local observables~\cite{GarnetChan2023,kechedzhi2024effective,tindall2024efficient}. 
These algorithms are effective because reproducing local observables with finite precision requires simulation of a limited effective circuit volume~\cite{kechedzhi2024effective}. 
However, reproducing higher-order out-of-time-ordered correlators (OTOC$^{(k)}$ with $k \geq 1$), where the butterfly and measurement operators are located at opposite ends of the system, requires simulating a large circuit volume that scales with the system size. This scaling suggests that algorithms optimized for local observables are likely inefficient for such tasks. Nonetheless, specialized algorithms can still reduce classical simulation costs for $k \geq 1$. In this section, we introduce classical sampling algorithms that exploit phase randomization within out-of-equilibrium dynamics to achieve these reductions.

\subsection{Introduction to tensor network contraction}\label{sec:tn}

The gold standard for classical simulation of random quantum circuits is tensor network contraction (TNC).
In a few words, a tensor network is a collection of multi-index arrays that are multiplied together
\cite{markov2008simulating,biamonte2017tensor}. 
For instance, a matrix is a tensor of two indices $M_{ij}$ and the multiplication of two matrices
can be written as $\sum_{j} M_{ij} M'_{jk} = M''_{ik}$. When the equation involves multiple tensors and summations,
the order (also called ``path'') in which tensors are multiplied is crucial \cite{markov2008simulating,boixo2017simulation}:
indeed, different paths
can have costs in floating point operations (FLOPs) that could differ by orders of magnitude~\cite{gray2021hyper}.
Finding optimal paths is known to be hard,%
\footnote{To be precise, finding optimal path is an NP-Hard problem \cite{arnborg1987complexity}.}
and exact approaches are limited to small TNCs.
Significant effort has gone into developing this over recent years~\cite{chen2018classical,gray2021hyper,huang2020classical,pan2022solving,kalachev2021multi,kalachev2021classical,Morvan2024_rcs}.
Advancements have not only improved contraction path optimizations, but also became aware of memory constraints~\cite{chen2018classical}, the advantage of caching intermediate results in memory~\cite{kalachev2021multi,kalachev2021classical,Morvan2024_rcs}, as well as the possibility to compute several circuit output amplitudes with a cost similar to that of a single one~\cite{pan2022solving,kalachev2021multi,kalachev2021classical,Morvan2024_rcs}. 
These improved techniques allow for the increasingly more precise estimation of the cost for executing tensor network based random circuit sampling simulations on the world's largest supercomputer, Frontier.\\

Given a circuit $C$, the exact expectation value can be computed by converting
the expectation value of $\langle Z \rangle = \bra{0}C  Z C^\dagger\ket{0}$ to a TNC. 
However, computing the exact expectation
value can be demanding, as the expected FLOPS to contract a TNC tends to grow exponentially with the depth of the TNC. 
To limit the contraction cost, we decomposed the expectation value as a sum of orthogonal projections on all qubits
at the measurement level excluding the measurement qubit and a few gates around it (up to $20$ open qubits in total):
\begin{align}
\label{eq:TN_projections}
    \langle Z \rangle &= \bra{0}C  Z \Biggl[\sum_x\Pi_x\Biggr] C^\dagger \ket{0}\nonumber\\
    &= \sum_x \Biggl[\bra{0}C\Pi_x\Biggr] Z \Biggl[\Pi_x C^\dagger \ket{0}\Biggr]\nonumber\\
    &= \sum_x \bra{\psi_x}  Z \ket{\psi_x}
\end{align}
with $\ket{\psi_x}$ a partial quantum state for a given projection $x$. The exact expectation value is obtained
by summing over all the possible projections $x$. However, an approximate expectation value can be obtained
by using a small number of projections~\cite{Mi_OTOC_2021}. Indeed, observing that 
$p_x = |\!\left\langle\psi_x | \psi_x\right\rangle\!|^2$ can be interpreted as probability, with 
$\sum_x p_x = 1$, it is possible to use rejection sampling to sample 
random projections $x$ and compute approximate
expectation values as:
\begin{align}
    \left\langle Z \right\rangle &= \sum_x \bra{\psi_x}  Z \ket{\psi_x}\nonumber\\
                                 &= \sum_x p_x \frac{\bra{\psi_x}  Z \ket{\psi_x}}{p_x}\nonumber\\
                                 &= \mathbb{E}\left[\frac{\bra{\psi_x}  Z \ket{\psi_x}}{p_x}\right].
\end{align}
That is, the expectation value $\langle Z\rangle$  is reduced to computing the expectation value over the
classical distribution induced by $p_x$ of $\psi_x = \frac{\bra{\psi_x}  Z \ket{\psi_x}}{p_x}$.
In general, the exact sampling from $p_x$ is impossible because it requires computing
all partial $\ket{\psi_x}$, hence the exact expectation value. Inspired by the \emph{frugal} sampling in \cite{markov2018quantum}, we perform rejection sampling
by exactly computing $p_x$ for a small number of randomly chosen projections $\{x_1,\,x_2,\,\cdots,\,x_k\}$ from the uniform distribution,
and then accepting/rejecting them with a probability $r_x = \frac{p_x}{\max_{\{x_1,\,x_2,\,\cdots,\,x_k\}}{p_x}}$.
While such sampling algorithm does not guarantee that the accepted projections will have the same
probability as been sampled from the exact probability distribution (if not in the limit of
$k \propto 2^n$, with $n$ being the number of qubits), we observe that it works well in practice.\\

\noindent\textbf{The tensor network contraction optimizer (\texttt{TNCO}) ---}
We developed a high-performance novel package,
\texttt{TNCO}, for optimizing the contraction path of a tensor network and estimating its runtime (including the constraints from finite random access memory on a supercomputer).
A tensor network contraction path (TNC) in \texttt{TNCO} is represented as a tree, with the root of the tree being the output tensor. If any of the intermediate tensors in the TNC cannot fit in memory, indices are \emph{sliced} until
the memory of all tensors (defined as $w = \sum_i \log_2 d_i$, with $d_i$ being the $i$-th tensor index) 
is not larger than a given target width. Unless specified otherwise, for each Frontier's node we are assuming the maximum width of $33$.%
\footnote{Each node in Frontier has $8$ GPUs, for a total amount of $512$ GB of
high-bandwidth memory.}
TNCs are updated by proposing local updates on the corresponding contraction tree,
which are accepted or rejected using Metropolis-Hastings. Representing TNCs as trees allows us to limit the optimization landscape to only topologically different configurations, avoiding the extra cost to explore symmetrical contractions. 

As observed in \cite{Morvan2024_rcs}, optimal contractions without memory constraints may not be well suited for slicing. To slice a TNC, \texttt{TNCO} simultaneously optimizes the contraction path and the location of slices. Furthermore, \texttt{TNCO} uses hyper-indices to connect diagonal gates, and provides paths caching the intermediate tensors: if an intermediate tensor fits in memory, the subtree of such tensor is pruned from the TNC
and the tensor is cached to be reused across multiple slices. To keep the cost of running \texttt{TNCO} limited, we do not account for them in the total memory count.
%we assume that the cached tensors have a negligible impact and, therefore, they are not accounted in the total
%memory count.%
\footnote{Since the total amount of required memory without caching is at most 4 times
the size of the largest intermediate tensor, limiting the width is effective to ensure that the full contraction
can be performed in memory. However, this is no longer true when caching is involved and the full contraction
might not fit in memory at fixed maximum tensor width.}

\subsection{Classical simulation of OTOC$^{(1)}$}\label{sec:classical_otoc}

In this section, we discuss classical methods used to simulate $\text{OTOC}^{(1)}$. We begin with subsections~\ref{sec:mc_intro}, \ref{sec:cachedMC} and \ref{sec:tnmc}, wherein we define a family of powerful {\it Monte Carlo} methods, which exploit having exactly two copies of the time-evolved butterfly operator.
%
%We continue then with the section \ref{sec:tn}, which covers the exact tensor network contraction, a prime method we benchmark the computational complexity against, which also applies to the case of $\text{OTOC}^{(2)}$.

\subsubsection{Introduction to Monte Carlo algorithms}\label{sec:mc_intro}
Let us split the time-evolution of the butterfly operator into $T$ pieces (either individual gates or groups of those), and write
\begin{gather}
    {U} = \prod_{t = 0}^{T - 1} {g}_t.
\end{gather}
Having two copies of a time-evolved butterfly operator opens a prospect to a powerful  Monte Carlo family of methods based on the {\it diagonal approximation}. Consider a time-evolved butterfly operator ${B}(T) = {U}^{\dagger} {B} {U}$ expanded in the Pauli basis ${B}(T) = \sum_P b_P {P}$, where $P$ are all possible $4^{n}$ Pauli strings, with $n$ being the number of qubits.\\

For the sake of discussion, let us temporarily consider the computation of a trace-OTOC. Plugging this Pauli decomposition into the expression for trace-OTOC, we obtain
\begin{align}
\label{eq:trace-otoc}
    \mathcal{C}^{(2)} &= \sum\limits_{P, P^\prime} b_P b_{P^\prime} \text{Tr} \left[{P} {M} {P}^\prime {M} \right] \nonumber\\&= \sum\limits_P |b_P|^2 (-1)^{\{P, M\}},
\end{align}
where $b_P$ are squared normalized, and $(-1)^{\{P, M\}} = \pm 1$ if $P$ and $M$ either commute or anti-commute.

The appearance of the positive measure $|b_P|^2$ allows one to sample the ``observable'' $\mathcal{O}({P}) = (-1)^{\{P, M\}}$ using Monte Carlo
\begin{gather}
    \mathcal{C}^{(2)} \approx \frac{1}{M} \sum_{P \sim |b_P|^2} \mathcal{O}({P}),
\end{gather}
where the $M$ samples are drawn from the $|b_P|^2$ distribution. Here, $P\sim|b_P|^2$ means sampling a Pauli string $P$ with the probability induced by $|b_P|^2$. Importantly, direct sampling from $|b_P|^2$ is hard. In the next sections, we will introduce several Monte Carlo methods, which only differ in the way they approximate the $|b_P|^2$ distribution in order to be able to sample from it. To introduce a common way of approximation, we will invoke the {\it path integral} formalism. 

To this end, we will view the time-dependent butterfly operator $|B(t)\rangle$ as a vector in the linear operator space satisfying the condition $|B(t = 0)\rangle = |B\rangle$. Action of a gate on the operator ${B}(t + 1) = {g}^{\dagger}_t \,{B}(t)\,{g}_t$ can be viewed as an action of an induced transfer matrix ${G}_t$ with elements 
\begin{gather}
    \left[{G}_t\right]_{pq} = \text{Tr}\left[{P} {g}^{\dagger}_t {Q} {g}_t\right]
\end{gather}
in the operator linear space $|B(t  + 1)\rangle = {G}_t |B(t)\rangle$.

For the sake of clarity, let us assume that $B$ is a Pauli string such that $P_0 = B$. In this notation, an element of the distribution reads
\begin{align}\label{eq:path_integral}
    |b_P|^2 &= \langle P | {G}_{T - 1} \cdots {G}_0 |B\rangle \langle B | {G}_0^{\dagger} 
        \cdots {G}^{\dagger}_{T - 1} |P\rangle \nonumber\\
        &= \sum\limits_{\boldsymbol Q, \boldsymbol{Q^\prime}} \langle P | {G}_{T - 1} |Q_{T - 1}\rangle \cdots \langle Q_0| {G}_0 |P_0\rangle \nonumber \\
        &\hspace{50pt}\times \langle P_0 | {G}_0^{\dagger} | Q^\prime_0\rangle \cdots \langle Q_{T - 1}^\prime| {G}^{\dagger}_{T - 1} |P\rangle
\end{align}
where we have inserted an identity resolution $\sum_Q |Q\rangle\! \langle Q|$ after each gate with $Q$ being a Pauli string, and a bold symbol $\boldsymbol Q = \{Q_0,\,Q_1,\,\ldots,\,Q_{T-1}\}$ denotes the collection of resolution insertions over all time stamps $t$.

The path integral representation Eq.\,\eqref{eq:path_integral} opens the prospect for a family of approximations, which we call diagonal approximations. For the rest of this Section, we describe multiple Monte Carlo algorithms based on the diagonal approximation.\\

\noindent\textbf{Vanilla Monte Carlo ---} The first example of such approximation is the {\it vanilla} Monte Carlo, where the condition $\boldsymbol Q = \boldsymbol Q^\prime$ is enforced, yielding to
\begin{gather}
\label{eq:projection_vanilla}
    |b_P|^2_{\text{vanilla}} = \sum\limits_{\boldsymbol Q} |\langle P | {G}_{T - 1} |Q_{T - 1}\rangle|^2 \ldots | \langle Q_0| {G}_0 |P_0\rangle|^2,
\end{gather}
which presents a direct receipt of sampling $P$ from $|b_P|^2_{\text{vanilla}}$: begin with $|P_0\rangle$, act a transfer matrix ${G}_0 |P_0\rangle$, sample $Q_0 \sim |\langle Q_0|{G}_0 |P_0\rangle|^2$, proceed to the next step for $Q_1$ and so on.

Observe that this approximation may be seen as ignoring all possible {\it cycles} (loops) in the butterfly operator evolution tree in the Pauli basis. Indeed, the information about two Pauli paths possibly rejoining later is neglected, and sampling takes place at every node of the tree.\\

\noindent\textbf{Cached Monte Carlo (CMC) ---} Looking at Eq.~\eqref{eq:path_integral}, it is immediate to realize that there are $n \cdot (T-1)$ locations in \emph{space} and \emph{time} where $\boldsymbol Q$ and $\boldsymbol Q^\prime$ can assume the same value, that is the locations where the two paths join. Therefore, the condition $\boldsymbol Q = \boldsymbol Q^\prime$ can be seen as a collections of pairs of two numbers $(x,\,t)$ that enforce the constraints $Q_{x, t} = Q^\prime_{x, t}$ for all $x$ and $t$. Consequently, the vanilla Monte Carlo can be improved by relaxing as many of such constraints as possible. This leads to the idea of the {\it cached} Monte Carlo (CMC). Practically, we work in a sparse representation of $|B(t)\rangle$ and begin with applying the induced gates to $|B\rangle$. When this representation overfills the memory at time $t$, we choose some qubit $x$, and enforce $Q_{x, t} = Q^\prime_{x, t}$. This enforcement creates a positive measure, and we immediately sample 
\begin{gather}
\label{eq:projection_cached}
    Q_{x, t} \sim \langle B(t-1) | Q_{x, t}\rangle\!\langle Q_{x, t}| \otimes \mathcal{I}_{n - 1} |B(t-1)\rangle,
\end{gather}
where $|Q_{x, t}\rangle\! \langle Q_{x, t}| \otimes \mathcal{I}_{n-1}$ is a projector onto a Pauli operator $Q_{x, t} \in \{I, X, Y, Z\}$ at a site $x$ and time $t$. That is, this sampling allows one to filter the sparse representation of $|B(t)\rangle$ by only keeping the Pauli strings which have the operator $q$ at a site $x.$ Having reduced the memory footprint, one can proceed with applying further gates until the procedure needs to be repeated.

Unlike the vanilla Monte Carlo, this procedure is capable of capturing {\it short} loops in the Pauli evolution tree, as it can act several transfer matrices sequentially without sampling, thus correctly accounting for interferences between different Pauli paths which result in the same Pauli string.\\

\noindent\textbf{Tensor-network Monte Carlo (TNMC) ---} Lastly, instead of sparsely representing the time-evolved operator $|B(t)\rangle$, one can directly sample
\begin{equation}\label{eq:projection_tn}
   P_{t+1}\sim |\langle P_{t+1}| G_t | P_t \rangle|^2
\end{equation}
by transforming a slice of the circuit into a tensor network and using sampling methods for shallow circuits such as  Ref.~\cite{bravyi2022simulate}.
This procedure is repeated over $K$ samples to target a desired Monte Carlo accuracy.
We call this application of tensor network contraction {\it tensor-network} Monte Carlo (TNMC). Observe that, if $T = 1$, there is no approximation and $P_N$ would be sampled with the exact distribution. Although, for $T$ sufficiently large, the resulting tensor network is shallow enough to be advantageous to use tensor network methods.\\

\noindent\textbf{Monte Carlo methods applied to state-OTOC ---} Let us consider a state-OTOC with a particular state $|\psi\rangle.$ In such case, starting from the initial butterfly vector $|B\rangle$, we still apply the induced transfer matrices ${G}_t$ and enforce diagonal constraints as in Eq.\,\eqref{eq:projection_vanilla}, \eqref{eq:projection_cached} or \eqref{eq:projection_tn}. In the last step, however, we compute the observable differently,
\begin{gather}
    O = \sum\limits_{P, P^\prime}b_P b_{P^\prime} \langle \psi|{P} {M} {P}^\prime {M}|\psi\rangle
\end{gather}
for the vanilla and cached Monte Carlo. For TNMC, we instead contract a shallow tensor network to obtain $\langle \psi| ({g}_{T-1} {P} {g}^{\dagger}_{T-1} {M})^2|\psi\rangle$, where ${P}$ was sampled in the last step, and ${g}_{T-1}$ is the remaining part of the full unitary thereafter.\\

\noindent\textbf{Relation to the Pauli insertion averaging ---} Finally, we show that a diagonal approximation of the form $Q_{x, t} = Q^\prime_{x, t}$ is related to the Pauli insertion averaging protocol which could be done experimentally or within state-vector simulations. To this end, we split the whole unitary in two parts: before the projection at time $t$, U(t,\,0), and after the projection $U(T,\,t)$. We then perform the Pauli expansion of the partially time-evolved butterfly operator ${B}(t) = {U}^{\dagger}(t,\,0) {B} {U}(t,\,0) = \sum_{P} b_P {P}$. Now, suppose that we also inserted a Pauli operator $Q_{x,t} \in \{{I}, {X}, {Y}, {Z}\}$ at a site $x$ at time $t$ and average over these options. The resulting expression reads
\begin{gather}
    \mathcal{C}^{(2)}_{\rm ins.} = 
    \frac{1}{4} \sum\limits_{{Q_{x,t}}, P, P^\prime} b_P b_{P^\prime} \langle{U}^{\dagger}(T, t) {Q_{x,t}} {P} {Q_{x,t}}{U}(T, t) \\ \nonumber \times  {M} {U}^{\dagger}(T, t) {Q_{x,t}} {P}^\prime {Q_{x,t}}{U}(T, t) {M}\rangle,
\end{gather}
where $\langle \ldots \rangle$ could denote either the trace or the state expectation. 
Importantly, the Pauli products are simplified as ${Q_{x,t}} {P} {Q_{x,t}} = {P} s({P}_x, {Q_{x,t}})$, where $s({P}_x, {Q_{x,t}}) = \pm 1$, depending on whether $P_x$ and $Q_{x,t}$ commute or anticommute. Because
\begin{gather}
    \frac{1}{4}\sum\limits_{Q_{x,t}} s({P}_x, {Q_{x,t}}) s({P}^\prime_x, {Q_{x,t}}) = \delta_{{P}_x, {P}^\prime_x},
\end{gather}
we get 
\begin{gather}
    \mathcal{C}^{(2)}_{\text{ins.}} = \sum\limits_{P, P^\prime | P_x = P^\prime_x} b_P b_{P^\prime} \langle{U}^{\dagger}(T, t) {P}{U}(T, t) M \nonumber \\  
     \times {U}^{\dagger}(T, t) {P}^\prime {U}(T, t) {M}\rangle,
\end{gather}
which means that experimentally (or in state-vector simulations) averaging over correlated Pauli insertion affects the OTOC computation exactly in the same way, as the point-like projection in Eq.\,\eqref{eq:projection_cached}. Clearly, several such projections (Pauli insertion averages) can be applied at the same time, for instance in a fixed-time cut manner as done in the main text.

In the next sections, we delve into the details of implementation, performance and complexity of those variations of the Monte Carlo approach.

%\subsubsection{The vanilla Monte Carlo algorithm}\label{sec:mc_algorithm}
%\input{mc_sampling}

\subsubsection{Cached Monte Carlo (CMC): details}\label{sec:cachedMC}

\begin{figure*}[t!]
    \centering
    \includegraphics[width=\textwidth]{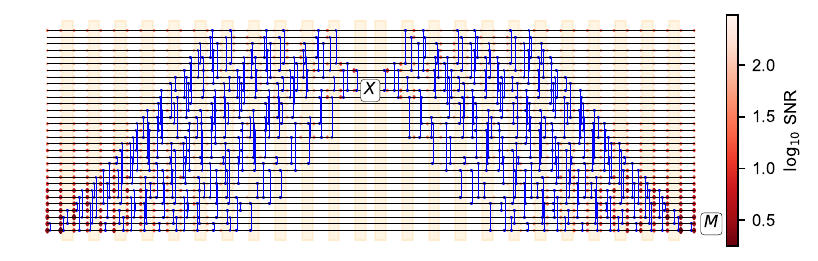}
    \caption{{\bf SNR of projecting in the Pauli basis.} A layout of a two-dimensional 31-qubit OTOC circuit. The shaded orange regions identify layers of two-qubit gates, and the blue lines representing two-qubit gates ate shifted horizontally within a layer for visibility. The application of a unitary ${U}$ is followed by the butterfly operator ${X}$, inverse unitary ${U}^{\dagger}$ and the measurement operator. The colorbar represents the SNR reduction upon insertion of a projection at a given space-time location $(q, t)$. The image is symmetric, since the Monte Carlo algorithm works in the Pauli space.}
    \label{fig:OTOC_projections_SNR_reduction}
\end{figure*}
The cached Monte Carlo scheme (CMC) is in principle heuristic (as we cannot bound the error), but that converges to the exact simulation in the limit of an exponentially-large cache size.
This flexibility allows us to test this method on the OTOC experiments presented in the main text with available classical hardware. For a 40-qubit experiment presented in the main text, the cached Monte Carlo gives the experiment-level $\text{SNR} \approx 5.4$, while for the 95-qubit experiment with two cuts we obtained an SNR above random guessing, but below the estimated SNRs of the experiment. In this section, we provide a few specific implementation details.

As discussed in Section\,\ref{sec:mc_intro}, the cached MC algorithm stores the non-zero elements of the coefficients $\bs{b}(t)$ of the Pauli, which we call ``cache'' $\mathcal{C}$, of size $|\mathcal{C}|$. If the cache overfills, the method uses a point-like projection Eq.\,\eqref{eq:projection_cached}. \\

\noindent\textbf{Computation of the expectation value ---} After applying all gates from ${U}$ and projecting along the way, the final cache $\mathcal{C}^{\{Q_{x,t}\}}$ depends on the set of projected Paulis $\{Q_{x,t}\}$ and the projection outcomes. Most often, we compute OTOC on top of a bitstring, which is an eigenstate of the measurement operator ${M}$. In particular, let us consider ${M} = {Z}$ and $|\psi\rangle = |00\ldots\rangle$. In such case, to avoid quadratic in $|\mathcal{C}|$ computation time, we first form the resulting ``wave function'' as 
\begin{gather}
    |\psi_{\{Q_{x,t}\}}\rangle = \sum\limits_{(b_P, P) \in \mathcal{C}} b_{P} \left({P} |0\rangle\right),
\end{gather}
and write the OTOC as
\begin{gather}
    \mathcal{C}^{(2)} \approx \frac{1}{M} \sum\limits_{{\{Q_{x,t}\}}} \langle \psi_{\{Q_{x,t}\}}| {M} |\psi_{\{Q_{x,t}\}}\rangle.
\end{gather}
where $M$ is the number of Monte Carlo samples. The required number of samples is smaller than in the vanilla Monte Carlo, since the norm of the expectation values $\langle \psi_{\{Q_{x,t}\}}| {M} |\psi_{\{Q_{x,t}\}}\rangle$ decays algebraically with the cache size. \\

\begin{figure}[b!]
    \centering
    \includegraphics[width=0.45\textwidth]{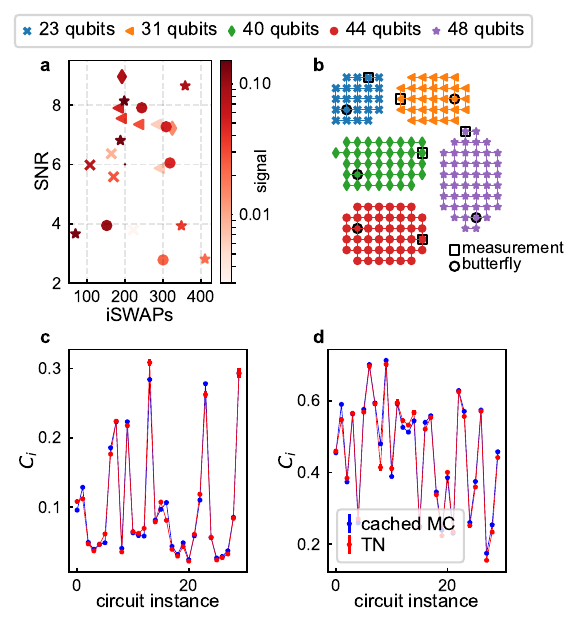}
    \caption{{\bf Benchmarking cached Monte Carlo on small systems. a,} The SNR between the cached MC and the exact simulation, {\bf b.} The circuit geometries. For each geometry, we considered a sequence of circuit depths to vary the signal (standard deviation of the OTOC values over circuit instances). The largest exactly-contracted circuit has 410 iSWAP gates on 48 qubits at 29 cycles. {\bf c-d.} 48-qubit comparison between the tensor-network contraction and the cached MC at depths 28 and 24, respectively.}
    \label{fig:small_geometries}
\end{figure}

\begin{figure}[t!]
    \centering
    \includegraphics[width=\columnwidth]{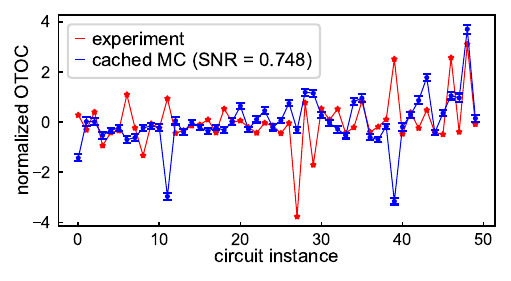}
    \caption{{\bf Comparison between the experiment and cached MC at 95-qubit circuits with 1000 iSWAPs.}}
    \label{fig:OTOC_CMC_large_systems}
\end{figure}

\noindent\textbf{The choice of the projected qubit $\boldsymbol{Q_{x,t}}$ ---}
We have empirically observed that the impact on the simulation SNR depends on the particular choice of the projected qubit, once the cache overfills. In particular, the spatial and temporal proximity of a projection to the measurement operator is an important indicator of the systematic bias introduced by the projection. Fig.~\ref{fig:OTOC_projections_SNR_reduction} illustrates the difference between in-bulk and idle projections, as well as the impact of proximity to the measurement operator. We observe that projections appearing closer to the measurement and butterfly operators lead to a greater reduction in SNR. We also observe that for the qubit lines far from the measurement operator, the in-bulk projections are more harmful than the idling projections.\\

\noindent\textbf{Small Pauli path truncation ---} An important part of the algorithm is to truncate small wave function elements $b_P \in \mathcal{C}$. Since the vector $b_P(t)$ is normalized, we truncate the cache elements with
\begin{gather}
\label{eq:truncation}
    |b_P|^2 < \frac{\texttt{tol}}{|\mathcal{C}|}
\end{gather}
and typically choose $0.001 < \texttt{tol} < 0.01$.\\

\noindent\textbf{Results on smaller systems ---} Before applying this algorithm to the full chip, we first assess its capacity on smaller systems where we can compute the OTOC circuits exactly, either using the tensor network contraction or the exact state-vector simulation. For these simulations, we set $|\mathcal{C}| = 50 \times 10^6$ and $\texttt{tol} = 0.001.$ We perform the exact simulations of the systems with 23, 31, 40, 44, and 48 qubits and with up to 410 iSWAP gates. For each geometry, we consider 30 random circuit instances. The system geometries and the resulting SNR as a function of the signal are shown in Fig.\,\ref{fig:small_geometries}. The observed SNRs are consistently well above random guessing.\\

\noindent\textbf{Application to the 95-qubit circuit ---} Finally, we apply the cached Monte Carlo algorithm to the 95-qubit OTOC$^{(1)}$ experiment with 1000 iSWAP gates and 29 cycles, as reported in the main text. %We will show that in the former case, the cached Monte Carlo algorithm obtains better-than-guessing SNR, while in the latter case the signals are uncorrelated.

%Pauli-injections in the experiment require the algorithm to project all qubits at those times, which allows one to free the cache without the SNR reduction. We report the comparison against the case with the Pauli-injection averaging in Fig.\,\ref{fig:OTOC_CMC_large_systems}. 
%We renormalize the OTOC as $\tilde{C}_i = (C_i - \bar{C}) / \text{std}\, C$, to ensure $\langle \tilde C_i \rangle = 0$ and $\text{std}\,\tilde{C}_i = 1$. Here, $\text{std}\, C$ is the standard deviation of $C_i$ computed over all circuit instances. After such rescaling, the SNR between the experiment and the cached MC is computed as $\text{SNR} = 1 / \text{std}\, (\tilde{C}^{\text{exp.}}_i - \tilde{C}^{\text{MC}}_i).$

In a simple model with the rescaled signals being uncorrelated and instances being drawn from the normal distribution $\mathcal{N}(0, 1)$, the SNR over $M$ circuit instances follows the inverse $\chi_{M - 1}$ distribution, which yields
\begin{gather}
    \langle\text{SNR}^{\text{uncorrelated}}\rangle_{\text{samples}} = \frac{\sqrt{M}}{2} \frac{\Gamma(M - 1)}{\Gamma(M - 1 / 2)} = \\ \nonumber = \frac{1}{\sqrt{2}} \left(1 + \frac{5}{4 M} + \mathcal{O}(M^{-2})\right).
\end{gather} 

%We see that in the case of the experiment with two cuts, the cached Monte Carlo gets to the SNR = 1.08, better than random guessing, as shown in Fig.\,\ref{fig:OTOC_CMC_large_systems}a. However, the space between cycles 7 and 26 has most of the gates, and the cached MC with $|\mathcal{C}| = 50 \times 10^6$ needs to make $\mathcal{O}(100)$ extra point-like projections.

In Fig.\,\ref{fig:OTOC_CMC_large_systems}b, we show the comparison between the cached MC and the experiment without any Pauli averaging. In this case, the SNR = 0.748 is close to the theoretical estimate for uncorrelated signals, $1/\sqrt{2} + 5 / (4 \sqrt{2} M) = 0.724$ at $M = 50$, indicating that the cached MC algorithm, limited by the memory of a single GPU, is unable to capture the correlations in this case.\\

\noindent\textbf{The cost of a perfect simulation without projections ---}
Here, we estimate the amount of memory required to perform this algorithm on the 95-qubit 1000-iSWAP circuit with only cache truncations of the type Eq.\,\eqref{eq:truncation}, but avoiding projections. To this end, we perform a simulation of the 95-qubit circuit with a cache size limit of $|\mathcal{C}| \leqslant 200 \times 10^6$. During the simulation, the cache becomes overfilled, and a qubit is chosen for projection to free space. We track how many Pauli strings are stored in the cache as a function of the number of non-projected qubits. We observe that the cache limit of $|\mathcal{C}| \leqslant 200 \times 10^6$ may already saturate at 15 non-projected qubits, which yields a cache size scaling of $|\mathcal{C}| \sim 3.57^{N_{\text{non-projected}}}$. Since these circuits involve at most 40 unprojected qubits, this would necessitate an impractically large cache size of $|\mathcal{C}| \leqslant 3.57^{40} \sim 10^{22}$.

\subsubsection{Tensor Network Monte Carlo (TNMC): details}\label{sec:tnmc}

\begin{figure*} [t!]
    \centering
    \includegraphics[width=1.0\textwidth]{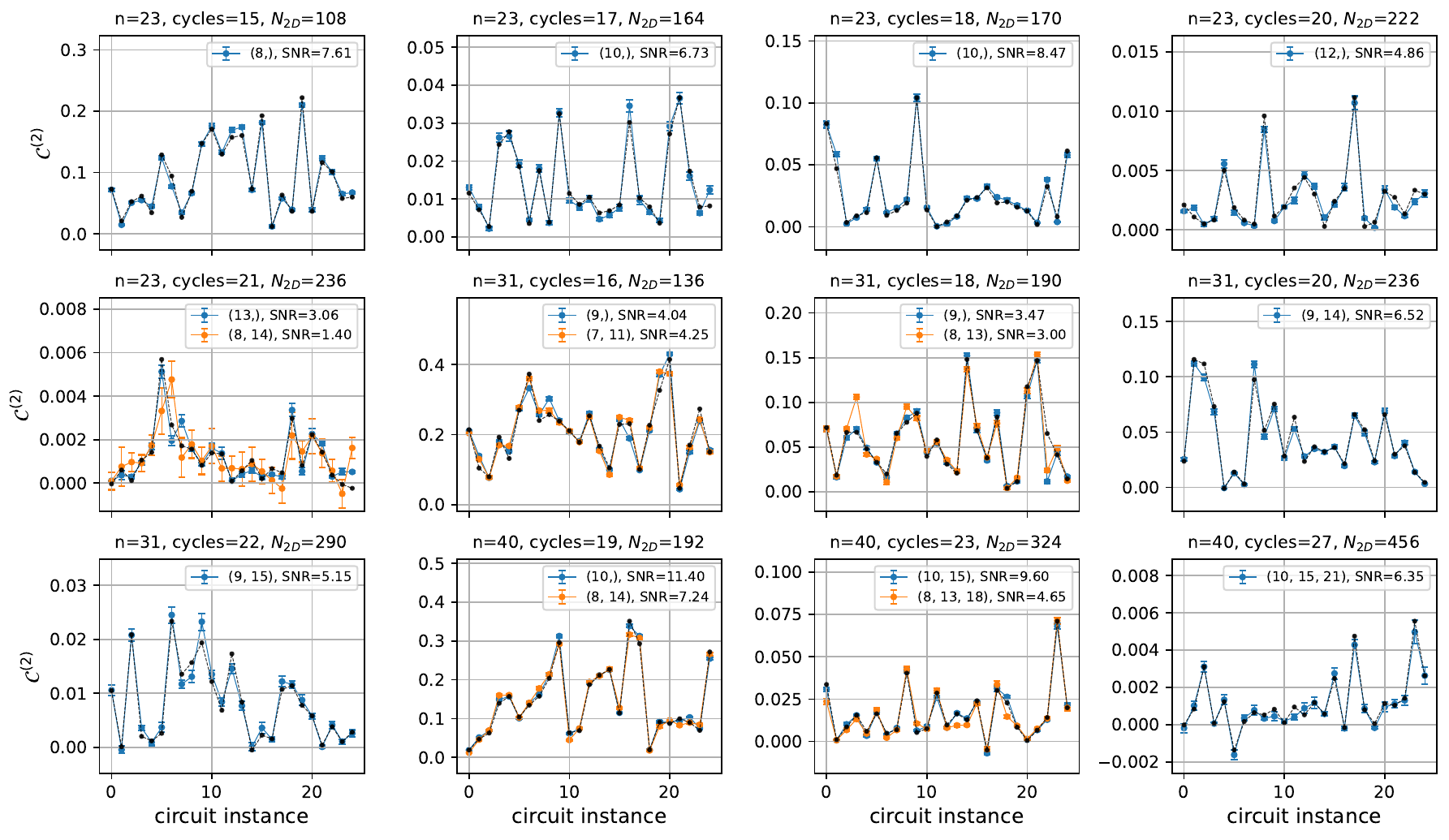}
    \caption{{\bf Performance of TNMC.} $\mathcal{C}^{(2)}_{\rm TNMC}$ as a function of circuit instance for different geometries of OTOC circuits of $n = $ 23, 31, and 40 qubits.
    The exact value of $\mathcal{C}^{(2)}$ is shown in black.
    The cut scheme used in each case and the SNR achieved is shown in the legend.
    In all cases we use $K = 5000$ Monte Carlo samples per point.}
    \label{fig:tnmc_small_geometries}
\end{figure*}

\begin{figure}[b!]
    \centering
    \begin{overpic}[width=1.0\columnwidth]{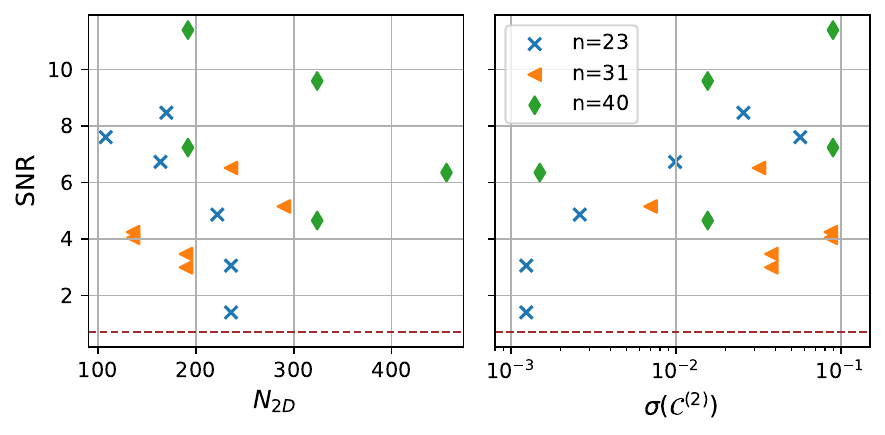}
      \put(5, 50){\textbf{a}}
      \put(53, 50){\textbf{b}}
    \end{overpic}
    \caption{{\bf SNR achieved by TNMC for different circuit geometries.}
    Scatter plot of the SNR as a function of the number of iSWAP-like gates $N_{\rm 2D}$ (\textbf{a}) and the signal size $\sigma(\mathcal{C}^{(2)})$ (\textbf{b}).
    We consider circuits of different number of qubits $n$ and different depths (see Figs.~\ref{fig:tnmc_small_geometries}).
    The baseline SNR of $1/\sqrt{2}\approx 0.71$ is shown as a dashed, brown line.}
    \label{fig:tnmc_snr_small_geometries}
\end{figure}

\begin{figure*} [t!]
    \centering
    \includegraphics[width=1.0\textwidth]{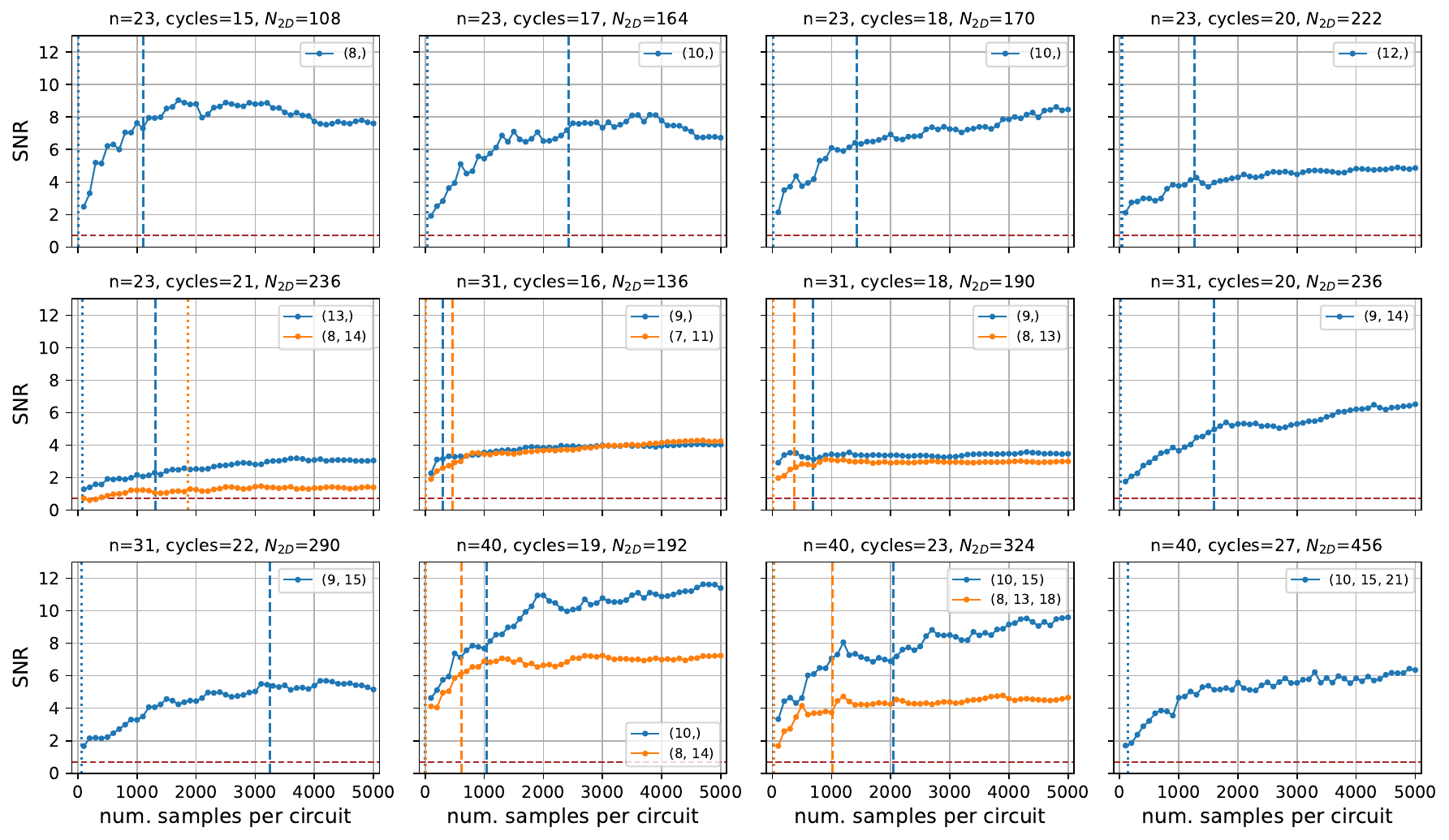}
    \caption{{\bf Convergence of TNMC with the number of samples.} Convergence of $\mathcal{C}^{(2)}_{\rm TNMC}$ as a function of the number of samples per circuit, $K$, for different circuit geometries of $n = $ 23, 31, and 40 qubits and varying depths.
    The cut scheme used in each case is shown in the legend.
    In addition, the values of $K_{\rm onset}$, after which the SNR is expected to grow, and $K_{\rm plateau}$, after which the SNR is expected to plateau, are shown as dotted and dashed vertical lines, respectively.
    The baseline SNR of $1/\sqrt{2}\approx 0.71$ is shown as a dashed, brown line.} 
    \label{fig:tnmc_convergence_small_geometries}
\end{figure*}

As discussed in Section~\ref{sec:classical_otoc}, Pauli string $P_{t+1}$ can be sampled by converting
the expression in Eq.~\eqref{eq:projection_tn} to a tensor network. This is done by a simple extension of the method of Ref.~\cite{bravyi2022simulate} from bitstrings to Pauli strings.
This sampling method has a complexity dominated by the evaluation of 
$\Tr [P_{t+1} G_t P_t ]$
by tensor network contraction (TNC). 
In addition, we focus on state-OTOC, which involves the computation of the remaining
\begin{align}
    \label{eq:remaining_tnmc}
    \langle \psi| ({g}_{T-1} {P_{T-1}} {g}^{\dagger}_{T-1} {M})^2|\psi\rangle 
\end{align}
as explained in Section~\ref{sec:classical_otoc}, in practice carried out via TNC.
Our implementation makes use of NVIDIA's \texttt{cuQuantum SDK} library for the optimization and computation of TNC on GPUs.\\

\noindent\textbf{Choosing circuit cuts ---}
Given an OTOC$^{(1)}$ with unitary $U$, TNMC ``cuts'' $U$ into $T$ shallower circuits at $T-1$ depth locations.
Once the circuit is cut, we need to sample $P_{T-1}$ and evaluate expression~\eqref{eq:remaining_tnmc} a number of times.
Sampling $P_{T-1}$ involves the computation of a number of TNCs that is linear in the number of gates in the circuit.
The aggregated cost of these contractions is dominated by the largest of the contraction costs of all $\Tr [P_{t+1} G_t P_t ]$ over subcircuits $g_t$, which scales exponentially in the size of $G_t$.
For the reasons presented, it is important to choose the location of the cuts carefully, while trying to minimize the number of cuts $(T-1)$ in order to achieve a good approximation.
In practice, given a certain value for $T$, we optimize the TNCs involved and choose the cutting scheme that minimizes computational cost.
If none of the options is computationally practical, then we explore schemes with larger values of $T$ until a viable cutting scheme is found.
For large circuits this might involve a large number of cuts, and hence a poor approximation of the OTOC$^{(1)}$ values.\\

\noindent\textbf{Performance of TNMC over small circuits ---}
In order to study the performance of TNMC, we focus first on small circuits for which it is possible to numerically compute the exact values of the OTOC$^{(1)}$, $C^{(2)}$.
We consider a set of circuit geometries involving $n =$ 23, 31, and 40 qubits.
For each value of $n$, we consider circuits of different number of cycles and different number of gates.
The largest example involves $N_{\rm 2D}=456$ iSWAP-like gates, i.\,e., almost half the number of two-qubit gates in the largest OTOC$^{(1)}$ circuits presented in this paper.
In each case, we have a set of 25 circuit instances.
For each circuit instance, we estimate $C^{(2)}$ with $K=5000$ TNMC samples.
Finally, for each circuit geometry, we consider either one or more values of $T$.

Fig.~\ref{fig:tnmc_small_geometries} shows the results of the TNMC computations over 12 circuit geometries.
In all cases we see a clear correlation between the approximate TNMC OTOC$^{(1)}$ values, $\mathcal{C}^{(2)}_{\rm TNMC }$, and the exact values $\mathcal{C}^{(2)}$.
The values of the SNR are typically relatively large, reaching a maximum of 11.40 for a 40 qubit circuit geometry with 19 cycles and $N_{\rm 2D} = 192$, requiring only a single cut after cycle 10 in $U$.
On the largest circuit geometry studied, with 27 cycles and $N_{\rm 2D} = 456$, TNMC with three cuts (on cycles 10, 15, and 21) reaches an SNR of 6.35.
Interestingly, there are non-negligible variations in the SNR achieved by TNMC for different circuit geometries and for different cut schemes.
In particular, in the case of 23 qubits and 21 cycles TNMC reaches lower values of the SNR, with a value as low as 1.40 with two cuts.

In an attempt to correlate the performance of TNMC to circuit parameters, we consider the SNR of all examples presented here as a function of $N_{2D}$, as well as a function of the typical size of the OTOC signal, i.\,e., the standard deviation over circuits $\sigma(C^{(2)})$.
We show this in Fig.~\ref{fig:tnmc_snr_small_geometries}.
Overall there is no clear correlation between either of these quantities.
When postselecting to either $n=23$ or $n=40$, there is a weak correlation indicating that the SNR decreases with increasing $N_{2D}$.
Similarly, there is weak evidence that the SNR increases with signal size $\sigma(C^{(2)})$.
So far, however, we have ignored the effect of having an insufficient number of Monte Carlo samples, $K$, for the estimation of $C^{(2)}_{\rm TNMC}$.\\

\noindent\textbf{Convergence of TNMC with the number of samples ---}
It is important to study the convergence of TNMC as a function of the number of Monte Carlo samples $K$.
Fig.~\ref{fig:tnmc_convergence_small_geometries} shows the SNR as a function of the number of samples used to estimate $\mathcal{C}^{(2)}_{\rm TNMC}$, $K$.
The behavior of this function is as expected.
At small $K$, the SNR is close to $1/\sqrt{2}$, shown as a dashed brown line.
This is the baseline SNR, observed between two uncorrelated sets of values.
Then, as $K$ grows we observe a steady increase in SNR with some statistical fluctuations.
Finally, at large $K$ the SNR plateaus at its final value.
While most examples considered show convergence (they reach their plateau value), some would need larger $K$ to converge.

\begin{figure*} [t!]
    \centering
    \begin{overpic}[width=0.8\textwidth]{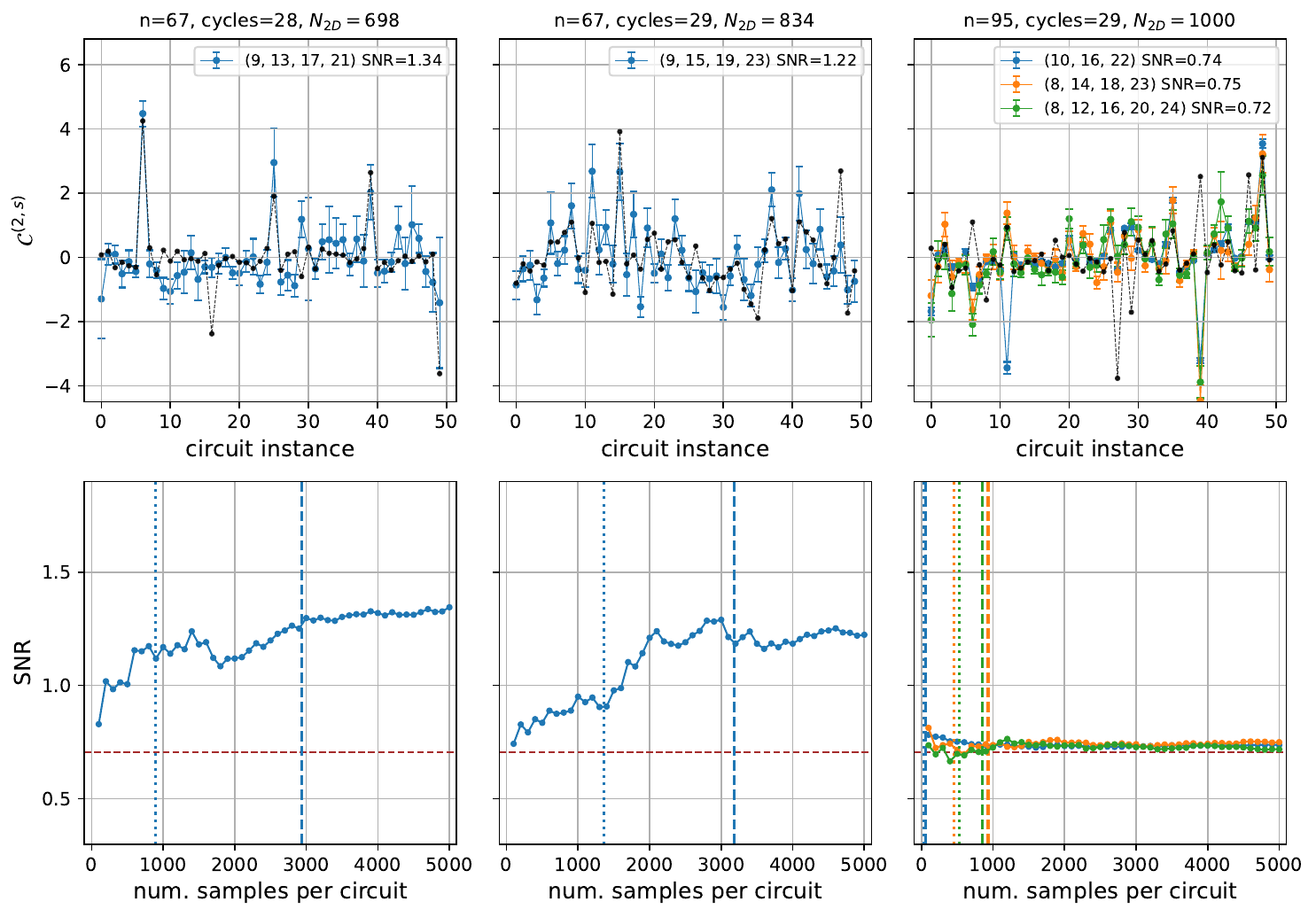}
      \put(4.5, 67.5){\textbf{a}}
      \put(36.5, 67.5){\textbf{b}}
      \put(68.5, 67.5){\textbf{c}}
      \put(4.5, 34){\textbf{d}}
      \put(36.5, 34){\textbf{e}}
      \put(68.5, 34){\textbf{f}}
    \end{overpic}
    \caption{{\bf Performance of TNMC over large experiments.} \textbf{a-c}, Standardized TNMC approximation of $\mathcal{C}^{(2, s)}_{\rm TNMC}$ as a function of circuit instance \emph{compared to experimental data} (black).
    We consider three experiments: (\textbf{a}) over 67 qubits and $N_{\rm 2D}=698$, (\textbf{b}) over 67 qubits and $N_{\rm 2D}$, and (\textbf{c}) over 95 qubits and $N_{\rm 2D}=1000$.
    The cut schemes and SNR \emph{with respect to experimental data} are shown in the legend.
    Note the presence of a significant correlation in the first case, with an SNR of 1.55, and the lack of a meaningful correlation in the second case.
    In both cases we use 5000 Monte Carlo samples per point.
    \textbf{d-f}, Convergence of TNMC with number of samples up to $K=5000$.
    The degradation of SNR compared to the values of Fig.~\ref{fig:tnmc_small_geometries} is due to the fact that here we compare against experimental data, which is itself imperfect, and to the increased size of the circuits in terms of qubits, depth, and $N_{\rm 2D}$.
    The SNR decreases as the circuits grow in volume.
    In addition, at $n=95$ and $N_{\rm 2D}=1000$ we see a slight increase in the SNR as the number of cuts decreases, although remaining small and close to $1/\sqrt{2}$.}
    \label{fig:tnmc_large_experimental}
\end{figure*}

\begin{table*}[t]
\centering
\begin{tabular}{|c|c|c|c|c|c|c|c|c|c|c|}
\hline
 \multirow{2}{*}{$\boldsymbol{n}$} &  \multirow{2}{*}{\textbf{Cycles}} &  \multirow{2}{*}{$\boldsymbol{N_{\rm 2D}}$} &  \multirow{2}{*}{\textbf{Cuts}} & \multicolumn{3}{c|}{\textbf{TNMC (Exact)}} &  \multicolumn{3}{c|}{\textbf{TNMC (Experiment)}} & \textbf{Experiment (Exact)} \\ \cline{5-11} 
 &  &  &  & $\boldsymbol{K_{\rm onset}}$ & $\boldsymbol{K_{\rm plateau}}$ & \textbf{SNR} & $\boldsymbol{K_{\rm onset}}$ & $\boldsymbol{K_{\rm plateau}}$ & \textbf{SNR} & \textbf{SNR} \\ \hline \hline
  \multirow{6}{*}{23} & 15 & 108 & (8) & 13 & 1105 & 7.61 & 13 & 479 & 4.74 & 7.21 \\ \cline{2-11}
  & 17 & 164 & (10) & 34 & 2425 & 6.73 & 34 & 575 & 3.08 & 5.26 \\ \cline{2-11}
  & 18 & 170 & (10) & 10 & 1432 & 8.47 & 10 & 495 & 5.34 & 6.67 \\ \cline{2-11}
  & 20 & 222 & (12) & 38 & 1270 & 4.86 & 48 & 493 & 2.55 & 3.57 \\ \cline{2-11}
  & \multirow{2}{*}{21} & \multirow{2}{*}{236} & (13) & 78 & 1316 & 3.06 & 78 & 709 & 2.46 & \multirow{2}{*}{2.45} \\
  & & & (8, 14) & 1860 & 7981 & 1.40 & 1861 & 4344 & 1.13 & \\ \hline
 \multirow{5}{*}{31} & \multirow{2}{*}{16} & \multirow{2}{*}{136} & (9) & 9 & 296 & 4.04 & 9 & 127 & 2.87 & \multirow{2}{*}{4.23} \\
 &  & & (7, 11) & 15 & 460 & 4.25 & 15 & 176 & 2.79 & \\ \cline{2-11}
 & \multirow{2}{*}{18} & \multirow{2}{*}{136} & (9) & 14 & 691 & 3.47 & 14 & 313 & 3.45 & \multirow{2}{*}{3.85} \\
 & & & (8, 13) & 17 & 375 & 3.00 & 17 & 197 & 2.40 &  \\ \cline{2-11}
 & 20 & 236 & (9, 14) & 15 & 1592 & 6.52 & 15 & 466 & 3.46 & 3.72 \\ \cline{2-11}
 & 22 & 290 & (9, 15) & 65 & 3250 & 5.15 & 65 & 455 & 2.19 & 3.40 \\ \hline
 \multirow{5}{*}{40} & \multirow{2}{*}{19} & \multirow{2}{*}{192} & (10) & 4 & 1043 & 11.40 & 4 & 188 & 5.41 & \multirow{2}{*}{4.88} \\
 & & & (8, 14) & 7 & 621 & 7.24 & 7 & 317 & 5.53 & \\ \cline{2-11}
 & \multirow{2}{*}{23} & \multirow{2}{*}{324} & (10, 15) & 15 & 2041 & 9.60 & 15 & 504 & 4.53 & \multirow{2}{*}{3.88} \\
 & & & (8, 13, 18) & 24 & 1013 & 4.65 & 24 & 291 & 2.75 & \\ \cline{2-11}
 & 27 & 456 & (10, 15, 21) & 138 & 7863 & 6.35 & 138 & 1623 & 2.78 & 2.59 \\ \hline
 67 & 28 & 698 & (9, 13, 17, 21) & - & - & - & 684 & 2347 & 1.34 & - \\ \hline
 67 & 29 & 834 & (9, 15, 19, 23) & - & - & - & 1357 & 3178 & 1.22 & - \\ \hline
 \multirow{3}{*}{95} & \multirow{3}{*}{29} & \multirow{3}{*}{1000} & (10, 16, 22) & - & - & - & 38 & 70 & 0.74 & - \\
 & & & (8, 14, 18, 23) & - & - & - & 279 & 602 & 0.75 & - \\
 & & & (8, 12, 16, 20, 24) & - & - & - & 506 & 775 & 0.72 & - \\ \hline
\end{tabular}
\caption{\textbf{Comparison between TNMC, experimental, and exact signals.}
Analysis of the performance of TNMC compared to experimental and exact OTOC signals.
We show the number of qubits $n$, the number of cycles in $U$, the number of iSWAP-like gates, and the cutting scheme used by TNMC.
The results include the SNR between the TNMC OTOC values and the exact ones (when available), as well as the SNR between TNMC and experimental values, and that between the experiment and the exact values.
In addition, we estimate $K_{\rm onset}$ at which some non-trivial SNR is expected and $K_{\rm plateau}$ at the final plateau in SNR is approached for TNMC.
We consider the same circuit geometries as in Figs.~\ref{fig:tnmc_small_geometries} and \ref{fig:tnmc_large_experimental}.
The last two rows correspond to large circuits for which only experimental values of the OTOC are available.}
\label{table:tnmc}
\end{table*}

Let us now define estimate the number of samples needed for convergence of the SNR in $K$.
Let us start by defining some quantities.
First, we refer to the TNMC estimate of the OTOC$^{(1)}$ for circuit instance $i$ as $\mathcal{C}^{(2)}_{\rm TNMC, i}$.
We also refer to the standard deviation of the distribution of Monte Carlo sample contributions to $\mathcal{C}^{(2)}_{\rm TNMC, i}$ as $\sigma_{\rm TNMC}(\mathcal{C}^{(2)}_{{\rm TNMC}, i})$.
This is not to be confused with $\sigma(\mathcal{C}^{(2)}_{\rm TNMC})$, which is the standard deviation of $\mathcal{C}^{(2)}_{\rm TNMC}$ across circuit instances.
In addition, we define the absolute value of the difference between the TNMC estimate and the true value of an OTOC as $\Delta_{\rm TNMC} \mathcal{C}^{(2)} = |\mathcal{C}^{(2)} - \mathcal{C}^{(2)}_{\rm TNMC}|$.
Finally, follow the convention of adding a superscript $s$ to denote standardization of the values of the OTOC to having mean 0 and standard deviation 1.
With this nomenclature at hand, let us focus on three helpful quantities, all of them relative to standardized signals:
\begin{enumerate}
    \item The signal size, defined as 
    \begin{align}
    \label{eq:tnmc_alpha}
    \alpha = \sigma(\tilde{\mathcal{C}}^{(2, s)})
    \end{align}
    By definition of standardization, $\alpha=1$.
    \item The typical spread of TNMC Monte Carlo sample contributions, characterized by the mean value over circuit instances of $\overline{\sigma_{\rm TNMC}(\mathcal{C}^{(2,s)}_{\rm TNMC})} $, i.e.,
    \begin{align}
    \label{eq:tnmc_beta}
    \beta = \overline{\sigma_{\rm TNMC}(\mathcal{C}^{(2,s)}_{\rm TNMC})}
    \end{align}
    \item The typical discrepancy between the TNMC estimate and the true value of the OTOC, characterized by the mean over circuit instances of $\Delta_{\rm TNMC} \mathcal{C}^{(2,s)} = |\mathcal{C}^{(2,s)} - \mathcal{C}^{(2,s)}_{\rm TNMC}|$, i.e.,
    \begin{align}
    \label{eq:tnmc_gamma}
    \gamma = \overline{\Delta_{\rm TNMC} \mathcal{C}^{(2,s)}}
    \end{align}
\end{enumerate}
We are now equipped to compare the typical statistical error in the Monte Carlo estimates of the OTOC to the signal size and to the typical difference between the estimated OTOC values and the exact ones.
The statistical error is typically of the order of $\beta / \sqrt{K}$, where $K$ is the number of Monte Carlo samples.
If $\beta / \sqrt{K} > \alpha = 1$, then the error in the estimates is too large to detect much correlation between $\mathcal{C}^{(2, s)}_{\rm TNMC}$ and $\mathcal{C}^{(2, s)}$.
We expect SNR to increase with $K$ once $K \gtrapprox (\beta/\alpha)^2 = \beta^2 \equiv K_{\rm onset}$.
Similarly, we expect to approach the plateau in SNR once the statistical error in $\mathcal{C}^{(2, s)}_{\rm TNMC}$ is smaller than the typical difference between both signals, i.\,e., when $\beta / \sqrt{K} \lessapprox \gamma$ or $K \gtrapprox (\beta / \gamma)^2 \equiv K_{\rm plateau}$.

In Fig.~\ref{fig:tnmc_convergence_small_geometries} we plot vertical lines at both $K_{\rm onset}$ (dotted) and $K_{\rm plateau}$ (dashed).
While both values are only ballpark indicators of the crossover between the behavior of the SNR as a function of $K$, we can use them as proxies to estimate the number of samples needed to measure correlation and final SNR value when comparing TNMC to exact numerics or experiments.\\

\noindent\textbf{Performance of TNMC over large circuits ---}
We now turn our attention to the performance of TNMC over circuits that are impractical to simulate exactly with classical computers.
In those cases we do not have exact values of the OTOC to compare against, but rather experimental values.

We consider two circuit geometries involving (1) 67 qubits and $N_{\rm 2D}=698$ gates, (2) 67 qubits and $N_{\rm 2D}=834$, and (3) 95 qubits and $N_{\rm 2D}=1000$ gates.
In the first case we run TNMC with 4 cuts, achieving an SNR of 1.34 when comparing to experimental data.
In the second case the SNR drops to 1.22 with 4 cuts, likely due to the increased circuit size.
In the last case we run TNMC with 3 (on NVIDIA servers), 4, and 5 cuts, achieving SNRs of 0.74, 0.75, and 0.72, respectively.
%It is interesting to observe the degradation of the SNR as the number of cuts increases.
In all 3 cases the SNR is not substantially higher than baseline $1/\sqrt{2} \approx 0.71$.
This data is presented in Fig.~\ref{fig:tnmc_large_experimental}.

In order to compare the performance of TNMC over small and large circuits on an equal footing, we compute SNR over all circuits with respect to experimental values of the OTOC.
We compile this and other data in Table~\ref{table:tnmc}.
It is interesting to compare the degradation of the SNR when computed with respect to experimental OTOCs instead of against exact values of the OTOCs.
Beyond 40 qubits we cannot compute the SNR with respect to $\mathcal{C}^{(2)}$.
We still see, however, the decrease of SNR as the circuit size increases, as well as a typically increasing SNR as the number of cuts decreases.
Finally, at $n=95$ and $N_{\rm 2D}=1000$ we cannot achieve a substantially large SNR.

One possible strategy to find statistically significant SNR on the $N_{\rm 2D}=1000$ circuits could be attempting to run TNMC with 2 cuts.
After optimization of the TNCs involved in this procedure, we estimate that a single sample with $T=2$ (optimally chosen at depths 12 and 20) would require $1.81 \times 10^{15}$ FLOP per sample.
This is equivalent to effectively 5.3 seconds for every 1000 samples on Frontier, which delivers a peak performance of 1.71 ExaFLOPS.
We constrain the memory footprint of each TNC to satisfy that of a single GPU.
Different independent contractions, such as those of different samples, can be carried out in parallel over different GPUs.
This is a somewhat challenging computational effort.

\noindent\textbf{Future TNMC improvements ---}
It is possible to improve the TNMC method and its implementation to achieve lower computational cost.
First, the current implementation cuts the OTOC unitary over all qubits at certain depth locations.
More flexible cutting schemes can certainly be beneficial.
Furthermore, the sampling procedure of Ref.~\cite{bravyi2022simulate} allows for more elaborate reorganization of the tensor network evaluations involved, leading to a lower cost per sample than the current implementation.
In addition, the evaluation of Eq.~\eqref{eq:projection_tn} and similar might admit approximations that lower the computational cost.
Finally, considering contractions with less stringent memory constraints can lower their cost dramatically.
This, however, requires a substantial effort in the underlying TNC implementation.
These potential improvements are left as directions of future research.\\

%\subsubsection{TN-MC performance on iSwap$^c$ circuits for $c<1$.}\label{sec:tnmc_iswapc}
%\input{tnmc_iswapc}

%\subsubsection{Exact tensor network contraction}%\label{sec:tn}
%
%\input{tn_exact}

\subsubsection{Clifford-based expansions (CBE)}\label{sec:clifford_expansion}
Clifford-base expansion (CBE) algorithms has been successfully used in the past to classically simulate OTOC$^{(1)}$ circuits \cite{Mi_OTOC_2021}. Similar to the cached Monte Carlo, the idea of (CBE) algorithms is to follow the evolution of the coefficients $b_P$, and keep track of only those coefficients which are non-zero.
Unlike cached Monte Carlo, coefficient are explored ``depth-first'': that is, a single branch of the Pauli-path is followed till all $u_t$ are applied, and only the final Pauli string is stored in memory.
Without any truncation, CBE algorithms provide the exact calculation of the correlation operator $\mathcal{C}^{(2k)}$. However, the number of $b_P$ coefficients exponentially grows with the number of non-Clifford gates.

In \cite{Mi_OTOC_2021}, the only non-Clifford gates used for OTOC$^{(1)}$ were 4 out of 8 single-qubit rotations, i.e. $\sqrt{W^{\pm1}}$ and $\sqrt{V^{\pm1}}$, with \mbox{$W = \sqrt{(X + Y)/\sqrt{2}}$} and \mbox{$V = \sqrt{(X - Y)/\sqrt{2}}$}. Because
\begin{align*}
    & W^\dagger X W = \frac{X + Y}{2} + \frac{Z}{\sqrt{2}} \\
    & V^\dagger X V = \frac{X - Y}{2} + \frac{Z}{\sqrt{2}} \\
    & W^\dagger Y W = \frac{X + Y}{2} - \frac{Z}{\sqrt{2}} \\
    & V^\dagger Y V = \frac{X - Y}{2} - \frac{Z}{\sqrt{2}} \\
    & W^\dagger Z W = \frac{Y - X}{\sqrt{2}} \\
    & V^\dagger Z V = -\frac{X + Y}{2},
\end{align*}
the expected number of paths to explore will be of the order of $\sim 3^{N_D/2} 2^{N_D/4}$, with $N_D$ being the number of non-Clifford rotations (see Fig.~S15 of \cite{Mi_OTOC_2021}). In our OTOC experiments, all gates are non-Clifford, making the exact simulation of $\mathcal{C}^{(2k)}$ unpractical.

In parallel to the cached Monte Carlo, approximation schemes can be introduced to CBE algorithms to reduce the amount of coefficients $b_P$ to keep track of. One possible approximation consists in only removing all paths with coefficients $b_P$ below a certain threshold. By properly tuning such a threshold, it is possible to reduce the amount of paths to explore. However, our numerical studies (not reported) show that the cached Monte Carlo largely outperforms CBE.

%\subsubsection{Reduced-density operator approaches}\label{sec:rdo}
%\input{reduced_density_operator}

\subsection{Classical simulation of higher-order OTOCs}\label{sec:classical_otoc2}

%{\color{green}{Tom: We need to add an intro paragraph here explaining that for OTOC$^{(2)}$, all known methods either cannot achieve significant SNR ($>1$), or have a formal exponential scaling cost. Then, mention that gate truncation is a strong contender for OTOC$^{(2)}$. Also, we can use this area to quickly go over the methods from OTOC that we have discounted for consideration from OTOC$^{(2)}$ due to poor performance.}}

In this section we expand our OTOC simulation efforts to target OTOC$^{(2)}$, both attempting to extend the methods attempted for OTOC$^{(1)}$, and trying new methods solely targeting $\mathcal{C}^{(4)}_{\rm off-diag}$.
Despite trying many approaches, we have been unable to find a polynomially-scaling method with significant SNR ($>1$) for OTOC$^{(2)}$, which marks a clear difference to the success of Monte Carlo methods on $\mathcal{C}^{(2)}$.
This underpins our claim that the experimental data reported in the main text is beyond-classical.
We begin in Sec.~\ref{sec:mc_otoc2} by attempting to extend the vanilla Monte Carlo to estimate OTOC$^{(2)}$.
Beyond small system sizes, we find that SNR above $1$ can not be achieved.
The same is true for the cached Monte Carlo, which we attempt in Sec.~\ref{sec:cached_mc_otoc2} and find evidence of a clear sign problem.
Indeed, this sign problem persists even when estimating the mean OTOC$^{(2)}$,  $\overline{C^{(4)}}$, as we show in Sec.~\ref{sec:permutationmontecarlo}, and to a much stronger degree than the sign problem in estimating $\sigma(C^{(2)})$ that was reported on in Ref.~\cite{Mi_OTOC_2021}.

We next turn to exponentially-scaling methods, or methods with a formal exponential cost that may be heuristically truncated (e.\,g. by bond dimension).
This includes many popular methods such as Clifford expansions \ref{sec:clifford_expansion}, Pauli-weight truncation algorithms \ref{sec:pauli_truncation}, and variational approaches such as neural quantum states \ref{sec:nn_wf} and matrix-product states \ref{sec:mps_mpo}. None of the above are able to achieve a significant SNR in OTOC$^{(2)}$ approximation with the resources at hand.

Therefore, we believe that tensor network contraction remains the fastest classical algorithm for OTOC$^{(2)}$ simulation. We detail methods for the optimization of exact tensor network contraction for OTOC$^{(2)}$ circuits in Sec.~\ref{sec:tn_estimates_otoc2}.

Then, in an attempt to reduce the contraction costs, in Sec.~\ref{sec:tngr} we detail a heuristic method to remove the least important gates at the cost of reducing the simulation SNR. It is challenging to estimate the resulting  tensor contraction cost after such gate removal protocol, since it is hard to extrapolate the SNR as a function of the removed gate count in beyond-classical circuits; our best estimates yield a range of between a few minutes and a month. However, as the true number of removable gates remains unknown, we keep our claim that our OTOC$^{(2)}$ experiment is beyond-classical.

\subsubsection{Vanilla Monte Carlo simulation of \texorpdfstring{OTOC$^{(2)}$}{OTOC-2}}\label{sec:mc_otoc2}

\begin{figure*}
\centering
\begin{overpic}[scale=0.7]{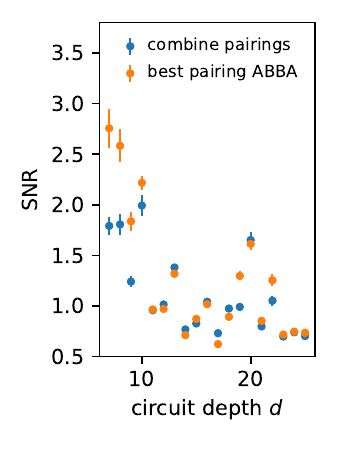}\put(10,98){\textbf{a}}\end{overpic}
\begin{overpic}[scale=0.7]{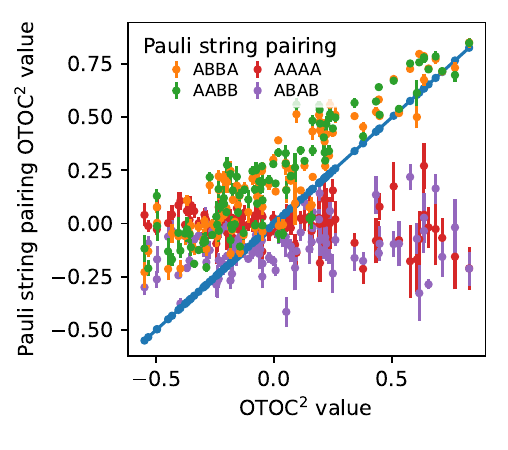}\put(10,87){\textbf{b}}\end{overpic}
\begin{overpic}[scale=0.7]{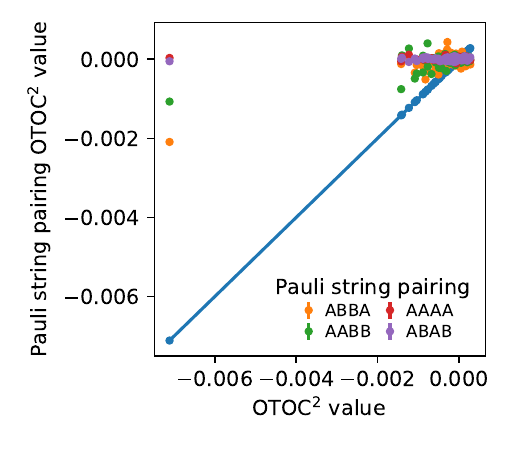}\put(10,87){\textbf{c}}\end{overpic}
\caption{{\bf a}, The results for the SNR when approximating the signal using the approximation introduced in Eq.\,\eqref{eq:diagonal_approx_otoc_sq}. We also consider the ABBA pairing. This pairing which yields the highest SNR in 100 random circuit instances. {\bf b-c}, the true $\mathrm{OTOC}^{(2)}$ expectation values for 100 random circuit instances plotted against the different pairings Eq.\,\eqref{eq:diagonal_approx_otoc_sq}. The circuit depth is 10 (center) and 20 (right), the system size is $26$.}
\label{fig:study_outliers}
\end{figure*}
\begin{figure}[t!]
    \centering
    \includegraphics[scale=0.8]{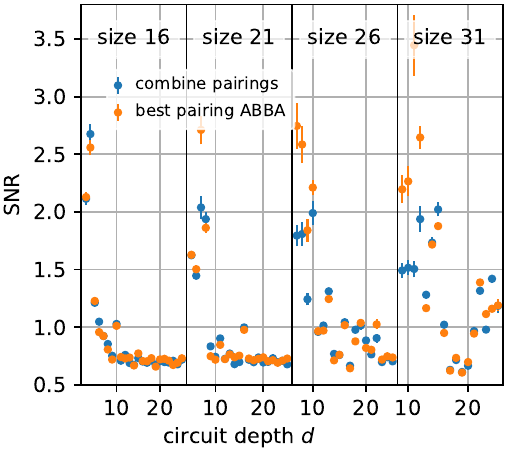}
    \caption{\textbf{SNRs when using the diagonal parts of the sum to estimate the OTOC$\boldsymbol{^{(2)}}$ signal, but only considering the results inside a $\boldsymbol{4\sigma}$ region around the mean to exclude outliers.} See also Fig.~\ref{fig:study_outliers}. The number of qubits, the sizes, are 16, 21, 26 and 31 from left to right.}
    \label{fig:signal_spoofing_system_sizes}
\end{figure}
\begin{figure}[t!]
    \centering
    \includegraphics[scale=0.8]{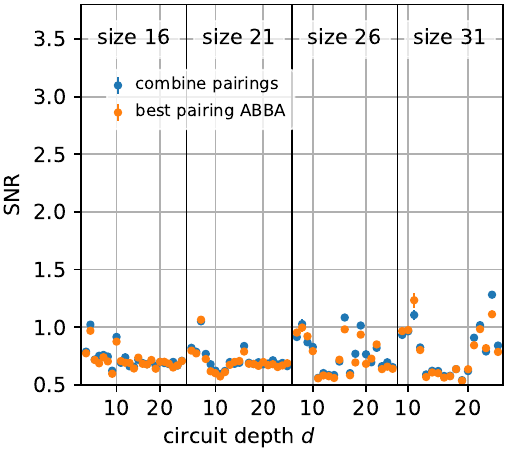}
    \caption{\textbf{SNR when using the diagonal pairings to estimate the subtracted OTOC$\boldsymbol{^{(2)}}$ signal, i.\,e. the true signal with the ABBA pairing subtracted.} We only consider the results inside a $4\sigma$ region around the mean. The number of qubits are 16, 21, 26 and 31 from left to right. 
    %The SNR for the 31 qubit case at depth 19 is surprisingly high. The reason being that here the ABBA signal is comparably large to the full signal, but uncorrelated. Hence, the subtracted signal actually becomes correlated with the ABBA part relative to the ratio of the signal strengths, which happens to be around $\sigma_\mathrm{subtracted}/\sigma_\mathrm{full}=1.4$ at circuit depth 19 for 31 qubits. %changed to Xiao's definition of SNR, text does not apply anymore
    }
    \label{fig:subtracted_signal_spoofing_system_sizes}
\end{figure}
In this section we adapt the methods of the previous section that were used for simulating the OTOC$^{(1)}$ to simulating the OTOC$^{(2)}$.
For simplicity, here we consider  splitting the circuit $U$ into only two parts $U=u_1 u_2$. Writing an observable $B$ as $B(t=1)=u_1^\dagger B u_1$, we obtain as an expression for the OTOC$^{(2)}$
\begin{align}
    \mathcal{C}^{(4)}=\big\langle [u_2^\dagger B(1) u_2 M]^4 \big\rangle.
\end{align}
Now, let us expand the time evolved butterfly in terms of Pauli strings $P$, yielding
\begin{align}
    &\mathcal{C}^{(4)}=\sum_{P,P^\prime,Q,Q^\prime} b_{P}(1)b_{P^\prime}(1)b_{Q}(1)b_{Q^\prime}(1)\nonumber\\ &\hspace{1cm}\big\langle u_2^\dagger P u_2 M u_2^\dagger P^\prime u_2 M u_2^\dagger Q u_2 M  u_2^\dagger Q^\prime u_2 \big\rangle. 
    \label{eq:diag_app_otoc_sq_otoc_sq}
\end{align}
As in Section~\ref{sec:mc_intro}, we can square the $b_{P}(1)$ terms to give a probability distribution $p_{P}(1)=|b_{P}(1)|^2$, which corresponds to the weight of the string $P$ in the decomposition of $B$.

%We wish to sample from this distribution and use this to evaluate Eq.~\eqref{eq:diag_app_otoc_sq_otoc_sq} (on the assumption that we can simulate half the circuit exactly).
To obtain a positive Monte Carlo measure, we need to disregard the terms in Eq.~\eqref{eq:diag_app_otoc_sq_otoc_sq} which do not allow for a probability interpretation in this basis.
%That is, we drop all terms which do not contain pairs of strings that can be combined to give a probability distribution as above.
This yields the approximation
\begin{gather}
    \mathcal{C}^{(4)}\approx\mathcal{C}^{(4)}_{\mathrm{MC}} = \sum_{P,P^\prime,Q,Q^\prime} \left(\delta_{P,P^\prime}\delta_{Q,Q^\prime}+\delta_{P,Q}\delta_{P^\prime,Q^\prime} \right. + \\ \nonumber \left. \delta_{P,Q^\prime}\delta_{P^\prime,Q}-2\delta_{P,P^\prime}\delta_{P^\prime,Q}\delta_{Q,Q^\prime}\right) b_{P}(1)b_{P^\prime}(1)b_{Q}(1)b_{Q^\prime}(1) \times \\\nonumber \big\langle  u_2^\dagger P u_2 M u_2^\dagger P^\prime u_2 M u_2^\dagger Q u_2 M  u_2^\dagger Q^\prime u_2 \big\rangle, \label{eq:diagonal_approx_otoc_sq}
\end{gather}
where we subtract the term where all Pauli strings are the same, as this is included across the first three terms.
The terms in the bracket give four contributions to the MC-approximated OTOC, which we label in order as ``AABB'', ``ABAB'', ``ABBA'', and ``AAAA''.
We can evaluate each contribution separately by Monte Carlo sampling from the $\bs{p}(1)$ vectors and evaluating the expectation value in Eq.~\eqref{eq:diagonal_approx_otoc_sq} as an OTOC$^{(2)}$ (possibly with different butterfly operators at different times).

We test this approximation numerically by considering a rectangular $l \times 5$ system, with the four corner qubits removed, for $l=4,\,5,\,6$ and $7$. In Fig.~\ref{fig:signal_spoofing_system_sizes}, we plot the SNR of the approximation defined in Eq.~\eqref{eq:diagonal_approx_otoc_sq} (using only a single cut), as a function of the depth.
For small depths, we observe that this approximation can be used to approximate the $\mathrm{OTOC}^{(2)}$.
At larger depths, there may be individual circuits with large signal sizes, yielding a large SNR despite the approximation being poor for the most circuits, see Fig.~\ref{fig:study_outliers}.

In Fig.~\ref{fig:signal_spoofing_system_sizes}, we plot the contribution to $\mathcal{C}^{(4)}_{\mathrm{MC}}$ from various pieces of Eq.~\eqref{eq:diagonal_approx_otoc_sq}.
We observe that keeping only the ABBA pairing works comparably to keeping a larger part of the sum.
Furthermore, we note that the two pairings ABBA and AABB have the largest contributions to the $\mathrm{OTOC}^{(2)}$ signal, see Fig.~\ref{fig:study_outliers}.

Due to the strong correlation between the whole sum $\mathrm{OTOC}^{(2)}$ and the pairing ABBA at small depths, we look into the sum in Eq.~\eqref{eq:diag_app_otoc_sq_otoc_sq} without the pairing term ABBA.
This can be done on the quantum device by measuring the $\mathrm{OTOC}^{(2)}$ and the ABBA term and taking the difference; we now study this option numerically.
Given the chance of outliers, we only consider circuits generating the subtracted $\mathrm{OTOC}^{(2)}$ signals inside a $4$-$\sigma$ range around the mean of the $\mathrm{OTOC}^{(2)}$, see Fig.~\ref{fig:study_outliers}.
The results for the different system sizes are shown in Fig.~\ref{fig:subtracted_signal_spoofing_system_sizes}.
We observe that the SNR drops, and the Monte Carlo is unable to estimate the residual correlation.
Though subtracting the ABBA part induces some correlation, we cannot observe any systematic behavior.
We conclude that these two parts of the sum are fundamentally different.
One part can be estimated, by restricting Pauli strings to be the same after half the iteration, while the other part is far less correlated with those terms.

Note that simulating the ABBA pairing is not necessarily computationally simple, and should scale exponentially with system size and circuit depth.
However, it is exponentially cheaper than simulating the whole circuit, and may be feasible with sufficient computational power.

\subsubsection{Cached Monte Carlo and the sign problem}\label{sec:cached_mc_otoc2}

We now attempt to go beyond the classical pairings shown in Section\,\ref{sec:mc_otoc2} and see if cached Monte Carlo may be used to compute the off-diagonal contributions to $\text{OTOC}^{(2)}$. To this end, let us discuss a natural extension of the projection procedure Eq.\,\eqref{eq:projection_cached} onto the OTOC$^{(2)}$ computation:
\begin{gather}
    \mathcal{C}^{(4)} = \langle 0|({B}(T) {M})^4 |0\rangle,
\end{gather}
which now contains four copies of ${B}(T)$. At first, one sparsely time-evolves a single copy of the butterfly operator until a cache overfills at time $t$. In a natural generalization of the Eq.\,\eqref{eq:projection_cached} we decompose each of the four caches (before the first projection, they are identical) as
\begin{gather}
    \label{eq:site_deco_2}
    {B}_k(t) = \sum\limits_{\alpha_k} z_{k, \alpha_k} {P}_i^{\alpha_k} \otimes {B}_{k, \bar{i}, \alpha_k},
\end{gather}
where, ${P}_i^{\alpha_k}$ is a Pauli operator acting on a site $i$, ${B}_{k, \bar{i}, \alpha_k}$ is the ``adjoint'' operator defined at all sites except for $i$, $0 \leqslant k < 4$ enumerates the four caches, and the prefactors are normalized as $\sum_{\alpha_k} z_{k, \alpha_k}^2 = 1.
$\footnote{In practice, such decomposition is done by selecting all Pauli strings from the cache $\mathcal{C}$ that have a Pauli $P_i$ at a site $i$.}
Plugging this decomposition in, we obtain
\begin{gather}
\label{eq:OTOC2_deco}
    \mathcal{C}^{(4)} = \sum\limits_{\alpha_1, \alpha_2, \alpha_3, \alpha_4} \underbrace{\left(\prod_{k=0}^3 z_{k, \alpha_k}\right)}_{W(\boldsymbol \alpha)} \times \\ \nonumber \times \langle 0| \prod_{k=0}^3 \left({U}^{\dagger}( t, T) P_i^{\alpha_k} \otimes {B}_{\bar{i}, \alpha_k} {U}(t, T){M}\right) |0\rangle,
\end{gather}

Unlike the $\text{OTOC}^{(1)}$ case, where the trace condition only allows equal Pauli operators, here, of those $4^4 = 256$ terms, $64$ satisfy
\begin{gather}
\label{eq:condition}
    \text{Tr}\, \prod_k P^{\alpha_k}_i \neq 0,
\end{gather}
including contributions with $\alpha_1 \neq \alpha_2 \neq \alpha_3 \neq \alpha_4$ (that is, all four different). Having four copies of the butterfly operator leads to the {\it sign problem} manifested in the following way. On average, $z_{k, \alpha_k} \approx 1/2,$ and a typical weight $W(\boldsymbol \alpha) \sim 1/16$. But since there are 64 non-zero contributions, $W(\boldsymbol \alpha)$ does not form a probability distribution.

Naturally, one could neglect more terms, by requesting pairwise matching $P_{0} = P_{1},\, P_2 = P_3$ (or any permutation thereof). Being restricted to these 16 contributions, $W_{\text{restricted}}(\boldsymbol \alpha)$ forms a probability distribution. This restriction represents the computation of the {\it diagonal} part of $\text{OTOC}^{(2)}$.

Without such a restriction, one needs to re-weight the Monte Carlo measure by $\sum_{\boldsymbol \alpha} W(\boldsymbol \alpha) \approx 4$. Such reweighting rescales an observable by a factor $\sim 4$ each time a projection is done, and leads to the exponential variance growth. This remanifestation of the sign problem for OTOC$^{(2)}$ prevents the cached Monte Carlo from accessing the non-diagonal part of $\text{OTOC}^{(2)}.$

\subsubsection{Sign problem for $\overline{\mathcal{C}^{(4)}}$-preserving dynamics}\label{sec:permutationmontecarlo}
\begin{figure*}
    \centering
    \begin{overpic}[width=0.3\textwidth]{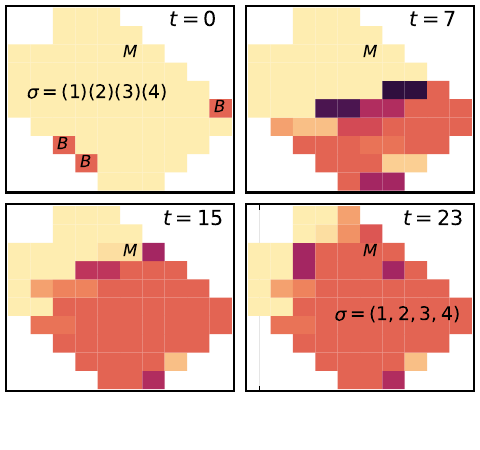}\put(4,97){\textbf{a}}\end{overpic}
    \begin{overpic}[width=0.32\textwidth]{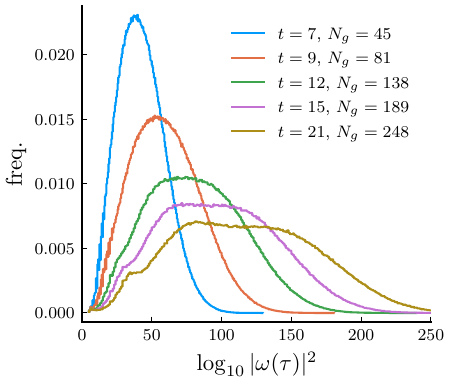}\put(9,90){\textbf{b}}\end{overpic}
    \begin{overpic}[width=0.32\textwidth]{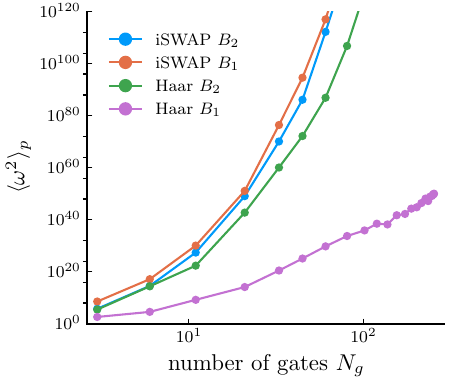}\put(9,90){\textbf{c}}\end{overpic}
    \caption{\textbf{Sign problem for computing $\boldsymbol{\overline{\mathcal{C}^{(4)}}}$.}  All parts of this figure use the $65$-qubit experimental $\mathcal{C}^{(4)}$ circuit layouts and either Haar-random two qubit gates or iSWAP-like gates, as indicated.
    {\bf a}, A snapshot of a typical permutation basis trajectory sampled using Eq.~\ref{eq:incrementalsampling}. Different colors correspond to different permutations -- a typical trajectory has a single large domain behind the operator front, which progresses outwards with the butterfly velocity. {\bf b}, The distribution of trajectory weights with iSWAP-like gates, cut at different depths $t$ after which the circuit has experienced $N_g$ total gates. In a large system, the distribution approaches log-normal with a mean and width that grows in time (not shown); the behavior here is similar but with distortions due to the boundaries of the circuit. {\bf c}, Variance of trajectory weights for a sample of $3 \cdot 10^6$ trajectories, using both iSWAP-like gates and Haar random two qubit gates, with two different choices for the basis in which the sampling takes place [see text]. In all cases, the weights grow tremendously quickly, indicating that sampling has a severe sign problem.
    }
    \label{fig:permutation_monte_carlo}
\end{figure*}

Unlike the case of $\mathcal{C}^{(2)}$, we find that Monte Carlo sampling is stymied by a sign problem even for the simpler problem of computing the average value $\overline{\mathcal{C}^{(4)}}$ over a Haar-random ensemble of gates. In this section we show that the computation of $\overline{\mathcal{C}^{(4)}}$ can be viewed as a globally truncated form of the dynamics similar to that achieved by vanilla Monte Carlo for $\mathcal{C}^{(2)}$, which, if not for the sign problem, would provide a starting point for approximating individual circuit $\mathcal{C}^{(4)}$ values. Thus, it is important to verify that this sign problem is actually a significant barrier. To quantitatively measure the severity of this sign problem, we consider sampling techniques that sample from a related positive distribution and absorb the sign into sample weights, as is done by quantum Monte Carlo practitioners. With this approach, we find that the growth of classical complexity is exponential in the volume of the operator front of $B(t)$, rather than the total volume of the operator $B(t)$, a reduction that is enabled by discarding irrelevant components of $4$ copies of the butterfly operator $B^{\otimes 4}(t)$.
However, this growth rate is still more than sufficient to render the circuits considered in this paper beyond any classical compute resources.

In this section we use a `folded' picture of the dynamics, in which $4$ copies of the butterfly operator $B$ are simultaneously evolved in an expanded Hilbert space of $4^4 = 256$ operators per site. 
Using a trace over $4$ copies of the original Hilbert space, we can rewrite the definition of $\mathcal{C}^{(4)}$ as
\begin{align*}
\mathcal{C}^{(4)} &= 2^{-N} \mathrm{Tr}\left[ \Pi_{(1,2,3,4)}^{\dagger} B^{\otimes 4}(t) M^{\otimes 4} \right] 
\\ &= 2^{-N} \mathrm{Tr}\left[\Pi_{(1,2,3,4)}^{\dagger} U(t)^{\otimes 4} B^{\otimes 4} U^{\dagger \otimes 4}(t) M^{\otimes 4} \right]
\end{align*}
Here, $\Pi_{(1, 2, 3, 4)}$ refers to a global permutation operator that permutes the $4$ copies of the Hilbert space, and we use the standard cycle notation for permutations --- thus, $(1, 2, 3, 4)$ refers to a permutation that sends copy $j$ to copy $(j+1) (\mathrm{mod} 4)$. We note that the $4!=24$ global permutations commute with the dynamical operator $U(t)^{\otimes 4}$ as it is the same on all $4$ copies.

Naively, this expanded Hilbert space greatly increases the complexity of dynamics algorithms, but it also enables truncating much of the information encoded in the combined $4$-copy operator $B^{\otimes 4}(t)$ while still approximately capturing the $\mathcal{C}^{(4)}$ front. 
To help explain the origins of this simplification, we note that this expression can be viewed as analogous to a time-ordered, equilibrium correlation function in the enlarged Hilbert space. Just as in that case, this correlation function undergoes a process of thermalization, in which the information content of $B^{\otimes 4}(t)$ becomes scrambled and inaccessible to local observables. For sufficiently deep circuits, $B^{\otimes 4}(t)$ locally approaches equilibrium. Thermalization of a Heisenberg-evolved operator effectively projects the operator onto the global symmetries of the evolution as equilibrium is approached; components of $B^{\otimes 4}(t)$ on these global symmetry operators are fixed in time, while all other components effectively decay away. The global symmetries of the $4$ copy evolution are in fact just the $24$ global permutation operators. 
%This thermalization of multiple tensor copies of a generic evolution has been previously studied under the name \emph{deep thermalization}.
The process of thermalization occurs behind the $\mathcal{C}^{(4)}$ operator front, as the equilibrium value of $\mathcal{C}^{(4)}$ is $0$. The benefit of exploiting thermalization is thus to discard information in $B^{\otimes 4}(t)$ behind the operator front.

% In the remainder of this section, we outline a Monte Carlo algorithm whose complexity growth is confined to the operator front, with the goal not of computing individual circuit $\mathcal{C}^{(4)}$ values but merely
In this section, we specifically want to compute the average value $\overline{\mathcal{C}^{(4)}}$ over a Haar random ensemble of single qubit gates inserted at every qubit and at every time step. 
This average is equivalent to projecting the dynamics of $B^{\otimes 4}(t)$ locally into the span of the permutation operators on the $4$-copies of each site. These projections are sufficient to truncate globally scrambled information in $B^{\otimes 4}(t)$ --- the average dynamics is non-unitary and explicitly drives the operator into a superposition of the $24$ global symmetries behind the front, with only information in the front being non-trivial. 

By construction, computing this average cannot reproduce the spread in $\mathcal{C}^{(4)}$ values for circuits that have the same two-qubit gates and only differ by single qubit gates. However, one can easily imagine improvements that can capture more variance between circuits by restricting the projections to a subset of space-time locations, at increased computational complexity. The cached Monte Carlo and TNMC algorithms for computing $\mathcal{C}^{(2)}$ can also be viewed as computing the result of projections at subsets of space-time locations, with those projections also resulting from averaging over an ensemble of single qubit gate insertions; just as those algorithms improve upon vanilla Monte Carlo, achieving some success in approximating $\mathcal{C}^{(2)}$, we expect that similar improvements can be made upon any algorithm that computes $\overline{\mathcal{C}^{(4)}}$. As these improvements use increased computational cost to capture more of the variance between circuits, we expect that the complexity of computing this average $\overline{\mathcal{C}^{(4)}}$ provides a lower bound for the difficulty of this entire class of algorithms. 
% We see below that a sign problem plagues computations of $\overline{\mathcal{C}^{(4)}}$, which we believe is a fundamental difference between $\mathcal{C}^{(2)}$ and $\mathcal{C}^{(4)}$.

Now let us explain the mechanics of computing
$$
\overline{\mathcal{C}^{(4)}} = 2^{-N} \mathrm{Tr}\left[ \Pi_{(1,2,3,4)}^{\dagger} \overline{B^{\otimes 4}(t)} M^{\otimes 4} \right] 
$$
by evolving the average $4$-copy operator $\overline{B^{\otimes 4}(t)}$. This calculation follows the strategy used for several random circuit calculations in the literature.
As the average over each single qubit gate insertion is independent, we can replace each two-qubit gate $u$ in the circuit, which appears in this operator evolution as a superoperator $S[u] = u^{\otimes 4} \cdot u^{\dagger \otimes 4}$, with a rotation averaged superoperator 
$$ 
T = \int dv_1 dv_2 dv_3 dv_4 S\left[(v_1 \otimes v_2) u (v_3 \otimes v_4) \right].
$$
As stated before, the operation of $T$ is precisely only non-zero on the span of the Haar-invariant operators, which are spanned by products of permutation operators on each site, and is zero on the orthogonal complement to this space. 

To proceed further, we pick a basis of operators for the span of the permutation operators, and expand our dynamics into a sum over basis trajectories. In our numerical experiments, the best choice of basis is the local permutation operators $\{\pi_\sigma\}$ which permute the $4$ copies of one site as the basis elements, which has the unique property that the fixed points of the average operator evolution can each be represented by a single computational basis element, rather than a large superposition. This basis is non-orthogonal and overcomplete, as only $14$ of the $24$ permutations are linearly independent. I will assume this choice below to simplify the exposition, but alternate choices proceed similarly.

Let us consider directly expanding the dynamics
as a sum over permutation basis trajectories

\begin{gather}
\mathcal{C}^{(4)} = 2^{-N} \sum_{\btau} T_{\btau} B_{\btau^0} M_{\btau^{f}}, \label{eq:c4traj}
\\ T_{\btau} \equiv \prod_{t=0}^{t_f - 1} \prod_{\langle i, j\rangle \in G(t)} \hat{T}^{\sigma_i^{t+1} \sigma_j^{t+1}}_{\sigma_i^{t}, \sigma_j^{t}}, \nonumber
\\ B(0)^{\otimes 4} = \sum_{\btau^0 = \{ \sigma_1^0, \ldots \sigma_N^0 \}} B_{\btau^0} \pi_{\sigma_1^0} \ldots \pi_{\sigma_N^0} \nonumber
\\ M_{\btau^f} = \mathrm{Tr} \left[ \Pi_{(1,2, 3,4)}^{\dagger} M^{\otimes 4} \pi_{\sigma_1^{t_f}} \ldots \pi_{\sigma_N^{t_f}} \right] \nonumber
\end{gather}
where the trajectory label $\btau = \{ \sigma_1^0, \sigma_2^0, \ldots  \sigma_N^{t_f}\}$ labels each assignment of permutations to all sites and times, $T_{\btau}$ labels the weight assigned to this trajectory, and $B_{\btau^0},M_{\btau^{t_f}}$ quantify the effect of the initial and final boundary conditions, respectively.

The final boundary condition produces trajectory weights that can be as large as $2^{3N}$ from the overlap between a permutation $\pi_{(1, 2, 3, 4)}$ on each site and $\Pi_{(1,2, 3,4)}^{\dagger}$. As $\mathcal{C}^{(4)}$ is a quantity bounded between $-1$ and $1$, this implies that vary rare trajectories could make order $1$ contributions, leading to difficulties in sampling. As such, we find that the following rescaling of the basis elements fixes the problem
$$
\pi_{\sigma} \rightarrow \frac{\pi_{\sigma}}{\mathrm{Tr}(\pi_{(1,2,3,4)}^{\dagger} \pi_{\sigma})},
$$
rendering each $M_{\btau^f}$ to be of order one. Moreover, the same replacement allows the $2^{-N}$ factor in Eq.~\ref{eq:c4traj} to be absorbed into the butterfly operator coefficients $B_{\btau^0}$, yielding order one numbers for those as well. Finally, because of the rescaling of the permutations, we have conservation of total output weight at each gate:
\begin{equation}
\sum_{\sigma_i^{t+1}, \sigma_j^{t+1}} \hat{T}^{\sigma_i^{t+1} \sigma_j^{t+1}}_{\sigma_i^{t}, \sigma_j^{t}} = 1.
\label{eq:pmcweightconservation_micro}
\end{equation}
This means that the total weight is conserved throughout the evolution, encoding the conservation law
$$
2^{-N} \mathrm{Tr}\left[ \Pi_{(1,2,3,4)}^{\dagger} B^{\otimes 4}(t) \right] = \frac{\mathrm{Tr}(B(t)^4)}{\mathrm{Tr}(I)} = \frac{\mathrm{Tr}(B(0)^4)}{\mathrm{Tr}(I)}.
$$
When $B(0)$ is a Pauli operator, this conserved value takes the value $1$.

In summary, we have \emph{almost} written the evolution of $\overline{B^{\otimes 4}(t)}$ as a Markovian evolution. However, the transfer matrix has signs, and the trajectory weights are both positive and negative, despite being conserved. We will use a reweighting scheme to try to sample anyway. To see how this works, it is useful to rewrite the observable as 
$$
\overline{\mathcal{C}^{(4)}} = \frac{\sum_{\btau} T(\btau) M(\btau)}{\sum_{\btau} T(\btau)}, 
$$
where $T(\btau) = T_{\btau} B_{\btau^0}$ and $M(\btau) = M_{\btau^f}$, and where the denominator \begin{equation}
   \sum_{\btau} T(\btau) = 1 \label{eq:pmcweightconservation},
\end{equation}
which can be derived from Eq.~\ref{eq:pmcweightconservation_micro}.
The sampling schemes under consideration sample trajectories according to a probability distribution $p(\btau)$ and reweight the trajectories
by $\omega(\btau) = T(\btau)/p(\btau)$, so that 
$$
\overline{\mathcal{C}^{(4)}} = \frac{\sum_{\btau} p(\btau) \omega(\btau) M(\btau)}{\sum_{\btau} p(\btau) \omega(\btau)} \approx \frac{\sum_{\text{sampled }\btau} \omega(\btau) M(\btau)}{\sum_{\text{sampled }\btau} \omega(\btau)}.
$$

The severity of the sign problem can be diagnosed by computing the sample variance of the denominator
$$
\text{Var}(\omega) = \sum_{\btau} p(\btau) \omega(\btau)^2 - \left(\sum_{\btau} p(\btau) \omega(\btau)\right)^2 = \langle \omega(\btau)^2 \rangle_{p} - 1,
$$
where $\langle \rangle_{p}$ indicates the average is taken with respect to the distribution $p$. As $\langle \omega \rangle_p = 1$ by construction, a number of samples $N_{s} \sim \text{Var}(\omega)$ is necessary to get an accurate estimate of the denominator. As an example, in the case where 
\begin{equation}
    p(\btau) = \frac{|T(\btau)|}{\sum_{\btau} |T(\btau)|}, \label{eq:abs_dist}
\end{equation}
one can show with Eq.~\ref{eq:pmcweightconservation} that
$$
\langle\omega^2\rangle_p = \frac{1}{\langle \text{sgn }T(\tau) \rangle_p^2};
$$
so an exponentially large $\langle\omega^2\rangle_p$ exactly corresponds to an exponentially small average sign, which is commonly used as an indicator of the sign problem --- see, for example Ref.~\cite{Troyer2005-kv}.

Instead of sampling from Eq.~\ref{eq:abs_dist}, we use the following incremental sampling algorithm to sample trajectories $\btau$ and corresponding weights $\omega(\btau)$ one gate at a time:
\begin{itemize}
    \item Start with an initial configuration $(\sigma^{0}_1, \sigma^{0}_2, \sigma^{0}_3, \ldots )$ and weight $\omega = 1$
    \item Stochastically apply each gate one at a time, replacing $\sigma_i, \sigma_j$  by $\sigma_i', \sigma_j'$ with probability
\begin{equation}
p^{\sigma_i' \sigma_j'}_{\sigma_i \sigma_j} = \frac{|\hat{T}^{\sigma_i' \sigma_j'}_{\sigma_i \sigma_j}|}{ \sum_{\sigma_i' \sigma_j'} |\hat{T}^{\sigma_i' \sigma_j'}_{\sigma_i \sigma_j}| }. \label{eq:incrementalsampling}
\end{equation}
\item  Update the weight variable $$\omega \leftarrow \omega \frac{\hat{T}}{p} = \omega\, \text{sgn} \left(\hat{T}^{\sigma_i' \sigma_j'}_{\sigma_i \sigma_j} \right)  \sum_{\sigma_i' \sigma_j'} |\hat{T}^{\sigma_i' \sigma_j'}_{\sigma_i \sigma_j}|.$$
\end{itemize}
At each step, the weight of the trajectory potentially changes sign and is multiplied by a factor $Z(\sigma_i, \sigma_j) = |\sum_{\sigma_i' \sigma_j'} \hat{T}^{\sigma_i' \sigma_j'}_{\sigma_i \sigma_j}| \geqslant 1$, where $\sigma_i$ and $\sigma_j$ are two permutations at the current location of a gate. One may expect then that the trajectory weights grow exponentially in magnitude with the number of applied gates. In practice, the growth is much slower than this implies for the following reason: $Z(\sigma_i, \sigma_j) = 1$ whenever $\sigma_i = \sigma_j$, which follows from the conservation law Eq.~\ref{eq:pmcweightconservation_micro}, and additionally $Z(\sigma_i, \sigma_j) = 1$ whenever $\sigma_i$ and $\sigma_j$ differ by a single transposition.
Furthermore, constant domains of a single permutation $\sigma$ are fixed under this update rule, and typical trajectories produce a large growing domain of a single permutation that expands outwards with the butterfly velocity, as illustrated in Fig.~\ref{fig:permutation_monte_carlo}~(a). Thus, only gates in the region of the operator front contribute to the growth of weights $\langle \omega(\tau)^2 \rangle_p$ and the severity of the sign problem. This suggests that our algorithm has successfully discarded irrelevant information in $B^{\otimes 4}(t)$ behind the front.

Numerical results for the growth of the sign problem for the $65$ qubit $\mathcal{C}^{(4)}$ circuits in this paper are shown in Fig.~\ref{fig:permutation_monte_carlo}~(b-c). 
Due to the repeated random multiplications that govern the growth of the weights, the weights follow a log-normal like distribution whose mean grows with depth of the circuit. The magnitude of this growth is rapid for $q=2$, producing astronomically sized weights. In Fig.~\ref{fig:permutation_monte_carlo}~(c), we compare the growth of the number of samples needed $N \sim \langle \omega^2 \rangle_p$ versus depth for the circuits used in this paper, which use iSWAP-like gates, as well as circuits with the same layout and Haar-random two qubit gates.
The circuits of this paper produce complexity growth that is far too rapid to be overcome with any existing classical compute resources --- indeed, the complexity of this algorithm far exceeds that of brute-force state vector calculations on $65$ qubits for these circuits. We see that averaging over a circuit ensemble with Haar random two-qubit gates decreases the difficulty compared to ensembles which fix the two qubit gate and only average over single qubit gates, but not sufficiently to make the problem tractable. In addition, we show comparisons between two different basis choices for the sampling; the $24$-dimensional permutation basis described above, labeled $B_1$ in the figure, and a basis using just the $14$ linearly independent permutations which appear most frequently in the trajectories, labeled $B_2$, which we show performs much worse for the Haar ensemble and similarly for the iSwap-like gates.

\subsubsection{Cost estimate of exact tensor network simulation of \texorpdfstring{OTOC$^{(2)}$}{OTOC-2}}
\label{sec:tn_estimates_otoc2}

\begin{figure*}
    \centering
    \begin{minipage}{0.46\textwidth}
      \begin{overpic}[width=0.85\textwidth]{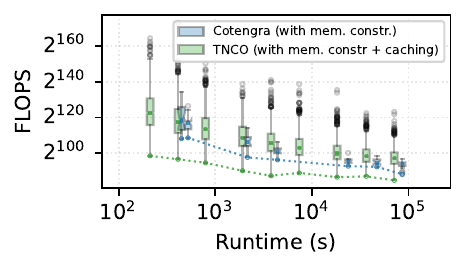}
        \put(10,55){\textbf{a}}
      \end{overpic}
      \begin{overpic}[width=0.85\textwidth]{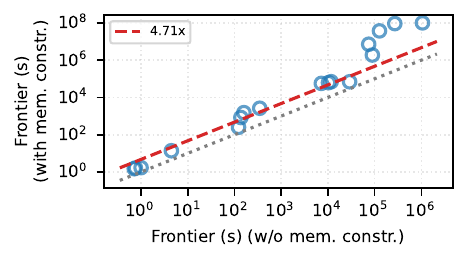}
        \put(10,55){\textbf{b}}
      \end{overpic}
    \end{minipage}
    \begin{minipage}{0.50\textwidth}
        \begin{overpic}[width=\textwidth]{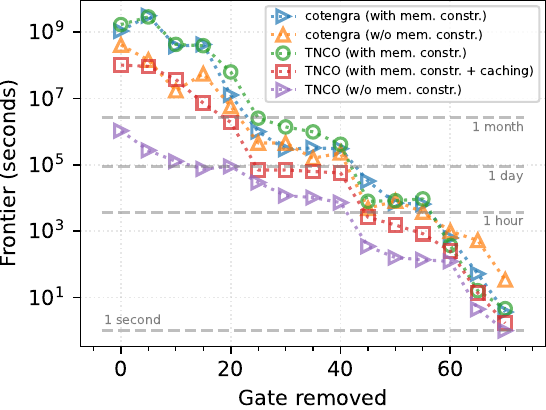}
          \put(10,77){\textbf{c}}
        \end{overpic}
    \end{minipage}
    \caption{\textbf{Optimization using tensor network contraction of the 65-qubit OTOC$\boldsymbol{^{(2)}}$ circuits.} \textbf{a}, Total time spent to optimize the tensor network contraction of the $65$-qubit
    OTOC$^{(2)}$ circuits with $N_{\rm 2D} = 1020$ on 
    $20$ \texttt{c2-standard-60} nodes on Google cloud, for zero gates removed. 
    \textbf{b}, Overhead by imposing memory constraints (each point correspond to a different circuit
    with a given number of gates removed). For both plots, a maximum width of $33$ and
    $20$ open qubits are assumed.
    \textbf{c}, Estimated time on Frontier for different numbers of gates removed.}
    \label{fig:tnc:frontier_time}
\end{figure*}

\begin{figure*}
    \centering
    \begin{overpic}[width=0.24\textwidth]{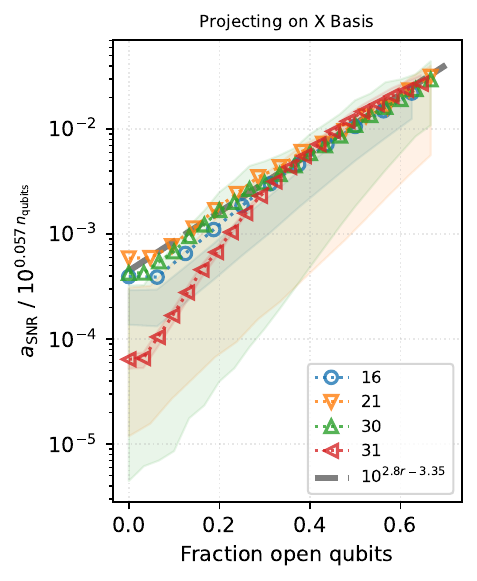}\put(10,97){\textbf{a}}\end{overpic}
    \begin{overpic}[width=0.24\textwidth]{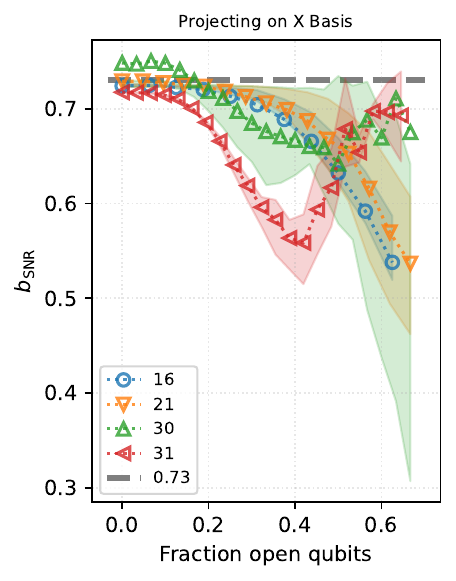}\put(10,94){\textbf{b}}\end{overpic}
    \begin{overpic}[width=0.24\textwidth]{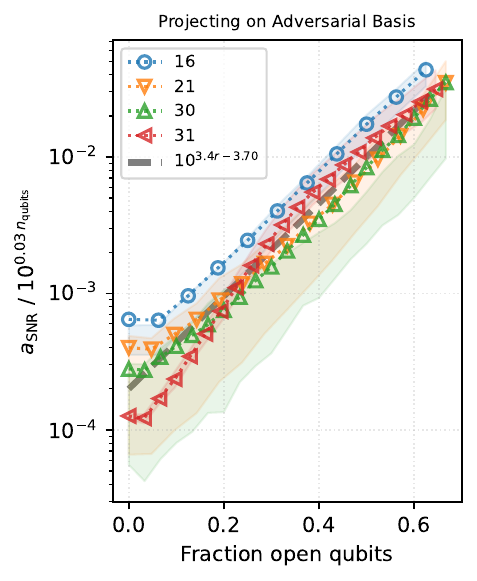}\put(10,97){\textbf{c}}\end{overpic}
    \begin{overpic}[width=0.24\textwidth]{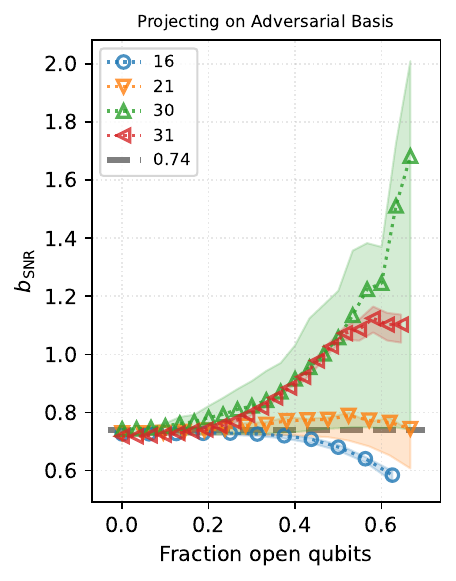}\put(10,94){\textbf{d}}\end{overpic}
    \caption{\textbf{Scaling coefficient for $\boldsymbol{{\rm SNR} = a_{\rm SNR} \sqrt{N_{\rm projs}} + b_{\rm SNR}}$, with $\boldsymbol{N_{\rm projs}}$ being the number of projections}. \textbf{a-b}, scaling for projecting on the $X$ basis. \textbf{c-d}, scaling for projecting on the the adversarial basis. The expected SNR for $65$ qubits and $20$ open qubits are respectively $16.67$ and $0.83$.
    In all plots, the shared area corresponds to the $25\%-75\%$ confidence interval.}
    \label{fig:tnc:snr_coeff}
\end{figure*}
\begin{table*}
\centering
\begin{tabular}{|cccccc|}
\hline
Gates Removed & SNR (MC) & $\log_2({\rm FLOPS})$ & Frontier Time & $\log_2({\rm FLOPS})^*$ & Frontier Time$^*$  \\
\hline
 0 & $\infty$ & 78.25 &       12 days, 5 hours & 84.83 &    3 years, 2 months\\
 5 &    61.17 & 76.29 &        3 days, 3 hours & 84.70 &   2 years, 10 months\\
10 &    55.20 & 75.20 &        1 day, 11 hours & 83.40 &     1 year, 2 months\\
15 &    39.83 & 74.43 &   20 hours, 45 minutes & 81.02 &    2 months, 22 days\\
20 &    35.45 & 74.69 &      1 day, 50 minutes & 79.09 &    21 days, 21 hours\\
25 &    24.36 & 73.08 &     8 hours, 6 minutes & 74.32 & 19 hours, 15 minutes\\
30 &    20.26 & 71.76 &    3 hours, 15 minutes & 74.33 & 19 hours, 24 minutes\\
35 &    18.72 & 71.57 &    2 hours, 51 minutes & 74.17 & 17 hours, 20 minutes\\
40 &    15.84 & 71.06 &                2 hours & 74.04 & 15 hours, 46 minutes\\
45 &    12.38 & 66.69 &  5 minutes, 48 seconds & 69.60 &           44 minutes\\
50 &    10.39 & 65.56 &  2 minutes, 39 seconds & 68.82 &           25 minutes\\
55 &     8.78 & 65.34 &  2 minutes, 17 seconds & 67.95 &           14 minutes\\
60 &     7.88 & 65.18 &   2 minutes, 2 seconds & 66.21 &            4 minutes\\
65 &     7.65 & 60.40 &           4.47 seconds & 62.03 &        13.78 seconds\\
70 &     4.19 & 58.25 &           1.01 seconds & 59.01 &         1.70 seconds\\
\hline
\end{tabular}
\caption{\textbf{Tensor network contraction estimates for the 65-qubit OTOC$\boldsymbol{^{(2)}}$ circuits with
$\boldsymbol{N_{\rm 2D} = 1020}$ by removing gates that minimize the SNR reduction.}
Columns with an asterisk
indicates contractions with a tensor width limited to $33$ and $20$ open qubits. 
Frontier time is computed assuming a theoretical maximum of $1.71\times 10^{18}$ FLOPS, with a $20\%$ of
efficiency.}
\label{tab:tnc:frontier_time}
\end{table*}

The results in this section form the basis for the classical cost estimates of exact simulation 
that were presented in the main text.
They are also used in conjunction with a gate-removal protocol in Section~\ref{sec:tngr} to develop heuristic
tensor-network-based methods for OTOC$^{(2)}$ simulation.
\\

\noindent {\bf Classical cost estimate of tensor networks for exact OTOC$^{\boldsymbol{(2)}}$ simulation ---} %\label{sec:tn_estimates} 
We observe that choosing the right basis to perform mid-circuit projections in TN contraction (as in Eq.\,\eqref{eq:TN_projections}) greatly affects 
the variance of sampling
\cite{Mi_OTOC_2021}. In particular, we (i) choose the basis orthogonal to the initial product state basis (in our case, $X$ basis orthogonal to the $Z$ basis).

Alternatively, one could (ii) choose in a basis induced by the last mid-circuit layer of single-qubit gates, hence simplifying the last layer of two-qubit gates and thus reducing the cost of contracting the tensor network.

%To better understand how the SNR would be affected by either selection of basis,  we performed a scaling analysis for OTOC$^{(2)}$ circuits up to $31$ qubits. 

To make sure we choose the best projection basis, we compare (i) and (ii) in small systems. Because of the rejection sampling algorithm, the SNR follows:
\begin{equation}
    {\rm SNR}_{\rm projs} = a_{\rm SNR} \sqrt{N_{\rm projs}} + b_{\rm SNR}.
\end{equation}
\noindent Fig.\,\ref{fig:tnc:snr_coeff} shows the coefficients $a_{\rm SNR}$ and $b_{\rm SNR}$ for both bases. The results are obtained by
averaging SNR fits over multiple circuit geometries (with different depth and number of
butterflies), with the shaded areas corresponding to the $25\%-75\%$ confidence
interval. 

By scaling the coefficient $a_{\rm SNR}$ and $b_{\rm SNR}$ with the number of total qubits, we can estimate the expected SNR for larger system sizes. For the case of $65$ qubits with $20$ qubits open and a single projection, we estimate (i) an $\text{SNR} = 16.67$ (\mbox{$1.55-21.60$} confidence interval) using the $X$ basis, and (ii) an $\text{SNR} = 0.83$ (\mbox{$0.73-0.87$}
confidence interval) using the adversarial basis. Therefore, we further work in the $X$ basis.

Fig.~\ref{fig:tnc:frontier_time} (right panel) shows the expected time to contract a single $65$-qubit OTOC$^{(2)}$
tensor network with either no memory constraints or with memory constraints and $20$ open qubits.
The Frontier time is calculated by assuming a peak of $1.71\cdot 10^{18}$ EXAFLOPS, with a $20\%$ of efficiency~\cite{Mi_OTOC_2021}.
The contraction path optimization has been performed on $20$ \texttt{c2-standard-60} Google Cloud nodes (see Fig.~\ref{fig:tnc:frontier_time}). 

Due to the contraction path structure (a single large tensor with many small tensors sequentially applied to it), the overhead of slicing is only $\sim\!\!5\times$ larger than without any memory constraints. Besides, caching becomes less relevant for a sufficiently large
number of gates removed. Finally, we compare \texttt{TNCO} to \texttt{Cotengra}, one of the widely used contraction path optimizers~\cite{gray2021hyper}.

\subsubsection{Reducing tensor network costs via gate removal}\label{sec:tngr}

It is known that one needs to only simulate the effective volume of a circuit in order to determine output from a given qubit~\cite{kechedzhi2024effective,patra2024efficient,tindall2024efficient}.
Claims comparing quantum and classical performance thus need to allow the classical computer to remove gates, or entire qubits.
Qubit removal is typically impossible for OTOC and 
%$\text{OTOC}^{(2)} - \text{cut}\,\text{OTOC}^{(2)}$
$\mathcal{C}^{(4)}_{\rm off-diag}$ 
circuits, as the long-distance nature of the signal allows the entire chip to be entangled.
However, gate removal is possible; for example, the gates outside the intersection of the butterfly and measurement light cones may be trivially removed (and for this work already are excluded from the gate counts $N_\text{2D}$ reported in the main text).
Removing any other gate comes with a potential systematic error, which can only be determined in exact simulation, or extrapolation till convergence. On the beyond-classical boundary, both options become challenging.

In this section, we outline an approximate method to remove gates from a circuit based on the Monte-Carlo approximation.
%This presents the strongest competition known for the simulation of $\Delta \mathcal{C}^{(4)}$ at the SNRs achieved in experiment.
This routine correctly determines the order in which the gates should be removed, however the estimates of the systematic error of the gate removal can differ by an order of magnitude when compared to the exact results. As such, we present a {\it range} of possible estimates for competing classical algorithms, as is reported in the main text.\\

\begin{figure}[t!]
    \centering
    \includegraphics[width=0.48\textwidth]{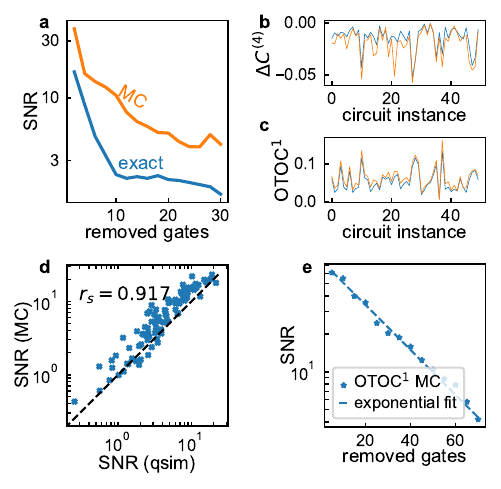}
    \caption{{\bf Monte Carlo-based gate removal. a,} In 50 instances of the 31-qubit circuit with 18 cycles, we computed SNR after removing each gate within the (i) exact simulation of $\mathcal{C}^{(4)}_{\rm off-diag}$ (qsim) and (ii) Monte Carlo simulation of OTOC$^1$ (MC). We ordered gates by their SNR impact increase, and show the SNR of cumulative removal. {\bf b-c,} At the point where $\text{SNR}_{\text{MC}} = 6$ (24 removed gates),
    $\text{SNR}_{\text{qsim}} = 2,$ and we compare of the signal without gate removal (blue) and with 24 gates removed (orange) for $\mathcal{C}^{(4)}_{\rm off-diag}$ (qsim, {\bf b}) and OTOC$^{(1)}$ (MC, {\bf c}). {\bf d,} The Spearman's rank correlation coefficient $r_s = 0.917$ indicates a high degree of correlation between (i) the ideal gate removal ordering and (ii) the Monte Carlo proxy. {\bf e,} The Monte Carlo proxy SNR as a function of the number of removed gates (ordered using the Monte Carlo ranking) in the 1020 OTOC$^{(2)}$ circuit. The exponential fit yields $\text{SNR}_{\text{MC}}(N_{\text{removed}}) \approx 74.0\, e^{-N_{\text{removed}} / 24.4}$.}
    \label{fig:MC_gate_removal}
\end{figure}

\noindent {\bf Heuristic classical algorithm for gate removal ---} 
Our algorithm proceeds by first ordering the gates by the reduction in SNR that results from their removal, and then trimming them one by one.
\footnote{In principle, this may not remove the optimal group of $G$ gates, but testing all possible combinations is exponentially hard in $G$.}
%
%However, in a two-dimensional grid with sides smaller than width 10, and with a discrete gate model instead of a continuous Hamiltonian, finite-size effects become significant.
%This makes it challenging to rank gates by their signal contribution without access to the exact simulation results.
We estimate the SNR from the real circuit using a proxy classical simulation.
Throughout this section and further, when a gate is removed, it is removed in all copies of $U$ and $U^{\dagger}$.
%We do not require this simulation to have a large SNR, but rather a similar relative sensitivity to the removal of a gate.

\begin{figure}
    \centering
    \includegraphics[width=\columnwidth]{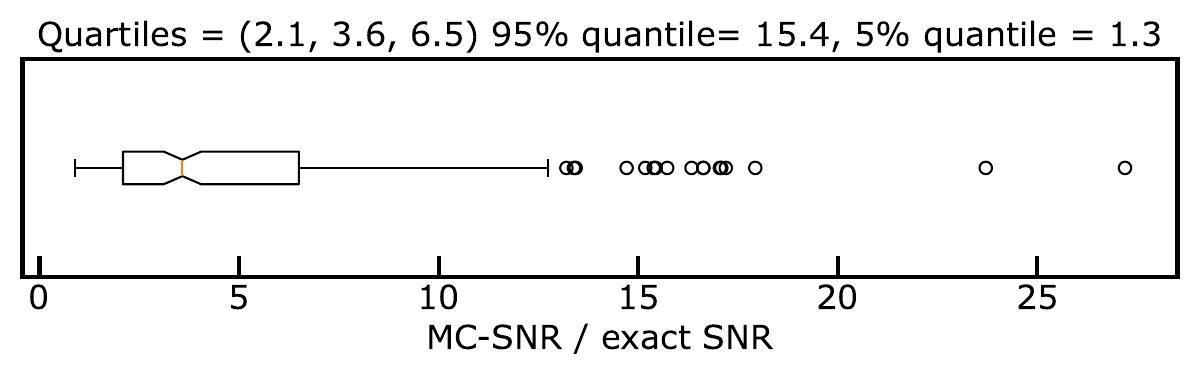}
    \caption{\textbf{Box plot of ratio between Monte Carlo SNR and exact SNR.} Positions of various quantiles are given in the title text.}
    \label{fig:SNR_scaling_boxplot}
\end{figure}

We test the gate removal using the vanilla Monte Carlo estimation (Section~\ref{sec:mc_intro}) of an OTOC$^{(1)}$ with the equivalent circuit $U$ (half of the initial OTOC$^{(2)}$ circuit). We benchmark using the Monte Carlo gate ranking in this OTOC$^{(1)}$ circuit as a proxy for the gate ranking in the ideal $\mathcal{C}^{(4)}_{\rm off-diag}$ circuits (with subtracted ABBA ``classical'' part) by simulating (i) $\mathcal{C}^{(4)}_{\rm off-diag}$ in the 31-qubit circuits with 18 cycles (50 instances) exactly and (ii) OTOC$^{(1)}$ within Monte Carlo and computing the SNRs of each individual gate removal. The comparison is shown in Fig.\,\ref{fig:MC_gate_removal}a. If both the gate removal order and cumulative removal SNR are computed within (ii), the SNR is highly overestimated, as compared to the (i) case. For instance, at the point where $\text{SNR}_{\text{MC}}$ crosses 6 (24 removed gates), $\text{SNR}_{\text{qsim}} = 2$, and in Fig.\,\ref{fig:MC_gate_removal}b-c we show the comparisons between the ideal (without removed gates) and the trimmed (24 removed gates) cases for (i) and (ii). While the Monte Carlo (ii) shows high agreement, the true signal (i) misses peaks, leading to the SNR of 2. We note that (i) uses the perfect gate ranking, but perhaps a non-optimal greedy cumulative removal starting with most insignificant gates. Finding the optimal strategy, however, would be exponentially hard in $N_{\text{removed}}$. Despite the perfect gate ranking used in (i), the SNR of
$\mathcal{C}^{(4)}_{\rm off-diag}$
is significantly lower than the Monte Carlo SNR obtained in (ii). Therefore, the SNR obtained in (ii) provides an upper bound on the desired SNR obtained in (i). In Fig.\,\ref{fig:SNR_scaling_boxplot}, we show the SNR ratios between (i) and (ii) for various system sizes and geometries to elaborate on the possible rescaling factor between the Monte Carlo proxy and the exact result.

Additionally, in Fig.\,\ref{fig:MC_gate_removal}d we look into the removal of individual gates. We observe that the Monte Carlo on average overestimates the SNR at the level of removal of individual gates, while the gate rankings are highly-correlated with the Spearman's rank correlation coefficient $r_s = 0.917$, which justifies using (ii) the gate ranking while trimming gates in the 65-qubit circuits.\\

\noindent\textbf{Application to OTOC$^{\boldsymbol{(2)}}$ circuits considered in this work ---}
We now turn to estimating the complexity of simulating the $1020$-gate data in Fig. 4f of the main text via tensor network methods, targeting the same SNR $\sim 3.5$ that was experimentally achieved.
In Fig.~\ref{fig:MC_gate_removal}e we plot the Monte-Carlo SNR estimates for this circuit with different numbers of gates trimmed, and observe that this follows a rough exponential decay law.
Following this ordering, it remains to estimate the two systematic biases described above; the bias from the truncation itself, and the residual sampling noise from the finite number of tensor network projections Eq.\,\eqref{eq:TN_projections}.
%As one can observe from the above results, we have accurate estimates of neither.
%In practice, one could simply start from a large truncation and re-add gates to the unitary until convergence, however as we are considering circuits at the beyond-classical boundary this is not feasible for our study.
%Given these fundamental limitations, we resorted to using small-size circuits to get reasonable assumptions and bounds for the classical simulations at scale, often favoring the classical side over the quantum one.

\begin{figure*}[th!]
    \centering
    \begin{overpic}[width=0.40\textwidth]{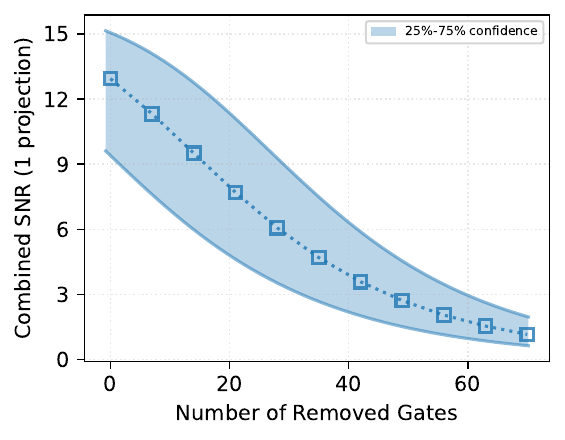}\put(5,80){\textbf{a}}\end{overpic}
    \hspace{3pt}
    \begin{overpic}[width=0.48\textwidth]{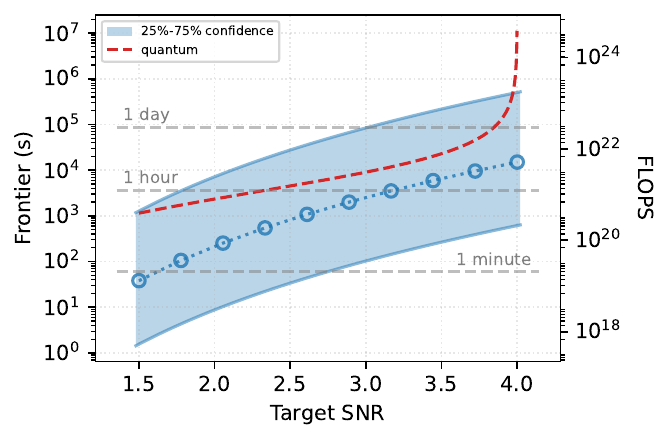}\put(5,67){\textbf{b}}\end{overpic}
    \caption{\textbf{SNR estimates from tensor network contraction calculations, after removing gates.} {\bf a}, Estimated SNR for the $65$-qubit OTOC$^{(2)}$ circuits with $N_{\rm 2D} = 1020$ by
    removing gates, for a single projection. {\bf b},
    The runtime on Frontier for different target SNR. The dashed-line corresponds to the time
    to reach the same target SNR using the quantum device.}
    \label{fig:tnc:combined_snr}
\end{figure*}

Combining the results from Fig.~\ref{fig:SNR_scaling_boxplot} (which shows the range of estimates for the multiplier between the Monte Carlo proxy SNR and our true SNR), and the results from Fig.~\ref{fig:tnc:snr_coeff},
we can extrapolate the optimal number of removed gates and number of projections to reach a target
SNR (see Fig.~\ref{fig:tnc:combined_snr}). 
We assume that the error induced by removing gates is uncorrelated with the error induced by removing gates. Fig.~\ref{fig:tnc:combined_snr}\,(left) shows the combined SNR by assuming a single projection. 
In the plot, we also assume that the multiplicative scaling factor between Monte Carlo SNR
and exact SNR is $3.6$ ($2.1$ --- $6.5$ confidence interval of $25\%-75\%$), as well as an SNR of $16.67$
for a single projection (see Section \ref{sec:tn_estimates_otoc2}). We combine these two estimates in Fig.\,\ref{fig:tnc:combined_snr}\,(right), showing the expected time to contract
a single tensor network on Frontier for a given target SNR.
% In Fig.~\ref{fig:tnc:combined_snr} we report the expected time on Frontier to contract a single
% OTOC$^{(2)}$ circuit. These estimates are obtained by combining the error
% of removing gates in the OTOC$^{(2)}$ circuit and the error of approximating the TNC.
The runtime on Frontier
is computed by dividing the FLOPS for the best tensor network contraction path we found using \texttt{TNCO} by the theoretical peak performance of Frontier ($1.71\cdot 10^{18}$ FLOPS), assuming a $20\%$ FLOPS efficiency~\cite{Mi_OTOC_2021}.

%More precisely, 
%we assume that the SNR by removing gates will decrease as:
%\begin{equation}
%    {\rm SNR} = \frac{259}{\alpha} \cdot e^{-N_{\rm removed} / 19},
%\end{equation}
%with $N_{\rm removed}$ the number of removed gates, and it will increase by using more projections as:
%\begin{equation}
%    {\rm SNR} = \mu \cdot \sqrt{N_{\rm projs}},
%\end{equation}
%with $N_{\rm projs}$ the number of projections. Table~\ref{tab:tnc:combined_snr}
%reports the combined percentiles and the expected FLOPS / seconds on Frontier to compute the expectation value 
%of OTOC$^{(2)}$ circuits.

\subsubsection{Pauli-weight truncation algorithms}
Another possible approach to the $\text{OTOC}^{(2)}$ computation is the {\it truncation in the Pauli space}~\cite{begusic2024, angrisani2025}. More specifically, one time-evolves the butterfly operator ${B}(t)$ in the Pauli space in sparse representation as in Eq.\,\eqref{eq:diag_app_otoc_sq_otoc_sq}, but keeps only the Pauli strings with a limited number of non-identity entries, that is $X$, $Y$ or $Z$. Importantly, unlike the applications considered in 
Refs.~\cite{begusic2024, angrisani2025}, $\text{OTOC}^{(2)}$ contains four copies of the butterfly operator. This makes it difficult to pre-select non-zero Pauli contributions based only on their Pauli weight $L$. In this section, we will argue that this approach is not applicable to the computation of higher-order OTOCs.\\

\noindent\textbf{Large-weight Pauli contributions to OTOC$\boldsymbol{^{(k)}}$ ---} In contrast to expectation values and correlation functions where low-weight Pauli strings provide the dominant contribution, in the case of circuit-to-circuit variance of OTOC$^{(k)}$ low weight Pauli strings do not contribute. Indeed a vanishing commutator $[B(t), M]=0$ implies OTOC$^{(k)}=1$ and zero circuit-to-circuit variance. Whereas non-vanishing commutator $[B(t), M]\neq0$ requires the expansion of $B(t)$ into Pauli strings to contain strings of sufficient weight to span the distance between $B$ and $M$ in space, corresponding to large weights for the geometries we consider. The distribution of Pauli strings over their Hamming weights is determined by the physics of operator spreading. For a random circuit a typical string has at each qubit operator $I, X, Y, Z$ with roughly equal probability. Therefore for a system of $n$ qubits  we expect a typical Pauli weight of
$\kappa n$, where $\kappa = \frac{3}{4}$ for OTOC. Below we provide a more accurate estimate specific to OTOC$^{(2)}$.
Including this large Pauli weight makes Pauli truncation algorithms impractical for OTOC$^{(k)}$.\\

\noindent\textbf{Counting of nonzero contributions for OTOC$\boldsymbol{^{(2)}}$ ---} To better understand why the pre-selection of non-zero Pauli contribution fails for OTOC$^{(2)}$, let us consider the following combinatorial argument. First, we estimate, Pauli strings of which weight $L$ give the largest number of contributions $K(L)$ to the expectation. Here, $K(L)$ will be the total number of ways how one could pick four Pauli strings with Pauli weight $L$, such that they lead to a non-zero trace. For simplicity, we will assume that only four Pauli-strings with the same Pauli weight would satisfy the non-zero trace condition, which will serve as a lower bound.

In the sector of Pauli strings of the Pauli weight of exactly $L$, at each site, we will consider having ${I}$ with the probability $1 - 3 p = 1 - L / N$, and each of the ${X},$ ${Y}$ and ${Z}$ with the probability $p = L / (3N)$. Here, $N$ is the number of sites. Sampling an operator at each of the $N$ sites, we obtain
\begin{gather}
    P(p) = \text{Prob}\left(\text{Tr}\, \left[P_1 P_2 P_3 P_4 \right] \neq 0 \right) = \\ \nonumber = 192 p^4 - 192 p^3 + 72 p^2 - 12 p + 1.
\end{gather}

\begin{figure}[t!]
    \centering
    \begin{overpic}[width=0.48\textwidth]{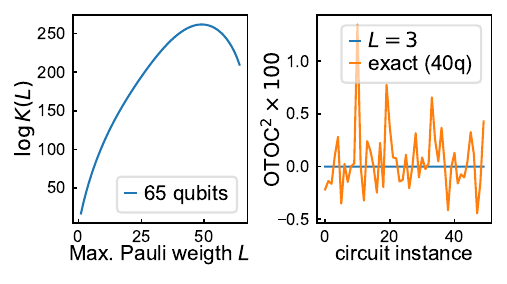}
      \put(5, 55){\textbf{a}}
      \put(52, 55){\textbf{b}}
    \end{overpic}
    \caption{{\bf Simulation of OTOC$\boldsymbol{^{(2)}}$ using Pauli weight truncation. a}, Combinatorial estimation of the number of contributions to $\text{OTOC}^{(2)}$ coming from each $L$-truncated Pauli sector. {\bf b}, Application of the Pauli truncation algorithm to the 40-qubit $\text{OTOC}^{(2)}$ circuit shown in the main text.}
    \label{fig:Pauli_weight}
\end{figure}

In total, there are ${N \choose L} 3^L$ different Pauli strings with the weight $L$, which gives $\log\,K(L)~=~ N \log P(p) + 4 L \log\,3 + 4 \log {N \choose L}$. The curve $\log\,K(L)$ for $N = 65$ (beyond-classical dataset) is shown in Fig.\,\ref{fig:Pauli_weight}a. The curve peaks at $L^* = 47$. At this Pauli weight, there are $R(L^*) \sim 10^{38}$ possible Pauli strings. \\

\noindent\textbf{Computation of OTOC$^{\boldsymbol{(2)}}$ from the butterfly operator representation ---} Given the Pauli-weight truncated representation of the butterfly operator (with $R$ terms), one requires to compute the $\text{OTOC}^{(2)}$ expectation. In the case higher-order OTOCs, the trace condition leaves numerous pairing options (64 in the case of OTOC$^{(2)}$) at each site. Therefore, we present an algorithm to directly compute $\text{OTOC}^{(2)}$ given some representation of ${B}$ in the Pauli basis. 

For simplicity, we consider the $|0\rangle = |00\ldots0\rangle$ initial state which is an eigenstate of the measurement operator, and 
\begin{gather}
    \mathcal{C}^{(4)} = \langle 0|{B}{M}{B}{M}{B}{M}{B} |0\rangle.
\end{gather}
First, we compute $|\psi\rangle = {M} {B}|0\rangle$ in $\mathcal{O}(R)$ time. Then, the expression reduces to
\begin{gather}
    \mathcal{C}^{(4)} = \sum\limits_{a, b} \psi^*(a) \psi(b) \langle a|{B} {M} {B}|b\rangle,
\end{gather}
where the sums in $a$, $b$ run over $\mathcal{O}(R)$ non-zero elements of the wave function $|\psi\rangle$. For each of the $\mathcal{O}(R^2)$ combinations of $a$ and $b$, we build $|\psi_a\rangle = {B}|a\rangle$, $|\psi_b\rangle = {B}|b\rangle$, and compute $\langle \psi_a|{M}|\psi_b\rangle$ in linear time. The total algorithm complexity is $\mathcal{O}(R^3)$.\\

\noindent\textbf{Application in a realistic simulation ---} We apply the Pauli truncation algorithm to the 40-qubit $\text{OTOC}^{(2)}$ circuits shown in the main text. For $L = 3$, the maximum cache size is $R = 3^4 \times {40 \choose 3} = 0.8 \times 10^6$, and we can run the $\mathcal{O}(R^3)$ computation algorithm. For $L = 4$, with the cache size of $R = 23 \times 10^{6}$, the simulation becomes infeasible. The results for the $L = 3$ case are shown in Fig.\,\ref{fig:Pauli_weight}b. For 40 qubits, we expect the most contributions to come from the sector with $L \sim 15$. Due to the excessive truncation, the result is vanishingly small. Likewise, renormalization of cache after each truncation leads to non-unitarity, and values of $\text{OTOC}^{(2)}$ uncorrelated to the ground truth.\\

\noindent\textbf{Noisy OTOC$\boldsymbol{^{(k)}}$ and Pauli-weight truncation  ---} Noise in practical quantum processors provides an avenue to reduce the cost of Pauli path algorithms for correlation functions. Noise affects Pauli strings exponentially in their Hamming weight. Therefore there is a natural criteria for truncation of Pauli strings with the largest weights reducing the cost of the computation. This approach however does not reduce the cost of computing OTOC$^{(k)}$ provided the noise in the device is sufficiently weak. Indeed for an output of the Pauli truncation algorithm to demonstrate substantial correlation with the noise-free OTOC$^{(2)}$ Pauli strings of typical weights need to be included. For the circuit geometries we consider in this paper the distribution of Pauli weights is relatively narrow around its typical value. Therefore exploiting the effect of noise does not lead to a substantial reduction in the cost of Pauli truncation algorithm compared to its noise free version.

%Scrambling by a random circuit increases operator entanglement whereas noise decreases it. As long as the %scrambling process is sufficiently fast compared to the rate of noise computation of OTOC$^{(k)}$ the distribution of Pauli strings with respect to their Hamming weight is approximately the same as in the noiseless case. Therefore the algorithms relying on truncation of strings with large weights do not reduce the cost of the classical computation.
\label{sec:pauli_truncation}

\subsubsection{Matrix product state methods}\label{sec:mps_mpo}
In this section, we discuss the simulation of $\text{OTOC}^{(2)}$ large-scale circuits using matrix product state (MPS) and matrix product operator (MPO) methods. These methods' computational complexity is controlled by the entanglement generated in the time-evolution. At a cost of imperfect fidelity, these methods compress a wave function by truncating small Schmidt coefficients. At the same time, the generated entanglement could be decreased by the gate removal procedure (see Section\,\ref{sec:tngr}), which also comes at the fidelity cost. In this section, we will demonstrate that the large entanglement generated in our experiments makes the application of such methods prohibitively expensive.

We first consider an MPS simulation. We enumerate the sites of a two-dimensional circuit in the {\it snake} manner, and form an MPS state
\begin{gather}
\label{eq:state_MPS}
    |\psi_\chi\rangle = {U}^{\dagger} {B} {U} {M} {U}^{\dagger} {B} {U}|0\rangle
\end{gather}
we then truncate the bond dimension to $\chi$ after each gate. Lastly, we measure $\mathcal{C}^{(4)}_\chi = \langle \psi_\chi | {M} |\psi_\chi\rangle.$

We consider the MPS simulation cost of the beyond-classical $\text{OTOC}^{(2)}$ circuits in the two limiting regimes: (i) no truncation, but the entanglement is lowered by removing as many gates as possible, (ii) no gates removed, but the Schmidt values are truncated. In both these cases, we consider a minimum $\text{SNR} \sim 3$.\\

\noindent\textbf{Entanglement in beyond-classical circuits ---} 
Determining the entanglement entropy for large quantum systems is a daunting task. Indeed, the complexity of computing the entanglement entropy by splitting a quantum system of $n$ qubits in two is proportional to $\mathcal{O}(2^{3n/2})$. 
In some cases, bounds on the entanglement entropy can be computed efficiently by exploiting the fact that Clifford gates form a 2-design. As shown in \cite{Morvan2024_rcs, kechedzhi2024effective}, if all 2-qubit gates are iSWAPs and all the 1-qubit gates are uniformly sampled from $Z^p \sqrt{X} Z^{-p}$, with $p\in\{-1,\,-1/4,\,-1/2,\,\ldots,\,3/4\}$, then the entanglement entropy of such circuits is bounded by the entropy of the closely related Clifford circuits where $p$ is sampled from $p\in\{-1,\,-1/2,\,0,\,1/2\}$ instead. 
More precisely,
the purity $\overline{ {\rm Trace} [\rho_L^2 ]}$ averaged over circuits, where $\rho_L$ is the reduced density matrix after tracing half of the system, is the same for the two ensembles (see Section~H5 of \cite{Morvan2024_rcs} for more details).
%Since the von Neumann entropy, $-{\rm Trace}[\rho_L \log_2 \rho_L]$, and the R\'enyi entropy, $-\log_2 {\rm Trace}[\rho_L^2]$, are the same for Clifford circuits, 
Using the Jensen's inequality, it is possible to bound the entanglement entropy for non-Clifford circuits as
\begin{equation}\label{eq:entropy_bound_clifford}
    -\overline{{\rm Trace}[\rho_L \log_2 \rho_L]} \geq
    -\log_2 \overline{ {\rm Trace} [\rho_L^2] }_{\rm Clifford},
\end{equation}
with $-\overline{ {\rm Trace} [\rho_L^2] }_{\rm Clifford}$ being the average R\'enyi entropy using Clifford gates only.

\begin{figure}
    \centering
    \includegraphics[width=0.45\textwidth]{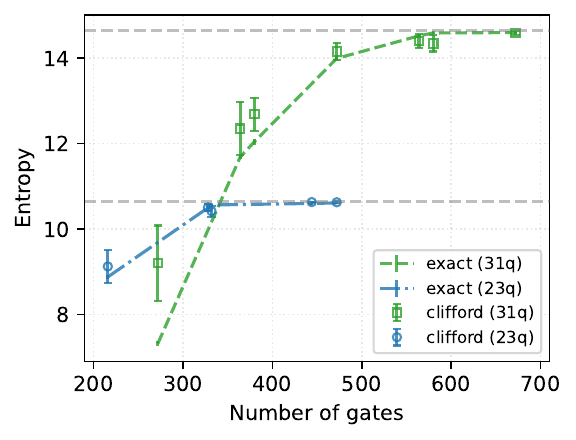}
    \caption{\textbf{\bf Entanglement entropy by varying the circuit depth for OTOC$\boldsymbol{^{(2)}}$ with 23 qubits and 31 qubits.}  The two sets of circuits use the same gate ensemble used experimentally. The horizontal gray-dashed lines correspond to the random Haar limit. The R\'enyi Clifford entropy well estimates the exact von Neumann entropy, including the saturation of the entropy at large depth.}
    \label{fig:cl_entropy_zero_removed}
\end{figure}

\begin{figure*}
    \centering
    \begin{overpic}[width=0.30\textwidth]{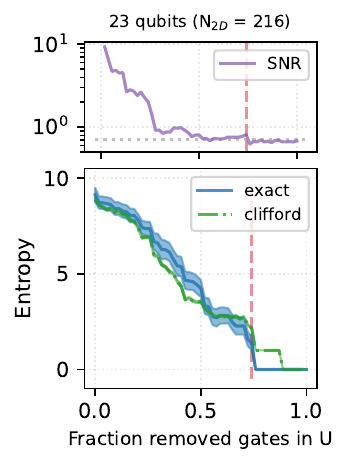}\put(5,100){\textbf{a}}\end{overpic}
    \begin{overpic}[width=0.30\textwidth]{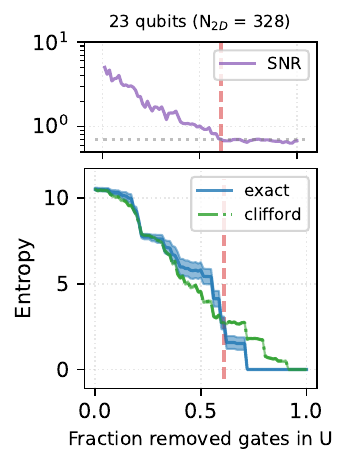}\put(5,100){\textbf{b}}\end{overpic}
    \begin{overpic}[width=0.30\textwidth]{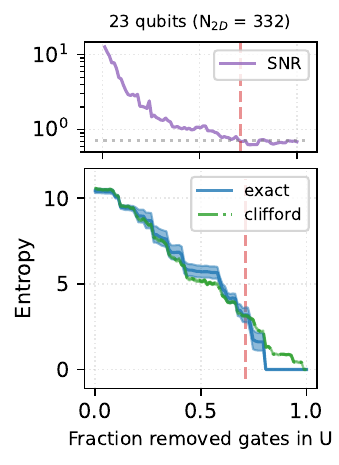}\put(5,100){\textbf{c}}\end{overpic}\\
    \begin{overpic}[width=0.30\textwidth]{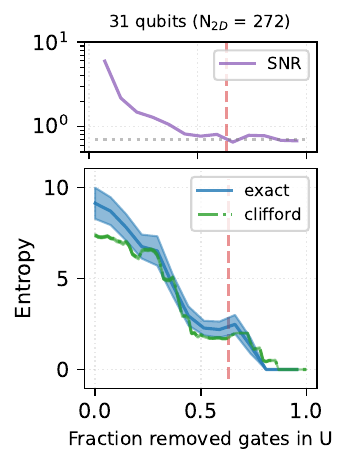}\put(5,100){\textbf{d}}\end{overpic}
    \begin{overpic}[width=0.30\textwidth]{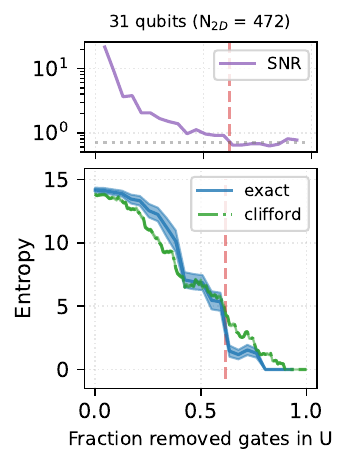}\put(5,100){\textbf{e}}\end{overpic}
    \begin{overpic}[width=0.30\textwidth]{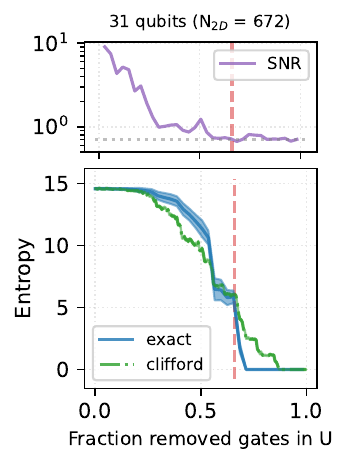}\put(5,100){\textbf{f}}\end{overpic}
    \caption{{\bf Comparison between the exact von Neumann entropy and the R\'enyi Clifford entropy for OTOC$\boldsymbol{^{(2)}}$ circuits with 23 qubits (a-c) and 31 qubits (d-f).} The two sets of circuits use the same gate ensemble used experimentally. The horizontal gray-dotted line corresponds to the baseline SNR of $1 / \sqrt{2}$, while the horizontal red-dashed line corresponds to the number of removed gates when the SNR drops below the baseline. As long as the SNR is larger than $1 / \sqrt{2}$ the R\'enyi Clifford entropy is a good estimator of the exact von Neumann entropy.}
    \label{fig:cl_entropy_fixed_depth}
\end{figure*}

\begin{figure}[t!]
    \centering
    \includegraphics[width=0.48\textwidth]{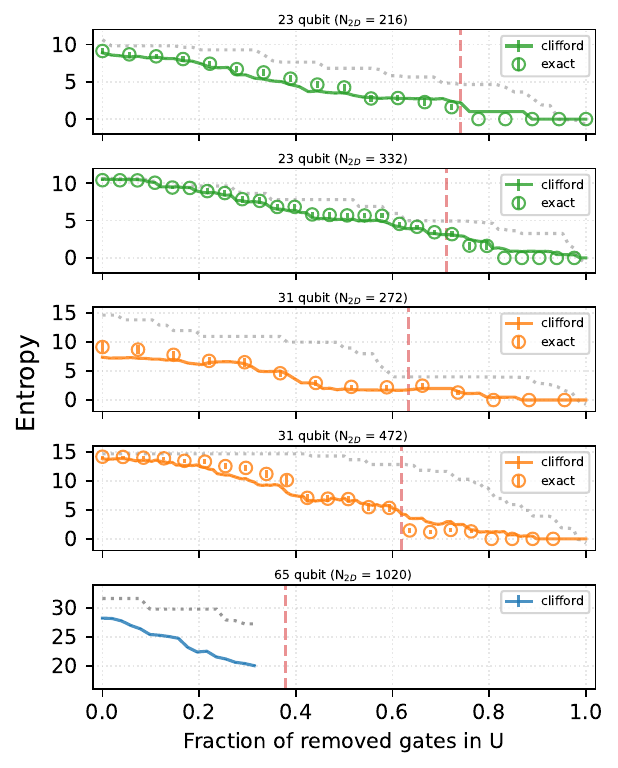}
    \caption{{\bf Entanglement entropy for OTOC$\boldsymbol{^{(2)}}$ circuits of different system sizes.} The plot compares the exact von Neumann entropy $-\overline{{\rm Trace}[\rho_L \log_2 \rho_L]}$ compared to the Rényi entropy $-\log_2 \overline{{\rm Trace} [\rho_L^2]}_{\rm Clifford}$. Vertical dashed-red lines correspond to the fraction of number of gates in $U$ for which the SNR drops below the baseline of $1/\sqrt{2}$, while dotted-gray lines correspond to the  maximum entanglement entropy for random Haar states. The SNR limit for $65$ qubit is computed by using estimates from cached Monte Carlo. For 65-qubit circuit with $N_{\textrm 2D} = 1020$, 70 removed gates correspond to the fraction $(4 \times 70) / 1020 \approx 0.274.$}
    \label{fig:clifford_entangl_scaling}
\end{figure}

Equation~\eqref{eq:entropy_bound_clifford} is valid for the ensembles described above~ \cite{Morvan2024_rcs, kechedzhi2024effective}, while the experimental circuits presented in this paper have a different gate ensemble. Nevertheless, small-scale numerical results in Fig.~\ref{fig:clifford_entangl_scaling} show that the Clifford purity in Eq.~\eqref{eq:entropy_bound_clifford} can be still used as an estimator for the experimental circuits.
Indeed, Fig.~\ref{fig:cl_entropy_zero_removed} compares the exact entanglement entropy to the Clifford R\'enyi entropy for OTOC$^{(2)}$ circuit with $23$ and $31$ qubits respectively, by varying the number of entangling gates. Both set of circuits use the same gate ensemble used in the experimental setup. More precisely, the figure shows that the Clifford R\'enyi entropy is a good estimator of the entanglement entropy, correctly predicting when the entanglement entropy saturates (indicated by the dashed-gray line). We also report the comparison by varying the number of removed gates in $U$, Fig.~\ref{fig:cl_entropy_fixed_depth}. The main observation is that the Clifford R\'enyi entropy is still a good estimator of the entanglement entropy even when gates are removed, as long as the final SNR is larger than the baseline $1/\sqrt{2}$.

Therefore, assuming that the Clifford estimates hold for larger systems, we expect an entanglement entropy of $\sim\!30$ for the 65 qubits and $N_{\rm 2D} = 1020$ entangling gates (see Fig.~\ref{fig:clifford_entangl_scaling}).
\\

\noindent\textbf{Simulation cost without truncation ---}
We begin with analyzing the case (i), where we decrease the maximum wave function entanglement entropy by removing gates from the circuit. As shown in Section\,\ref{sec:tngr}, removing $\sim\!70$ gates gives the SNR of at most 3. After such removal, as shown in Fig.\,\ref{fig:clifford_entangl_scaling}, the Rényi entropy is $\sim\!21$, which leads to the required bond dimension of $\chi_{\text{min}} = 2^{21}$, and the respective FLOP count of at least $8 \chi_{\text{min}}^3 = 2^{66} = 7.3 \times 10^{19}$, assuming single complex precision. With the Frontier FLOPs of $2 \times 10^{18}$ and assuming effective performance of 20\,\%, we obtain a single SVD run time of at least 3 minutes, making this approach not faster than the direct tensor network contraction.\\

\begin{figure}[t!]
    \centering
    \includegraphics[width=0.48\textwidth]{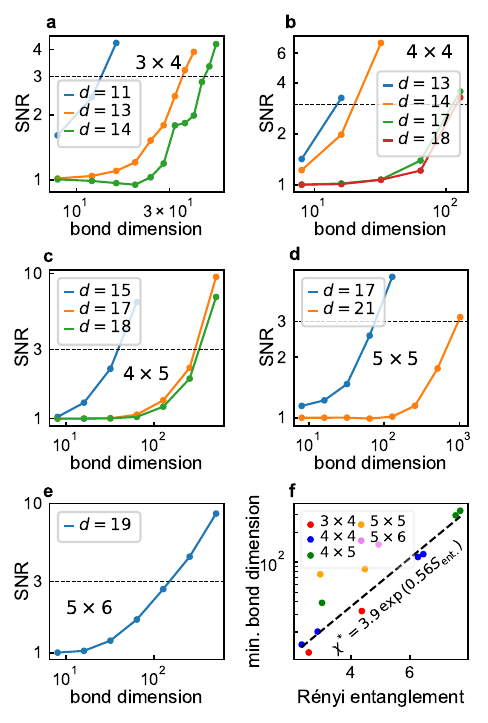}
    \caption{{\bf MPS simulation of OTOC$\boldsymbol{^{(2)}}$ circuits of size $\boldsymbol{l_x \times l_y}$. a-e,} SNR as a function of the maximum bond dimension for several depths. We extract $\chi^*_{\text{SNR} = 3}$ using a polynomial extrapolation. {\bf Fitting, f} Fitting $\chi^*_{\text{SNR} = 3}$ as a function of $S_{\text{Rényi}}^{\text{max}}$ using an exponential fit. }
    \label{fig:MPS_extrapolations}
\end{figure}

\noindent\textbf{Simulation cost with truncation ---} 
Finally, we consider the case where the truncation is the sole source of finite SNR. We first estimate the minimum bond dimension $\chi^*_{\text{SNR} = 3}$ required to reach $\text{SNR} \sim 3$, by linking $\chi^*_{\text{SNR} = 3}$ to the maximum Rényi entropy $S_{\text{Rényi}}^{\text{max}}$ generated during the composition of a state Eq.\,\eqref{eq:state_MPS}. We do that by running MPS simulations in smaller systems with up to 30 qubits. The results are shown in Fig.\,\ref{fig:MPS_extrapolations}. The panels Fig.\,\ref{fig:MPS_extrapolations}a-e show SNR as a function of maximum bond dimension in the systems of various size and depth. Fig.\,\ref{fig:MPS_extrapolations}f shows the minimum required bond dimension $\chi^*_{\text{SNR} = 3}$ as a function of the maximum Rényi entropy $S_{\text{Rényi}}^{\text{max}}$. Fitting as an exponential gives 
\begin{gather}
    \chi^*_{\text{SNR} = 3} = 3.9\,e^{0.56\, S_{\text{Rényi}}^{\text{max}}},
\end{gather}
where we note that larger systems' points mostly lie above the fit, which signals that the fit likely presents a conservative estimate. As shown in Fig.\,\ref{fig:clifford_entangl_scaling}, the Rényi entropy in the beyond-classical OTOC$^{(2)}$ circuits without gate removal is $\approx 29$, which gives $\chi^*_{\text{SNR} = 3} \sim 30 \times 10^{6}$, and therefore per-SVD $2.8 \times 10^{23}$ FLOPS, resulting into 8 days on Frontier assuming $\sim\! 20\,\%$ FLOPS efficiency, making this approach impractical.\\

\begin{figure}[t!]
    \centering
    \includegraphics[width=0.48\textwidth]{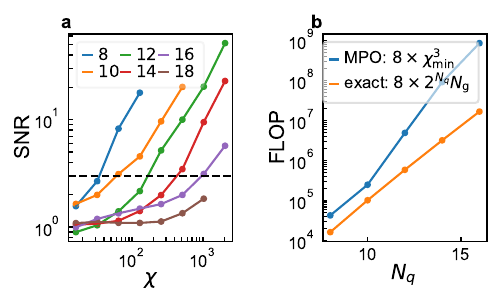}
    \caption{{\bf MPO simulation of one-dimensional OTOC$\boldsymbol{^{(2)}}$ circuits. SNR as a function of the maximum bond dimension a,} In one-dimensional geometries with $N_q = 8,\,10,\ldots, 18$, MPO simulation SNR as a function of the maximum virtual MPO bond dimension $\chi$. {\bf FLOP extrapolation b,} We compare the lower bound on the MPO simulation complexity $8 \chi_{\text{min}}^3$ to the cost of a state-vector simulation, $8 \times 2^{N_q} \times N_{\text{gates}}.$}
    \label{fig:MPO_extrapolations}
\end{figure}

\noindent\textbf{OTOC$\boldsymbol{^{(2)}}$ simulation in the MPO formalism ---} 
Finally, we test the prospects of an MPO simulation. We time-evolve the butterfly operator in the form of an MPO, and then compute (contract) an observable using several copies of the MPO. To obtain a reliable scaling, here we study a family of one-dimensional systems with the butterfly operator placed two steps behind the front with the butterfly velocity $v_B = 3/5$. The results are shown in Fig.\,\ref{fig:MPO_extrapolations}, where we compare the cost of an MPO simulation to the cost of the direct state-vector simulation. For all considered system sizes, a state-vector simulation requires less computation. Therefore, due to the large operator entanglement entropy, the MPO approach does not give advantage over the direct state-vector simulation of OTOC$^{(2)}$, which is impractical in the large-scale circuits.

\subsubsection{Neural quantum states}\label{sec:nn_wf}
\begin{figure}[t!]
    \centering
    \includegraphics[width=0.48\textwidth]{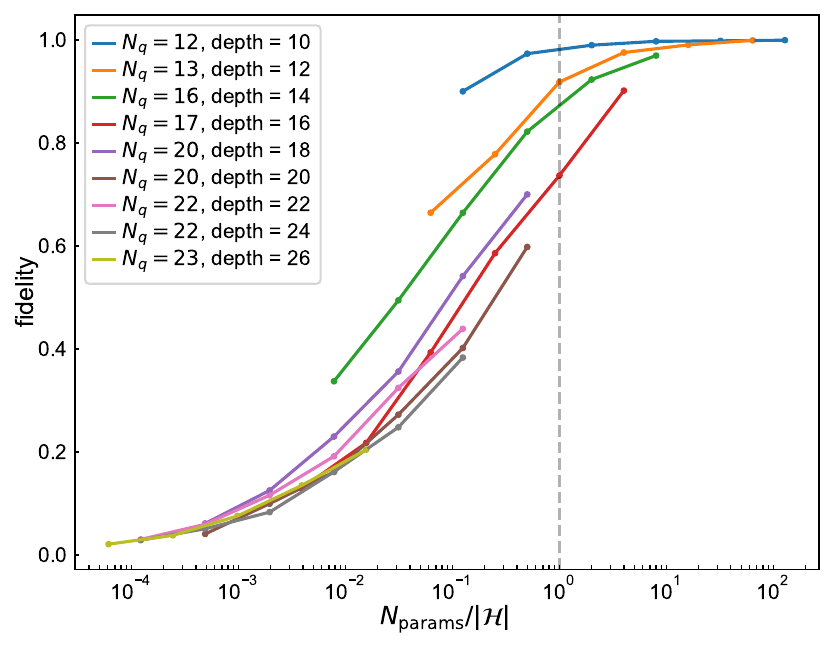}
    \caption{{\bf Neural Quantum States (NQS) applied to a typical OTOC$^{(2)}$ wave function used in this work.}  To obtain each point, we select the best run of ten random initializations, and average all curves over 5 random circuit instances. The curves show that the overlap between the NQS wave function and the state-vector (exact) wave function, as a function of the number of parameters in the neural network, divided by the Hilbert space size.}
    \label{fig:NQS_results}
\end{figure}

In this section, we discuss the prospect of applying the {\it Neural Quantum States} (NQS) method~\cite{Carleo_2017}, wherein a quantum state is variationally represented in form of a neural network, $\Psi(\boldsymbol{\theta}, s) \to \psi(s)$, a function of trained parameters $\boldsymbol \theta$, which, given a bit string $s$, produces the respective element of the wave function $\psi(s)$. This approach has been applied to time-evolution in cases of either adiabatic evolution~\cite{mauron2025challengingquantumadvantagefrontier,carleo2024simulatingadiabaticquantumcomputation,Schmitt_2022} or special regimes of the transverse-field Ising model~\cite{sinibaldi2025timedependentneuralgalerkinmethod,Schmitt_2020,vandewalle2024manybodydynamicsexplicitlytimedependent,chen2025convolutionaltransformerwavefunctions}.

\begin{figure}[t!]
    \centering
    \includegraphics[width=0.48\textwidth]{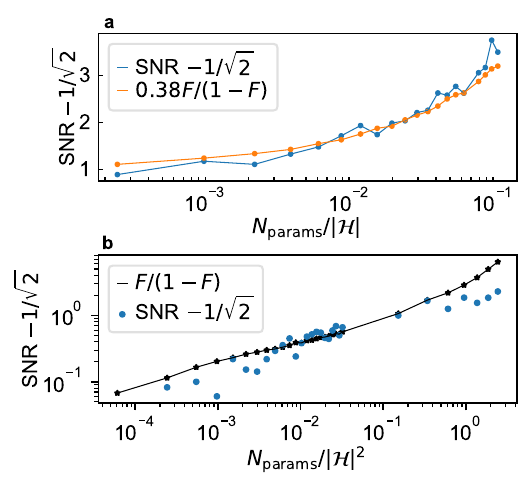}
    \caption{{\bf Fidelity and SNR comparison. a,} In a 16-qubit one-dimensional system with 36 layers of random $SU(4)$ two-qubit gates, we plot the SNR deviation from benchmark $\text{SNR} - 1 /  \sqrt{2}$ (computed over 30 random circuit instances) and $\alpha F / (1 - F)$, where we fit $\alpha \sim 0.38$, as a function of the parameter count. We observe that a large SNR in the OTOC$^{(2)}$ observable is only achievable at a significant fidelity. {\bf b,} In a 16-qubit one-dimensional chain, in the OTOC$^{(1)}$ circuits, we use NQS in the Pauli basis to fit the time-evolved $B(t)$ operator and plot $F / (1 - F)$ (where $F$ is the fidelity of the operator representation). We also plot the SNR deviation from the benchmark $\text{SNR} - 1 /  \sqrt{2}$ (computed over 30 random circuit instances). }
    \label{fig:SNR_Fidelity}
\end{figure}

In this section, we fit an exact state-vector as a neural network with a variable number of parameters. This allows us to assess the {\it expressivity} of the NQS ansatz without delving into particular details of the time-evolution implementation. At first, we use a standard architecture with $N_h = 2, 3, 4$ dense hidden layers and \texttt{ReLu} nonlinearities, which has two outputs predicting $\log |\psi(s)|$ and $\text{arg}\,\psi(s)$~\cite{Westerhout_2020}. This choice of ansatz is governed by the random circuit nature of time-evolution (as opposed to the Hamiltonian evolution) and the absence of any spatial symmetries, which disfavors convolutional neural networks (CNNs). Later, we consider CNN and the ansatz with complex weights and biases, which avoids the amplitude-phase separation.

\begin{figure*}[t!]
    \centering
    \includegraphics[width=\textwidth]{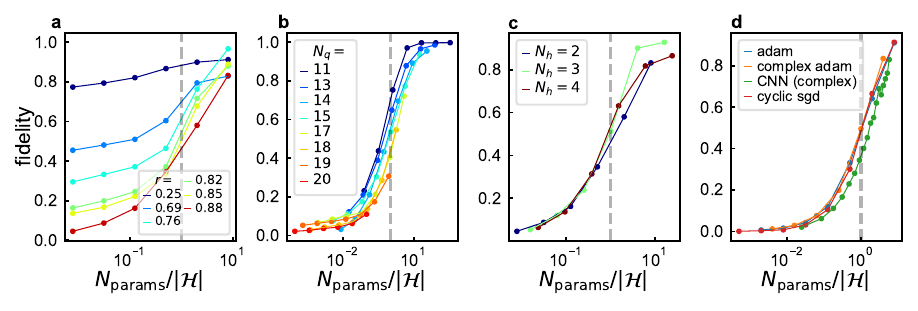}
    \caption{{\bf NQS expressivity tests in various scenarios a,} In a 16-qubit one-dimensional chain with $SU(4)$ random two-qubit gates at various depths, we plot fidelity (best in 10 runs) as a function of $N_{\text{params}} / |\mathcal{H}|$ (two hidden layers). Here, the relative entropy is defined as $r = S^{\text{ent.}}_{\text{Rényi}} / \text{max}\,S^{\text{ent.}}_{\text{Rényi}}$.
    {\bf b,} For one-dimensional chains with $N_q$ qubits, we plot fidelity (best in 10 runs) as a function of $N_{\text{params}} / |\mathcal{H}|$ (two hidden layers). For each system size, we choose the depth such as $0.85 \leqslant r \leqslant 0.90$, in order to replicate the case for beyond-classical dataset. 
    %{\bf c,} In a $4 \times 4$-qubit grid and various depths, we compare the fidelities (best in 10 runs) obtained in (i) sequential fitting approach (solid), or (ii) fitting the final wave function (dashed). 
    {\bf d,} In a 16-qubit chain and depth 70, comparison of architectures with $N_h = 2, 3, 4$ hidden layers. {\bf e,} In a 16-qubit chain at depth 48, we contrast using the Adam optimizer, stochastic gradient descent with Nesterov momentum and cyclic learning rate. We also attempt to make all the parameters of the neural network complex to avoid the amplitude-phase separation and predict the entries of the wave function directly (complex Adam) and do the same in a CNN (complex). All experiments collapse on the same curve.
    }
    \label{fig:nqs_largepanel}
\end{figure*}

In Fig.\,\ref{fig:NQS_results}, we present the case $N_h = 2$, and we fit the final state-vector directly. We consider a family of 31-qubit circuits used in the main text, and control their size and entanglement by truncating the depth at a particular value. We optimize the neural network parameters $\boldsymbol \theta$ to obtain the maximum possible overlap with the exact wave function using the ADAM optimizer~\cite{kingma2017adammethodstochasticoptimization} (the variations, such as AdaBound~\cite{doeschl2024neuralnetworkquantumstates} yield practically similar results). We normalize the $x$--axis for all curves by the maximum Hilbert space size of a particular system. To obtain the best possible fit, we train a neural network starting from 10 different random initial conditions.

Fig.\,\ref{fig:NQS_results} shows that the curves feature near-``universal'' behavior and have an inflection point around the Hilbert space size, $N_{\text{params}} \sim 0.1 |\mathcal{H}|$, with fidelity $\sim 0.5$, slowly decreasing with the system size. If we assume that the fidelity $F$ contributes to the signal, while the infidelity \mbox{$1-F$} contributes to the noise, we can approximately write $\text{SNR} \propto F / (1-F)$, which we illustrate in Fig.\,\ref{fig:SNR_Fidelity}a. This means that achieving the experimental $\text{SNR} \sim 3$ would require to train a neural network with the parameters count of order the full Hilbert space size, which we deem impractical. 

This result is expected since, as Section\,\ref{sec:mps_mpo} shows, the wave functions produced in our $\text{OTOC}^{(2)}$ circuits have entanglement close to the maximum (with the Rényi entropy being $\sim\!90\,\%$ of the maximum, see Fig.~\ref{fig:clifford_entangl_scaling}). Therefore, unlike the case of low-entanglement or low-energy evolution, their accurate representation requires a number of parameters of the order of $|\mathcal{H}|$. This intuition aligns with previous studies showing an exponential scaling (in evolution time) of the required number of parameters~\cite{Lin_2022,Passetti_2023}.

For the case of $\text{OTOC}^{(1)}$, we check if NQS is capable of efficiently compressing the time-evolved butterfly operator $B(t)$ in the Pauli basis, instead of the state in the computational basis, probing the {\it operator entanglement}. The result is shown in Fig.\,\ref{fig:SNR_Fidelity}b. We observe that obtaining an accurate $B(t)$ representation requires achieving substantial fidelity, which is even more difficult in the Pauli basis. This agrees well with the fact that cached Monte Carlo simulations (sec.\,\ref{sec:cachedMC}) require numerous projections, and therefore the $B(t)$ operator evolution has large {\it quantum magic}, hindering the efficient compression of $B(t)$ with NQS~\cite{Liu_2022,Garcia_2023}.

The most prominent example of successful application of the NQS method to the time-evolution is the transverse-field Ising model initiated in the product state polarized along the transverse field~\cite{sinibaldi2025timedependentneuralgalerkinmethod,Schmitt_2020,vandewalle2024manybodydynamicsexplicitlytimedependent,chen2025convolutionaltransformerwavefunctions} for different field strengths $h = 2 h_{\text{c}},\, h_{\text{c}},\, 0.1 h_{\text{c}}$ (here, $h_{\text{c}} \approx 3.044$ is the critical value). Importantly, the evolution for $h = 2 h_{\text{c}}$ and $h = h_{\text{c}}$ reaches only $\sim 0.08$ and $< 0.2$ of the maximum von Neumann entanglement entropy, respectively. Hence, this allows for an efficient compression of a wave function in terms of either neural networks, or matrix-product states. In particular, the case $h = 0.1 h_{\text{c}}$ is near-classical and, therefore, it allows for a near-exact projected entangled pair state description with a very small bond dimension of $D = 2$~\cite{Czarnik_2019}, which a neural network can learn as well. 
It is also important to notice that MPS simulations may provide a better accuracy than NQS simulations for transverse-field Ising models at the finite energy density phase transition that require a large bond dimension, as shown in \cite{haghshenas2025digitalquantummagnetismfrontier}.
%Moreover, an application of NQS to simulate the states of transverse-field Ising model requiring higher bond dimension, at finite energy density phase transition, showed lower accuracy than regular MPS methods~\cite{haghshenas2025digitalquantummagnetismfrontier}.

\begin{figure}[b!]
    \centering
    \includegraphics[width=0.48\textwidth]{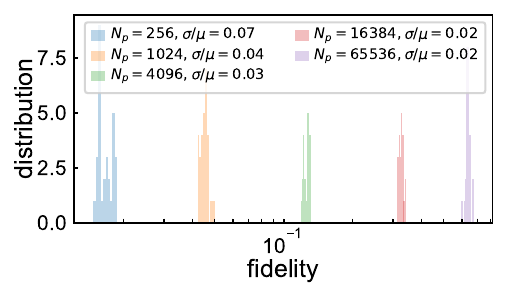}
    \caption{{\bf Distributions of fidelities.} Here, we vary the number of hidden features and launch 30 NQS optimizations starting from different initial random parameter initializations. We observe low variance-to-mean $\sigma / \mu \ll 1$ ratios.}
    \label{fig:NQS_dist}
\end{figure}

By varying the circuit depth and thus tuning the entanglement, we observe a potential reduction of the number of parameters required by the NQS simulations, which is large in the low-entanglement regime, and disappears once a wave function becomes generic enough, as we show in Fig.\,\ref{fig:nqs_largepanel}a. However, if we focus on the depths where the Rényi entropy is over 85\,\% of the maximum, the inflection point is always at around $N_{\text{params}} \sim |\mathcal{H}|$ (see Fig.\,\ref{fig:nqs_largepanel}b), signalling that our beyond-classical OTOC$^{(2)}$ wave functions can not be efficiently expressed in terms of a NQS.

%We contrast our approach of fitting the final wave function directly, against refitting our wave function after each circuit moment, mimicking a time-dependent variational algorithm. Fig.\,\ref{fig:nqs_largepanel}c shows that the latter approach gives no advantage over direct fitting. 
We now discuss different neural network architectures. First, we tested the effect of varying the number of hidden NQS layers, with results shown in Fig.\,\ref{fig:nqs_largepanel}c. We clearly observe that the representation quality depends only on the number of parameters, $N_{\text{params}}$. The use of a convolutional neural network did not improve the scaling, which is expected due to the absence of any spatial symmetry, as shown in Fig.\,\ref{fig:nqs_largepanel}d. Furthermore, avoiding the amplitude-phase separation and considering a neural network with complex parameters does not improve the scaling either (Fig.\,\ref{fig:nqs_largepanel}d). 

Finally, we discuss the potential effect of local minima in the optimization. Fig.\,\ref{fig:nqs_largepanel}d shows no practical difference between choosing the Adam or stochastic gradient descent with Nesterov momentum ($\eta = 0.9$) and cyclic learning rate, which is believed to avoid sharp local minima better than Adam~\cite{liu2025epochalsawtootheffectunveiling}. Furthermore, we consider the distribution of fidelities in multiple optimizations starting from different random initializations in Fig.\,\ref{fig:NQS_dist}. We observe very small variance-to-mean ratio, i.\,e. small dependence of the outcome on the initial network initialization.

\section{Theory of OTOCs and higher-order OTOCs}\label{sec:otoc_theory}

In this section, we develop theoretical models of OTOC$^{(k)}$ that describe the generic physics underlying the experiments performed in the main text.
The first is an analysis of Haar random circuit ensembles, Section~\ref{sec:permutation_basis}, using a mapping to a statistical theory derived by exact averages over the gate ensemble, as has been done previously for OTOC in Refs.~\cite{Mi_OTOC_2021, Khemani2018-od, Nahum2018-pw, von-Keyserlingk2018-ez}. In this framework, we analyze circuit-to-circuit variance of OTOC using perturbation theory in $1/d$, where $d$ is the Hilbert space dimension of a qudit, and verify it with a tensor network simulation for large 1D chains of up to 1000 qudits, see~\ref{sec:permutation_basis}. We demonstrate that the magnitude of the variance scales as a slow polynomial of inverse circuit depth. The theoretical and numerical prediction of OTOCs with $k>1$ and their circuit-to-circuit variances for $k\geq2$ is challenging as the respective statistical models suffer from a sign problem, Section~\ref{sec:permutationmontecarlo}. This complexity also increases with $k$. In Section~\ref{sec:meanotoc2} we use tensor network simulations for large 1D chains to study the circuit average of OTOC$^{(2)}$, which demonstrates a time dependence consistent with the spectral gap closing phenomena discussed in the main text. For circuit-to-circuit variance of OTOC$^{(2)}$ we rely on exact numerical simulations up to $n=36$ qubits ($d=2$). We find that the circuit variance for $k=2$ decays slowly with depth. Additionally, we numerically study the decay of $\mathcal{C}^{(2)}_{\mathrm{off-diag}}$ and $\mathcal{C}^{(4)}_{\mathrm{off-diag}}$, which are constructed as in the main text by subtracting from $\mathcal{C}^{(2)}$ or $\mathcal{C}^{(4)}$ respectively the average of the same over inserting Pauli strings midway through the circuit. For $\mathcal{C}^{(2)}$, the Pauli insertions induce pairings of the forward and backwards trajectory of the operator; nonetheless, we show that this captures in large 1D circuits most of the full signal, with the remainder $\mathcal{C}^{(2)}_{\mathrm{off-diag}}$ approaching $0$ polynomially but with a faster power than $\mathcal{C}^{(2)}$. For $\mathcal{C}^{(4)}$, two separate sets of uncorrelated Pauli insertions are introduced for the first and second copies of the evolution in the circuit. This allows capturing part of the $\mathcal{C}^{(4)}$ signal, but there is a sizable remainder due to contributions that satisfy the trace condition with pairing of trajectories, $\mathcal{C}^{(4)}_{\mathrm{off-diag}}$. The decay of $\mathcal{C}^{(4)}_{\mathrm{off-diag}}$ is large 1D circuits is quite slow, but our numerics don't suffice to resolve whether the asymptotic fraction of the full $\mathcal{C}^{(4)}$ signal that this represents saturates or drops to $0$.

%However, the Haar theory becomes challenging to analyze beyond small $k$, as the number of degrees of freedom grows rapidly with $k$ and because the mappings produce dynamical rules with sign problems, as shown in Section~\ref{sec:permutationmontecarlo}.

%it is both necessary for classical hardness against a standardized OTOC signal in the main text, and will required for learning molecular Hamiltonians from nuclear magnetic resonance experiments in Section~\ref{sec:nmr}.

In Section~\ref{sec:phasetransition}, we develop a theory of $\mathrm{OTOC}^{(k)}$ for {\it all} $k$ by relating these objects to the spectrum of the {\it correlation operator} (see the main text and Section~\ref{sec:phasetransition} for the definition). 
First, we introduce an exactly-solvable random-matrix model that has no spatial structure but has a tunable parameter that controls the degree of scrambling. 
Variation of this parameter captures the evolution between the limits of no dynamics and of a fully scrambling dynamics.
We show that the spectrum of the correlation operator undergoes a gap closing transition with the increase of the scrambling parameter; we analyze the properties of this transition and find its effect on $\mathrm{OTOC}^{(k)}$ values.
Additionally, we use exact numerics to argue that the picture of the spectral phase transition holds also in local circuits, with the gap closing as one tunes across the operator front.

The combination of the circuit models we analyze in this Section demonstrate a set of features generic for  out-of-equilibrium quantum dynamics: large circuit-specific signal OTOC$^{(k)}$ which can be efficiently measured on a quantum processor and absence of known efficient classical simulation algorithms for $k>2$. This motivates circuit-specific OTOC$^{(k)}$ values as a resource for Hamiltonian learning with the potential for quantum advantage in  practical settings. 
 % intro to all theory

%%%% KK start

\subsection{Statistical theory of Haar averaged random circuits}\label{sec:permutation_basis}

In circuits with Haar random gate ensembles, averages over the Haar ensemble can be exactly mapped to ``classical'' statistical lattice models which can be described as $S_k$ ferromagnets; see the review Ref.~\cite{Fisher2023-sr} for an overview of the many previous works involving these mappings, and Refs.~\cite{Mi_OTOC_2021, Nahum2018-pw, Khemani2018-od, von-Keyserlingk2018-ez, Bertini2020-he} for previous applications to the study of OTOCs.

% For the purposes of this Supplementary Information, we will reintroduce this formalism in a manner appropriate for computing the mean and fluctuations of OTOC$^{k/2}$ for arbitrary $k$ in circuits with Haar random gates. This will be used in Section~\ref{sec:otoc_scaling} to probe the scaling of OTOC fluctuations with the system size, and again in Section~\ref{sec:permutation_based_algorithms} to motivate classical algorithms for OTOC simulation.

% \subsubsection{OTOC circuit-to-circuit variance} \label{sec:variance}

In this section we analyze OTOCs for qudits subject to a circuit $U(t)$ of depth $t$ that consists of a bricklayer pattern of Haar random two-qubit gates, which take the form
\begin{align}
    \mathcal{C}^{(2k)}_{B,M}(t) = \frac{1}{d^{n}}\textrm{Trace} \left[\left(B(t) M\right)^{2k}\right]
\end{align}
for systems of $n$ qudits of dimension $d$, where $B$ and $M$ are the butterfly and measurement operators respectively.
% In the language of the main text, $O(t)\equiv A_0(t)$, $M\equiv B_x$, and $\mathcal{C}^{(k)}\equiv C^{(k)}_{A,B}(x,t)$; we include the additional indices here for convenience.
We consider average values over the circuit ensemble, which we denote with the horizontal bar. We rewrite the previous equation as
\begin{gather}\label{eq:meanotoc}
\overline{\mathcal{C}^{(2k)}_{B,M}(t)} = \frac{1}{d^n}\textrm{Trace} \left[ \pi_{(1,2,\ldots k)} \overline{B(t)^{\otimes 2k}} M^{\otimes 2k}\right],
\end{gather}
a correlation function in an enlarged Hilbert space that consists of $2k$ copies of the original. $\pi_{\sigma}$ denotes a permutation operator that acts on the $2k$ copies of Hilbert space by sending the $j$-th copy to the $\sigma(j)$-th copy, where $\sigma$ is a permutation from the symmetric group $S_{2k}$. We adopt the standard cycle notation for permutations~\cite{Diaconis} --- thus, $(1, 2, \ldots 2k)$ refers to the permutation that sends $j$ to $(j+1)\,(\text{mod }2k)$.

% A cyclic permutations of $l$ elements defined by a generator $g(i)$ acting on the index $i$, is denoted by $\left(g^0(1) g^1(1) g^2(1)\text{...} g^l(1)\right)$. Every permutation can be represented as a product of cycle permutations

Additionally, we consider the variances
\begin{gather}
    \Delta^{(2k)}_{B,M}(t) =  \overline{\mathcal{C}^{(2k)}_{B,M}(t) \mathcal{C}^{(2k)}_{B,M}(t)} - \overline{\mathcal{C}^{(2k)}_{B,M}(t)} \; \overline{\mathcal{C}^{(2k)}_{B,M}(t)}. % \textrm{Tr}\left\{\left(C_{0} D_{x} (t) \right){}^2\right\}}
    \label{eq:DeltaACBD}
\end{gather}
The first term can be rewritten as above using a Hilbert space with $4k$ copies of the original,
\begin{multline}
\overline{\mathcal{C}^{(2k)}_{B,M}(t) \mathcal{C}^{(2k)}_{B,M}(t)} 
\\=  \frac{1}{d^{2n}}\textrm{Trace} \left[ \pi_{(1,2,\ldots 2k)(2k+1, \ldots 4k)} \overline{B(t)^{\otimes 4k}} M^{\otimes 4k}\right].
\end{multline}

In the average expressions, each gate $U_{ij}$ acting on a pair of sites $i, j$ in the circuit appears in the average $r = 2k$ times as $U_{ij}$ and $2k$ times as $U^{\dagger}_{ij}$; in the second moments, $r = 4k$ times each.
Generally, in an expression involving $r$~copies of $U_{ij}$ and $r$~copies of $U_{ij}^{\dagger}$ of a Haar random gate, the average 
\begin{gather}
    \mathcal{P}_r \equiv \overline{\underset{s=1}{\overset{r}{\otimes }}U_{i_s j_s} \left(U_{i_s' j_s'}\right){}^{\dagger }} \label{eq:P_k}
\end{gather}
can be evaluated exactly as a projection onto the space of Haar invariant operators. 
A convenient albeit non-orthogonal basis for the invariant subspace~\cite{collins2022weingarten} is the set of permutation operators, $\sigma, \tau$, permuting the $n$~copies, $|\sigma \rangle = \prod_{s=1}^{r}\delta_{i_s, \sigma(j^\prime_s)}$. We rewrite the circuit average $\mathcal{P}_r$ in this subspace as
\begin{gather}
    \mathcal{P}_r = \sum_{\sigma,\tau} |\tau\tau\rangle \textrm{Wg}(\tau\sigma^{-1})\langle\sigma \sigma |, \label{eq:P_kviaWg}.
\end{gather}
Here $\textrm{Wg}(\tau\sigma^{-1})$ is the Weingarten function corresponding to the two-qudit Hilbert space dimension $d^2$, which can be computed by inverting the overlap matrix of the permutations~\cite{Weingarten1978-ro, collins2022weingarten}.
 Overlaps between permutations are given by $\langle \alpha|\beta\rangle = d^{r - |\alpha^{-1}\beta|}$, where $|\alpha^{-1}\beta|$ is the minimum number of elementary transpositions comprising the permutation $\alpha^{-1}\beta$. 
Dynamics is generated by applying two-qudit projectors $\mathcal{P}_r$, which by Eq.~(\ref{eq:P_kviaWg}) has matrix elements
\begin{gather} \label{eq:unscaleddynamics}
    \mathcal{P}_r |\mu\nu\rangle = \sum_\tau T_{\tau\tau,\mu\nu} \ket{\tau\tau}, \\
    T_{\tau\tau,\mu\nu} =\sum_\sigma \textrm{Wg}(\tau\sigma^{-1})\langle\sigma | \mu \rangle \langle \sigma | \nu\rangle.
%\approx \left(\frac{d}{d^2+1}\right)^2 1/d^2,
\end{gather}
Specifically, for each Haar random two-qudit gate between qudits at sites $i, j$, $ \mathcal{P}_r$ is applied to sites $i, j$ starting with $r$ copies of the initial operator $B(t=0)$.

The projectors force the operator dynamics to take place entirely within the subspace spanned by products of permutations on every site; thus we can write the average evolved operator $\overline{B(t)^{\otimes r}}$ in the form
\begin{gather}
\sum_{\sigma_i \in S_r} B_{\sigma_1, \ldots, \sigma_n}(t) \ket{\sigma_1, \sigma_2, \ldots, \sigma_n}
\label{eq:permutation-wavefunction}
\end{gather}
for $t \geq 1$.
In this basis, uniform two-qudit operators are stationary $\mathcal{P}_r |\sigma \sigma \rangle = |\sigma \sigma \rangle$. Operators with globally constant permutation factors $|\sigma \sigma ...\sigma\rangle$ are thus stationary states of the dynamics; for this reason these models are referred to as $S_r$ ferromagnets~\cite{Fisher2023-sr, Zhou2019-zu}.
Non-trivial dynamics occurs only along domain walls between permutations.

% In the following we will adopt the cycle notation for permutations. A cyclic permutations of $l$ elements defined by a generator $g(i)$ acting on the index $i$, is denoted by $\left(g^0(1) g^1(1) g^2(1)\text{...} g^l(1)\right)$. Every permutation can be represented as a product of cycle permutations~\cite{Diaconis}.

%Note that cycle permutations do not necessarily commute and therefore the order of application is important.

Additionally, as a consequence of the unitarity of the circuits, these dynamics preserve $r!$ conserved charges $$\langle \omega \omega ...\omega | B^{\otimes r}(t) \rangle \text{ for each } \omega \in S_r.$$
For reasons that will become clear when we analyze the dynamics in the following sections, we find it convenient to use a change of basis that is a simple rescaling of the permutations $\hat{\sigma} = \sigma/\langle \omega | \sigma \rangle,$ where $\omega = (1, 2, \ldots 2k)$ for the mean OTOC$^{(k)}$ calculation and $\omega = (1, 2, \ldots 2k)(2k+1, 2k+2, \ldots 4k)$ for the second moment OTOC$^{(k)}$ calculation. Quantities in this rescaled basis are decorated with hats below to distinguish them.
This results in a equivalent dynamical rule
\begin{gather} \label{eq:scaleddynamics}
    \mathcal{P}_r |\hat{\mu}\hat{\nu}\rangle = \sum_\tau \hat{T}_{\tau\tau,\mu\nu} \ket{\hat{\tau}\hat{\tau}}, \\
    \hat{T}_{\tau\tau,\mu\nu} =\sum_\sigma \textrm{Wg}(\tau\sigma^{-1})\frac{\langle\sigma | \mu \rangle \langle \sigma | \nu\rangle \langle \omega | \tau \rangle^2}{\langle \omega | \mu \rangle \langle \omega | \nu \rangle},
\end{gather}
which by construction has the additional property that it conserves the total weight in the rescaled basis
\begin{multline*}
    \langle \omega \ldots \omega |\sum_{\sigma_i \in S_r} \hat{B}_{\sigma_1 \ldots \sigma_N}(t) | \hat{\sigma}_1 \ldots \hat{\sigma}_n \rangle = \sum_{\sigma_i \in S_r} \hat{B}_{\sigma_1 \ldots \sigma_n}(t).
\end{multline*}

Below, we analyze how the domain walls evolve in the calculations for mean OTOC$^{(1)}$, second moment of OTOC$^{(1)}$, and mean OTOC$^{(2)}$. We attack the problem both perturbatively in $1/d$ and use tensor network algorithms to directly compute the dynamics of the permutation wavefunction Eq.~(\ref{eq:permutation-wavefunction}) for various values of $d$. These results are used to validate the perturbative results and show how they extend to small values of $d$.

\subsubsection{Circuit average OTOC\,$^{(1)}$}

% The second term in Eq.~(\ref{eq:DeltaACBD}) is a product of two independent averages of $ \overline{C_{A,B}(x,t)}$. Such that each of the factors can be calculated using $k=2$ in Eq.~(\ref{eq:P_kviaWg}). 

%The final state corresponds to $|S...S\rangle$, which has the overlaps $\langle S|I \rangle = d$ and $\langle S|S \rangle = d^2$. 

A precise quantitative description of dynamics of the mean OTOC$^{(k)}$ for $k=1$ is well known~\cite{Mi_OTOC_2021, Nahum2018-pw, Khemani2018-od}; the circuit average dynamics is Markovian and can be described as a stochastic growth of a domain. We review the key features of these results here as they are relevant for understanding the following sections.

For simplicity, we will analyze the case where the butterfly operator $B$ is chosen as a non-identity Pauli operator on the left-most site; after averaging the first Haar random gate, the state takes the form
\begin{multline}\label{eq:init_state}
\overline{B^{\otimes 2}(t=1)} = \frac{1}{d^4-1} \left( d^2 \ket{S S I \ldots I} -  \ket{I I I \ldots I} \right),
\end{multline}
where $I=(1)(2)$ and $S=(12)$ are the identity and swap permutation operators, respectively. 

The non-trivial matrix elements in Eq.~(\ref{eq:unscaleddynamics}) are $$T_{SS,IS}=T_{II,IS}=T_{SS,SI}=T_{II,IS} = \frac{d}{d^2+1}.$$
Thus, applying $\mathcal{P}_2$ causes hopping of the domain wall; a single domain wall cannot be destroyed without meeting another domain wall or the boundary, and no new domain walls are created.
Thus, the first component of Eq.~(\ref{eq:init_state}) evolves into a superposition of states with a single domain wall between the $S$ domain and $I$ domain. The hopping occurs in both directions equally according to the dynamical rule Eq.~(\ref{eq:unscaleddynamics}).

The desired quantity $\overline{\mathcal{C}^{(2)}}$ is expressed as the overlap of this evolved operator with the final state $d^{-n} \bra{S S \ldots S} M^{\otimes 2}$. 
Aside from the location of the measurement operator $M$, the final state weights components of $\overline{B^{\otimes 2}(t)}$ by a factor $d^{\ell}$, where $\ell$ is the size of the domain of $S$. Thus, configurations with 
large domains of $S$ dominate contributions to $\overline{\mathcal{C}^{(2)}}$.

% However, large domains of $S$ have smaller weights in the expansion of $\overline{A^{\otimes 2}(t)}$, leading to a competition between effects.

The combined effect of the dynamics and the final state reweighting can be incorporated by a change of basis to the rescaled operators $\hat{I}=\frac{I}{d}, \hat{S}=\frac{S}{d^2}$.
After this change of basis, the dynamical rule Eq.~(\ref{eq:unscaleddynamics}) is modified into Eq.~(\ref{eq:scaleddynamics}) with nontrivial matrix elements $\hat{T}$
\begin{gather}
    \hat{T}_{SS, IS} = \hat{T}_{SS, SI} = \frac{d^2}{d^2+1}, \, \nonumber
    \\ \hat{T}_{II, IS} = \hat{T}_{II, SI} = \frac{1}{d^2+1}. \label{eq:T_biased_walk}
\end{gather}
This dynamics is Markovian --- the total weight is conserved and all matrix elements are positive. The operator coefficients in Eq.~(\ref{eq:permutation-wavefunction}) after this change of basis can be reinterpreted as probabilities. Unlike the dynamics under $T$, the dynamics under $\hat{T}$ breaks the symmetry between the two permutations; $\hat{S}$ domains tend to grow and $\hat{I}$ domains tend to shrink.

% A pair of qubits in different domains $I, S$ both flip to the $S$ domain with probability $p = d^2/(d^2+1)$ and to the $I$ domain with probability $1-p$. 
% The exponential weighting of $S$ domains by the final state has been reincorporated throughout the dynamics, causing a preferential growth of the $S$ domain and an apparent breaking of the symmetry between the two permutations. 
Let us now specialize to the case where the measurement operator $M$ chosen to be a non-identity Pauli operator on a single site $x$; we make the $x$ dependence of the correlator explicit by writing the OTOC in this case as $\overline{\mathcal{C}^{(2)}}(x, t)$.
In this case, the overlap with the final state can be rewritten as the probability to find the permutation $S$ at site $x$, so that $\overline{\mathcal{C}^{(2)}}(x, t)$ precisely measures the progress of the growth of the $S$ domain across the system. 

In one dimension, for a butterfly operator that starts at the left of the system, the growth of the $S$ domain is described entirely by the location of the right end point of the domain wall, which undergoes a biased random walk with probability  $p=d^2/(d^2+1)$ to the right and $1-p$ to the left. 
% 
% Under this rescaling, the overlap with the final state treats all configurations equally when $B_x$ is the identity. As Eq.~\ref{eq:meanotoc} yields $\overline{\mathcal{C}^{(2)}} = 1$ when $B_x$ is the identity operator, there is a conserved total weight which we will refer to as a quasi-probability, as the individual term weights total to $1$ but are not guaranteed to be positive. 
% 
% Thus, we can express the state as a single domain wall described by a probability $W(x,t)$ for the domain to be located at $x,t$ on the dual lattice,
% \begin{gather}
% W(x,t) = \sum_{x^\prime}\langle x|T^t |x^\prime \rangle W(x^\prime,0)
% \end{gather}
% with the transfer matrix $\langle x|T |x^\prime \rangle$. For Haar two-qubit gates the transfer matrix reads,
% \begin{gather}
% \langle x| T  |x'\rangle = p \delta _{x,x'+1}+(1-p)\delta _{x,x'-1}, \label{eq:T_biased_walk}
% \end{gather}
% where $p = d^2/(d^2+1)$. This is nothing but a transfer matrix of a biased random walk, which can be iterated exactly. The result takes the form,
% \begin{gather} 
% \overline{\mathcal{C}}\approx \sum _{u}p^{t-u} (1-p)^{u}  B_u^t,  
% %\omega(u,v,t) = \left(B_v^t B_u^t- B_{v-1}^{t}B^t_{u-1}\right)
% \end{gather}
% where the sum is over $u\in [\frac{t-x+3}{2}, t] $ %and $v\in [0, \frac{t+x-1}{2}]$
% and $B_v^t$ is a binomial coefficient.  
%
The time dependence of $\overline{\mathcal{C}}$ has the shape of a wavefront that propagates to the right with the average velocity $v_B$ and broadens diffusively with width $\sqrt{2Dt}$. The values of $v_B$ and $D$ for Haar random 1D circuits are directly computable from the random walk hopping probability $p$:
\begin{gather}
v_B = 2p-1 = \frac{d^2-1}{d^2+1}
\\ D = 2 p ( 1 - p) = \frac{2 d^2}{(d^2+1)^2}.
\label{eq:haarvbD}
\end{gather}

% Asymptotically, $\overline{\mathcal{C}^{(2)}}(x, t)$ becomes the cumulative probability that the random walker has reached site $x$ by time $t$. 
As the probability distribution for the location of the random walker becomes a normal distribution at long times, with mean $x_0 + v_B t$ and standard deviation $\sqrt{2 D t}$, asymptotically $\overline{\mathcal{C}^{(2)}}(x, t)$ becomes an error function. Plotted against a rescaled coordinate $$
u = \frac{ x - (x_0 + v_B t)}{\sqrt{2 D t}},
$$
the mean OTOC quickly converges to a scaling function that describes the asymptotic behavior, as shown in Fig.~\ref{fig:c4_permutations}(a). In higher dimensions, the results are qualitatively similar, however, the width of the front scales with different exponents. It has been argued~\cite{Nahum2018-pw} that these exponents correspond to Kardar-Parisi-Zhang universality class~\cite{Kardar1986-yl}.

% \subsubsection{OTOC variance using 4 copy}

\subsubsection{OTOC\,$^{(1)}$ variance}

\begin{figure*}
    \centering
    \begin{overpic}[width=0.32\textwidth]{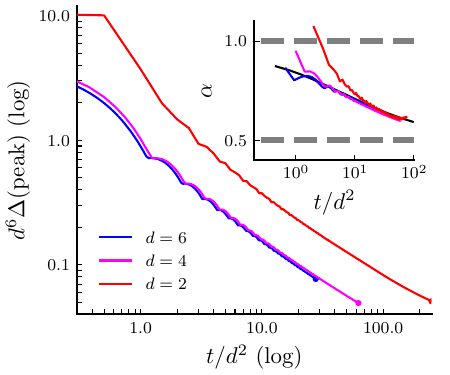}\put(2,88){\textbf{a}}\end{overpic}
    \begin{overpic}[width=0.32\textwidth]{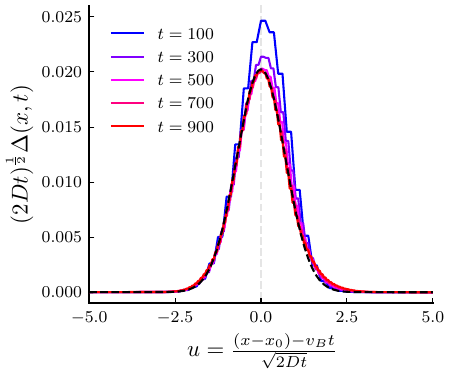}\put(2,88){\textbf{b}}\end{overpic}
    \begin{overpic}[width=0.32\textwidth]{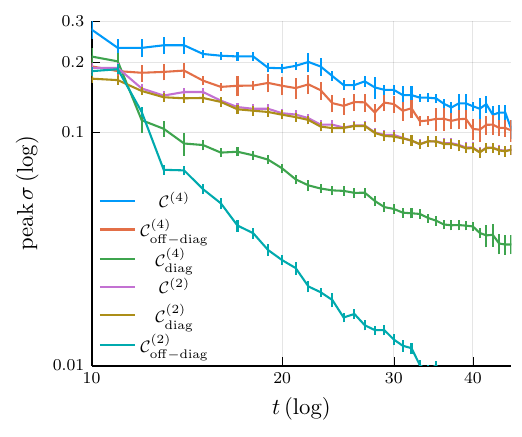}\put(2,88){\textbf{c}}\end{overpic}
    \caption{\textbf{Fluctuations of OTOCs for a 1D brickwork circuit with Haar random two-qubit gates.} The butterfly operator $B$ is chosen as a Pauli $Z$ on the left-most site, while the measurement operator is chosen as a Pauli $Z$ on a site $x$ that varies across the 1D chain. The signal propagates from left-to-right with the butterfly velocity $v_B$ in a diffusively growing front. {\bf a}, Peak variance of $\mathcal{C}^{(2)}$ (labeled $\Delta = \sigma^2(\mathcal{C}^{(2)})$) plotted versus time, rescaled for comparison between different values of qudit dimension $d$. Exponent of the decay $\alpha$, extracted from the average slope on the log-log plot over a sliding time window, drifts from $1/t$ to $1/\sqrt{t}$ (inset), consistent with the crossover form predicted in Eq.~(\ref{eq:varOTOC}). As $d$ is increased, the curves collapse under the rescaling. {\bf b}, Spatial distribution of the variance of $\mathcal{C}^{(2)}$. Asymptotically, when plotted against the rescaled front coordinate $u$ and rescaled by $\sqrt{2Dt}$ the variance collapses onto a curve proportional to a Gaussian with standard deviation $1/\sqrt{2}$ (black dashed curve). {\bf c}. Peak standard deviation of $\mathcal{C}^{(2)}$ and $\mathcal{C}^{(4)}$, as well as the diagonal and off-diagonal components thereof, measured numerically using state vector calculations on up to $36$ qubits with a sample of 200 random circuits. The variance of $\mathcal{C}^{(4)}$ undergoes a similar power-law decay as that of $\mathcal{C}^{2)}$. For each circuit, we additionally compute $\mathcal{C}^{(2)}_{\mathrm{diag}}$ and $\mathcal{C}^{(4)}_{\mathrm{diag}}$ by averaging over random Pauli strings inserted at a depth half way through the circuit, as done in the main text, and the corresponding off-diagonal components by subtracting. While $\mathcal{C}^{(2)}_{\mathrm{off-diag}}$ and $\mathcal{C}^{(4)}_{\mathrm{off-diag}}$ both appear to decay as power-laws, $\mathcal{C}^{(2)}_{\mathrm{off-diag}}$ decays more rapidly.}
    \label{fig:c2_variance}
\end{figure*}

% By independently averaging over the operators at the measurement and butterfly qudits, we can remove the specific dependence on the choice of $A$ and $B$. This allows us to replace the inital and final state in our calculation with
% \begin{gather}
%  \overline{A^{\otimes 2}} = \overline{B^{\otimes 2}} = \frac{d^2}{d^4-1}\left(|S_1S_2\rangle -\frac{1}{d^2} |I_1 I_2 \rangle\right), \label{eq:init_state}
% \end{gather}
% where $I$ and $S$ are identity and swap operators, respectively. 

%Similar average over butterfly operators $\overline{B^{\otimes 2}}=\overline{D^{\otimes 2}}$.% reads,
%\begin{gather}
%   \overline{B^{\otimes 2}} = \left(\langle II | -\frac{1}{d} \langle S S |\right).
%\end{gather}
% We denote OTOC variance with this initial/final state as $\Delta$. 

In contrast to the average, the second moment of OTOC$^{(1)}$ (first term of Eq.~(\ref{eq:DeltaACBD})) requires a circuit average of four copies of operator evolution, i.e. in Eq.~(\ref{eq:P_k}) we set $r=4$. In this case Eq.~(\ref{eq:P_kviaWg}) projects onto a subspace spanned by $r!=24$ states labeled by permutations of four elements. 

Once again using a non-identity Pauli operator on the left-most site as the initial state $B(t=0)$ of our evolution, after the first layer of Haar random gates the operator is projected into the permutation space
$$
\overline{B^{\otimes 4}(t=1)} = \sum_{\alpha \in S_4} c_{\alpha} \ket{\alpha \alpha I \ldots I},
$$
creating an initial superposition of domains of different types. As in the case of the mean OTOC, the evolution proceeds by creating a operator front that propagates ballistically with speed $v_B$ and broadens diffusively. Behind the front, where the dynamics is fully scrambled, one always finds a uniform domain of a single permutation or a superposition of uniform domains $\sum_{\alpha} \ket{\alpha \alpha \alpha \ldots I I I}$, where the $\ldots$ conceals a potentially complicated superposition of permutations within the front region. Physically, this uniform domain indicates that far behind the front the dynamics is indistinguishable from global scrambling using $r$th moment quantities.

% contains a term corresponding to a domain configuration $|\sigma_0^\prime, \sigma_1^\prime\sigma_2...\sigma_n\rangle$, where $\sigma_0^\prime, \sigma_1^\prime = (12)(34)$ and the rest are identity $\sigma_i = I, i= 2,...,n$. The domain $\sigma_0, \sigma_1$ is separated from the rest of the chain by a double domain wall. 
% The final state corresponds to $\mathcal{N}^{-1}\langle \sigma_0...\sigma_{x-1}\sigma^\prime_{x}\sigma_{x+1}...\sigma_{n} |$ where
% $\sigma_i = (12)(34), i\neq x$. Here the factor $\mathcal{N}^{-1}$ accounts for normalization of trace. 

%As in other sections of this SI, it is convenient to rescale weight including the effect of the final state,
%\begin{gather}
%    \tilde{J}_{\tau\tau,\mu\nu} = J_{\tau\tau,\mu\nu} \frac{\langle \sigma \sigma | \tau\tau \rangle }{\langle \sigma \sigma | \mu \nu \rangle }.
%\end{gather}
%Dynamics generated by the random circuit converges to stationary states which are uniform in space, i.e. single domain, as was shown after Eq.~(\ref{eq:P_kviaWg}). 

The desired second moment quantity is computed as the overlap with $d^{-2n} \bra{\omega \ldots \omega} M^{\otimes 4}$, where $\omega = (12)(34)$.
Ignoring momentarily the effect of the butterfly operator $M$, 
the overlap of a uniform domain $\alpha$ of size $l$ with the final state $\omega$ is exponential 
\begin{gather}
d^{-2n}\langle \omega | \alpha \rangle^l = d^{(2-|\omega^{-1}\alpha|) l}.
\end{gather}
The effect of the final state competes with the weight of the domain wall which in the leading order corresponds to $(d/(d^2+1))^{t |\alpha|}$ for each moment in time. Domains for which these two effects balance each other out, i.e. that satisfy,
\begin{gather}
2-|\omega^{-1}\alpha| - |\alpha| = 0,
\end{gather}
contribute significantly to the variance of OTOC. Only four types of domains satisfy this condition: $\alpha \in \{I, (12), (34), (12)(34)\}$. The contribution of all other domains to the OTOC variance is exponentially suppressed by its boundary. Moreover, the entropy of domain configurations is not sufficient to compensate for this effect: the number of configurations of a domain with perimeter $l$ on a square lattice scales as $2^l$, which is insufficient to overcome the weight of its boundary $\left(\frac{d}{d^2+1}\right)^l$, for $d\geq2$.

Using the rescaled dynamics Eq.~(\ref{eq:scaleddynamics}), the effect of the final state is incorporated throughout the dynamical evolution. In this dynamics, the suppressed domains are unstable and with weight $1 - \mathcal{O}(d^{-2})$ transition into non-suppressed domains, which are stable. As a result, after an initial transient dynamics a typical component of the permutation wavefunction is described as having large domains of only $I, (12), (34)$ and $(12)(34)$. For the choice of a butterfly operator that is a non-identity Pauli operator, the situation is even simpler: a typical configuration has far behind the front $(12)(34)$, a single domain wall transitioning to a domain of $(12)$ or $(34)$ in the front, and a second domain wall to the $I$ domain ahead of the front. The two domain walls each undergo random walks and interact only when colliding.

In the limit $d\rightarrow\infty$ the two domain walls may be treated as independent, in which case their joint probability distribution is simply a product. Independent domain walls are described by the transfer matrix in Eq.~(\ref{eq:T_biased_walk}). The leading order correction corresponds to a collision of the two domain walls, where the following process is possible,
\begin{gather}
(12)(34) I \rightarrow \sigma \sigma \rightarrow (12)(34) I. \label{eq:single_step_bubble}
\end{gather}
Here arrows correspond to applying a layer of two-qudit gates. For independent domain walls this process has the weight $4p^2(1-p)^2$. The leading order correction corresponds to generating intermediate domains  $\sigma =\{ (14)(23), (13)(24)\}$ which introduces correlations between the domain walls. The respective correction to OTOC variance is of the order of $d^{-4}$. 

%A careful reader may notice that the intermediate states $\sigma =\{ (14)(23), (13)(24)\}$ differ from $(12)(34)$ only by relabelling of the copies. 

Consider two domain walls propagating independently with a single collision point, where the process in Eq.~(\ref{eq:single_step_bubble}) occurs. The probability of moving for time $t$ without changing the relative distance between the pair of domain walls solves $(p^2+(1-p)^2)^t \approx \exp(-2t/d^2)$. Therefore for very short times $t<d^2$ the collision of domain walls is unlikely. In this case the OTOC variance is dominated by a boundary term at the point of measurement. At longer times, $t>d^2$, domain walls collide in the bulk. The leading order calculation corresponds to a single collision. 

These considerations produce the following result for the maximum of the variance $\Delta$ over the location of the measure and butterfly qudits as a function of the circuit depth $t$:
\begin{gather}
\textrm{max}\Delta \propto \left\{
    \begin{array}{cc}
    \frac{1}{d^4} \frac{1}{t}, \; t<d^2, \\
    \frac{1}{d^5} \frac{1}{\sqrt{t}}, \;  t\geq d^2. \\   
    \end{array}
    \right. \label{eq:varOTOC}
\end{gather}
This result is valid for $d\gg1$, but by numerical simulation we show that this result is valid as well for relatively small $d$. The result of these numerics is detailed in Fig.~\ref{fig:c2_variance}. As both lines of Eq.~(\ref{eq:varOTOC}) are consistent with a single scaling form $d^6 \Delta = F(t/d^2)$, we confirm the $d$ dependence by plotting $d^6 \Delta$ versus the rescaled time $t/d^2$. This results in a collapse of the data for curves with $d \geq 4$. While $d = 2$ does not collapse onto the same curve as $d \geq 4$, in either case the peak variance is indeed governed at short times by $1/t$ decay and at long times by $1/\sqrt{t}$ decay. 

This numerics uses an MPS representation of the permutation wavefunction to exactly compute the dynamics in the permutation subspace, and it uses \emph{time-evolving block decimation}~\cite{Vidal2004-wr} to evolve this wavefunction. Specifically, we use the rescaled dynamics Eq.~\ref{eq:scaleddynamics} as the evolution rule, which results in improved convergence for computing the mean and second moment of OTOC.
The algorithm uses a truncation parameter $\varepsilon \lesssim 10^{-12}$ which governs the level of compression in the standard manner, by discarding small singular values $s_a$ whose total $\sum_a s_a^2 \leq \varepsilon$, with the result checked for convergence over several orders of magnitude in $\varepsilon$.

% The computation confirms a slow polynomial $t^{-\frac{1}{4}}$ decay of the OTOC fluctuations, which are concentrated in the operator front, as was argued in Ref.~\cite{Nahum2018-pw} using a conjectural hydrodynamic description of the OTOC.

\begin{figure*}
    \centering
    \begin{overpic}[width=0.32\textwidth]{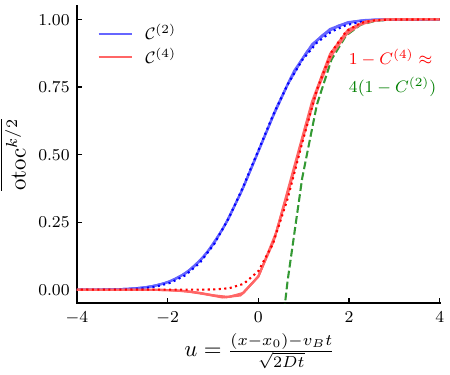}\put(2,88){\textbf{a}}\end{overpic}
    \begin{overpic}[width=0.32\textwidth]{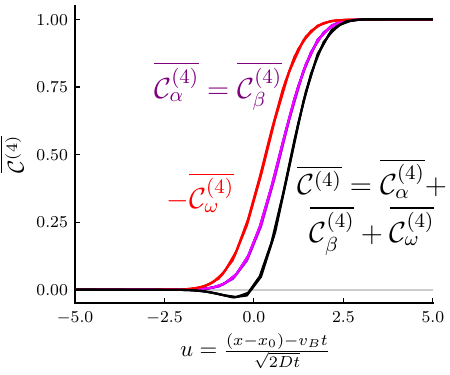}\put(2,88){\textbf{b}}\end{overpic}
    \begin{overpic}[width=0.32\textwidth]{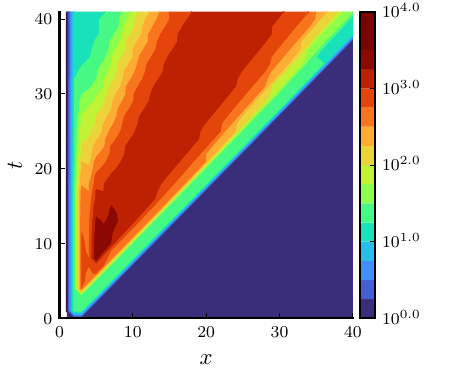}\put(2,88){\textbf{c}}\end{overpic}
    \caption{{\bf a}, Comparing the mean value of OTOC$^{(k/2)}$ for $k=2$ and $k=4$  for a 1D brickwork circuit with Haar random two-qubit gates. The butterfly operator $O$ is chosen as a Pauli $Z$ on the left-most site, while the measurement operator is chosen as a Pauli $Z$ on a site $x$ that varies across the 1D chain. The signal propagates from left-to-right with the butterfly velocity $v_B$ in a diffusively broadening front; plotted against the rescaled coordinate $u$, the signal collapses into a universal shape. The scaling limit of $k=2$ is well approximated by an error function (dashed lines), while $k=4$ deviates from an error function. In the leading edge of the front, the mean OTOC$^{(k/2)}$ satisfies $1 - \mathcal{C}^{(k)} \approx (\frac{k}{2})^2 (1 - \mathcal{C}^{(2)})$. This property is a manifestation of the correlation operator spectrum when it consists of small arcs, as shown in Fig.~\ref{fig:rmt_numerics}. In the middle of the front, $\mathcal{C}^{(4)}$ oscillates, qualitatively similarly to what is seen in the random matrix model Fig.~\ref{fig:rmt_higher_order_otocs}. {\bf b}, In the permutation representation, a superposition of three growing domains is created, with domains of $\alpha=(12)(34)$, $\beta=(14)(23)$, and $\omega=(1234)$ which in the scaling limit comprise the entire signal. The signal is separated into pieces $\mathcal{C}^{(4)}_{\alpha}$ by projecting the left-most site into the permutation $\alpha$, with the total well approximated by $\mathcal{C}^{(4)}_\alpha$ + $\mathcal{C}^{(4)}_\beta$ + $\mathcal{C}^{(4)}_\omega$, the later of which is negative. {\bf c}, Shown here is the MPS bond dimensions in the mean OTOC$^{(2)}$ calculation. The peak bond dimension saturates in time and follows the position of the front, allowing for an efficient simulation of the mean OTOC$^{(2)}$ in 1D. However, the front is a complex superposition of permutation configurations, requiring an bond dimension of over $1500$ to represent accurately. A two dimensional extended mean OTOC$^{(2)}$ front (not shown) is far more complex and beyond the capabilities of our MPS simulations to capture.}
    \label{fig:c4_permutations}
\end{figure*}

\subsubsection{Mean OTOC\,$^{(2)}$}\label{sec:meanotoc2}

As in the case of the variance of OTOC$^{(1)}$, for the analysis of mean OTOC$^{(2)}$ ($\overline{C^{(4)}}$) we use $r=4$ in Eq.~(\ref{eq:P_kviaWg}). The averaged Haar gates project onto a subspace spanned by $r!=24$ states labeled by permutations of four elements, and the dynamical rule Eq.~(\ref{eq:unscaleddynamics}) is identical. However, the final state changes from $\omega = (12)(34)$ to $\omega = (1234)$; the components of the permutation wavefunction with large overlap on that state undergo an evolution that is qualitatively different. As in the previous sections, the effect of this distinct final state can be incorporated throughout the evolution using the rescaled dynamics Eq.~(\ref{eq:scaleddynamics}), but now using overlaps with the cycle $\omega=(1234)$ for rescaling. 

The analysis of stable and unstable domains from the variance of OTOC$^{(1)}$ needs to be modified as follows: the overlap of a domain $\alpha$ with the new final state $d^{-n} \bra{\omega \ldots \omega}$ instead results in an exponential weight
\begin{gather}
d^{-l}\langle \omega | \alpha \rangle^l = d^{(3-|\omega^{-1}\alpha|) l}.
\end{gather}
Domains for which the dynamical weight of the domain wall balances the effect of the final state are ones that satisfy
\begin{gather}
3-|\omega^{-1}\alpha| - |\alpha| = 0.
\end{gather}

Many more permutations satisfy this than the $4$ in the previous section. They can be characterized easily by noticing that $|\sigma^{-1} \tau|$ is a distance metric on the \emph{Cayley graph}, a graph with permutations as nodes and edges for permutations connected by multiplication of a single transposition. Thus, by the triangle rule, every permutation satisfies
\begin{gather}
|\omega^{-1}\alpha| + |I^{-1}\alpha| \geqslant |I^{-1}\omega| = 3,
\end{gather}
with the equality satisfied by permutations on a shortest path between $\omega=(1234)$ and the identity $I$ on the Cayley graph. By manual inspection of the Cayley graph, this results in the following list of fourteen stable permutations:
\begin{multline}
I, (12), (13), (14), (23), (24), (34), (123), (124), (134), \\ (234), (12)(34), (14)(23), (1234).
\end{multline}

In addition, we find that there is a selection rule in the dynamics; after an initial transient time, permutations that have zero overlap with $B(t=0)^{\otimes 4}$ will not appear behind the front.
As a result, an initial butterfly operator chosen as a non-identity Pauli evolves into a superposition of only three distinct permutation domains corresponding to the permutations $\alpha=(12)(34)$, $\beta=(14)(23)$, and $\omega=(1234)$, which one can write schematically as
$$
 \overline{B^{\otimes 4}(t)} \sim \ket{\hat{\alpha} \hat{\alpha} \hat{\alpha} \ldots \hat{I}  \hat{I}} +  \ket{\hat{\beta} \hat{\beta} \hat{\beta} \ldots \hat{I}  \hat{I}} -  \ket{\hat{\omega} \hat{\omega} \hat{\omega} \ldots \hat{I}  \hat{I}},
$$
using as before the notation $\hat{\sigma}$ to represent a permutation rescaled by its overlap with the final boundary condition $\bra{\omega}$.
We note that each of these expanding domains becomes a complex superposition of all permutation types --- primarily the 14 stable permutations listed above --- in the region of the front. The total weight in the collections of terms with a $\alpha, \beta$, and $\omega$ domain quickly converges exactly to $1$, $1$, and $-1$ respectively, which can be derived analytically by considering the average output of a globally Haar random unitary acting on the chosen butterfly operator. 

% To leading order in $1/d$, the structure of the $\omega$ domain wall is dominated by basis states with $3$ domain walls, transition from $(1234)$ to $I$ in steps of single transpositions. The single transposition domain walls undergo free random walks, and interact upon contact. By contrast, the $\alpha$ and $\beta$ domain walls consist of both configurations with $2$ and $4$ domain walls --- the later configurations first transition into an $\omega$ domain and then down to $I$ in three steps. We observe that the $4$ domain wall configurations are created with processes that result in both positive and negative weights and that significant interference effects play a role in their dynamics. The persistent negative weights that appear in these evolutions are a consequence of the non-Markovianity in the dynamics and have an important consequence for the complexity of Monte Carlo sampling algorithms that will be explored further in App.~\ref{sec:permutationmontecarlo}.

The value of OTOC$^{(2)}$ for the measurement operator $M$ that consists of a non-identity Pauli at site $x$ can be expressed as the total quasi-probability of finding one of $9$ permutations at site $x$ --- specifically, the $9$ permutations $\tau$ such that
$$
{\rm Trace}{\left[ C^{\dagger} Z^{\otimes 4} \hat{\tau} \right]} = 1.
$$
In the 1D chain, we can compute the resulting dynamics with MPS, as in the previous section. In Fig.~\ref{fig:c4_permutations}(a), we compare the mean OTOC$^{(1)}$ result with mean OTOC$^{(2)}$. While both transition from $1$ to $0$ across the operator front, the mean OTOC$^{(2)}$ is sharper and takes a scaling form that deviates from an error function. In Fig.~\ref{fig:c4_permutations}(b), we see that the resulting OTOC$^{(2)}$ signal can be split into three parts, corresponding to the $\hat{\alpha}, \hat{\beta},$ and $\hat{\omega}$ domains behind the front. Each of these fronts grows with the same butterfly velocity $v_B$ that is relevant for $k=2$. In Fig.~\ref{fig:c4_permutations}(c), we show that the front is a complex superposition of permutations requiring a large bond dimension to represent accurately with an MPS. For truncation error $\varepsilon = 10^{-12}$, the maximum bond dimension saturates around $1750$; computations with smaller bond dimensions show significant error buildup in the mean OTOC$^{(2)}$ values over time. For two dimensional systems, MPS requires a cost exponential in width to capture the front, which results in unfeasibly large bond dimensions for computing mean OTOC$^{(2)}$ even in narrow strip geometries.

% but they spread with distinct diffusion constants $D \neq D_{\omega} \neq D_{\alpha}=D_{\beta}.$

% First, there is a hierarchy of favored domains which determine the direction of growth of domain walls. 
% Second, domain walls between permutations that differ by more than one transposition tend to split into multiple, elementary domain walls. In isolation, these elementary domain walls act as biased random walkers, but they have an effective contact interaction which changes their dynamics. (Reference figure, Cayley graph of S4) 
% Additionally, certain domain walls are unstable; they strongly favor production of an additional domain, creating more domain walls. (Reference figure, Cayley graph of S4) 
% Lastly, domains are often produced in superposition with both positive and negative weights, and interference effects between oppositely-signed processes plays a crucial role in the dynamics.

% \subsubsection{Estimating average OTOC$^{(2)}$ via permutation Monte Carlo}\label{sec:average_otoc2_permutation}
% \input{permutation_mc} % permutation basis analysis of Haar random mean otoc^1, var otoc^1, and mean otoc^2 using stat. mech. model, perturbation theory and MPS.

\subsection{Criticality in the correlation operator spectrum}\label{sec:phasetransition}

\subsubsection{Overview: spectrum and moments of the correlation operator \label{sec:rmt_intro}}
The goal of this supplement is to elucidate spectral properties of the ``correlation operator'' $C(t) = M B(t)$. 
We will show that the spectrum of $C(t)$ undergoes a dynamical phase transition as a function of time $t$.
On one side of the transition, the spectrum is gapped, while on the other side it is gapless.
We will demonstrate how the phase transition is reflected in out-of-time-ordered correlators (OTOCs).

Explicitly, the correlation operator between local observables $M$ and $B$ is given by
\begin{equation}\label{eq:rmt_corr_op_def}
C(t) = M U^\dagger(t) B U(t).
\end{equation}
Here $U(t)$ is a unitary matrix, e.g., generated by an application of a random circuit of depth $t$ (comprising single-qubit and local two-qubit operations) to $n$ qubits. Hereafter, we will focus on the case in which $M$ and $B$ are Pauli matrices acting on two different qubits, $M = Z_M$ and $B = Z_B$. This choice makes the correlation operator unitary, $C^\dagger C = \openone$. It also leads to the ``parity'' symmetry $C^\dagger = Z_{M}C Z_{M}$.

\begin{figure}[t]
  \begin{center}
    \includegraphics[scale = 1.1]{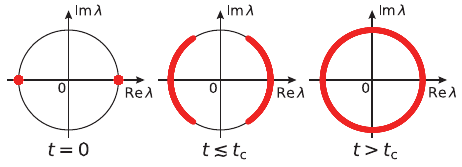}
    \caption{\textbf{Sketch of the evolution of the correlation operator spectrum with $\boldsymbol{t}$ across the phase transition point $\boldsymbol{t_\star}$.} (See Figs.~\ref{fig:rmt_rmt_results} and \ref{fig:rmt_numerics} for fine details). At $t = 0$, there are only two distinct eigenvalues, $\lambda = \pm 1$; each has a macroscopic degeneracy $D_{\rm H} / 2$ determined by the Hilbert space dimension $D_{\rm H}$. At finite $t < t_\star$, the spectrum broadens into two arcs. However, it remains gapped. The gap vanishes at $t \geq t_\star$.}
    \label{fig:rmt_sketch}
  \end{center}
\end{figure}

To explain what makes the spectral structure of $C(t)$ interesting, we first note that, due to the unitarity, all eigenvalues $\lambda$ of $C(t)$ lie on a unit circle in the complex plane, $\lambda = e^{i\varphi}$. 
The distribution of eigenvalues evolves with $t$. It is in this evolution that the phase transition occurs, see Fig.~\ref{fig:rmt_sketch}. 
Hints towards the existence of the phase transition can be seen from the comparison of the two extreme cases. 
At $t = 0$, unitary $U = \openone$. 
This turns $C(0)$ into a product of commuting operators $Z_M$ and $Z_B$, which can be straightforwardly diagonalized.
There are only two distinct eigenvalues: $\lambda = \pm 1$.
Each of the two has a macroscopic degeneracy $D_{\rm H} / 2$ determined by the Hilbert space dimension $D_{\rm H} = 2^n$.

In the opposite limit, $t \rightarrow \infty$, $U(t)$ becomes a fully scrambling unitary. This randomizes $C(t)$, and the eigenvalues of the latter spread uniformly across the circle. The mean angular spacing between the levels can be estimated as $\delta \varphi \sim 1 / D_{\rm H} = 2^{-n}$; it is exponentially small in the system size. Thus, for all practical purposes the spectrum can be viewed as a gapless one.

On the other hand, the spectrum cannot immediately turn gapless with the increase of $t$ from zero. Indeed, at small but finite $t$, perturbation theory is justified. It leads to the splitting of the degeneracy but only by a finite value. The spectrum, therefore, consists of two non-overlapping arcs, see Fig.~\ref{fig:rmt_sketch}.
The size of the arcs increases with $t$ until, eventually, the gap closing occurs. 
This defines a critical point $t_\star$ of the spectral phase transition.

To elucidate the evolution of the spectrum across $t_\star$, we introduce the spectral density
\begin{equation}
    S (\varphi) = \frac{1}{D_{\rm H}}\sum^{D_{\rm H}}_{i = 1} \delta(\varphi - \varphi_i)
\end{equation}
[in passing, we note that $S (\varphi) = S (-\varphi)$ due to the parity symmetry].
Below, we will develop a theory for the spectral density in an analytically-solvable random matrix theory model, see Section~\ref{sec:rmt_model}; we will also present numerical results on quantum circuits composed of random two-qubit gates, see Section~\ref{sec:rmt_numerics}.

The reconstruction of the spectrum occurring across $t = t_\star$ is manifested in the dependence of higher-order OTOCs, ${\cal C}^{(2k)}(t)$, on $k$. To see this, note that the higher-order OTOCs are nothing but the Fourier harmonics of the spectral density $S(\varphi)$ 
%\sm{Is it possible to avoid the use of OTOC$^{(k)}$ in formulas? In this case, OTOC$^{(k)}$ would be ${\rm Tr} [\mathcal{C}^{(2k)}]$, since Tr is the normalized trace}:
\begin{equation}\label{eq:rmt_OTOCs}
    {\cal C}^{(2k)}(t) = \tfrac{1}{D_{\rm H}} \mathrm{Trace}[C^{2k}(t)] =\hspace{-0.15cm} \int_{-\pi}^{\pi} \hspace{-0.25cm}d\varphi\, S(\varphi) \cos(2k\varphi).
\end{equation}
The asymptotic behavior of the harmonics at $k \gg 1$ reveals the presence of discontinuities in the function $S(\varphi)$; it is thus different below and above $t_\star$.
In the gapped phase, $t < t_\star$, $S(\varphi)$ is supported only within the spectral arcs [see Fig.~\ref{fig:rmt_sketch}], and its derivatives are discontinuous at the arcs edges. This translates into a slow power-law decay of the harmonics with $k$,
\begin{equation}\label{eq:rmt_n_below}
    {\cal C}^{(2k)}(t) \propto \frac{\cos(\omega k - \xi)}{k^{\alpha}},\quad\quad t < t_\star.
\end{equation}
The decay is accompanied by oscillations [$\omega$ and $\xi$ depend smoothly on $t_\star - t$]. A power-law dependence persists to the critical point,
\begin{equation}\label{eq:rmt_n_at}
    {\cal C}^{(2k)}(t) \propto \frac{(-1)^k}{k^{\alpha_{\rm c}}},\quad\quad t = t_\star,
\end{equation}
although the exponent changes to $\alpha_{\rm c} \neq \alpha$.
The behavior is different in the gapless phase, $t > t_\star$, where $S(\varphi)$ becomes a smooth function. The smoothness results in an exponential dependence 
\begin{equation}\label{eq:rmt_n_above}
    {\cal C}^{(2k)}(t) \propto (-1)^k \exp(-\nu k),\quad\quad t > t_\star,
\end{equation}
with $\nu \equiv \nu(t - t_\star) > 0$.
The change in the asymptotic behavior across $t_\star$ is a direct signature of the phase transition that can be observed by measuring OTOCs of different order.\\

The remainder of the supplement is organized in the following way. We describe the random-matrix theory model of the spectral phase transition in Section~\ref{sec:rmt_model}. In Section~\ref{sec:rmt_otocs}, we find OTOCs in this model. In particular, we compare the asymptotic behavior of the high-order OTOCs below, at, and above the critical point, see Section~\ref{sec:rmt_crit_otoc}. Section~\ref{sec:rmt_derivations} contains detailed derivations of the central equations of the random-matrix theory.
Lastly, in Section~\ref{sec:rmt_numerics}, we demonstrate the existence of the phase transition in a realistic random quantum circuit.

\subsubsection{Random matrix model of the spectral phase transition\label{sec:rmt_model}}
In this section, we present a solvable random-matrix based model allowing one to capture the evolution of spectrum of the correlation operator $C = Z_M U^\dagger Z_B U$ with the increase of randomness in $U$. We will use this model to elucidate the critical behavior in the spectral density $S(\varphi)$, and show how the criticality is reflected in the higher-order OTOCs.

Specifically, we consider a model in which the evolution operator $U$ is a random $D_{\rm H} \times D_{\rm H}$ unitary matrix. To fully define the model, we need to describe an ensemble from which $U$ is sampled. A natural choice would be to first generate a random Hamiltonian $H$ from a Gaussian unitary ensemble (GUE), 
\begin{subequations}\label{eq:rmt_gue}
\begin{align}\label{eq:rmt_average_is_zero}
    &\langle H_{ij} \rangle = 0,\\
    &\langle H_{ij} H_{kl} \rangle = \frac{\gamma }{D_{\rm H}} \delta_{il}\delta_{jk},\quad i,j,k,l = 1,\dots, D_{\rm H},\label{eq:rmt_correlator}
\end{align}
\end{subequations}
and then introduce $U = \exp(-2iH)$ as the respective evolution operator [we normalized the correlator of $H$ by the Hilbert space dimension to make the thermodynamic limit $D_{\rm H} \rightarrow \infty$ well-defined].
%corresponding to this Hamiltonian. 
There is a correspondence between the model parameter $\gamma$ and the evolution time $t$ of Eq.~\eqref{eq:rmt_corr_op_def} and Fig.~\ref{fig:rmt_sketch}; {we will clarify this correspondence in Section~\ref{sec:rmt_numerics}.}
In principle, the described choice of $U$ would allow one to capture the crossover between $U = \openone$ and a fully scrambled $U$ by varying $\gamma$. The problem with this approach is the exponential dependence of $U$ on $H$; it makes finding the spectral density $S(\varphi)$ of $C$ analytically a difficult task. 
Instead, we will consider an ensemble of unitaries obtained as a Cayley transform of GUE:
\begin{equation}\label{eq:rmt_ensemble}
    U = \frac{\openone - i H}{\openone + i H}.
\end{equation}
For small $\gamma$, $U$ defined in this way is close to $U = \exp(-2iH)$. Although the respective ensembles become significantly different for $\gamma \gtrsim 1$, we will see that the phase transition occurs already at $\gamma = 1/3$. 
The main advantage of the ensemble in Eq.~\eqref{eq:rmt_ensemble} is that for such an ensemble it is possible to find the spectral density $S(\varphi)$ analytically \cite{khindanov2024}. 

To access $S(\varphi)$, we introduce the Green's function of the correlation operator:
\begin{equation}\label{eq:rmt_G}
    G(\lambda) = \frac{1}{\lambda \openone - C}.
\end{equation}
The Green's function contains the full spectral information about $C$; in particular, the spectral density can be represented as
\begin{equation}\label{eq:rmt_density}
    S(\varphi) = \frac{1}{2\pi D_{\rm H}}\sum_{s = \pm 1} s\lambda\, \bigl\langle\mathrm{Trace} [G(\lambda(1+s\,0^+))]\bigr\rangle\bigr|_{\lambda = e^{i\varphi}},
\end{equation}
where $0^+$ is an infinitesimal positive number and $\langle \dots \rangle$ denotes the averaging over the realizations of $H$.

\begin{figure*}[t]
  \begin{center}
    \includegraphics[scale = 1.0]{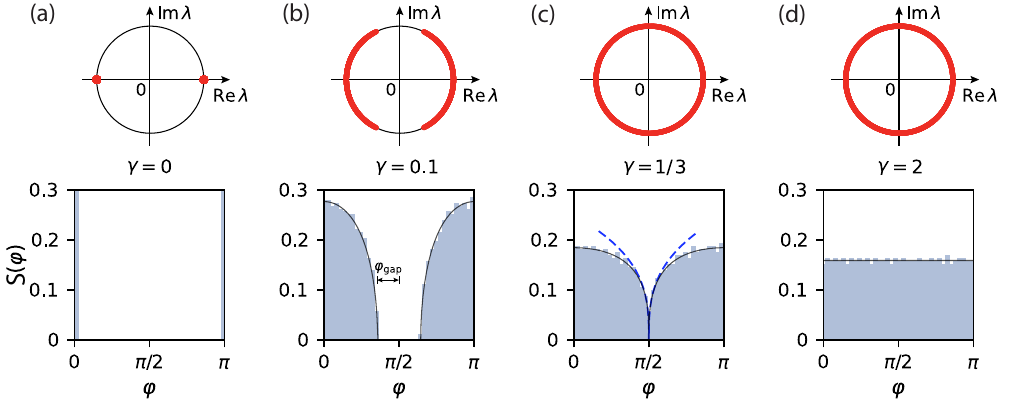}
    \caption{
    \textbf{Evolution of the correlation operator spectrum in the random-matrix theory model.} {\bf a}, At $\gamma = 0$, there are only two distinct eigenvalues $\lambda = \pm 1$. Each has a macroscopic degeneracy of $D_{\rm H} / 2$, where $D_{\rm H}$ is the Hilbert space dimension. {\bf b},~At a finite $\gamma$, the spectrum broadens into two arc (histogram is produced by numerically diagonalizing the random matrix $C$ for $D_{\rm H} = 2048$; solid black curve is obtained via a numeric solution of Eqs.~\eqref{eq:rmt_G_av} and \eqref{eq:rmt_selfenergy}]. {\bf c}, The arcs merge at the critical point $\gamma = \gamma_\star = 1 / 3$. Dashed blue curve depicts the asymptotic result for $S(\varphi)$ near $\varphi = \pi / 2$ [see Eq.~\eqref{eq:rmt_critical_dos}). {\bf d}, Spectral density becomes uniform at $\gamma = 2$. 
    As we will see, the evolution of the correlation operator spectrum in realistic quantum circuits with local, random gates is qualitatively similar to the one presented here (cf.~Fig.~\ref{fig:rmt_numerics}). 
    \label{fig:rmt_rmt_results}}
  \end{center}
\end{figure*}

The introduction of ensemble averaging deserves a comment. On the one hand, it is this ensemble averaging that makes the problem tractable. The averaging washes out fine features of the level distribution, and leads to a well-behaved, smooth density function $S(\varphi)$. On the other hand, it raises a question of the importance of sample-to-sample fluctuations. In fact, for our random-matrix theory model, sample-to-sample fluctuations are negligible; this is in full similarity to standard random-matrix theory behavior where the fluctuations are small in $1 / D_{\rm H}$. 

How does one carry out an ensemble averaging in Eq.~\eqref{eq:rmt_density}? Were $C$ linear in $H$, one could expand the right-hand side of Eq.~\eqref{eq:rmt_G} in powers of $C$, and straightforwardly perform the averaging term by term using Eq.~\eqref{eq:rmt_gue}.
However, $C$ depends on $H$ in a non-linear way; this makes a direct realization of such a program laborious. 

A key property of ensemble~\eqref{eq:rmt_ensemble} is that for this ensemble the task of finding $\langle {\rm Trace} [G(\lambda)] \rangle$ can actually be mapped onto inverting and averaging an auxiliary matrix linear in $H$. The mapping procedure closely follows Ref.~\onlinecite{khindanov2024}, and we explain it in detail in a later section [Section~\ref{sec:rmt_derivations}]. For now, we only present a final result 
for $\langle {\rm Trace} [G(\lambda)] \rangle$
valid in the limit of $D_{\rm H} \gg 1$. Using a variant of a self-consistent Born approximation tailored for the problem at hand, we will show that
\begin{equation}\label{eq:rmt_G_av}
    \langle {\rm Trace} [G(\lambda)] \rangle = \frac{D_{\rm H}}{2}\sum_{r = \pm 1} \frac{1}{\lambda -  \Bigl(\tfrac{\varepsilon + i}{\varepsilon - i}\Bigr)^2 r}.
\end{equation}
Here $\varepsilon \equiv \varepsilon(\lambda)$ is the ``self-energy'' function. The self-energy satisfies the self-consistency relation; in a parameterization $\lambda = e^{i\varphi}$ natural for evaluating $S(\varphi)$ [cf.~Eq.~\eqref{eq:rmt_density}], this relation reads
\begin{align}\label{eq:rmt_selfenergy}
    \varepsilon(\varphi) = \frac{\gamma}{2}\Bigl(&\frac{\varepsilon(\varphi)-\cot(\varphi / 2)}{1-\varepsilon^2(\varphi)+2\varepsilon(\varphi)\cot(\varphi / 2)}\notag\\
    &\quad\quad + \frac{\varepsilon(\varphi)+\tan(\varphi / 2)}{1-\varepsilon^2(\varphi)-2\varepsilon(\varphi)\tan(\varphi / 2)}\Bigr).
\end{align}
One immediate consequence of this equation is that the average spectral density satisfies $S(\varphi + \pi) = S(\varphi)$. 
{This feature results from  Eq.~\eqref{eq:rmt_average_is_zero}}. The other consequence of Eq.~\eqref{eq:rmt_selfenergy} is that $S(\varphi) = S(-\varphi)$, which is nothing but the exact parity symmetry discussed after Eq.~\eqref{eq:rmt_corr_op_def}.

Let us now use Eqs.~\eqref{eq:rmt_density}--\eqref{eq:rmt_selfenergy} to describe the evolution of the spectral density $S(\varphi)$ with $\gamma$, and elucidate the critical behavior in it. \\

%\subsubsection{Weak randomness, $\gamma \ll 1$}
\noindent\textbf{Weak randomness, $\boldsymbol{\gamma \ll 1}$ ---} To start off, it is useful to find $S(\varphi)$ in the limit of weak randomness, $\gamma \ll 1$. Without randomness, $D_{\rm H} / 2$ eigenvalues of $C$ are equal to $+1$, while the remaining $D_{\rm H} / 2$ are equal to $-1$. The introduction of randomness lifts this macroscopic degeneracy; however, as long as the randomness is weak, the eigenvalues remain concentrated near $\lambda = \pm 1$. Keeping the symmetry of $S(\varphi)$ in mind, we focus on the modification of the eigenvalue density $S(\varphi)$ near $\varphi = 0$.

For $\gamma \ll 1$ and $\varphi \ll 1$, the self-energy function is itself small, $\varepsilon(\varphi) \ll 1$ (as will become apparent momentarily). This allows one to simplify Eq.~\eqref{eq:rmt_selfenergy}, and bring it to a compact form
\begin{equation}
    \varepsilon(\varphi) = -\frac{\gamma}{2} \frac{1}{2\varepsilon(\varphi)+\varphi / 2}.
\end{equation}
A solution of the equation for $\varepsilon(\varphi)$ is straightforward:
\begin{equation}
    \varepsilon(\varphi) = -\frac{\varphi}{8} \pm \frac{i}{8} \sqrt{16 \gamma - \varphi^2}.
\end{equation}
Combining this result with Eqs.~\eqref{eq:rmt_density} and \eqref{eq:rmt_G_av}, we find a Wigner semicircle distribution of levels:
\begin{equation}\label{eq:rmt_small_tau}
    S(\varphi \ll 1) = \frac{1}{16\pi\gamma} \sqrt{16\gamma - \varphi^2}\cdot\Theta(16\gamma - \varphi^2)
\end{equation}
[here $\Theta(x)$ is a step function]. In addition to a semicircle centered at $\varphi = 0$, $S(\varphi)$ has a semicircle around $\varphi = \pi$, of the same width.

The above shows that the eigenvalues of $C$ form two disjoint arcs on a unit circle, see Fig.~\ref{fig:rmt_rmt_results}(b). Outside of the arcs, the density of levels is negligible; the spectrum has a gap.
The arcs expand with the increase of $\gamma$ [see Eq.~\eqref{eq:rmt_small_tau}]. 
The expansion culminates at a critical value $\gamma_\star$, at which the arcs merge and the spectrum becomes gapless [Fig.~\ref{fig:rmt_rmt_results}(c)]. 

Due to the symmetry of $S(\varphi)$, the closing of the gap occurs at $\varphi = \pm \pi / 2$. 
One can find the critical point $\gamma_\star$ by examining the structure of solutions of Eq.~\eqref{eq:rmt_selfenergy} for $\varepsilon(\pm \pi/2)$. 
The gap closing corresponds to the appearance of complex (as opposed to purely real) solutions for this quantity. 
From this condition, we find $\gamma_\star = 1/3$. With that, we proceed to the critical behavior of $S (\varphi)$.\\

\noindent\textbf{Critical behavior of the spectral density ---}
At $\gamma = \gamma_\star = 1 / 3$ and $\varphi = \pi / 2$, the self-energy function vanishes, $\varepsilon = 0$.
To study the critical behavior of the spectral density $S(\varphi)$, we find how $\varepsilon$ deviates from zero with $\gamma - \gamma_\star$ and $\varphi - \pi/2$.
The structure of these deviations directly translates into a respective structure of $S (\varphi)$. Indeed, it follows from Eqs.~\eqref{eq:rmt_density} and \eqref{eq:rmt_G_av} that the spectral density satisfies
\begin{equation}\label{eq:rmt_se_to_rho_crit}
    S(\varphi) = -\frac{2}{\pi}\mathrm{Im}\,\varepsilon(\varphi),
\end{equation}
as long as the self-energy is small, $|\varepsilon|\ll 1$.

 Performing an expansion in the deviations $\gamma - \gamma_\star \ll 1$ and $\varphi - \pi/2 \ll 1$ in Eq.~\eqref{eq:rmt_selfenergy}, we obtain the following equation for $\varepsilon(\varphi)$:
\begin{equation}\label{eq:rmt_cubic}
    (\varphi - \pi / 2) + 9 \varepsilon(\varphi) (\gamma - \gamma_\star) + 17 \varepsilon^3(\varphi) = 0.
\end{equation}
This is a cubic equation, and it can be solved explicitly.  At this point though, it is more instructive to examine its general scaling properties.

\begin{figure}[t]
  \begin{center}
    \includegraphics[scale = 1]{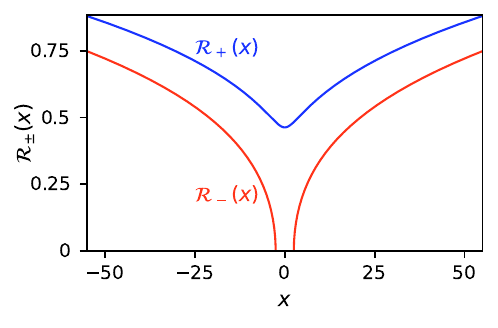}
    \caption{\textbf{Scaling functions $\boldsymbol{{\cal R}_{\pm}(x)}$ characterizing the behavior of the spectral density in the vicinity of the critical point}. See Eqs.~\eqref{eq:rmt_scaling}--\eqref{eq:rmt_scaling_below}.}
    \label{fig:rmt_scaling}
  \end{center}
\end{figure}

To implement the scaling analysis, we note that Eq.~\eqref{eq:rmt_cubic} can be made ``dimensionless'' by introducing rescaled variables $\tilde{\varepsilon} \equiv |\gamma - \gamma_\star|^{1/2} \varepsilon$ and $\tilde{\varphi} \equiv |\gamma - \gamma_\star|^{3/2}(\varphi - \pi/2)$. In terms of these variables, the equation acquires the form
\begin{equation}\label{eq:rmt_cubic_rescaled}
    \tilde{\varphi} \pm 9\tilde{\varepsilon} + 17 \tilde{\varepsilon}^3 = 0,
\end{equation}
where the $+$ and $-$ signs correspond to $\gamma > \gamma_\star$ and $\gamma < \gamma_\star$, respectively. Clearly, the solution for $\tilde{\varepsilon}$ depends only on a rescaled angle $\tilde{\varphi}$.
This observation stipulates a scaling relation for the self-energy: $\varepsilon(\varphi) = (\gamma - \gamma_\star)^{1/2} {\cal E}\bigl([\varphi - \pi / 2] / (\gamma - \gamma_\star)^{3/2}\bigr)$.  Thus, $S(\varphi)$ obeys a scaling law too [cf.~Eq.~\eqref{eq:rmt_se_to_rho_crit}]:
\begin{equation}\label{eq:rmt_scaling}
    S(\varphi) = \begin{cases}
    (\gamma - \gamma_\star)^{1/2}{\cal R}_+\bigl(\tfrac{\varphi - \pi / 2}{(\gamma - \gamma_\star)^{3/2}}\bigr),\quad\quad \gamma > \gamma_\star,\\
    (\gamma_\star - \gamma)^{1/2}{\cal R}_-\bigl(\tfrac{\varphi - \pi / 2}{(\gamma_\star - \gamma)^{3/2}}\bigr),\quad\quad \gamma < \gamma_\star.
    \end{cases}
\end{equation}
The scaling functions are different above and below the phase transition point; they follow from the solution of the cubic equation, Eq.~\eqref{eq:rmt_cubic_rescaled}. Above the transition, $\gamma > \gamma_\star$, we obtain
\begin{align}\label{eq:rmt_scaling_above}
    {\cal R}_{+} (x) = \frac{\sqrt{3}}{34^{1/3}\pi}\Bigl[&\Bigl(\sqrt{x^2 + \tfrac{108}{17}} + x\Bigr)^{1/3} \notag\\
    &+\Bigl(\sqrt{x^2 + \tfrac{108}{17}} - x\Bigr)^{1/3}\Bigr].
\end{align}
Notice that ${\cal R}_{+} (x) > 0$ for all $x$; this feature reflects that the spectrum is gapless at $\gamma > \gamma_\star$.

The structure of the scaling function changes below the transition point due to the gap opening.
The magnitude of the gap can be found by a direct inspection of Eq.~\eqref{eq:rmt_cubic_rescaled}. 
A simple analysis of this equation shows that there exists a parameter region $\tilde{\varphi} \in [-\sqrt{108/17} , \sqrt{108/17}]$ in which all three solutions for $\tilde{\varepsilon}$ are real.
In this region, the spectral density vanishes [cf.~Eq.~\eqref{eq:rmt_se_to_rho_crit}]. 
The scaling function describing the gapped spectrum at $\gamma < \gamma_\star$ reads
\begin{align}\label{eq:rmt_scaling_below}
    {\cal R}_-(x) &= \frac{\sqrt{3}}{34^{1/3}\pi}\Bigl[\Bigl(|x| + \sqrt{x^2 - \tfrac{108}{17}}\Bigr)^{1/3} \notag\\
    &-\Bigl(|x| - \sqrt{x^2 - \tfrac{108}{17}}\Bigr)^{1/3}\Bigr]\cdot\Theta\Bigl(|x| - \sqrt{\tfrac{108}{17}}\Bigr).
\end{align}
Eqs.~\eqref{eq:rmt_scaling}--\eqref{eq:rmt_scaling_below} give a complete description of the behavior of spectral density $S(\varphi)$ in the vicinity of the critical point. The behavior of scaling functions ${\cal R}_{\pm}(x)$ is demonstrated in Fig.~\ref{fig:rmt_scaling}.

Let us use the derived scaling functions to provide a set of explicit results for the dependence of the spectrum of $C$ on $\gamma - \gamma_\star$ and $\varphi$.
First, by combining Eqs.~\eqref{eq:rmt_scaling} and \eqref{eq:rmt_scaling_below}, we obtain the dependence of the spectral gap on $\gamma_\star - \gamma$:
\begin{equation}\label{eq:rmt_gap}
    \varphi_{\rm gap} = \sqrt{\tfrac{108}{17}} (\gamma_\star - \gamma)^{3/2}.
\end{equation}
Near the edge of the gap, $0 < |\varphi - \pi/2| - \varphi_{\rm gap} \ll \varphi_{\rm gap}$, the spectral density has a square root behavior,
\begin{equation}\label{eq:rmt_cusps}
S(\varphi) = \frac{2}{\pi}\left(\frac{|\varphi - \tfrac{\pi}{2}| - \varphi_{\rm gap}}{3\sqrt{51(\gamma_\star -\gamma)}}\right)^{1/2}\hspace{-0.2cm}\cdot \Theta\bigl(|\varphi - \tfrac{\pi}{2}| - \varphi_{\rm gap}\bigr).
\end{equation}
Eq.~\eqref{eq:rmt_gap} and Eq.~\eqref{eq:rmt_cusps} elucidate how the arcs in the spectrum of $C$ approach each other as $\gamma \rightarrow \gamma_\star$.

At the critical point, $\gamma = \gamma_\star$, the gap is absent but $S(\pi / 2) = 0$. It is natural to ask how $S(\varphi)$ deviates from zero when $\varphi$ is tuned away from $\pi / 2$. By assessing the asymptotic behavior of scaling functions ${\cal R}_\pm(x)$ at $|x| \gg 1$, we find \begin{equation}\label{eq:rmt_critical_dos}
    S(\varphi) = \frac{\sqrt{3}}{17^{1/3}\pi} |\varphi - \pi / 2|^{1/3}.
\end{equation}
In Fig.~\ref{fig:rmt_rmt_results}(c), we compare this prediction to an explicit diagonalization of matrix $C$, as well as to a direct numerical solution of Eqs.~\eqref{eq:rmt_density}--\eqref{eq:rmt_selfenergy}. The three results are in a perfect agreement.

Lastly,  we describe how $S(\varphi = \pi / 2)$ departs from zero with the increase of $\gamma$ above $\gamma_\star$. Setting $\varphi = \pi / 2$ in Eq.~\eqref{eq:rmt_scaling}, we arrive to
\begin{equation}\label{eq:rmt_buildup}
    S(\pi / 2) = \frac{1}{\pi} \sqrt{\tfrac{36}{17}}(\gamma - \gamma_\star)^{1/2}\cdot \Theta(\gamma - \gamma_\star),
\end{equation}
where a specific numeric coefficient follows from Eq.~\eqref{eq:rmt_scaling_above}.

Eqs.~\eqref{eq:rmt_gap}--\eqref{eq:rmt_buildup} define a set of critical exponents: $3/2$ for the dependence of the gap on $\gamma_\star - \gamma$; $1/2$ for the behavior of $S(\varphi)$ near the gap edge; $1/3$ for the $\varphi$-dependence of $S(\varphi)$ at the critical point; and $1/2$ for the build up of the spectral density at $\varphi = \pm \pi / 2$ with $\gamma - \gamma_\star$. \\

\noindent\textbf{Spectral density at $\boldsymbol{\gamma = 2}$}
The spectral density of $C$ becomes more and more uniform with the increase of $\gamma$ above its critical value $\gamma_\star = 1/3$. This process completes at $\gamma = 2$, where $S(\varphi)$ becomes constant. To see this, we note that at $\gamma = 2$ Eq.~\eqref{eq:rmt_selfenergy} admits a $\varphi$-independent solution $\varepsilon(\varphi) = \pm i$. Its use in Eqs.~\eqref{eq:rmt_density} and \eqref{eq:rmt_G_av} leads to $S(\varphi) = 1/(2\pi)$. A uniform $S(\varphi)$ can be rationalized by noting that, at $\gamma = 2$, the ensemble defined by Eq.~\eqref{eq:rmt_ensemble} mimics the scrambling property $\langle U \rangle = 0$ of the circular unitary ensemble (i.e., the Haar-random distribution).

\subsubsection{OTOCs in the random-matrix model\label{sec:rmt_otocs}}
We succeeded in showing the existence of a phase transition that occurs in the spectrum $S(\varphi)$ of the correlation operator $C$ with a variation of $\gamma$. It is important to realize, though, that $S(\varphi)$ itself cannot be accessed directly on a quantum computer; one can only measure its moments, i.e., OTOCs [see Eq.~\eqref{eq:rmt_OTOCs}]. In this section, we find OTOCs in the random-matrix theory model of Section~\ref{sec:rmt_model}. We will obtain explicit expressions describing the evolution of the low-order OTOCs with $\gamma$; we will also elucidate signatures of the spectral phase transition in the dependence of ${\rm OTOC}^{(k)} \equiv {\cal C}^{(2k)}$ on $k$. 
Throughout this section, we focus on ensemble-averaged quantities; this is justified by the fact that the spectral density $S(\varphi)$ is self-averaging in our model. \\
%\sm{Same here. It would be better to avoid the use of OTOC$^{(k)}$ in formulas, because OTOC$^{(k)}$ and $\mathcal{C}^{(2k)}$ are used interchangeably in the rest of the SM.}

\noindent\textbf{Derivation of the $\boldsymbol{\gamma}$-dependence of the low-order OTOCs ---}
\begin{figure}[t]
  \begin{center}
    \includegraphics[scale = 1]{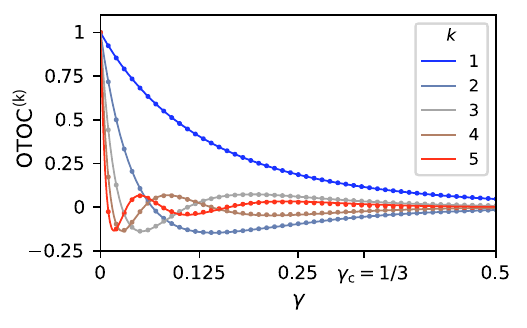}
    \caption{\textbf{Dependence of $\bm{{\cal C}^{(2k)}}$  on $\boldsymbol{\gamma}$ for $\boldsymbol{k = 1,}$$\dots$$\boldsymbol{, 5}$.} Solid curves are produced using analytical formulas (Eqs.~\eqref{eq:rmt_OTOC}--\eqref{eq:rmt_OTOC3}, and their generalizations for $k = 4$ and $5$). Dots depict the results obtained by a direct diagonalization of a matrix $C$ for $D_{\rm H} = 1024$.}
    \label{fig:rmt_higher_order_otocs}
  \end{center}
\end{figure}
As follows from Eq.~\eqref{eq:rmt_G}, individual OTOCs can be extracted from the Laurent expansion of the Green's function:
\begin{equation}
    \tfrac{1}{D_{\rm H}}\,\langle {\rm Trace}[G(\lambda)] \rangle = \sum_{k=0}^{\infty}\frac{1}{\lambda^{2k+1}}{\cal C}^{(2k)}.
\end{equation}
[the terms with odd powers of $\lambda$ cancel due to the symmetry of the spectrum $S(\varphi) = S(\varphi + \pi)$]. 
We carry out the Laurent expansion with the help of Eqs.~\eqref{eq:rmt_G_av} and \eqref{eq:rmt_selfenergy}. First, we iteratively solve Eq.~\eqref{eq:rmt_selfenergy} for the self-energy at $\lambda \rightarrow \infty$; this allows us to represent it as a series $\varepsilon(\lambda) = \varepsilon_0 + \varepsilon_2 / \lambda^2 + \varepsilon_4 / \lambda^4 + \dots$. For example, $\varepsilon_0$ is given by
\begin{equation}
    \varepsilon_0 = -\frac{i}{2}\Bigl(\sqrt{1 + 4\gamma} - 1\Bigr).
\end{equation}
The expression for $\varepsilon_2$ and higher-order coefficients are cumbersome, and we refrain from presenting them here. Then, we obtain the OTOCs by substituting the found $\varepsilon(\lambda)$ into Eq.~\eqref{eq:rmt_G_av}, and expanding the latter in $1/\lambda$. With due diligence, this procedure can be used to find ${\cal C}^{(2k)}$ with any given $k$. For $k = 1, 2$ and $3$, we find
\begin{widetext}
\begin{align}\label{eq:rmt_OTOC}
    {\cal C}^{(2)}(\gamma) &= \left(\frac{3-\sqrt{4\gamma +1}}{1+\sqrt{4\gamma + 1}}\right)^4,\\
    {\cal C}^{(4)}(\gamma) &= \left(\frac{3-\sqrt{4\gamma +1}}{1+\sqrt{4\gamma + 1}}\right)^6\frac{(4\gamma - 22)\sqrt{1+4\gamma} + 26 - 24 \gamma}{\sqrt{1+4\gamma}(1+\sqrt{1+4\gamma})^2},\label{eq:rmt_OTOC2}\\
    {\cal C}^{(6)}(\gamma) &= 8\left(\frac{3-\sqrt{4\gamma +1}}{1+\sqrt{4\gamma + 1}}\right)^8\frac{
    (241 + 240 \gamma - 78 \gamma^2 + 8 \gamma^3)
    \sqrt{1+4\gamma} 
    -239 - 814 \gamma + 736 \gamma^2 - 96 \gamma^3
    }{(1+4\gamma)^{3/2}(1+\sqrt{1+4\gamma})^4}.\label{eq:rmt_OTOC3}
\end{align}
\end{widetext}
As expected, ${\cal C}^{(2)} = {\cal C}^{(4)} = {\cal C}^{(6)} = 1$ at $\gamma = 0$. In the regime where $U$ is fully scrambling, $\gamma = 2$, we obtain ${\cal C}^{(2k)} = 0$; this is in agreement with a uniform distribution $S(\varphi)$ found above. Analytical expressions for higher-order OTOCs ($k = 4, 5,\dots$) 
quickly become unwieldy, so we do not present formulas with $k \geq 4$ explicitly.

Fig.~\ref{fig:rmt_higher_order_otocs} shows the dependence of ${\cal C}^{(2k)}$ with \mbox{$k=1,\dots,5$} on $\gamma$. In contrast to the lowest-order ${\cal C}^{(2)}$, OTOCs with $k \geq 2$ are non-monotonic, sign-alternating functions of $\gamma$.\\

\paragraph{Signature of the phase transition in the OTOCs.\label{sec:rmt_crit_otoc}}
Fig.~\ref{fig:rmt_higher_order_otocs} may create an impression that the phase transition occurring at $\gamma_\star = 1/3$ is undetectable on the level of OTOCs. This is not correct. As discussed in Section~\ref{sec:rmt_intro}, the phase transition can be revealed by assessing the asymptotic dependence of OTOCs on $k$. Here we find the exponents $\alpha$ and $\alpha_{\rm c}$ characterizing this dependence in our random-matrix theory model [cf.~Eqs.~\eqref{eq:rmt_n_below} and \eqref{eq:rmt_n_at}]. We also find signatures of the criticality in the dependence of $\omega$ and $\nu$ [see Eqs.~\eqref{eq:rmt_n_below} and \eqref{eq:rmt_n_above}] on $\gamma - \gamma_\star$.

To set the stage, it is useful evaluate ${\cal C}^{(2k)}$ deep in the gapped phase, $\gamma \ll \gamma_\star$. In this case, the distribution of levels is a Wigner semicircle [see Eq.~\eqref{eq:rmt_small_tau}]. For such a simple distribution, all of the OTOCs can be found analytically. Using Eq.~\eqref{eq:rmt_small_tau} in Eq.~\eqref{eq:rmt_OTOCs} and computing the resulting integral, we obtain
\begin{equation}\label{eq:rmt_bessel}
    {\cal C}^{(2k)}(\gamma) = \frac{J_1(8k\sqrt{\gamma})}{4k\sqrt{\gamma}}, 
\end{equation}
where $J_1(x)$ is the Bessel function. At large $k$, this expression turns into
\begin{equation}\label{eq:rmt_otoc_vs_n_deep}
    {\cal C}^{(2k)}(\gamma) \approx \frac{\cos\bigl(8\sqrt{\gamma} k - 3\pi / 4\bigr)}{8\sqrt{\pi}\gamma^{3/4}k^{3/2}}. 
\end{equation}
The dependence of OTOC on $k$ is oscillatory; the frequency of oscillations $\omega = 8\sqrt{\gamma}$ depends on $\gamma$. 
Eq.~\eqref{eq:rmt_otoc_vs_n_deep} fixes $\alpha = 3/2$ for an exponent describing the power-law decay of the higher-order OTOCs in the gapped phase [cf.~Eq.~\eqref{eq:rmt_n_below}].

We note that the value $\alpha = 3/2$ persists beyond the applicability of Eq.~\eqref{eq:rmt_otoc_vs_n_deep}; in fact, it remains the same throughout the gapped phase. This value reflects the character of the edge singularity in the spectral density at $\gamma < \gamma_\star$, $S(\varphi) \propto (|\varphi - \pi / 2| - \varphi_{\rm gap})^{1/2}\cdot \Theta(|\varphi - \pi / 2| - \varphi_{\rm gap})$.

The character of the edge singularity changes at the critical point, $\gamma = \gamma_\star$. There, $S(\varphi) \propto |\varphi - \pi / 2|^{1/3}$ [see Eq.~\eqref{eq:rmt_critical_dos}]. The exponent governing the dependence of OTOCs on $k$ jumps to $\alpha_{\rm c} = 4 / 3$. 

In the gapless phase, $\gamma > \gamma_\star$, $S(\varphi)$ becomes a smooth function. This results in the exponential decay of its harmonics with $k$, ${\cal C}^{(2k)} \propto {\rm exp} (-\nu k)$. Near the critical point, the dependence of the coefficient $\nu$ on $\gamma - \gamma_\star$ follows from scaling law \eqref{eq:rmt_scaling}. Using this law together with Eq.~\eqref{eq:rmt_OTOCs}, we obtain $\nu \propto (\gamma - \gamma_\star)^{3/2}$. 
A similar argument applied the gapped side of the transition gives $\omega - \pi \propto (\gamma_\star - \gamma)^{3/2}$.

\begin{figure*}[t]
  \begin{center}
    \includegraphics[scale = 1]{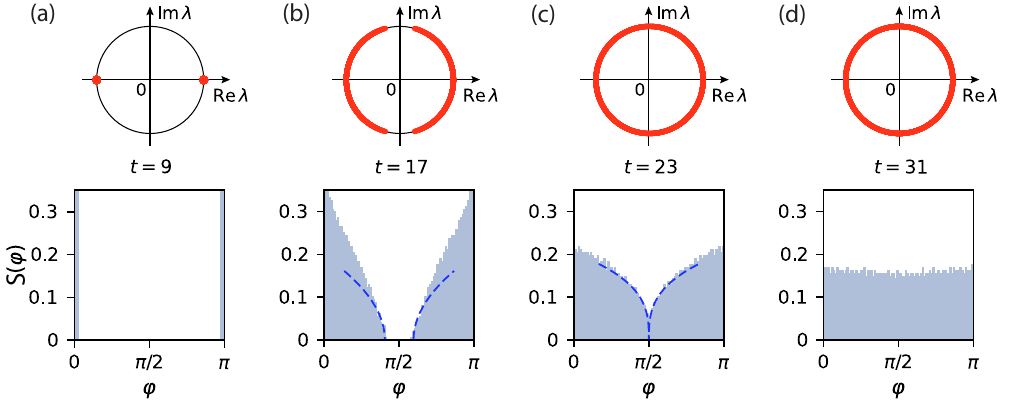}
    \caption{
    \textbf{Evolution of the correlation operator spectrum in a quantum circuit composed of Haar random two-qubit gates (the system size is $\boldsymbol{n = 12}$).}
    The behavior of the spectral density $S(\varphi)$ is qualitatively similar to the respective behavior in the random-matrix model (cf.~Fig.~\ref{fig:rmt_rmt_results}). In panel ({\bf b}), the dashed line shows the dependence $S(\varphi) \propto (|\varphi - \pi / 2| - \varphi_{\rm gap})^{1/2}\cdot\Theta(|\varphi - \pi / 2| - \varphi_{\rm gap})$ expected from the random-matrix theory [Eqs.~\eqref{eq:rmt_small_tau} and \eqref{eq:rmt_cusps}]. The dashed line in panel ({\bf c}) depicts $S(\varphi) \propto |\varphi - \pi / 2|^{1/3}$ expected at the critical point (Eq.~\eqref{eq:rmt_critical_dos}). 
    \label{fig:rmt_numerics}}
  \end{center}
\end{figure*}

\subsubsection{Derivation of Eqs.~\texorpdfstring{\eqref{eq:rmt_G_av}}{} and~\texorpdfstring{\eqref{eq:rmt_selfenergy}}{}\label{sec:rmt_derivations}}
Let us now explain how we derived Eqs.~\eqref{eq:rmt_G_av} and \eqref{eq:rmt_selfenergy} that were central for the analysis of spectral phase transition. 

The object needed to compute $S(\varphi)$ is the trace of the ensemble-averaged Green's function:
\begin{equation}\label{eq:rmt_what_needed}
    \langle {\rm Trace}_{\rm \,H} [G(\lambda)] \rangle = \Bigl\langle {\rm Trace}_{\rm \, H} \Bigl[\frac{1}{\lambda - Z_{M} U^\dagger Z_{B} U}\Bigr] \Bigr\rangle
\end{equation}
[we explicitly indicated that the trace is evaluated over the $D_{\rm H}$-dimensional Hilbert space; this is done to distinguish $\mathrm{Trace}_{\rm\, H}$ from the traces over the auxiliary spaces which we introduce below]. At face value, the ensemble averaging in Eq.~\eqref{eq:rmt_what_needed} is hard to carry out. The reason is the non-linear dependence of $Z_{M} U^\dagger Z_{B} U$ on $H$ [cf.~Eq.~\eqref{eq:rmt_ensemble}]. A method for simplifying
problems like this one was recently developed in Ref.~\onlinecite{khindanov2024}. It was shown that evaluation of $\langle {\rm Trace}_{\rm \, H} [G(\lambda)] \rangle$ for ensemble \eqref{eq:rmt_ensemble} can be reduced to a simpler problem involving inversion and averaging a matrix {\it linear} in $H$. The latter problem is amenable to standard diagrammatic technique \cite{AGD}.

The coveted linear-in-$H$ matrix is constructed in several steps. We begin by noting that the denominator in the expression for the evolution operator $U$ [see Eq.~\eqref{eq:rmt_ensemble}] appears when one inverts an auxiliary matrix 
\begin{equation}\label{eq:rmt_L}
    \check{L} = 
    \begin{pmatrix}
    \openone & \openone\\
    H & i\openone
    \end{pmatrix}.
\end{equation}
In fact, $U$ can be represented as a matrix element of $\check{L}^{-1}$:
\begin{equation}\label{eq:rmt_U_L}
    U = \bm{v}^T \check{L}^{-1} \bm{u}, \quad \bm{u} = 
    \begin{pmatrix}
    \sqrt{2}\\
    0
    \end{pmatrix},\quad
    \bm{v} = 
    \begin{pmatrix}
    \tfrac{1}{\sqrt{2}}\\
   \tfrac{-1}{\sqrt{2}}
    \end{pmatrix}.
\end{equation}
We will denote the auxiliary two-dimensional vector space in which $\check{L}$ acts by a subscript $\rm A$. Eq.~\eqref{eq:rmt_U_L} allows us to express the correlation operator as a trace over the auxiliary space:
\begin{equation}\label{eq:rmt_C_L}
    C \equiv \mathrm{Trace}_{\rm\, A}\bigl[Z_M\check{\mu}_M(\check{L}^\dagger)^{-1}Z_B\check{\mu}_B\check{L}^{-1} \bigr],
\end{equation}
where $2\times 2$ matrices $\check{\mu}_M$ and $\check{\mu}_B$ are given by
\begin{equation}
    \check{\mu}_M = \bm{u}\bm{u}^T,\quad\quad \check{\mu}_B = \bm{v}\bm{v}^T.
\end{equation}
Due to the structure of these matrices, not only $C$ but also any power of $C$ can be represented as a trace:
\begin{equation}\label{eq:rmt_C_L_n}
    C^k \equiv \mathrm{Trace}_{\rm\, A}\bigl[\bigl(Z_M\check{\mu}_M(\check{L}^\dagger)^{-1}Z_B\check{\mu}_B\check{L}^{-1}\bigr)^{k} \bigr],
\end{equation}
Our key observation is that the object under the trace in Eqs.~\eqref{eq:rmt_C_L} and \eqref{eq:rmt_C_L_n} emerges naturally when one inverts the matrix
\begin{equation}\label{eq:rmt_matrix_w_no_name}
    \sqrt{\lambda} \begin{pmatrix} 
    \check{0} & \check{L}\\
    \check{L}^\dagger & \check{0}
    \end{pmatrix}_{\rm K}\, - 
    \begin{pmatrix} 
    Z_M \check{\mu}_M &\check{0}\\
    \check{0} & Z_B \check{\mu}_B
    \end{pmatrix}_{\rm K}.
\end{equation}
This matrix acts in a further-enlarged Hilbert space; we indicated the new two-dimensional subspace by a subscript $\rm K$. As follows from Eq.~\eqref{eq:rmt_L}, the matrix in Eq.~\eqref{eq:rmt_matrix_w_no_name} is linear in $H$. It is this matrix whose inverse will be the main object in the calculation. We denote the inverse by
\begin{equation}\label{eq:rmt_G_nice}
    \hat{\cal G} (\lambda) = \Bigl[\sqrt{\lambda} \begin{pmatrix} 
    \check{0} & \check{L}\\
    \check{L}^\dagger & \check{0}
    \end{pmatrix}_{\rm K}
    -
    \begin{pmatrix} 
    Z_M \check{\mu}_M & \check{0}\\
    \check{0} & Z_B \check{\mu}_B
    \end{pmatrix}_{\rm K}\Bigr]^{-1}.
\end{equation}

To establish a direct connection between $\hat{\cal G}(\lambda)$ and the correlation operator $C$, it is useful to evaluate 
\begin{equation}
    \mathrm{Trace}_{\rm \,K, A} \Bigl[ \begin{pmatrix} 
    Z_M \check{\mu}_M & \hspace{-0.2cm}\check{0}\\
    \check{0} &\hspace{-0.2cm} Z_B \check{\mu}_B
    \end{pmatrix}_{\rm K} \hat{\cal G}(\lambda)\Bigr].
\end{equation}
Performing a series expansion of $\hat{\cal G}(\lambda)$ at $\lambda \rightarrow \infty$, we obtain
\begin{align}
    &\mathrm{Trace}_{\rm \,K, A}\Bigl[\begin{pmatrix} 
    Z_M \check{\mu}_M & \hspace{-0.2cm}\check{0}\\
    \check{0} &\hspace{-0.2cm} Z_B \check{\mu}_B
    \end{pmatrix}_{\rm K} \hat{\cal G}(\lambda)\Bigr]\notag \\
    &=2\sum_{k=1}^{\infty} \frac{1}{\lambda^k}  \mathrm{Trace}_{\rm \, A}\bigl[\bigl(Z_M\check{\mu}_M(\check{L}^\dagger)^{-1}Z_B\check{\mu}_B\check{L}^{-1}\bigr)^{k} \bigr] \notag\\&= 2\sum_{n=1}^{\infty} \frac{1}{\lambda^k}C^k = 2(\lambda G(\lambda) - \openone).
\end{align}
To transition from the second to the third lines, we used representation~\eqref{eq:rmt_C_L_n} for the $k$-th power of the correlation operator; the last equality stems from the definition of the Green's function $G(\lambda) = 1 / [\lambda \openone - C]$. Finally, we substitute the obtained result in Eq.~\eqref{eq:rmt_what_needed}. This leads to
\begin{align}\label{eq:rmt_simple_rep}
    &\langle {\rm Trace}_{\rm \,H} [G(\lambda)] \rangle\\
    &=\frac{1}{\lambda}\Bigl(D_{\rm H} + \frac{1}{2}\mathrm{Trace}_{\rm\, K, A, H}\Bigl[\begin{pmatrix} 
    Z_M \check{\mu}_M & \hspace{-0.2cm}\check{0}\\
    \check{0} &\hspace{-0.2cm} Z_B \check{\mu}_B
    \end{pmatrix}_{\rm K} \bigl\langle\hat{\cal G}(\lambda)\bigr\rangle\Bigr]\Bigr).\notag
\end{align}
The main advantage of representing ${G}(\lambda)$ in terms of $\hat{{\cal G}}(\lambda)$ is that the linear dependence of $\hat{\cal G}^{-1}$ on $H$. This fact allows us to carry out the ensemble-averaging using the diagrammatic techniques \cite{AGD}. We explain how this is done in the next paragraph.\\

\noindent\textbf{Ensemble averaging ---}
In this paragraph, we use the diagrammatic technique to perform the averaging of $\hat{\mathcal{G}}(\lambda)$.
A first step is to split $\hat{\mathcal{G}}^{-1}(\lambda)$ into two parts:
\begin{equation}
	\hat{\mathcal{G}}^{-1}(\lambda) = \hat{\mathcal{G}}^{-1}_0(\lambda)+\sqrt{\lambda} H\hat{\Pi}_{x}.
\end{equation}
Here $\hat{\mathcal{G}}^{-1}_0(\lambda)\equiv\hat{\mathcal{G}}^{-1}(\lambda)|_{H=0}$ is the bare Green's function and
\begin{equation}
	\hat{\Pi}_{x}=
	\begin{pmatrix}
		0 & 0 & 0 & 0 \\
		0 & 0 & 1 & 0 \\
		0 & 1 & 0 & 0 \\
		0 & 0 & 0 & 0 \\
	\end{pmatrix}_{\rm K,A}.
	\label{eq:rmt_pi_x}
\end{equation}
The object that we need to compute to find the spectral density is the average Green's function [see Eq.~\eqref{eq:rmt_simple_rep}]
\begin{equation}
	\bigl\langle\hat{\cal G}(\lambda)\bigr\rangle=\left\langle\frac{1}{\hat{\mathcal{G}}^{-1}_0(\lambda)+\sqrt{\lambda}H\hat{\Pi}_{x}} \right\rangle.
	\label{eq:G_def}
\end{equation}
To implement the diagrammatic calculation, we introduce diagrams for the bare and the dressed Green functions:
\begin{equation}
\begin{tikzpicture}
  \begin{feynman}
    \vertex (i);
    \vertex [right=1.2cm of i] (o);
    \diagram*{
      (o) --[fermion] (i)      
    };
  \end{feynman}
\end{tikzpicture}
~\equiv~\hat{\mathcal{G}}_0,\quad\quad
\begin{tikzpicture}
  \begin{feynman}
    \vertex (i);
    \vertex [right=1.2cm of i] (o);
    \diagram*{
      (o) --[fermion, line width=0.6mm] (i)      
    };
  \end{feynman}
\end{tikzpicture}
~\equiv~\langle\hat{\mathcal{G}}\rangle,
\end{equation}
We will represent averaging over the statistics of Gaussian variables $H_{ij}$ by a dashed line:
\begin{equation}
\begin{tikzpicture}
  \begin{feynman}
    \vertex (i1);
    \node[right=0.4cm of i1,draw,fill=white,circle] (v1){$H$};
    \vertex [right=0.8cm of v1] (o1);
    \vertex [right=0.8cm of o1] (i2);
    \node[right=0.4cm of i2,draw,fill=white,circle] (v2){$H$};
    \vertex [right=0.8cm of v2] (o2);
    \diagram*{
      (v1) --[fermion, edge label=$i$] (i1),        
      (o1) --[fermion, edge label=$j$]  (v1)
    };
    \diagram*{
      (v2) --[fermion, edge label=$k$] (i2),        
      (o2) --[fermion, edge label=$l$]  (v2),
      (v2) --[scalar, half right,edge label=$\frac{\gamma \lambda}{D_{\rm H}}\delta_{il}\delta_{jk}$] (v1)
    };
  \end{feynman}
\end{tikzpicture}~,
\end{equation}
see Eq.~\eqref{eq:rmt_correlator} [we included the vertex factors of $\sqrt{\lambda}$ into the disorder line].
As usual, the average Green's function satisfies the Dyson equation; in the diagrammatic form, this equation reads:
\begin{equation}
\vcenter{\hbox{\begin{tikzpicture}
  \begin{feynman}
    \vertex (i);
    \vertex [right=1.2cm of i] (o);
    \diagram*{
      (o) --[fermion, line width=0.6mm] (i)      
    };
  \end{feynman}
\end{tikzpicture}}}
~=~
\vcenter{\hbox{\begin{tikzpicture}
  \begin{feynman}
    \vertex (i);
    \vertex [right=1.2cm of i] (o);
    \diagram*{
      (o) --[fermion] (i)     
    };
  \end{feynman}
\end{tikzpicture}}}
~+~
\vcenter{\hbox{\begin{tikzpicture}
  \begin{feynman}
    \vertex (i);
    \node[right=1.2cm of i,draw,fill=white,circle] (v){$\hat{\Sigma}$};
    \vertex [right=1.45cm of v] (o);
    \diagram*{
      (v) --[fermion] (i),        
      (o) --[fermion, line width=0.6mm]  (v)
    };
  \end{feynman}
\end{tikzpicture}}}~,\label{eq:rmt_Dyson}
\end{equation}
where self-energy $\hat{\Sigma}$ includes all possible pairings of $H$ in the Taylor expansion of Eq.~\eqref{eq:G_def}. 
In the limit of large Hilbert space dimension, $D_{\rm H} \gg 1$, the self-energy can be found with the help of the self-consistent Born approximation ~\cite{AGD, khindanov2024}:
\begin{equation}
\vcenter{\hbox{\begin{tikzpicture}
  \begin{feynman}
    \node[right=1.5cm of i,draw,fill=white,circle] (v){$\hat\Sigma$};
    \diagram*{
      (v)
    };
  \end{feynman}
\end{tikzpicture}}}
~=~
\begin{tikzpicture}
  \begin{feynman}
  \tikzfeynmanset{every vertex={empty dot}}
    \vertex (i);
    \vertex [right=1.5cm of i] (o);
    \diagram*{
      (o) --[fermion, line width=0.6mm] (i);    
    };
    \diagram*{
      (i) --[scalar, half left] (o)
    };
  \end{feynman}
\end{tikzpicture}~.\label{eq:rmt_SCBA_self_energy}
\end{equation}
This approximation amounts to a resummation of all diagrams in which the disorder lines do not cross. It is justified by the fact that the crossing diagrams are suppressed by a factor $1 / D_{\rm H} \ll 1$.

Eq.~\eqref{eq:rmt_Dyson} and Eq.~\eqref{eq:rmt_SCBA_self_energy} are self-consistent: the average Green's function is determined by the self-energy [Eq.~\eqref{eq:rmt_Dyson}], while the self-energy itself depends on $\langle \hat{\cal G} \rangle$ via Eq.~\eqref{eq:rmt_SCBA_self_energy}. In the next paragraph, we perform the analysis of these equations and use them to derive the self-consistency relation, Eq.~\eqref{eq:rmt_selfenergy}.\\

\noindent\mbox{\textbf{Derivation of the self-consistency relation ---} Al-}
gebraically, the obtained system of self-consistent equations [Eqs.~\eqref{eq:rmt_Dyson} and \eqref{eq:rmt_SCBA_self_energy}] has the following form:
\begin{subequations}
\begin{align}
&\langle\hat{\mathcal{G}}\rangle=\hat{\mathcal{G}}_0+\hat{\mathcal{G}}_0\hat\Sigma\langle\hat{\mathcal{G}}\rangle, \label{eq:rmt_system_av_GF}\\
	&\hat\Sigma=\gamma\lambda\,\openone\otimes\hat{\Pi}_{x} \tfrac{1}{D_{\rm H}} \mathrm{Trace}_{\,\rm H}[\langle\hat{\mathcal{G}}\rangle]\hat{\Pi}_{x}\label{eq:rmt_system_selfenergy}
\end{align}
\end{subequations}
Below, we explain how this system can be solved.

\begin{figure}[t!]
  \begin{center}
    \includegraphics[scale = 1]{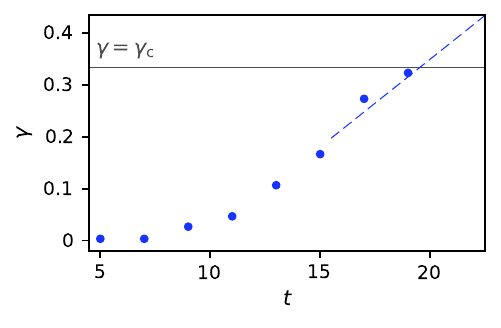}
    \caption{\textbf{Correspondence between parameter $\gamma$ of the random-matrix theory.} See Eq.~\eqref{eq:rmt_average_is_zero}, and circuit depth $t$. To establish the correspondence, we numerically obtain the gap at every $t$, and then find the value of the random-matrix parameter $\gamma$ that leads to the same gap. Dashed line shows the extrapolation of $\gamma (t)$ into the domain $\gamma > \gamma_\star$ in which there is no gap.}
    \label{fig:rmt_correspondence}
  \end{center}
\end{figure}

To begin with, notice that the presence of matrices $\hat{\Pi}_{x}$ in Eq.~\eqref{eq:rmt_system_selfenergy} enforces the form of the self-energy with at most four (out of $16$) non-zero components in ${\rm K}$ and ${\rm A}$ subspaces:
\begin{equation}
	\hat\Sigma=
	\begin{pmatrix}
		0 & 0 & 0 & 0 \\
		0 & \zeta_{11} & \zeta_{12} & 0 \\
		0 & \zeta_{21} & \zeta_{22} & 0 \\
		0 & 0 & 0 & 0 \\
	\end{pmatrix}_{\rm K,A}.
\end{equation}
\begin{figure*}[t!]
  \begin{center}
    \includegraphics[scale = 1]{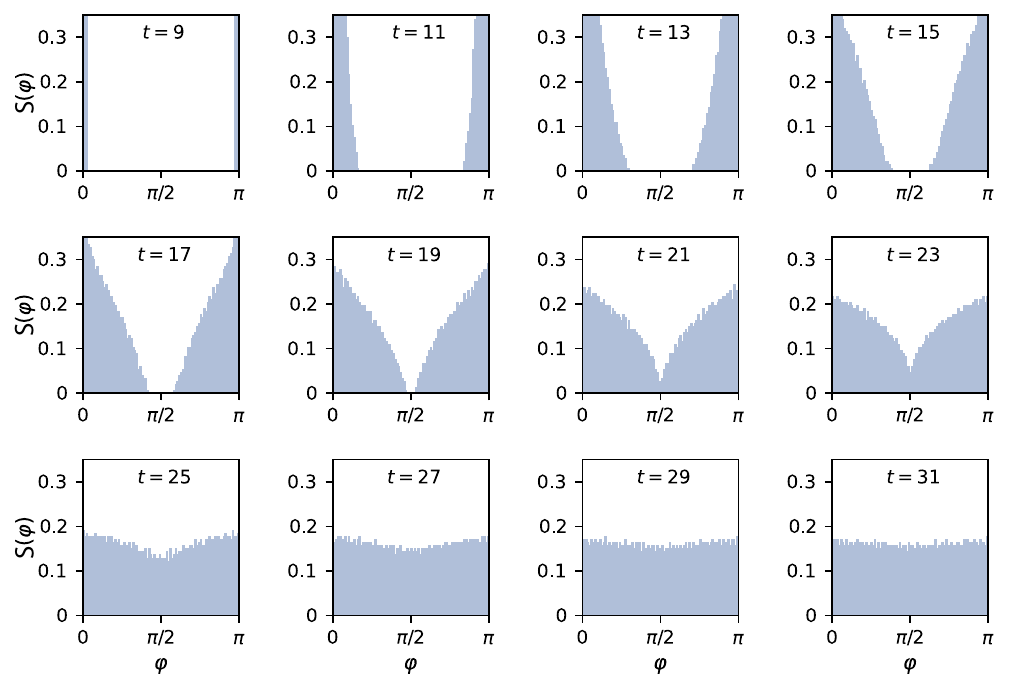}
    \caption{
    Full time dependence of the correlation operator spectrum in a quantum circuit composed of Haar random two-qubit gates [the circuit instance is the same as in Fig.~\ref{fig:rmt_numerics}].
    \label{fig:rmt_evolution}}
  \end{center}
\end{figure*}
This highlights a special role played by the inner $2\times 2$ block of this matrix, and suggests projecting Eq.~\eqref{eq:rmt_system_av_GF} onto this block as well. The projection is implemented by an operator:
\begin{equation}
	\hat\Pi=
	\begin{pmatrix}
		0 & 0 & 0 & 0 \\
		0 & 1 & 0 & 0 \\
		0 & 0 & 1 & 0 \\
		0 & 0 & 0 & 0 \\
	\end{pmatrix}_{\rm K,A}.
    \label{eq:Pi}
\end{equation}
We introduce a $2 \times 2$ matrix $g$ as the inner block of $\langle \hat{\cal G}\rangle$,
\begin{equation}
\hat\Pi \langle\hat{\mathcal{G}}\rangle \hat\Pi=
	\begin{pmatrix}
		0 & 0 & 0 & 0 \\
		0 & g_{11} & g_{12} & 0 \\
		0 & g_{21} & g_{22} & 0 \\
		0 & 0 & 0 & 0 \\
	\end{pmatrix}_{\rm K,A},
\end{equation}
and similarly $g_0$ as the inner block of $\hat{\cal G}_0$. We can represent the projected version of Eq.~\eqref{eq:rmt_system_av_GF} in terms of $g$, $g_0$, and $\zeta$ as
\begin{equation}\label{eq:rmt_g_small}
    g^{-1}=g_0^{-1}-\zeta.
\end{equation}
Eq.~\eqref{eq:rmt_system_selfenergy} can also be recast through $\zeta$ and $g$:
\begin{equation}\label{eq:rmt_zeta_small}
  \zeta = \gamma\lambda \openone\otimes \sigma_x\tfrac{1}{D_{\rm H}} {\rm Trace}_{\,\rm H}[g]\sigma_x.
\end{equation}
Here and below, $\sigma_i$ with $i=x,y,z$ are Pauli matrices in the inner subspace.
To finish defining the system of equations for $g$ and $\zeta$, we need to specify $g_0^{-1}$. A straightforward algebra following from the definition of $\hat{\cal G}$ (see Eq.~\eqref{eq:rmt_G_nice}) yields
\begin{align}\label{eq:rmt_g0_small}
    g_0^{-1} &= \sigma_y \sqrt{\lambda}\,\frac{Z_M Z_B + \lambda \openone}{Z_M Z_B - \lambda \openone}\notag\\
     &+\frac{1+\sigma_z}{2} \frac{2Z_M \lambda}{Z_M Z_B - \lambda \openone} + \frac{1-\sigma_z}{2} \frac{2Z_B \lambda}{Z_M Z_B - \lambda \openone}.
\end{align}
With that, we have all of the components needed to solve Eqs.~\eqref{eq:rmt_g_small} and \eqref{eq:rmt_zeta_small}.

A crucial observation that allows us to make further analytical progress is that the system defined by Eqs.~\eqref{eq:rmt_g_small} and \eqref{eq:rmt_zeta_small} admits a solution for the self-energy with a very simple matrix structure,
\begin{equation}\label{eq:rmt_ansatz}
    \zeta = -i\sqrt{\lambda}\,\varepsilon\,\sigma_y.
\end{equation}
We introduced a factor of $\sqrt{\lambda}$ here for notational convenience; parameter $\varepsilon \equiv \varepsilon (\lambda)$ depends on $\lambda$. The reason why ansatz \eqref{eq:rmt_ansatz} goes through Eqs.~\eqref{eq:rmt_g_small} and \eqref{eq:rmt_zeta_small} is that this ansatz preserves the structure of $g$ in which the diagonal terms are proportional to $Z_{M}$ and $Z_{B}$. Because ${\rm Trace}_{\,\rm H} [Z_{M,B}] = 0$, these terms would nullify on the right-hand side of Eq.~\eqref{eq:rmt_zeta_small}; only the terms $\propto \sigma_y$ would survive the trace. This is compatible with the left-hand side being $\propto \sigma_y$.

Combining Eqs.~\eqref{eq:rmt_g_small}, \eqref{eq:rmt_g0_small}, and \eqref{eq:rmt_ansatz}, and using them in Eq.~\eqref{eq:rmt_zeta_small}, we obtain a self-consistency relation for $\varepsilon$:
\begin{equation}\label{eq:rmt_selfcons_lambda}
    \varepsilon = \frac{\gamma}{2}\sum_{Z = \pm 1}\frac{\varepsilon - i \frac{\lambda - Z}{\lambda + Z}}{1-\varepsilon^2 + 2i\varepsilon \frac{\lambda - Z}{\lambda + Z}}.
\end{equation}
By substituting $\lambda = e^{i\varphi}$ into this equation, we arrive to Eq.~\eqref{eq:rmt_selfenergy} of Section~\ref{sec:rmt_model}.

We are now ready to obtain an expression for $\langle\hat{\cal G}(\lambda)\rangle$, and thus complete the derivation of $\langle{\rm Trace}_{\,\rm H}[G(\lambda)]\rangle$ [we remind one that the relation between the two Green's functions is given by Eq.~\eqref{eq:rmt_simple_rep}]. In fact, $\langle\hat{\cal G}(\lambda)\rangle$ can be straightforwardly expressed in terms of the already found $g_0$ and $\zeta$ [Eqs.~\eqref{eq:rmt_g0_small}--\eqref{eq:rmt_selfcons_lambda}]. To do that, let us first note an alternative, useful representation of Eq.~\eqref{eq:rmt_system_av_GF}:
\begin{equation}\label{eq:rmt_alternative}
\langle\hat{\mathcal{G}}\rangle=\hat{\mathcal{G}}_0+\hat{\mathcal{G}}_0\hat{\Pi}\hat\Sigma(\openone - \hat{\mathcal{G}}_0\hat\Sigma)^{-1}\hat{\Pi} \hat{\mathcal{G}}_0.
\end{equation}
To insert projectors $\hat{\Pi}$ here, we employed the property $\hat\Sigma = \hat{\Pi}\hat\Sigma = \hat\Sigma \hat{\Pi}$ that follows directly from Eq.~\eqref{eq:rmt_system_selfenergy}. Obviously, the matrix $\hat{\Pi}\hat\Sigma(\openone - \hat{\mathcal{G}}_0\hat\Sigma)^{-1}\hat{\Pi}$ differs from zero only in its inner block. Thus, we can represent it as
\begin{equation}
	\hat{\Pi}\hat\Sigma(\openone - \hat{\mathcal{G}}_0\hat\Sigma)^{-1}\hat{\Pi} =
	\begin{pmatrix}
		0 & 0 & 0 & 0 \\
		0 & \theta_{11} & \theta_{12} & 0 \\
		0 & \theta_{21} & \theta_{22} & 0 \\
		0 & 0 & 0 & 0 \\
	\end{pmatrix}_{\rm K,A}.
\end{equation}
A simple calculation gives
\begin{equation}\label{eq:rmt_theta}
    \theta = \zeta (\openone - g_0 \zeta)^{-1} = -i\sqrt{\lambda} \varepsilon \sigma_y(\openone +i\sqrt{\lambda} \varepsilon g_0\sigma_y)^{-1}.
\end{equation}
The use of Eqs.~\eqref{eq:rmt_alternative}--\eqref{eq:rmt_theta} in Eq.~\eqref{eq:rmt_simple_rep} leads to Eq.~\eqref{eq:rmt_G_av} of Section~\ref{sec:rmt_model}. We have now derived all of the equations needed to compute the spectral density $S(\varphi)$ in the limit $D_{\rm H} \gg 1$. 

\subsubsection{Numerical results on \texorpdfstring{$1{\rm D}$}{1D} random circuits \label{sec:rmt_numerics}}
Finally, we verify numerically that the phenomenology of spectral phase transition also applies to realistic quantum circuits composed of random, local gates. To this end, we consider a chain of $n$ qubits. As butterfly and measurement operators we use $Z_1$ and $Z_n$, respectively. We generate the evolution $U$ by a brickwork circuit composed of two-qubit Haar random gates, and track how the spectrum of $C = Z_{n}U^\dagger Z_1 U$ changes with the increase of the circuit depth $t$. 

The results of the numerical simulation are presented in Figs.~\ref{fig:rmt_numerics} and \ref{fig:rmt_evolution}. Even for a modest system size, $n = 12$, they demonstrate a qualitative agreement with the general expectations of Section~\ref{sec:rmt_intro}, as well as with predictions of the random-matrix theory model [cf.~Fig.~\ref{fig:rmt_rmt_results} and Fig.~\ref{fig:rmt_numerics}].
To provide the correspondence between $\gamma$ and $t$, we match the values of the gap at different times to $\varphi_{\rm gap}(\gamma)$ of the random-matrix theory model, see Fig.~\ref{fig:rmt_correspondence}.

%\bibliography{rmt_references.bib}

% t_c -> t_\star
% Remove subscript H in Eq 26
% B -> O % correlation operator spectrum, labeled as sec:phasetransition

% Symbol definitions
% General symbols
\def\Pauli{P}
\def\PauliI{I}
\def\PauliX{X}
\def\PauliY{Y}
\def\PauliZ{Z}
\def\StateVec#1{\left|#1\right\rangle}
\def\DensityMatrix{\rho}
\def\LocalHilbertDim{q}
\def\HilbertDim{D_{\mathrm{H}}}
\def\CircuitU{U}
\def\NumCycles{d}
\def\ExpValue#1{\left\langle#1\right\rangle}
\def\NumQubits{n}
\def\ProbVar{p}
\def\CoordX{x}
\def\CoordU{u}
\def\TimeVar{t}

% OTOCs
\def\MsmtOp{M}
\def\BflyOp{O}
\def\BflyOpTime{\BflyOp(\TimeVar)}
\def\CorrOp{C}
\def\CorrOpTime{C(\TimeVar)}
\def\MsmtQubit{Q_{\rm m}}
\def\BflyQubit{Q_{\rm b}}
\def\OTOCOp#1{\mathcal{C}^{(#1)}}
\def\OTOC{{\rm OTOC}}
\def\AveOTOCOp#1{\overline{\mathcal{C}}^{(#1)}}
\def\StdOTOCOp#1{\sigma\!\big({\mathcal{C}}^{(#1)}\big)}
\def\RescaledOTOCOp#1{{\mathcal{C}}^{(#1, s)}}
\def\SNR{{\rm SNR}}
\def\BflyVel{v_B}
\def\FrontDiffusionConst{D}

% Experimental details
\def\DecoherenceTime{T}
\def\MaxQubitFreq{f}
\def\ISWAP{G}
\def\CouplingStrenght{g}

% Noise and error mitigation
\def\NoiseDampingFactor{F_{\rm noise}}
\def\NumShots{K}
\def\ShotNoiseError{\epsilon_{\rm stats}}
\def\SystematicError{\epsilon_{\rm sys}}
\def\TotalError{\epsilon_{\rm tot}}

% Classical computation details
\def\TensorWidth{{\rm W}}
\def\DeltaOTOCOp#1{\Delta\OTOCOp{#1}}
\def\DeltaRescaledOTOCOp#1{\Delta\RescaledOTOCOp{#1}}

% Pauli pathing model
\def\PauliPathA{\bs{\alpha}}
\def\PauliPathB{\bs{\beta}}

% Permutation model
\def\PermutationA{\sigma}
\def\PermutationB{\tau}
\def\SymmetricGroup{S_k}
\def\PermutationOp#1{\pi_{#1}}
\def\Weingarten{{\rm Wg}}
\def\Projection#1{\mathcal{P}_{#1}}
\def\ProjDecomposition{T}
\def\RescaledPermutationA{\hat\PermutationA}
\def\RescaledPermutationB{\hat\PermutationB}
\def\RescaledProjDecomposition{\hat\ProjDecomposition}
\def\ProbDensity#1#2{W(#1,\,#2)}
\def\TEBDSingValue{s_a}

% Correlation operator spectrum
\def\EigenValue{\lambda}
\def\EigenPhase{\varphi}
\def\SpectralDensityFn#1{\rho(#1)}
\def\CriticalPoint{t_*}
\def\Hamiltonian{H}
\def\HamModelParam{\gamma}
\def\CriticalHamModelParam{\HamModelParam_*}
\def\GreenFn#1{G(#1)}
\def\SelfEnergy{\varepsilon}
\def\StepFn{\Theta}
\def\RescaledSelfEnergy{\tilde\SelfEnergy}
\def\RescaledHamModelParam{\tilde\HamModelParam}
\def\ScalingFn{\mathcal{R}_{\pm}}
\def\EigenValueGap{\EigenValue_{\rm gap}}
\def\BesselFn#1{J_1(#1)}

%\section{Table of symbols used in text}
\newpage
%\phantomsection
%\refstepcounter{section}
%\addcontentsline{toc}{section}{\thesection. Table of symbols used in text}
\begin{onecolumngrid}
\begin{center}
%\textbf{\thesection. TABLE OF SYMBOLS USED IN TEXT}
\section{Table of symbols used in text}
\end{center}
\end{onecolumngrid}

\begin{twocolumngrid}
\end{twocolumngrid}

\vspace{1pt}

\begin{longtable*}{c|l}
        \hline
        \textbf{Symbol}  & \textbf{Meaning} \\
        \hline
        \endhead
        \multicolumn{2}{r}{\textbf{General symbols}} \\
        \hline
        $\PauliI$, $\PauliX$, $\PauliY$, $\PauliZ$ & Pauli operators \\
        $\Pauli$ & Generic Pauli operator $\Pauli=\{\PauliI,\,\PauliX,\,\PauliY,\,\PauliZ\}^{\otimes \NumQubits}$ \\
        $\StateVec{\psi}$ & Quantum state vector \\
        $|P \rangle\rangle$ & Vector in superoperator formalism (corresponding to operator $P$) \\
        $\DensityMatrix$ & Density matrix \\
        $d$ & Local Hilbert space dimension \\
        $D_H$ & Hilbert space dimension, $\HilbertDim=d^{\NumQubits}$ \\
        $\CircuitU$ & Circuit unitary \\
        $u_t$ & Circuit sub-block unitary: $U=\prod_tu_t$. \\
        $t_f$ & Number of cycles or layers in a circuit \\
        $\Pi$ & Projector \\
        $\ExpValue{\cdot}$ & Expectation value \\
        $\overline{\cdot}$ & Average over circuits \\
        $\NumQubits$ & Number of qubits \\
        $q_m,\, q_b$ & Measurement and butterfly qubits \\
        $N_{\rm 2D}$ & Number of two-qubit gates within the lightcones of $q_m$ and $q_b$\\
        $\ProbVar$ & Probability variable \\
        $\CoordX,\, \CoordU$ & Spatial coordinates\\
        $\TimeVar$ & Time \\
        $\mathrm{Trace}$ & Trace, $\mathrm{Trace}[U] = \sum_j U_{jj}$ \\
        $\mathrm{Tr}$ & Normalized trace, $\mathrm{Tr}[P]=\frac{1}{2^n}{\rm Trace}[P]$\\
        $\hat{z}$ & Rescaled value of quantity $z$.\\
        $\widetilde{z}$ & Noisy value of quantity $z$.\\
        \hline
        \multicolumn{2}{r}{\textbf{OTOC-related symbols}} \\
        \hline
        $\MsmtOp$ & Measurement operator \\
        $B$ & Butterfly operator \\
        $B(t)$ & Butterfly operator evolved under unitary; $B(t)=U^{\dag}BU$  \\
        $\CorrOpTime$ & Correlation operator $C(t)=MB(t)$\\
        $OTOC^{(1)}$, $\OTOC$ & Out-of-time-ordered correlator ($\mathcal{C}^{(2)}=\mathrm{Trace}[\rho MB(t)MB(t)]$)\\
        $\OTOCOp{4}$, $\OTOC^{(2)}$ & OTOC squared ($\mathcal{C}^{(4)}=\mathrm{Trace}[\rho MB(t)MB(t)MB(t)MB(t)]$)\\
        $\OTOCOp{2k}$, $\OTOC^{(k)}$ & $k$-th order OTOC ($\mathcal{C}^{(2k)}=\mathrm{Trace}[\rho(MO(t))^{2k}]$) \\
        $\AveOTOCOp{2k}$ & OTOC$^{(k)}$ averaged over a circuit ensemble \\
        $L$ & Loschmidt echo. \\
        $\StdOTOCOp{2k}$ & Standard deviation of OTOC$^{(k)}$ over a circuit ensemble \\
        $\OTOCOp{2k}_{\mathrm{diag}}$ & OTOC$^{(k)}$ averaged over a set of inserted Pauli gates\\
         $\mathcal{C}^{(2k)}_{\rm off-diag}$ & Difference between OTOC$^{(k)}$ with and without averaging over Pauli insertion: \\
        & e.g. $\mathcal{C}^{(2k)}_{\rm off-diag}=\mathcal{C}^{(2k)}-\mathcal{C}^{(2k)}_{\mathrm{diag}}$.\\
        $\RescaledOTOCOp{2k}$ & Rescaled OTOC $\mathcal{C}^{(2k, s)} = (\mathcal{C}^{(2k)} - \overline{\mathcal{C}}^{(2k)}) / \sigma \left(\mathcal{C}^{(2)} \right)$ \\
        $\mathcal{C}^{(4,s)}_{\rm off-diag}$ & Rescaled difference between OTOC$^2$ values: $\mathcal{C}^{(4,s)}_{\rm off-diag}=\left(\mathcal{C}^{(4)}_{\rm off-diag} - \overline{\mathcal{C}^{(4)}}_{\rm off-diag}\right)/\sigma(\mathcal{C}^{(4)}_{\rm off-diag})$.\\
        $\SNR$ & Signal-to-noise ratio \\
        $\OTOCOp{2k}_{\mathrm{<method>}}$ & $<$method$>$ approximation of the OTOC$^{(k)}$.\\
        $\mathrm{SNR}_{\mathrm{<method>}}$ & SNR of $<$method$>$ approximation of the OTOC$^{(k)}$.\\
        $\BflyVel$ & Butterfly velocity \\
        $\FrontDiffusionConst$ & Front diffusion constant \\
        $b_{j}(t)$, $b_{\alpha}(t)$ & coefficient of butterfly operator in some basis decomposition\\
        $\alpha, \beta$ & Pauli path.\\
        & \\
        \hline
        \multicolumn{2}{r}{\textbf{Continued on next page}}\\
        \multicolumn{2}{r}{\textbf{Experimental details}} \\
        \hline
        $\DecoherenceTime_1$, $\DecoherenceTime_2$ $\DecoherenceTime_{\phi}$ & Decoherence times \\
        $\tau$ & Gate time \\
        $\MaxQubitFreq_{\max}$, $\MaxQubitFreq_{10}$ & Maximum allowed qubit frequency, idle frequency \\
        $\ISWAP_{ij}$ & iSWAP-like gate \\
        $\tau_r$ & Resonance time during $G_{ij}$ \\
        $\CouplingStrenght$, $\CouplingStrenght_{\max}$ & Coupling strength, maximum coupling strength during $G_{ij}$.\\
        \hline
        \multicolumn{2}{r}{\textbf{Noise and error mitigation}}\\
        \hline
        $\epsilon_{X}$ & Error probability per event $X$ (e.g. a gate). \\
        $\Gamma_{12}$ & Leakage rate.\\
        $\NoiseDampingFactor$ & Noise damping factor \\
        $\NumShots$ & Number of shots (experiment repetitions) \\
        $\ShotNoiseError$ & Statistical shot noise \\
        $\SystematicError$ & Systematic error \\
        $\TotalError$ & Total error \\
        $\SNR_{\mathrm{sys}}$ & Systematic error contribution to the SNR.\\
        \hline
        \multicolumn{2}{r}{\textbf{Classical computation details}} \\
        \hline
        ${\rm FLOP}$ & Floating point operations. \\
        ${\rm FLOPS}$ & Floating point operations per second. \\
        $\TensorWidth$ & Max tensor width during contraction.\\
        $\rho_{A}$ & Reduced density operator on subsystem $A$.\\
        $N_{\mathrm{projs}}$ & Number of projections in final step of tensor network contraction \\
        $\mathrm{SNR}_{\mathrm{projs}}, a_{\mathrm{SNR}}, b_{\mathrm{SNR}}$ & Coefficients of convergence of tensor network contraction with projection count.\\
        $\mathcal{C}$ & Cache of Pauli paths in cached-MC algorithm.\\
        $b_{\alpha}$ & operator amplitude in cached-MC algorithm.\\
        $\Lambda$ & Pauli transfer matrix. \\
        $c$ & Exponent on iSWAP gate when considering other types of random circuits. \\ 
        \hline
\end{longtable*}

\bibliography{refs}
\bibliographystyle{naturemag}

%\appendix